%% file: paper.tex
\def\elsevier{-2}
\ifnum\elsevier>0 
   \documentclass{elsart}
   \newcommand{\cal}[1]{\mathcal{#1}}
   
\else
   \documentclass[10pt]{article}
   \usepackage{a4wide}
\fi

\usepackage{epsfig,verbatim}
\usepackage{amsmath,amssymb}
\usepackage{graphicx,lscape,ltxtable}
\usepackage{color,afterpage,float}

\def\elscolor{-2}
\def\elsitalic{-2}

\newcommand{\bds}[1]{\boldsymbol{#1}}
\newcommand{\lra}[1]{\langle #1 \rangle }

\newcommand{\td}[1]{\tilde{#1}}
\newcommand{\mc}[1]{\mathcal{#1}}
\newcommand{\mb}[1]{\mathbf{#1}}
\newcommand{\ch}[1]{../../BibTex/#1}

\DeclareGraphicsExtensions{.eps}
\graphicspath{{Figures/}}

\begin{document}

\ifnum\elsevier>0 

\begin{frontmatter}
\title{Mean-field/PDF numerical approach for polydispersed turbulent
  two-phase flows}
\author[ademe]{E. Peirano\thanksref{ca}}
\author[edf]{S. Chibbaro}
\author[imp]{J. Pozorski}
\author[edf]{J-P. Minier}
\address[ademe]{ADEME, Renewable Energies Department, 500 route des
  Lucioles, 06560 Valbonne, France}
\address[edf]{Electricit\'e de France, Div. R\&D, MFTT, 6 Quai Watier, 78400 Chatou,
              France}
\address[imp]{Institute of Fluid-Flow Machinery, Polish Academy of
  Sciences, Fiszera 14, 80952 Gda\'{n}sk, Poland}
\thanks[ca]{Corresponding author. Tel.: +33(0)4 93 95 79 34, Fax: +33
  (0)4 93 65 31 96, E-mail: eric.peirano{@}ademe.fr.}
\begin{abstract}
The purpose of this paper is to give an overview in the realm of
numerical computations of polydispersed turbulent two-phase flows,
using a mean-field/PDF approach. In this approach, the numerical
solution is obtained by resorting to a hybrid method where the mean
fluid properties are computed by solving mean-field (RANS) equations
with a classical finite volume procedure whereas the local
instantaneous properties of the particles are determined by solving
stochastic differential equations (SDEs). The
fundamentals of the general formalism are recalled and particular
attention is focused on a specific theoretical issue: the treatment of
the multiscale character of the dynamics of the discrete particles,
that is the consistency of the system of SDEs in asymptotic
cases. Then, the main lines of the particle/mesh algorithm are given
and some specific problems, related to the integration of the SDEs,
are discussed, for example, issues related to the specificity of the
treatment of the averaging and projection operators, the time
integration of the SDEs (weak numerical schemes consistent with all
asymptotic cases), and the computation of the source terms. Practical
simulations, for three different flows, are performed in order to
demonstrate the ability of both the models and the numerics to cope
with the stringent specificities of polydispersed turbulent two-phase
flows.
\end{abstract}
\begin{keyword}
dispersed two-phase flows, turbulence, PDF methods, particle/mesh
algorithms.
\end{keyword}
\end{frontmatter}

\else

\title{Mean-field/PDF numerical approach for polydispersed turbulent
  two-phase flows}
\author{E. Peirano$^{1,*}$, S. Chibbaro$^{2}$, J. Pozorski$^{3}$ and
  J-P. Minier$^{2}$ \vspace{2mm} \\
\small $^{1}$ADEME, Renewable Energies Department, 500 route des Lucioles,
  06560 Valbonne, France \\ \small Tel.: +33(0)4 93 95 79 34, Fax:
  +33(0)4 93 65 31 96, \small E-mail: eric.peirano@ademe.fr
  \vspace{2mm} \\
\small $^{2}$Electricit\'e de France, Div. R\&D, MFTT, 6 Quai Watier, 78400
  Chatou, France \\ \small  E-mail: chibbaro@chi80bk.der.edf.fr, 
  Jean-Pierre.Minier@edf.fr \vspace{2mm} \\
\small $^{3}$ Institute of Fluid-Flow Machinery, Polish Academy of Sciences,
  Fiszera 14, 80952 Gda\'{n}sk, Poland \\ \small E-mail: jp@imp.gda.pl
  \vspace{2mm} \\
\small $^{*}$ Corresponding author }
\date{}
\maketitle
\begin{abstract}
The purpose of this paper is to give an overview in the realm of
numerical computations of polydispersed turbulent two-phase flows,
using a mean-field/PDF approach. In this approach, the numerical
solution is obtained by resorting to a hybrid method where the mean
fluid properties are computed by solving mean-field (RANS) equations
with a classical finite volume procedure whereas the local
instantaneous properties of the particles are determined by solving
stochastic differential equations (SDEs). The
fundamentals of the general formalism are recalled and particular
attention is focused on a specific theoretical issue: the treatment of
the multiscale character of the dynamics of the discrete particles,
that is the consistency of the system of SDEs in asymptotic
cases. Then, the main lines of the particle/mesh algorithm are given
and some specific problems, related to the integration of the SDEs,
are discussed, for example, issues related to the specificity of the
treatment of the averaging and projection operators, the time
integration of the SDEs (weak numerical schemes consistent with all
asymptotic cases), and the computation of the source terms. Practical
simulations, for three different flows, are performed in order to
demonstrate the ability of both the models and the numerics to cope
with the stringent specificities of polydispersed turbulent two-phase
flows. \newline \textbf{Key words:} dispersed two-phase flows,
turbulence, PDF methods, particle/mesh algorithms.
\end{abstract}

\fi

\tableofcontents
%===============================================================================

%===============================================================================
\section{Introduction}
%===============================================================================
Two-phase flows are relatively easy to observe: to get a first
inkling, one can think of throwing small light particles (which then
play the role of tracer particles) into a turbulent flow such as a
rapid river or a plume coming out of a chimney. The small solid
particles reveal the intricate and complex features of turbulent
flows: understanding and modelling these features, i.e. single-phase
turbulent flow modelling, is the subject of extensive research
\cite{Sre_99}. If one introduces larger and larger particles in the
flow, more complex phenomena take place: the behaviour of the particles
will reflect the interplay between the main physical mechanisms,
such as
particle inertia and turbulence of the carrying flow. Then,
eventually, when particles become large-enough, the effect of the
fluid may become negligible with respect to particle inertia. Thus,
turbulent fluid-particle flow modelling appears as a link between
subjects such as turbulence and granular flows \cite{Gen_99}. 
In general, two-phase flows are even more complex
since, in the case of air and water for example, different
configurations of the interface between the two phases may be
present. Yet, in the present study, attention will be focused on the
motion of particles embedded in a turbulent fluid,
i.e. \textit{polydispersed turbulent two-phase flows}, where the
geometrical configuration does not change.

Polydispersed turbulent two-phase flows are found in numerous
environmental and industrial processes, very often in contexts that
involve additional issues, for example chemical and combustion
ones. Therefore, modelling these flows raises very difficult
theoretical questions and, at the same time, one has to provide
answers to what we can refer to as engineering concerns. As a 
consequence, a theoretical and numerical model represents an attempt
to find a \textit{satisfactory} compromise between these sometimes
conflicting expectations. Before trying to outline what is meant by
satisfactory, let us describe the characteristics of the polydispersed
turbulent two-phase flows we consider here. 

In the present study, only non-reacting incompressible fluid-particle
flows, with no collisions between particles, are investigated
(particle dispersion and turbulence modulation induced by the presence
of the particles are the physical mechanisms under consideration).
This is not a strict limitation of the approach that will be adopted
since, as mentioned below, the probability density function (PDF)
formalism that shall be followed is precisely well-suited for the
extension to more complex physics, such as combustion. However, for
the sake of simplicity, we limit ourselves to the core physics
embodied by particle dynamics. In addition, only the case of solid
heavy particles is treated, i.e. the density of the particles is much
greater than that of the fluid, $\rho_p \gg \rho_f$. This hypothesis
simplifies the equation of motion of the discrete particles in a
turbulent flow which, retaining only drag and
gravity forces, can be written as:
\begin{equation} \label{exa_part_eqns}
\left\{\begin{split} 
& \frac{d{\bf x}_p(t)}{dt} = {\bf U}_p(t), \\
& \frac{d{\bf U}_p(t)}{dt} = \frac{1}{\tau_{p}}({\bf U}_{s}(t)-{\bf
  U}_{p}(t)) \;  + {\bf g}.
\end{split}\right.
\end{equation}
In these equations, ${\bf U}_{s}(t)={\bf U}(t,{\bf x}_{p}(t))$ is the
fluid velocity ``seen'', i.e. the fluid velocity sampled along the
particle trajectory ${\bf x}_{p}(t)$, where ${\bf U}(t,{\bf x})$ is
the local instantaneous (Eulerian) fluid velocity field. The particle
relaxation time, $\tau_p$, is defined as
\begin{equation} \label{definition taup}
\tau_{p}=\frac{\rho_p}{\rho_f}\frac{4 d_p}{3 C_D |{\bf U}_r|},
\end{equation}
where the local instantaneous relative velocity is
${\bf U}_r(t)={\bf U}_p(t)-{\bf U}_s(t)$. The drag coefficient,
$C_D$, is a non-linear function of the particle-based Reynolds number,
$Re_p=d_p |{\bf U}_r|/\nu_f$, which means that $C_D$ is a non-linear
function of the particle diameter, $d_p$, \cite{Cli_78}. This last
point represents a major theoretical difficulty for a statistical
treatment since we do not consider mono-dispersed two-phase flows
(where $d_p$ is constant), but \emph{polydispersed} two-phase flows
where $d_p$ covers a range of possible values (from very light
particles acting as fluid tracers to high-inertia particles in the
ballistic regime, where the effect of the fluid on the particle
dynamics can be neglected). In the particle dynamical equations, it is
important to note that we are dealing with the instantaneous fluid
velocities, ${\bf U}(t,{\bf x}_{p}(t))$. Yet, for high-Reynolds
turbulent flows, which are the most common ones, such an information
is not available due to the very large number of degrees of freedom of
the turbulent flows \cite{Mon_75}. A modelling step is necessary and
most models adopt a statistical approach where only some limited
information is sought for the fluid fields whereas particles are
tracked individually. In practical models, this information consists
in, for the fluid, the first two velocity moments, as in
$R_{ij}-\epsilon$ models \cite{Sto_96}, or even filtered velocity
fields as in LES calculations, e.g. \cite{Boi_00,Bel_04}. In the
present work, $R_{ij}-\epsilon$ models (RANS equations) will be used
in practical computations, but the PDF approach (for the particles) to
come is fully compatible with other approaches for the fluid,
for example LES.

As indicated above, in order to track the particles, a
\textit{satisfactory} model must be built for the evaluation of the
particle properties, 
cf. Eqs.~(\ref{exa_part_eqns}). By \textit{satisfactory}, it is meant
a model which has the following properties:
\begin{enumerate}
\item[(i)] the model treats the important phenomena, such as
  convection and the polydispersed nature of the particles, without
  approximation,
\item[(ii)] the approach is naturally set into a general formalism
  which allows additional variables, for more complex physics such as
  combustion issues, to be directly introduced,
\item[(iii)] the complete theoretical model must be tractable in
  complex geometries and applicable for engineering problems.
\end{enumerate}
The first two issues have been addressed in a previous review
work~\cite{Min_01} where a PDF approach
has been developed. In practice, the PDF approach has the form of a
particle stochastic method where the velocity of the fluid seen, 
${\bf U}_{s}(t)$, is
modelled as a stochastic diffusion process, i.e. the dynamics of the
particles are calculated from
SDEs (Stochastic Differential Equations), the so-called Langevin
equations~\cite{Min_01}. In polydispersed two-phase flows, a particle
point of view seems rather natural, given the physics considered. Yet,
the particles which are to be simulated represent samples of the pdf
and should not be confused with real particles. Within the PDF
formalism, this particle point of view is helpful to build the
theoretical model and, at the same time, represents directly a
discrete formulation of the model. However, in order to devise a
consistent framework, it is important to separate the two steps by
formulating the model in continuous time \emph{before} addressing the
questions of numerical methods for practical computations.

The purpose of the present review is to address point (iii) above, and
to discuss the general numerical methodology used for particle
stochastic or PDF models for polydispersed turbulent two-phase
flows. More specifically, it is aimed at providing answers to several
interrogations:
\begin{enumerate}
\item[(a)] what do the stochastic particles represent?
\item[(b)] how do we compute the stochastic differential equations?
\item[(c)] what are the various difficulties and sources of numerical
  errors in the complete numerical method?
\end{enumerate} 
More than trying to present definitive answers to the questions of
what numerical scheme should be used, the objective is to propose a
general numerical approach and to show how PDF models, Langevin
stochastic equations, particle/mesh and dynamical Monte Carlo methods
are closely connected and actually represent different translations of
the same idea. Within that context, a major goal is to emphasise that,
although numerical schemes are separated from the construction of the
theoretical model, they cannot be addressed only from a mathematical
point of  view. Indeed, it is important that they reflect the physical
properties of the continuous stochastic model, namely the multiscale
character of the Langevin equations presented in Section
\ref{sec:pdf_models}.

The paper is organised as follows. The mathematical background on PDF
equations and stochastic diffusion processes is recalled in Section
\ref{sec:formalism}. General and  state-of-the-art Langevin models for
polydispersed turbulent two-phase flow modelling are briefly presented
in Section \ref{sec:pdf_models}. A central point is the analysis of the
multiscale properties of the Langevin equations, and the expression of
the various physical limits when characteristic timescales become
negligible with respect to the observation timescale, cf. Section
\ref{sec:scale}. This analysis serves as a guideline for the
development of the numerical model in Section \ref{sec:num0}, that
contains both particle/mesh and time-integration issues. 
%modif EP (début)
\ifnum\elscolor>0 \color{blue} \fi
\ifnum\elsitalic>0 \itshape \fi
Then, a discussion is given on specific issues related to two-way
coupling, Section \ref{sec:twc}. Several numerical applications
representative of practical concerns are proposed in Section
\ref{sec:exemples}. 
\normalfont
\color{black}
%modif EP (fin)

%===============================================================================
\section{General formalism} \label{sec:formalism}
%===============================================================================
The general formalism on which the derivation of the system of
equations (RANS equations for the fluid and SDEs for the discrete
particles) relies, is based on the Lagrangian point of view: the system
(the fluid-particle mixture) is treated as an ensemble of fluid and
discrete particles. The discretisation of a continuous medium (the
fluid) with particles is not a natural step, but it is a practical
way, in the frame of the probabilistic
formalism briefly outlined here, to treat important physical phenomena
without approximation \cite{Min_01}. In the present formalism, a fluid
particle is an independent sample of the flow, with a given
pdf. Physically, a fluid particle can be seen as a small element of
fluid whose characteristic length scale is much larger than the
molecular mean free path and much smaller than the Kolmogorov length
scale. The fluid particles have a mass $m_f$, a volume $V_f$ and a
velocity that equals the fluid velocity field at the location of the
particle, ${\bf U}_f(t)={\bf U}(t,{\bf x}_f(t))$.

%-------------------------------------------------------------------------------
\subsection{Statistical approach} \label{sec:stat}
%-------------------------------------------------------------------------------
Let us consider an ensemble composed by $N_f$ fluid particles and
$N_p$ discrete particles interacting through forces that can be
expressed as functions, or functionals, of the variables attached to
each particle, e.g., $l$ variables for the fluid particles and
$q$ variables for the discrete particles (these variables can be, for
example, position, velocity, \dots). All available information is then
contained in the following state vector:
\begin{equation}
\left. \begin{split}
{\bf Z}(t)=\{Z_{f,1}^1(t), \ldots, Z_{f,l}^1(t)\,;\,\ldots\,;\,
& Z_{f,1}^{N_f}(t),\ldots, Z_{f,l}^{N_f}(t) ; Z_{p,1}^1(t), \ldots \\
& \ldots, Z_{p,q}^1(t)\,;\,\ldots\,;\, Z_{p,1}^{N_p}(t),
\ldots,Z_{p,q}^{N_p}(t)\},
\end{split}\right.
\end{equation}
where $Z_{f,j}^i(t)$ represents the variable $j$ attached to the fluid
particle labelled $i$ and $Z_{p,j}^i(t)$ represents the variable $j$
attached to the discrete particle labelled $i$. The dimension of the
state vector is then $l N_f + q N_p$. Let us suppose that the
dynamical behaviour of the closed system can be described in terms of
ordinary differential equations (the Navier-Stokes equations, in
Lagrangian form, for the fluid particles and for the discrete
particles, the equation of motion of a single particle in a turbulent
fluid-particle mixture), i.e.
\begin{equation}
\frac{d{\bf Z}(t)}{dt}={\bf A}(t,{\bf Z}(t)).
\end{equation}
Here, it is assumed that, in the Navier-Stokes equations, the local
instantaneous pressure gradient, the viscous forces and the source
term (due to the force exerted by the discrete particles on the fluid)
can be expressed as functionals of the state vector ${\bf
  Z}$(t). In sample space, this system of ODEs (Ordinary Differential
Equations) corresponds to the
Liouville equation~\cite{Gar_90}
\begin{equation} \label{eq:closed}
\frac{\partial{p(t;{\bf z})}}{\partial{t}} + 
\frac{\partial}{\partial{{\bf z}}}
({\bf A}(t,{\bf z})\; p(t;{\bf z}))=0,
\end{equation} 
where $p(t;{\bf z})$, the associated pdf, represents the probability
to observe at time $t$ the system in state ${\bf z}$. In the present
paper, we distinguish between physical space, ${\bf Z}$, and sample
space, ${\bf z}$. A distinction is also made, for the pdf, between
parameters and variables by separating them with a semi-colon,
i.e. $(t;{\bf z})$. 

In practice, the number of degrees of freedom of such a system is huge
(turbulent flow with a large number of particles) and one has to
resort to a contracted description in order to come up with a model
that can be simulated with modern computer technology. For
single-phase turbulent reactive flows, a one-point pdf, $p(t;{\bf z}_f^i)$, is
often retained \cite{Pop_85,Pop_94}. For the description of the
dynamics of the discrete particles in turbulent dispersed two-phase
flows, a  one-point pdf, $p(t;{\bf z}_p^j)$, is also
encountered~\cite{Sto_96,Min_01,Ree_92}. In this work,  as we shall
see in Section \ref{sec:model}, both approaches are gathered in the
form of a two-point pdf, $p(t;{\bf z}_f^i,{\bf z}_p^j)$ and the
associated reduced state vector (henceforth denoted by superscript
$r$) is
\begin{equation}
{\bf Z}^r(t)=\{Z_{f,1}(t), \ldots, Z_{f,l}(t),Z_{p,1}(t), \ldots,
Z_{p,q}(t)\}.
\end{equation}
The time evolution equations, in physical space, for this
sub-system have the form
\begin{equation}
\frac{d {\bf Z}^r(t)}{dt} = {\bf A}(t,{\bf Z}^r(t),{\bf Y}(t)),
\label{eq:reduced_a}
\end{equation} 
where there is a dependence on the external variable ${\bf Y}(t)$
(related to the particles not contained in ${\bf Z}^r(t)$ as only
pairs of particles, a fluid one and a discrete one, are under
consideration). In sample space, the marginal pdf $p^r(t;{\bf z}^r)$
verifies
\begin{equation}
\frac{\partial{p^{r}(t;{\bf z}^r)}}{\partial{t}} 
+ \frac{\partial}{\partial{{\bf z}^r}}
\left[ \langle {\bf A}\,|\,{\bf z}^r\, \rangle \, 
p^{r}(t;{\bf z}^r)\right] =0, 
\label{eq:reduced}
\end{equation}
where the conditional expectation is given by
\begin{equation}
\begin{split}
\langle {\bf A}\, |\, {\bf z}^r \,\rangle = 
\int {\bf A}(t,{\bf z}^r,{\bf y})\,& p(t;{\bf y}\,|\,{\bf z}^r) 
\,d{\bf y} \\ & = \frac{1}{p^r(t;{\bf z}^r)}
\int {\bf A}(t,{\bf z}^r,{\bf y})\,
p(t;{\bf z}^r,{\bf y})\,d{\bf y}.
\end{split}
\end{equation}
Eq. (\ref{eq:reduced}) is now unclosed, showing that a reduced
description of a system implies a loss of information and thus the
necessity to introduce a model.

A practical way to close the system is to resort to stochastic
differential equations (SDEs), as it shall be briefly explained in
Section \ref{sec:SDEphy}. Further detailed explanations for this move
can be found in Refs.~\cite{Min_01} and~\cite{Pop_00}. The stochastic
differential equations treated in this work have the following form
(${\bf Z}^r(t)$ is called a diffusion process)
\begin{equation} \label{sde}
dZ^r_i(t) = A_i(t,{\bf Z}^r(t))\,dt + B_{ij}(t,{\bf Z}^r(t))\,dW_j(t),
\end{equation}
where ${\bf W}(t) = (W_1(t),\ldots, W_d(t))$ is a set of independent
Wiener processes~\cite{Arn_74} and $d=l+q$ is the dimension of the
reduced state vector. These equations are often referred to as
\textit{Langevin equations} in the physical literature
\cite{Gar_90}. In Eq. (\ref{sde}), ${\bf A} = (A_i)$ is called the
drift vector and $ {\bf B} = (B_{ij})$ is the diffusion matrix. SDEs
require a strict mathematical definition of the stochastic integral as
%modif EP (début)
\ifnum\elscolor>0 \color{blue} \fi
\ifnum\elsitalic>0 \itshape \fi
it shall be explained in Section \ref{sec:intsto}. If one adopts the
It\^{o} definition of the stochastic integral (see Section
\ref{sec:intsto}) in Eq. (\ref{sde}), it can be shown, see
\cite{Gar_90} for example, that the corresponding equation in sample
space for $p^r(t;{\bf z}^r)$ is the Fokker-Planck equation
\normalfont
\color{black}
%modif EP (fin)
\begin{equation} \label{fokker-planck}
\frac{\partial p^r(t;{\bf z}^r)}{\partial t} =
-\frac{\partial}{\partial z_i^r}[\,A_i(t,{\bf z}^r)\, p^r(t;{\bf z}^r)\,] +
\frac{1}{2} \frac{\partial^2}{\partial z_i^r\partial z_j^r}
               [\,D_{ij}(t,{\bf z}^r)\, p^r(t;{\bf z}^r)\,],
\end{equation}
where $D_{ij} = B_{il}B_{jl} = (BB^T)_{ij}$ is a positive-definite
matrix. In a weak sense (when one is only interested in statistics of
the process), one can speak of an equivalence between SDEs and
Fokker-Planck equations. As we shall see below, this correspondence is
the cornerstone of the proposed numerical approach: the pdf can be
obtained by simulating the motion of stochastic particles,
i.e. Eq. (\ref{sde}). In other words, real particles are
replaced by stochastic particles which, if the model is suitable,
reproduce the same statistics as the real ones. Indeed, in many
problems of practical concern, the dimension of the reduced state
vector is large and properties of the coefficients ${\bf A}$ and 
${\bf B}$ make the direct solution of the above PDE (Partial
Differential Equation), i.e. the Fokker Planck equation,
numerically difficult. Instead, it is more appropriate to calculate
$p^r(t;{\bf z}^r)$ (or any moment extracted from it) from
Eq. (\ref{sde}). Practically, this is done by resorting to a dynamical
Monte Carlo method, that is by simulating a large number $N$ of
independent realisations of  ${\bf Z}^r(t)$. Then, at each time step,
the \textit{discrete} pdf,
%modif EP (début)
\ifnum\elscolor>0 \color{blue} \fi
\ifnum\elsitalic>0 \itshape \fi
$p_N^r(t;{\bf z}^r)$, can be computed from the set of $N$ independent
samples $\{{\bf Z}^{r,n}(t)\}$ as
\begin{equation}
p_N^r(t;{\bf z})=\frac{1}{N}\sum_{n=1}^N
\delta(z_1^r-Z_1^{r,n}(t))\,\delta(z_2^r-Z_2^{r,n}(t))\dots
\delta(z_d^r-Z_d^{r,n}(t)),
\end{equation}
where $d$ is the dimension of the reduced state vector and $n$ stands
for the sample index. The question to be answered is: in what sense
does the ensemble $\{{\bf Z}^{r,n}(t)\}$, from which the discrete pdf
$p_N^r(t;{\bf z}^r)$ is extracted, represent the underlying pdf
$p^r(t;{\bf z}^r)$? The answer to this question can be given in a weak
sense, that is when \textit{convergence in distribution} is ensured, i.e. 
$p_N^r(t;{\bf z}^r) \to p^r(t;{\bf z}^r)$ when $N \to \infty$.

A sequence of random variables $\{X^n\}$ converges in distribution to
$X$ if and only if, for any bounded continuous function $g$ on
${\mathbb R}$, one has $\lra{g(X^n)} \to \lra{g(X)}$ when $n \to
\infty$. In calculations, the mathematical expectations
$\lra{\,\cdot\,}$ is estimated by the ensemble average
$\lra{\,\cdot\,}_N$ over $N$ independent samples.
The \textit{law of large numbers} tells us that $\lra{g(X)}_N$ is an
unbiased estimation of $\lra{g(X)}$, that is
$\lra{g(X)}_N \to \lra{g(X)}$ when $N \to \infty$.  
Then, according to the \textit{central limit theorem}, the error
$\epsilon_N = \lra{g(X)}_N-\lra{g(X)}$, which is a random variable,
converges in distribution to a Gaussian random variable of zero mean
and standard deviation $\sigma[g(X)]/\sqrt{N}$ provided that the
variance of $g(X)$, $\sigma^2[g(X)]$, is finite.
\normalfont
\color{black}
%modif EP (fin)

In the following, the PDF approach shall always be understood as the
numerical solution of the set of SDEs equivalent, in a weak sense as
explained above, to the corresponding Fokker-Planck equation.

%-------------------------------------------------------------------------------
\subsection{Stochastic integrals and calculus} \label{sec:intsto}
%-------------------------------------------------------------------------------
Stochastic differential equations require a strict mathematical
treatment. The mathematical specificities of SDEs have far reaching
consequences for the derivation of accurate numerical schemes,
cf. Section \ref{sec:cond}. Some basic explanations are now
given to highlight the important points that are needed for practical
purposes. As a matter of fact, Eq.~(\ref{sde}) is just a shorthand
notation for
\begin{equation} \label{eq:ito}
Z^r_i(t)= Z^r_i(t_0) + \int_{t_0}^{t} A_i(s,{\bf Z}^r(s))\,ds +
\int_{t_0}^{t} B_{ij}(s,{\bf Z}^r(s))\,dW_j(s),
\end{equation}
where the first integral on the RHS (Right-Hand Side) is a classical
Riemann-Stieltjes one. In the second integral, integration is
performed with a measure, $d{\bf W}(t)$, that has non-conventional
properties. The Wiener process can be defined~\cite{Arn_74} as
\textit{the only stochastic process with independent Gaussian
  increments and continuous trajectories} (an increment, over a time
step $dt$, is defined as $dW_j(t)=W_j(t+dt)-W_j(t)$). The Wiener
process has the following properties \cite{Arn_74,Kle_99}:
\begin{enumerate}
\item[(i)] the trajectories of $W_j(t)$ are continuous, yet nowhere
  differentiable (even on small time intervals, $W_j(t)$ fluctuates
  enormously),
\item[(ii)] each increment is a Gaussian random variable:
  $\lra{dW_j(t)^{2p+1}}=0$ for the odd moments,
  $\lra{dW_j(t)^2}= dt$ and
  $\lra{dW_j(t)^{2p}}=o(dt)$, $\forall\; p>1$, for the even
  moments. Increments over small time steps are stationary and
  independent, $\lra{dW_j(t)}=0$, $\forall\; t$, and
  $\lra{dW_j(t)\,dW_j(t')}=0$ with $t \neq t'$,
\item[(iii)] the trajectories are of unbounded variation in every
  finite interval.
\end{enumerate}
The last property is the reason why the treatment of stochastic
integrals differs from that of classical (Riemann-Stieltjes)
ones (the Wiener process is not of finite
variation~\cite{Kle_99}). Without going too deep into mathematical
details, property (iii) 
simply implies that speaking of a stochastic integral without
specifying in what sense it is considered lacks rigour (in this work,
all stochastic integrals will be considered in the It\^o sense). In
classical integration, the limit of the following sum ($\tau_k \in
[t_{k},t_{k+1}]$)
\begin{equation}
\int_{t_0}^{t} B_{ij}(s,{\bf Z}^r(s))\,dW_j(s) = \lim_{N \to +\infty}
\sum_{k=0}^{N} B_{ij}(\tau_k,{\bf Z}^r(\tau_k))(W_j(t_{k+1})-W_j(t_{k})),
\end{equation}
should be independent of the choice
of $\tau_k$. This is not true in the above integral, because of
property (iii)\footnote{Here, the different modes of convergence of the
  random variable -the stochastic integral- are not dealt with. For further
  information, see \cite{Arn_74,Oks_95} for example.}. 
As a consequence, a choice has to be made for the sake
of consistency. Two main choices (there exist others) are
encountered in the literature, the It\^o and the Stratonovich
definitions.
In the It\^o definition, $\tau_k=t_k$ and the following limit is
under consideration
\begin{equation}
\lim_{N \to +\infty} \sum_{k=1}^{N} B_{ij}(t_k,{\bf
  Z}^r(t_k))(W_j(t_{k+1})-W_j(t_{k})).
\end{equation}
This choice has a major drawback, i.e. the rules of ordinary
differential calculus are no longer valid. However, this drawback is
balanced by \textit{the zero mean and isometry properties} which
are of great help when deriving weak numerical schemes, see Section
\ref{num-sch},
\begin{equation} \label{eq:iso_prop}
\left.\begin{split}
& \lra{\int_{t_0}^{t_1} X(s)\,dW(s)}=0, \\
& \lra{\int_{t_0}^{t_2} X(s)\,dW(s)\;\int_{t_1}^{t_3} Y(s)\,dW(s)}=
\int_{t_1}^{t_2} \lra{X(s)Y(s)}\,ds.
\end{split}\right.
\end{equation}
where $\lra{\,\cdot\,}$ is the mathematical expectation ($t_0\leq t_1 \leq
t_2 \leq t_3$, $X(t)$ and $Y(t)$ are two stochastic processes). These
properties no longer hold in the case of the Stratonovich
interpretation but the rules of ordinary differential calculus remain
valid. In the Stratonovich interpretation of the stochastic integral,
the basic idea is to choose $\tau_k$ as the middle point of the
intervals, i.e. $2\,\tau_k=t_k+t_{k+1}$. There are, as a
matter of fact, several possible choices, the most commonly
encountered in the mathematical literature being
\begin{equation}
\left.\begin{split}
& \int_{t_0}^{t} B_{ij}(s,{\bf Z}^r(s)) \circ dW_j(s) = \\
& \lim_{N \to +\infty}
\sum_{k=0}^{N} \frac{1}{2}[B_{ij}(t_k,{\bf
  Z}^r(t_k))+B_{ij}(t_{k+1},{\bf Z}^r(t_{k+1}))](W_j(t_{k+1})-W_j(t_{k})),
\end{split}\right.
\end{equation}
where $\circ$ indicates that the stochastic integral is treated in the
Stratonovich sense.

The distinction between the It\^o and the Stratonovich interpretations
is critical, especially when ensuring consistency in the derivation of
weak numerical schemes. There is actually an equivalence between the
two interpretations. In can be shown \cite{Arn_74,Oks_95} that
Eq. (\ref{sde}) written in the Stratonovich sense 
\begin{equation} 
dZ^r_i(t) = A_i(t,{\bf Z}^r(t))\,dt + B_{ij}(t,{\bf Z}^r(t)) \circ dW_j(t),
\end{equation}
is equivalent to the following SDE, written in the It\^o sense,
\begin{equation}
\begin{split}
dZ^r_i(t) = A_i(t,{\bf Z}^r(t))\,dt + B_{kj}(t,{\bf
  Z}^r(t))& \frac{\partial B_{ij}(t,{\bf Z}^r(t))}{\partial z_k}\,dt \\
& + B_{ij}(t,{\bf Z}^r(t))\,dW_j(t).
\end{split}
\end{equation}
This result can explain why, in some works, computations performed
with identical models can lead to contradictory results (the
difference between the results is a drift term which is, most of the
time, not negligible). Let us stress, once again, that even though the
purpose of the present paper is not to present mathematical
subtleties, a good understanding of stochastic calculus is needed when
deriving weak numerical schemes for SDEs encountered in fluid
mechanics.

As in the present work the It\^o interpretation is retained,
let us briefly present the basics of stochastic calculus. It has been
mentioned that, in the frame of the It\^o interpretation, the rules of
ordinary differential calculus are no longer valid. As a matter a
fact, this non-trivial consequence can be understood by going back to
the properties of the Wiener process. For the second order moment of
the increments of $W(t)$, one has $\lra{dW(t)^2}=dt$. This non trivial
result (in classical calculus one would expect a second order term) is
inherent to the nature of the Wiener process (it is a
non-differentiable process). As a consequence, the rules of classical
calculus must be modified when considering terms of at least order
$2$. This is the well-known It\^o formula. For any stochastic process
${\bf X}(t)$ which verifies the following SDE
\begin{equation} 
dX_i(t) = A_i(t,{\bf X}(t))\,dt + B_{ij}(t,{\bf X}(t))\,dW_j(t),
\end{equation}
the SDE verified by any smooth function $f(t,{\bf X}(t))$ is
\begin{equation}
\begin{split}
df(t,{\bf X}(t)) = \frac{\partial f}{\partial t}(t,{\bf X}(t))\,dt +
& dX_i(t) \frac{\partial f}{\partial x_i}(t,{\bf X}(t)) \\
& + \frac{1}{2}({\bf B}{\bf B}^T)_{ij}(t,{\bf X}(t))\frac{\partial^2
  f}{\partial x_i \partial x_j}(t,{\bf X}(t)) dt,
\end{split}
\end{equation}
where the last term on the RHS is a new term with respect to
classical differential calculus.

The main tools of the general formalism and the basics of stochastic
calculus have now been introduced. This short presentation is a brief
summary and a reader willing to derive weak numerical schemes for
SDEs may refer to Refs.~\cite{Arn_74,Oks_95} for the
mathematical background, Ref.~\cite{Gar_90} for the physical background and
more importantly Ref.~\cite{Klo_92} for the derivation of numerical
schemes. A key point, which is not
developed here (the main point being that stochastic integrals require
a careful treatment), is the derivation of stochastic Taylor series, a
tool which is needed when attempting to develop weak numerical schemes for
SDEs. There exists a comprehensive book on these techniques, see
Ref. \cite{Klo_92}.

%===============================================================================
\section{PDF models} \label{sec:pdf_models}
%===============================================================================
The purpose of this section is to put forward the system of equations
which is solved in the present mean-field/PDF approach, and to reveal
the existing link between the physics of the problem and the tools
that were presented in Section \ref{sec:formalism}. Once this is done,
attention shall be focused on a central specific
theoretical issue: the treatment of the multiscale character of the
dynamics of the discrete particles, cf. Section \ref{sec:scale}.

%-------------------------------------------------------------------------------
\subsection{Derivation of a PDF model} \label{sec:SDEphy}
%-------------------------------------------------------------------------------
If the trajectories of a pair of particles (a fluid particle and a
discrete particle) can be modelled by writing a system of equations
given by Eq.~(\ref{sde}), two issues must be solved:
\begin{enumerate}
\item[(i)] what is the dimension of the reduced state vector (what
  variables should be retained)?
\item[(ii)] and what is the form of the coefficients (the drift vector
and the diffusion matrix)?
\end{enumerate}
A physical answer can be given to issue (i) when there is a separation
of (time) scales. Let $dt$ be the reference timescale at which the
physical phenomena are observed. The separation of scales is
defined in terms of slow and fast variables. A slow variable is a
variable whose integral timescale, $T$, is much larger than $dt$ and
vice-versa for a fast variable whose integral timescale is $\tau$,
i.e. $\tau \ll dt \ll T$. The answer comes from the application of
ideas known in synergetics, the so-called slaving principle
\cite{Hak_89}. In this equilibrium hypothesis, the fast variables are
assumed to relax 'very rapidly' to their equilibrium values which can
be expressed as a function of the values taken by the slow modes. A
practical application of this principle is the fast-variable
elimination technique where the fast variables are replaced by models
which represent their equilibrium values and usually involve
white-noise terms \cite{Gar_90}. The fast-variable elimination
technique can be used in the derivation of one-point PDF models for
single-phase turbulent flows.

In this work, answers to issues (i) and (ii) are given separately for
the fluid and the particles in the form of one-point PDF models. In the
one-point PDF approach for the fluid (see Section \ref{sec:pdf_f}),
single-phase turbulence is under consideration whereas in one-point
pdf models for the discrete particles, the fluid-particle mixture is
under investigation with known properties for the fluid (see Section
\ref{sec:pdf_p}). The two-point description briefly introduced in
Section \ref{sec:formalism} will be addressed in Section
\ref{sec:model}.

%--------------------------------------------------------------------
\subsubsection{One-point PDF models for single-phase turbulent flows} \label{sec:pdf_f}
%--------------------------------------------------------------------
When the Reynolds number is sufficiently large, for a reference time
scale $dt$ in the inertial range ($\tau_{\eta} \ll dt \ll T_L$, where
$T_L$ is the integral Lagrangian timescale and $\tau_{\eta}$ is the
Kolmogorov timescale), the Kolmogorov theory \cite{Mon_75} tells us
that, for Lagrangian statistics, the covariance matrix of velocity
increments has the form
\begin{equation} \label{eq:cov}
\lra{dU_{f,i}(t)\,dU_{f,j}(t+dt)}=C_0\,\lra{\epsilon}\,dt\,\delta_{ij},
\end{equation}
where $\epsilon(t,{\bf x})$ is the local instantaneous energy
dissipation rate and $C_0$ is a constant.
Equation (\ref{eq:cov}) implies that one has for the autocorrelation
coefficients $R_U = 1-[(C_0\,dt)/(2\,T_L)] \simeq 1$ (velocity) 
and $R_A = \tau_{\eta}/T_L \ll 1$ (acceleration) \cite{Min_01}. This shows that,
for $dt$ in the inertial range,  the velocity of a fluid particle,
${\bf U}_f(t)$, is a slow variable whereas the acceleration, ${\bf
  A}_f(t)$, is a fast variable. This suggests, according to the
slaving principle, that ${\bf A}_f(t)$ should be eliminated and
replaced by a function of the slow modes, position and velocity,
i.e. ${\bf Z}^r(t)=\{{\bf x}_f(t),{\bf U}_f(t)\}$. The Kolmogorov
theory gives answers to issues (i) and (ii): the
dimension of the reduced state vector is $2$ and the diffusion matrix
should be given by $B_{ij}= \sqrt{C_0\,\lra{\epsilon}}\;
\delta_{ij}$. The use of a SDE is not enforced by Eq. (\ref{eq:cov})
but the linear dependence in time of the covariance matrix is a strong
indication. For further justifications concerning the use of SDEs for
modelling purposes, see Ref. \cite{Min_01}.

Using different arguments, several researchers \cite{Pop_00,Min_97}
have shown that a Langevin equation model for single-phase turbulence is
\begin{equation} \label{eq:SDEf}
dU_{f,i}(t) = -\frac{1}{\rho_f}\frac{\partial \lra{P}}{\partial x_i}\,dt +
G_{ij}(U_{f,j}-\lra{U_{j}})\,dt + \sqrt{C_0\,\lra{\epsilon}}\,dW_i(t),
\end{equation}
where $P(t,{\bf x})$ is the local instantaneous pressure field. All
mean fields, i.e. $\lra{P}$, $\lra{{\bf U}}$ and
$\lra{\epsilon}$ are evaluated at time $t$ for ${\bf x}={\bf
  x}_f(t)$. The return-to-equilibrium matrix, $G_{ij}$, depends on mean
fields and is usually written $G_{ij}=-\delta_{ij}/T_L + G_{ij}^a$
where $T_L$ is a timescale given by $1/T_L =
(1/2+3C_0/4)\lra{\epsilon}/k$ ($k$ is the turbulent kinetic
energy). The anisotropy matrix, $G_{ij}^a$, also depends on mean
fields only and can take different forms \cite{Pop_00}.   

Equation (\ref{eq:SDEf}) has two noteworthy properties:
\begin{enumerate}
\item[-] the coefficients of this SDE depend not only on time and
  ${\bf Z}^r(t)$ (as in Eq.~(\ref{sde})) but also on the expected values
  of functionals of the state vector. This dependence of the
  coefficients has important consequences, not only for the
  mathematical formalism, but for the numerical algorithm, see Section
  \ref{sec:num}. These equations are often called Mac-Kean SDEs in the
  mathematical literature.
\item[-] the model is not self-contained since external fields are
  needed to compute the drift vector and the diffusion matrix,
  i.e. the state vector should rather be written ${\bf
    Z}^r(t)=\{{\bf x}_f(t),{\bf U}_f(t),\varepsilon(t)\}$. In complete
  models, a specific SDE is written for
  $\varepsilon(t)=\epsilon(t,{\bf x}_f(t))$ and the mean
  pressure field is given by a Poisson equation, see
  Ref. \cite{Pop_00} for example.
\end{enumerate}

%--------------------------------------------------------------
\subsubsection{One-point PDF models for the discrete particles} \label{sec:pdf_p}
%--------------------------------------------------------------
Let us suppose that the discrete particles are moving in a turbulent
flow whose mean fields are known (only one-way coupling is
considered for the moment, i.e.\ the presence of particles
does not modulate the turbulence). These mean fields are most commonly
the two first velocity moments, $\lra{{\bf U}(t,{\bf x})}$ and
$\lra{{\bf U}(t,{\bf x})\otimes{\bf U}(t,{\bf x})}$, and $\lra{P(t,{\bf x})}$ and
$\lra{\epsilon(t,{\bf x})}$.

In the case of discrete particles, the choice of a suitable state
vector is more difficult than in the fluid case since there are no
general results indicating a clear separation of scales. However, an
extension of Kolmogorov theory \cite{Min_01,Min_04} shows that a
linear dependence for the covariance matrix of the increments of the
fluid velocity seen, i.e.\
$\lra{dU_{s,i}(t)\,dU_{s,j}(t+dt)}$, can be obtained under some
specific hypotheses, for $dt$ in the inertial range. Without being a
formal proof, this result suggests to include the fluid velocity seen
in the state vector, ${\bf Z}^r(t)=\{{\bf x}_p(t),{\bf U}_p(t),{\bf
  U}_s(t)\}$, and to model the increments of ${\bf U}_s(t)$ with a
Langevin equation. This choice differs from the one inherent to
kinetic models where ${\bf Z}^r(t)=\{{\bf x}_p(t),{\bf U}_p(t)\}$
\cite{Ree_92,Ree_93}. The existing correspondence between these two
approaches has been discussed elsewhere \cite{Min_01}: it can be shown
that including the fluid velocity seen in the state vector presents
several advantages from the modelling point of view. Furthermore,
kinetic models can be retrieved from the Langevin models for ${\bf
  U}_s(t)$ \cite{Min_01}. Issue (i) has now been addressed, the
dimension of the state vector is $3$. Let us move to issue (ii), that
is to write a SDE for the increments of ${\bf U}_s$(t). 

From a physical point of view, the problem of modelling particle
dispersion (i.e. deriving a model for ${\bf U}_s(t)$, see
Eq. (\ref{exa_part_eqns})) is more complicated than the diffusion one
(fluid  particles, cf. Section \ref{sec:pdf_f}) since two
additional physical mechanisms have to be
accounted for: particle inertia characterised by the timescale
$\tau_p$ and external force fields (gravity in our case),
Eq. (\ref{exa_part_eqns}). Two main approaches can be found in the
literature:
\begin{enumerate}
\item[-] approaches based on paths (trajectories). A two-step
  construction is considered: a Lagrangian step and an Eulerian
  step. The Lagrangian step corresponds to the trajectory, over a time
  interval $dt$, of a fluid particle located at time $t$ in the
  vicinity of the discrete particle (this step is directly given by
  Eq. (\ref{eq:SDEf})). The Eulerian step corresponds to a spatial
  correction which gives, from the location of the fluid particle at
  $t+dt$, the fluid velocity seen by the discrete
  particle at time $t+dt$. This modelling point of view
  has two major drawbacks: it leads to an artificial decrease of
  the integral timescale of ${\bf U}_s(t)$ (denoted $T_{L,i}^*$ in the
  present paper) and there is no clear separation between the effects
  of $\tau_p$ and ${\bf g}$~\cite{Min_01}. 
\item[-] approaches based on the physical effects \cite{Poz_98}. A
  two-step construction is also considered by decoupling the two
  physical mechanisms: the first step corresponds to the effects of
  $\tau_p$ in the absence of external forces 
  (in that case $T_{L,i}^*$ varies between two limit
  values, $T_E$ -the integral Eulerian time scale- when $\tau_p \to
  +\infty$ and $T_L$ when $\tau_p \to 0$). The second step corresponds
  to the effects of gravity alone which induce a mean drift and result
  in a decorrelation of ${\bf U}_s(t)$ with respect to 
  ${\bf U}_f(t)$. This effect is called the crossing trajectory effect
  (CTE) and is related to the mean relative velocity 
  $\lra{{\bf U}_r}=\lra{{\bf U}_p-{\bf U}_s}$. 
\end{enumerate}
In the present work, the derivation of a model for ${\bf U}_s(t)$ is
carried out by resorting to an approach based on the physical effects
where the influence of the first step is neglected. This assumption
allows us to extend Kolmogorov theory since the increments of the
fluid velocity seen are only governed by mean quantities~\cite{Min_01}.

%modif EP (début)
\ifnum\elscolor>0 \color{blue} \fi
\ifnum\elsitalic>0 \itshape \fi
Assuming, for the sake of simplicity, that the mean drift (the mean
relative velocity $\lra{{\bf U}_r}$) is aligned
with one of the coordinate axes (the general case is discussed
in Ref.~\cite{Min_01}), it can be shown~\cite{Min_01,Min_04} that a
possible model for the increments of the fluid velocity seen is (the
summation convention by repeated indices does not apply to the third
and fourth term on the RHS)
\begin{equation} \label{eq:dUs}
\left.\begin{split}
d& U_{s,i}(t) =-\frac{1}{\rho_f}\frac{\partial \lra{P} }{\partial x_i}\, dt
+ \left( \lra{U_{p,j}} - \lra{U_{j}} \right)
\frac{\partial \lra{U_{i}}}{\partial x_j}\, dt \\
& -\frac{1}{T_{L,i}^*}\left( U_{s,i}-\lra{U_{i}} \right)\, dt
+ \sqrt{ \lra{\epsilon}\left( C_0b_i \tilde{k}/k
+ \frac{2}{3}( b_i \tilde{k}/k -1) \right) }\, dW_i(t).
\end{split}\right.
\end{equation}
The CTE has been modelled by changing the timescale, compared to the
fluid case, in the drift term (third term on the RHS) and by adding a
mean drift term (second term on the RHS). The time scale is modified
according to Csanady's analysis~\cite{Csa_63}
\begin{equation} \label{eq:Csanady}
T_{L,i}^{*}= T_L \left( 1 + \beta_i^2 \displaystyle \frac{\vert
      \lra{{\bf U}_r}\vert^2}{2k/3} \right)^{-1/2},
\end{equation}
where $\beta_1=\beta$, if axis $1$ is aligned with the mean drift,
with $\beta=T_L/T_E$, and in the transversal directions (axes labelled
2 and 3) $\beta_i=2 \beta$. In the diffusion matrix, a new kinetic
energy has been introduced
\begin{equation} \label{eq:ktilde}
\tilde{k}= \frac{3}{2}
\frac{\sum^3_{i=1}b_i\lra{u_{i}^2}}{\sum^3_{i=1}b_i},
\end{equation}
where ${\bf u}(t,{\bf x})={\bf U}(t,{\bf x})-\lra{\bf U}$ and $b_i= T_L/T_{L,i}^*$.

Equation (\ref{eq:dUs}) has two noteworthy properties:
\begin{enumerate}
\item[-] it is consistent, by construction, with Eq. (\ref{eq:SDEf})
  when  $\tau_p \to 0$, that is when the discrete particles behave
  like fluid particles,
\item[-] it is a Mac-Kean SDE even though the mean fields of the fluid
  are known (they are given by solving RANS equations). Indeed, it is
  necessary to compute the mean velocity of the particles 
  $\lra{{\bf U}_p}$ to calculate not only the mean drift term (second
  term on the RHS) but also the integral time scale of 
  ${\bf U}_s(t)$, $T_L^*$ ($\lra{{\bf U}_s}$ is also needed for the
  computation of this time scale).
\end{enumerate}
Moreover, it must be emphasised that Eq. (\ref{eq:dUs}) is a possible
choice among others and that the exact form of a Langevin equation for
${\bf U}_s(t)$ still remains an open issue, see for exemple
Refs. \cite{Oll_04,Rey_00} for models suited for homogeneous turbulence. There exists an
alternative to Eq. (\ref{eq:dUs}) in the literature \cite{Sim_93}
where the coefficients are slightly different (the drift vector and
the diffusion matrix), the main difference being the form of the mean
drift term which is written in terms of instantaneous velocities rather
that mean velocities, i.e. $(U_{p,j}-U_{s,j})(\partial \lra{U_i}/\partial x_j)$.
This difference has been discussed elsewhere \cite{Min_01}. This form
of the mean drift term does not change the methodology which is
presented in the rest of the paper, but it modifies the structure of
the system of SDEs, i.e. ${\bf U}_s(t)$ depends explicitly on the
particle velocity, ${\bf U}_p(t)$. This matter will be discussed in
Section \ref{sec:twc}.
\normalfont
\color{black}
%modif EP (fin)

When two-way coupling occurs, that is when the mass of particles per
unit volume of fluid is sufficient to influence the characteristics of
turbulence, Eq. (\ref{eq:dUs}) can be supplemented by an acceleration
term ${\bf A}_{p \rightarrow s}$ which accounts for the influence
of the discrete particles on the statistics of the fluid velocity
sampled along the trajectory of a discrete particle, i.e.
\begin{equation}
dU_{s,i}(t) = A_{s,i}\,dt + A_{p \rightarrow s,i}\,dt +
B_{s,ij}\,dW_j(t),
\end{equation}
where the drift vector ${\bf A}_s$ and the diffusion matrix ${\bf
  B}_s$  are directly given by Eq. (\ref{eq:dUs}).

For ${\bf A}_{p \rightarrow s}$, the underlying force corresponds to
the exchange of momentum between the fluid and the particles (drag
force). The acceleration acting on the fluid element surrounding a
discrete particle can be obtained as the sum of all elementary
accelerations (due to the neighbouring particles)~\cite{Min_01}
\begin{equation} \label{eq:Ap_s}
{\bf A}_{p \rightarrow s} = - \frac{\alpha_p \rho_p}{\alpha_f \rho_f}
\frac{{\bf U}_s-{\bf U}_p}{\tau_p},
\end{equation}
i.e., at the discrete particle location ${\bf x}_p$, the
elementary acceleration $({\bf U}_p-{\bf U}_s)/\tau_p$ is multiplied
by the probable mass of particles divided by the probable mass of
fluid (since the total force is distributed only on the fluid
phase). In Eq. (\ref{eq:Ap_s}), it is implicitly assumed that all
particles under consideration have the same
acceleration. Moreover, $\alpha_f(t,{\bf x})$ and $\alpha_p(t,{\bf x})$
represent the probability of presence of the fluid and the particles,
respectively ($\alpha_f+\alpha_p=1$).

The complete set of SDEs which describes the one-point dynamical
behaviour of the discrete particles is
\begin{equation} \label{eq:SDEp}
\left\{\begin{split}
& dx_{p,i}(t)= U_{p,i}\,dt, \\
& dU_{p,i}(t)= \frac{U_{s,i}-U_{p,i}}{\tau_p}\,dt + g_i\,dt, \\
& dU_{s,i}(t) = A_{s,i}\,dt + A_{p \rightarrow s,i}\,dt +
  B_{s,ij}\,dW_j(t),
\end{split}\right.
\end{equation}
where ${\bf A}_s$ and ${\bf B}_s$ are calculated by resorting to Eqs.
(\ref{eq:dUs})-(\ref{eq:ktilde}). This set of SDEs is under
investigation in the present paper and the numerical methods needed to
solve it will be discussed in Section \ref{sec:num0}.

%-------------------------------------------------------------------------------
\subsection{Mean-field/PDF approach} \label{sec:model}
%-------------------------------------------------------------------------------
As specified in the Introduction, a mean-field/PDF approach is
adopted here for the computations of polydispersed turbulent two-phase 
flows, i.e., the fluid is described by solving RANS equations
whereas the dynamics of the discrete particles are simulated with
SDEs. The general formalism presented in Section \ref{sec:formalism}
and the models presented in Sections \ref{sec:pdf_f} and
\ref{sec:pdf_p} are sufficient to derive the mean-field/PDF model
which is used in the present work. As a matter of fact, the set of
SDEs, from which the pdf of the discrete phase can be extracted, has
already been presented, cf. Eqs. (\ref{eq:SDEp}). Only the
derivation of the set of mean-field (RANS) equations for the
continuous phase has to be put forward. This derivation can be done
using different techniques \cite{Sim_96,Pei_98a,Pei_02}. An
interesting technique, which is in line with the
tools introduced in Section \ref{sec:formalism}, is to resort to a
two-point description. This new approach hardly provides further
physical information but it allows both the fluid and the
particles to be described under the same formalism. This description
is briefly outlined here, supplementary information is given in
Appendix \ref{sec:appen}.

%---------------------------------------------------------------------
\subsubsection{PDF models for polydispersed turbulent two-phase flows} 
\label{sec:two_point}
%---------------------------------------------------------------------
The path which is adopted is to gather the preceding results that
have just been presented for the time increments of the fluid velocity
seen along discrete particle trajectories, Eq.~(\ref{eq:SDEp}), and for
the time increments of the fluid velocity along fluid particle
trajectories, Eq.~(\ref{eq:SDEf}). The system of
SDEs is, however, supplemented by one term (an acceleration)
${\bf A}_{p \rightarrow f}$ which reflects the influence of the
discrete particles on the fluid. The time rate of change of
${\bf Z}^r(t)=\{{\bf x}_f(t),{\bf U}_f(t),{\bf x}_p(t),{\bf
  U}_p(t),{\bf U}_s(t)\}$ is given by
\begin{equation} \label{eq:SDEfp}
\left\{\begin{split}
& dx_{f,i}(t)= U_{f,i}\,dt, \\
& dU_{f,i}(t) = A_{f,i}\,dt + A_{p \rightarrow f,i}\,dt +
  B_{f,ij}\,dW_j(t),\\
& dx_{p,i}(t)= U_{p,i}\,dt, \\
& dU_{p,i}(t)= ((U_{s,i}-U_{p,i})/\tau_p)\,dt + g_i\,dt, \\
& dU_{s,i}(t) = A_{s,i}\,dt + A_{p \rightarrow s,i}\,dt +
  B_{s,ij}\,dW_j^{\prime}(t),
\end{split}\right.
\end{equation}
where the drift vector ${\bf A}_f$ and the diffusion matrix 
${\bf B}_f$ are given by Eq.~(\ref{eq:SDEf}). The form of 
${\bf A}_{p \rightarrow f}$ is discussed in Appendix
\ref{sec:appen}. By assuming that the trajectories of a pair of
particles can be obtained in such a way, i.e. Eqs.~(\ref{eq:SDEfp}),
one should be aware that several assumptions have been made. Further
explanations on these assumptions are given in Refs.~\cite{Min_01} and
\cite{Pei_02}.

%-----------------------------------------------------------
\subsubsection{Mean-field/PDF model}
%-----------------------------------------------------------
It can be shown that the mean-field (RANS) equations for the fluid can
be extracted from Eqs.~(\ref{eq:SDEfp}), see Appendix
\ref{sec:appen}. The complete set of equations is given in Table
\ref{tab:RANS}. Depending on the closures chosen for the
return-to-equilibrium matrix, $G_{ij}$, different RSM equations can be
obtained. In practical computations, these mean-field equations are
supplemented with a PDE for $\lra{\epsilon}$ obtained by the same path
(as mentioned in Section \ref{sec:pdf_f}, a SDE can be written for
$\varepsilon(t)$).

The mean-field/PDF model has now been presented, that is the mean
field equations for the fluid and the set of SDEs for the discrete
particles. In practical computations, the mean fluid properties are
computed with classical finite volume techniques whereas the dynamics
of the stochastic particles are calculated by resorting to stochastic
calculus, i.e. by integrating in time a set of SDEs (weak
numerical schemes). The finite volume algorithms are well known and no
discussion on these methods is given here. However, some specific
issues related to the computation of the source terms in
Eqs.~(\ref{eq:feq_Uf}) and (\ref{eq:feq_ufuf}) will be addressed in
Section \ref{sec:source}. These source terms are computed from the
stochastic particles (the numerical solution of the SDEs) and, as we
shall see in Section \ref{sec:num}, information has to be exchanged
between the stochastic particles and the mesh on which the mean-fluid
properties are calculated. Before carrying on with the derivation of
weak numerical schemes for the set of SDEs, a very important issue has
to be put forward: the multiscale character of the model for the
discrete phase.

%modif EP (début)
\ifnum\elscolor>0 \color{blue} \fi
\ifnum\elsitalic>0 \itshape \fi
The discussions to come on the multiscale character of the set of
SDEs, Section \ref{sec:scale}, and on the derivation of weak numerical
schemes for these equations, Section \ref{sec:cond}, are presented in
the case of one-way coupling. This is not due to a limitation of the
method but it simply reflects the current status of the present
work. Indeed, the methodology, which is presented in Sections
\ref{sec:scale} and \ref{sec:num0}, remains valid in the case of
two-way coupling. The presentation of the general algorithm,
Section \ref{sec:num}, and the discussion on issues related to
projection and averaging in particle-mesh method, Section
\ref{sec:PA}, are put forward in the case of two-way
coupling. Specific issues related to two-way coupling will be
discussed in Section \ref{sec:twc}, namely the computations of the
source terms in the set of PDEs (RANS equations) and the extension of
the material presented in Sections \ref{sec:scale} and \ref{sec:cond}
to two-way coupling.
\normalfont
\color{black}
%modif EP (fin)

%===============================================================================
\section{Multiscale properties of the SDEs} \label{sec:scale}
%===============================================================================
There are three different timescales describing the dynamics of the
discrete particles, cf. Eq. (\ref{eq:SDEp}): $dt$, the
timescale at which the process is
observed, $T_{L,i}^*$ the integral timescale of the fluid velocity
seen, ${\bf U}_s(t)$, and $\tau_p$ the particle relaxation time. Once
again, it must be recalled that these SDEs have a physical meaning
only in the case where $dt \ll T_{L,i}^*$ and $dt \ll \tau_p$. What
happens if one of these conditions or both are not verified? It is in
fact possible to show that the system of SDEs converges towards
several limit cases which are consistent with the physics. The
mathematical details are not given here, see Ref. \cite{Min_03} for
further explanations. Nevertheless, the fast-variable
elimination technique is now presented in a simple case in order
to help the reader understand the form of two of the limit cases of
system (\ref{eq:SDEp}).

%-------------------------------------------------------------------------------
\subsection{Fast-variable elimination technique} \label{sec:fast}
%-------------------------------------------------------------------------------
Let us consider a model problem to illustrate this technique. A good
historical example is the treatment of Brownian motion. There exist
two points of view to address this problem: Einstein's point of view
where only position is retained in the state vector, ${\bf
  Z}(t)=\{{\bf x}(t)\}$, and Langevin's approach where the state
vector is composed by position and velocity, ${\bf Z}(t)=\{{\bf
  x}(t),{\bf U}(t)\}$. Let us consider one-dimensional Brownian motion
for the sake of simplicity. Langevin's model reads
\begin{equation} \label{eq:langevin}
\left\{\begin{split}
& dx(t) = U(t)\, dt, \\
& dU(t) = -(U(t)/T)\,dt + B\,dW(t),
\end{split}\right.
\end{equation}
this system being valid when $dt \ll T$ ($T$ is the integral time
scale of $U(t)$, $B$ is the diffusion coefficient). What happens if
this condition is not verified, i.e. when $T \to 0$? The velocity,
$U(t)$, becomes a fast variable and, according to the slaving
principle, it should be eliminated and its influence should be
expressed as a function of the slow modes (here position,
$x(t)$). This new description should then be consistent with
Einstein's model, $dx(t) = \sqrt{2 D}\;dW(t)$ where $D$ is the
diffusion coefficient.

By applying the rules of stochastic calculus (see Section
\ref{sec:intsto}), one can show that
\begin{equation}
U(t) \simeq B\,T \;\eta(t) \quad \text{with} \quad 
\eta(t)=\frac{1}{T}\exp(-t/T)\int_{-\infty}^{t}\exp(s/T)\,dW(s),
\end{equation}
where the influence of initial conditions have been neglected (one
integrates from $-\infty$). Here, $\eta(t)$ is a Gaussian random
variable \cite{Kle_99} (it is a stochastic integral of a deterministic
function) with $\lra{\eta(t)}=0$ (zero mean property) and (isometry
property)
\begin{equation}
\lra{\eta(t)\eta(t^{\prime})}=\frac{1}{2T}\exp(-|t-t^{\prime}|/T)
\xrightarrow[T \to\, 0]{} \delta(t-t^{\prime}).
\end{equation}
Therefore, $\eta(t)$ is a Gaussian white noise and consequently
$dW(t)=\eta(t)\;dt$. The limit system for Langevin's model,
Eqs. (\ref{eq:langevin}), when $T \to 0$ and $B\,T$ remains finite, is
then given by
\begin{equation}
dx(t) = \sqrt{2 D}\;dW(t),
\end{equation}
where $D=(B\,T)^2/2$. The velocity, whose integral timescale becomes
zero and whose variance becomes infinite (Gaussian white noise) has been
eliminated, that is the process is observed at a timescale $dt$
large in comparison to the velocity fluctuations (characteristic time
scale, $T$). Its influence is, however, left in the diffusion
coefficient of the reduced model (Einstein's model) provided that the
product $B\,T$ remains finite. As we shall see shortly in Section
\ref{sec:limcases}, this model problem is going to be helpful to
understand two of the four limit cases to come.

The above analysis of the model problem has been presented in the
\textit{continuous sense}, that is when both variables are continuous
functions of time. In practical applications, i.e. in
numerical computations or in experiments, both variables would have
been observed at discrete times. In numerical calculations, one
computes the solutions at discrete times where the time step
corresponds to the observation timescale $dt$. In experiments, the
variables of interest are measured at a given sampling frequency $f$
and one has $dt=1/f$. The development of the numerics is slightly
anticipated and the limit case $T \to 0$ is now
investigated in the \textit{discrete sense}, that is when $T \ll
\Delta t$. A good understanding of the subtle difference between the
continuous case and the discrete case will be helpful in the
development of weak numerical schemes to come, see Section
\ref{num-sch}.

In the \textit{discrete sense}, the limit case does not correspond to
$T \to 0$ but rather to $T \ll \Delta t$. Applying the rules of
stochastic calculus, the solution to system
(\ref{eq:langevin}) at $t=t_0+\Delta t$ is given by
\begin{equation} \label{eq:eqI}
\left\{\begin{split}
& x(t) = x(t_0) + U(t_0)\,T\,[1-\exp(-\Delta t/T)] + I_x(t), \\
& U(t) = U(t_0)\exp(-\Delta t/T) + I_U(t),
\end{split}\right.
\end{equation}
where the stochastic integrals are defined as
\begin{equation} \label{eq:IxIu}
\left\{\begin{split}
& I_x(t) = B\,T\int_{t_0}^{t}dW(s)-B
T\exp(-t/T)\int_{t_0}^{t}\exp(s/T)\,dW(s), \\
& I_U(t) = B\exp(-t/T)\int_{t_0}^{t}\exp(s/T)\,dW(s).
\end{split}\right.
\end{equation}
Here, $(I_x,I_U)$ is a vector composed by two dependent, centred
(zero mean property, see Section \ref{sec:intsto}) Gaussian random
variables (Gaussian since $I_x$ and $I_U$ are stochastic integrals of
deterministic functions \cite{Kle_99}). It can be shown that a
centred Gaussian vector can be expressed as the product of two
matrices by resorting to the Choleski algorithm: these matrices are
the covariance matrix and a vector composed of independent standard
Gaussian random variables (${\cal N} (0,1)$, i.e. zero mean and
variance equal to unity). This decomposition is well suited to
numerical applications since a set of independent standard Gaussian
random variables can easily be generated on computers by using suitable
random number generators. By applying the Choleski algorithm to
$(I_x,I_U)$, one can write
\begin{equation} \label{eq:xi}
\left\{\begin{split}
& I_x(t) = \left(\lra{I_x\,I_U}/\sqrt{\lra{I_U^2}}\right)\,\xi_U + 
\sqrt{\lra{I_x^2}-\lra{I_x\,I_U}^2/\lra{I_U^2}}\,\xi_x, \\
& I_U(t) = \sqrt{\lra{I_U^2}}\,\xi_U,
\end{split}\right.
\end{equation}
where $\xi_x$ and $\xi_U$ are two independent standard Gaussian random
variables. The components of the covariance matrix are given by
\begin{equation}
\left\{\begin{split}
& \lra{I_x^2} = (B\,T)^2 \{  \Delta t-T[1-\exp(-\Delta
t/T)][3-\exp(-\Delta t/T)]/2 \},\\ 
& \lra{I_U^2} =  B^2T[1-\exp(-2\Delta t/T)]/2,\\
& \lra{I_x\,I_U} = \{B\,T[1-\exp(-\Delta t/T)]\}^2/2.
\end{split}\right.
\end{equation}
Therefore, in the \textit{discrete sense}, the limit system to
Langevin's model becomes
\begin{equation} \label{eq:lim_Gauss}
\left\{\begin{split}
& x(t) =
x(t_0)+U(t_0)\,T+B\,T\left(\sqrt{\frac{T}{2}}\,\xi_U+\sqrt{\Delta t}\,\xi_x\right), \\
& U(t) = \sqrt{\frac{B^2T}{2}}\,\xi_U.
\end{split}\right.
\end{equation}
In the discrete sense, the velocity $U(t)$ does not 'disappear' (in the
continuous case, it becomes Gaussian white noise). This result is
physically sound since the velocity is only observed at time steps
which are large compared to its memory (integral timescale). Finally,
the above results are consistent with the observation of Einstein,
that is, for long diffusion times, one has  $\lra{x(t)^2}\sim (BT)^2\, t$.

%-------------------------------------------------------------------------------
\subsection{Limit cases} \label{sec:limcases}
%-------------------------------------------------------------------------------
From now on, the summation rule by repeating indices is dropped to
avoid confusion, as in Eq. (\ref{eq:dUs}) for example. The system of
SDEs describing the dynamics of the discrete particles reads
(\textit{from now on $B_{s,ij}$ is denoted $B_{ij}$ for the sake of
  simplicity})
\begin{equation}
\left\{
\begin{split} \label{eq:sysEDS}
dx_{p,i}(t) & = U_{p,i}\, dt, \\
dU_{p,i}(t) & =\frac{1}{\tau_p}(U_{s,i}-U_{p,i})\, dt + \mc{A}_i\,dt,\\
dU_{s,i}(t) & =-\frac{1}{T_{L,i}^*}U_{s,i}\,dt+C_i\, dt + \sum_j
B_{ij}\, dW_j(t),
\end{split} \right .
\end{equation}
%modif EP (début)
\ifnum\elscolor>0 \color{blue} \fi
\ifnum\elsitalic>0 \itshape \fi
where $C_i$ is a term that includes all mean contributions: the mean
pressure gradient, $-(\partial \lra{P}/\partial x_i)/\rho_f$, the mean drift term,
$(\lra{U_{p,j}}-\lra{U_j})(\partial \lra{U_i}/\partial x_j)$, and the mean
part of the return-to-equilibrium term, $\lra{U_i}/T^*_{L,i}$. As
explained above, two-way coupling is left out of the present
analysis. $\mc{A}_i$ is an acceleration (gravity in the
\normalfont
\color{black}
%modif EP (fin)
present work, but it can be extended for practical reasons to the case
of other external force fields).
Once again, system (\ref{eq:sysEDS}) has a
physical meaning only in the case where $dt \ll T_{L,i}^*$ and $dt \ll
\tau_p$. When these conditions are not satisfied, it is possible to show
that, in the {\em continuous sense} (time and all coefficients are
continuous functions which can go to zero), the system converges
towards several limit systems \cite{Min_03}.
\newline \textbf{Case 1}: when $\tau_p \rightarrow 0$, the particles
behave as fluid particles and one has
\begin{equation} \label{eq:limc1}
\text{system (\ref{eq:sysEDS}}) \; \xrightarrow[\tau_p \to 0]{}
\left\{\begin{split}
& dx_{p,i}(t) = U_{p,i}\, dt, \\
& U_{p,i}(t) = U_{s,i}(t), \\
& dU_{s,i}(t) = -\frac{1}{\rho_f} \frac{\partial \lra{P}}{\partial x_i}\,dt 
 -\frac{1}{T_L}(U_{s,i}-\lra{U_i})\,dt + \sqrt{C_0 \lra{\epsilon}}\,dW_i(t), 
\end{split}\right.
\end{equation}
that is, the model is consistent with a known turbulent fluid PDF model
\cite{Pop_94} as explained in Section \ref{sec:pdf_p}. This shows that
the model is a coherent generalisation of the fluid one, which can be
recovered as a limit case.
\newline \textbf{Case 2}: when $T_{L,i}^* \rightarrow 0$ and
$B_{ij}T_{L,i}^* \to {\rm cst}$, the fluid velocity seen becomes a fast
variable. It is then eliminated and one can write
\begin{equation} \label{eq:limc2}
\text{system (\ref{eq:sysEDS}}) \;
\xrightarrow[T_{L,i}^* \to 0]{(B_{ij} T_{L,i}^* \to {\rm cst})}
\left\{\begin{split}
dx_{p,i}(t) = & U_{p,i}\,dt, \\
dU_{p,i}(t) = & \frac{1}{\tau_p}(\lra{U_i}-U_{p,i})\,dt +
 \mc{A}_i\,dt + \\ & \hskip 2.0cm \sum_j \frac{B_{ij}
   T_{L,i}^*}{\tau_p}\,dW_j(t).
\end{split}\right.
\end{equation}
This result can be understood from the model problem of the previous
Section. ${\bf U}_s(t)$ has been eliminated but its influence is left in the
diffusion coefficient $B_{ij}T_{L,i}^*/\tau_p$. In this case, the equations
are equivalent to a Fokker-Planck model for particles of significant inertia.
\newline \textbf{Case 3}: When $\tau_p,\,T_{L,i}^* \rightarrow 0$ and at the
same time $B_{ij}T_{L,i}^* \to {\rm cst}$ , the fluid velocity seen becomes a fast
variable and the discrete particles behave as fluid particles. It can
be shown that
\begin{equation} \label{eq:limc3}
\text{system (\ref{eq:sysEDS}}) \;
\xrightarrow[\tau_p,T_{L,i}^* \to 0]{(B_{ij} T_{L,i}^* \to {\rm cst})}
\left\{\begin{split}
& dx_{p,i}(t) = \lra{U_i}\,dt + \mc{A}_i\,dt + \\ 
& \hskip 3.5cm \sum_j (B_{ij} T_{L,i}^*)\,dW_j(t).
\end{split}\right.
\end{equation}
We retrieve a pure diffusive behaviour, that is the equations of
{\it Brownian motion}, cf. Section \ref{sec:fast}.
\newline \textbf{Case 4}: at last, when $T_{L,i}^* \rightarrow 0$ with
no condition on $B_{ij}T_{L,i}^*$, the velocity of the fluid seen is
no longer random and the system becomes deterministic. The flow is
laminar and it can be proven that
\begin{equation} \label{eq:limc4}
\text{system (\ref{eq:sysEDS}}) \;
\xrightarrow[T_{L,i}^* \to 0]{}
\left\{\begin{split}
& dx_{p,i}(t) = U_{p,i}\, dt, \\
& dU_{p,i}(t) = \frac{1}{\tau_p}(\lra{U_i}-U_{p,i})\, dt 
  + \mc{A}_i\,dt, \\
& U_{s,i}(t) = \lra{U_i}.
\end{split}\right.
\end{equation}

Limit cases 1 to 3 reflect the multiscale character of the
problem. When the timescales go to zero (with a condition on their
products with the coefficients of the diffusion matrix), a hierarchy
of stochastic differential systems is obtained. Moreover, the
elimination of the fast variables (the velocities ${\bf U}_p(t)$ and
${\bf U}_s(t)$) does not mean that these variables do not (physically)
exist anymore: they simply become Gaussian
white noise. As we shall see in Section \ref{num-sch}, both 
${\bf U}_p(t)$ and ${\bf U}_s(t)$ become independent Gaussian random
variables, in the discrete sense, in limit case (iii) (in limit case
(ii), only ${\bf U}_s(t)$ becomes a Gaussian random variable). These
results are in line with the previous model problem,
cf. Eq. (\ref{eq:lim_Gauss}) in Section \ref{sec:fast}.

The existence of limit systems is a key point in the development of
weak numerical schemes to integrate in time the set of SDEs
describing the dynamics of the discrete particles, i.e. Eqs.
(\ref{eq:sysEDS}). As we shall see in the next Section, in numerical
computations, $dt$ the observation timescale of the process, is the
time step. A suitable weak numerical scheme should therefore be
consistent with all limit cases since, as we shall see, it is not
possible to control the time step to enforce the conditions necessary
for the validity of Eqs. (\ref{eq:sysEDS}). Before we carry
on to the time integration of Eqs. (\ref{eq:sysEDS}), let us give a
general overview of the numerical procedure which is needed to solve
the whole set of equations (the mean-field (RANS) equations for the
fluid and the SDEs for the discrete particles).

%===============================================================================
\section{Numerical approach} \label{sec:num0}
%===============================================================================
The mean-field/PDF model used in the present paper for practical
computations has now been given, see Table \ref{tab:RANS}. It
consists in a set of PDEs describing the dynamics of mean-fluid
quantities and a set of SDEs from which the joint pdf of the variables
of interest for the discrete particles can be extracted. In this
approach, the numerical solution is obtained by resorting to a hybrid
method where the mean-fluid properties are computed by solving the
mean-field (RANS) equations with a classical finite volume procedure
whereas the local instantaneous properties of the discrete particles
are determined by solving the set of SDEs, Eqs.
(\ref{eq:SDEp}). Therefore, the mean fluid properties are computed on
a mesh whereas the statistics of the discrete phase are calculated
from particles moving in the computational domain.

A closer look at the equation system presented in Table \ref{tab:RANS}
shows that the set of equations has the following properties:
\begin{enumerate}
\item[(i)] in the set of PDEs from which the mean-fluid properties are
  computed, mean fields involving discrete particles properties are
  needed in order to compute the source terms, see Eqs.
  (\ref{eq:feq_Uf}) and (\ref{eq:feq_ufuf}).
\item[(ii)] in the set of SDEs, the knowledge of statistical moments
  (for the fluid and the discrete particles),
  such as mean values and variances, at the locations of the
  stochastic particles, is required in order to compute the time
  evolution of the discrete particle properties and, thus, of
  the statistics derived from them. Indeed, from a mathematical
%modif EP (début)
\ifnum\elscolor>0 \color{blue} \fi
\ifnum\elsitalic>0 \itshape \fi
  standpoint, the set of SDEs can formally be written as (from now on,
  the notation is slightly changed: ${\bf Z}(t)$ denotes the state
  vector for the discrete particles, i.e. the superscript $r$ is
  dropped for the sake of clarity)
\begin{equation} \label{EqGeneric}
d{\bf Z}(t)={\bf A}(t,{\bf Z}(t),p(t;{\bf z}),{\bf Y}(t))\,dt
+ {\bds \sigma}(t,{\bf Z}(t),p(t;{\bf z}),{\bf Y}(t))\,d{\bf W}(t),
\end{equation} 
where $p(t;{\bf z})$ stands for the pdf of ${\bf Z}(t)$ and
${\bf Y}(t)$ represents external mean fields, i.e.
the fluid mean fields defined at particle locations. For each time
$t$, $p(t;{\bf z})$ has to be calculated in order to compute the
coefficients of the SDE: in the present approach, the pdf (or
any necessary moment extracted from it) is computed out of all
stochastic particles that are tracked (all values taken by 
${\bf Z}(t)$), cf. Section \ref{sec:stat}.
\normalfont
\color{black}
%modif EP (fin)
Thus, as mentioned in Sections \ref{sec:pdf_f} and
\ref{sec:pdf_p}, a kind of integro-differential equation, no longer
local in the space of ${\bf Z}(t)$, is obtained; it is called a Mac-Kean
SDE and is inherently difficult to solve~\cite{Ott_96,Tal_95}. In
other words, we are dealing with a system where particles
interact weakly (or indirectly) through the mean fields that they
create (it is the leading idea of one-point PDF models).
\end{enumerate}
As far as property (ii) is concerned, in practice, probabilistic
expectations of particle properties at a given point are approximated
by spatial averages over nearby particles, i.e. the
statistics extracted from the stochastic particles (which are needed
to compute the coefficients of system (\ref{eq:SDEp})) are not
calculated for each particle (this would cost too much CPU time) but
are evaluated at each cell centre of the mesh (generated for the
solution of the set of PDEs) following a given spatial average
(\textit{averaging operator}). These moments can then be evaluated for
each particle by interpolation or projection (\textit{projection operator}). The same
projection operator is used to compute the statistics of the fluid at
the locations of the stochastic particles. The source terms in the
PDEs, cf. property (i), are directly computed by resorting to
the averaged particle quantities.

This is the principle of particle-mesh methods: exchange data between
particles and mesh points. In the present work, the main advantage of
such methods is of course the reduction of CPU time but the use of
projection and averaging operators has some drawbacks, i.e.
it creates additional numerical errors and, in the general algorithm,
each particle has to be located in the mesh.

%-------------------------------------------------------------------------------
\subsection{Particle-mesh methods} \label{sec:num}
%-------------------------------------------------------------------------------
Historically, particle-mesh methods have been widely used in other
areas of physics like the dynamics of plasmas, astrophysical
simulations, electrostatics, etc.~\cite{Hoc_88}. In these
applications, the system of equations which is solved is deterministic
and the mesh is uniform, most of the time for unbounded domains or
bounded domains with periodic boundary conditions. These models are
often referred to as the particle-in-cell (PIC)
approach~\cite{Hoc_88,Bir_69,Oro_93}.

In fluid dynamics, apart from calculations of dispersed two-phase
flows, particle-mesh methods are mainly encountered in computations of
single-phase turbulence with stand-alone particle
models~\cite{Min_99b,Pop_95,VSl_98}, in hybrid particle/field
models~\cite{Mur_99}, in calculations of turbulent reactive flows with
PDF methods \cite{Pop_85}, and in numerical simulations of non-Newtonian
polymeric fluids~\cite{Ott_96}. In these applications, the situation
is different from the classical PIC approach. Indeed, the systems of
equations are stochastic and the domains are bounded with various
boundary conditions. Let us present the main lines of the
particle-mesh algorithm used in the present work. It is implemented in
the ESTET software \cite{Min_01rr}.

%----------------------------------
\subsubsection{General algorithm} \label{sec:algo}
%----------------------------------
Let $\{{\bf Y^{[x]}}\}$ stand for the set of the fluid mean fields at
the different mesh points and let $\{{\bf Y}^{(N)}\}$ be the fluid mean
fields interpolated at particle locations. Let $\{{\bf Z}^{(N)}\}$ denote
the set of variables attached to the stochastic particles and 
$\{{\bf Z^{[x]}}\}$ the set of statistics, defined at cell centres,
extracted from $\{{\bf Z}^{(N)}\}$. Time is discretised with a constant
time step $\Delta t=t_{n+1}-t_n$ and space with a uniform mesh of cell
size $\Delta x$.

The first step (operator $F$) of the algorithm is to solve the PDEs
describing the fluid,
\begin{equation} 
\{{\bf Y^{[x]}}\}(t_n) \text{\, and \,} \{{\bf Z^{[x]}}\}(t_n)
\xrightarrow{F} \{{\bf Y^{[x]}}\}(t_{n+1}).
\end{equation}
The $F$ operator corresponds to a classical finite volume
RANS solver and it gives the evolution in time of the
statistical moments of the fluid (the particle properties are needed
to compute source terms when two-way coupling is accounted for).

The second step (projection, operator $P$) consists in calculating
mean-particle properties and mean-fluid properties at particle
locations,
\begin{equation} 
\{{\bf Z^{[x]}}\}(t_n) \text{\, and \,} \{{\bf Y^{[x]}}\}(t_n)
\xrightarrow{P}
\{{\bf Z}^{(N)}\}(t_n) \text{\, and \,} \{{\bf Y}^{(N)}\}(t_n),
\end{equation}

Then, the stochastic differential system can be integrated in time
(operator $T$),
\begin{equation} 
\{{\bf Z}^{(N)}\}(t_n)\text{\, and \,} \{{\bf Y}^{(N)}\}(t_n)
\xrightarrow{T} \{{\bf Z}^{(N)}\}(t_{n+1}).
\end{equation}

Finally, from the new computed set of variables, at particle
locations, new statistical moments are evaluated at cell centres,
\begin{equation} 
\{{\bf Z}^{(N)}\}(t_{n+1}) \xrightarrow{A} \{{\bf Z^{[x]}}\}(t_{n+1}),
\end{equation}
and so on. The general algorithm is therefore defined by iterating the
four operators, $F \rightarrow P \rightarrow T \rightarrow A$. The
purpose of the present section, Section \ref{sec:num0}, is to discuss
different implementation aspects of the general algorithm and more
especially issues related to:
\begin{enumerate}
\item[(i)] the specificity of the treatment of the averaging and
projection operators, Section \ref{sec:PA},
\item[(ii)] the time integration of the SDEs (operator $T$),
  i.e. the determination of a suitable weak numerical scheme for
  system (\ref{eq:SDEp}) which is consistent with the multiscale
  character of the physics (asymptotic cases), Section \ref{sec:cond}.
\end{enumerate}
Before discussing these issues, some last clarifications are given on
the nature of the numerical errors generated by the particle-mesh
algorithm sketched above.

%--------------------------------------------------------
\subsubsection{Numerical errors in particle-mesh methods} \label{SS_Num_Err}
%--------------------------------------------------------
The numerical particle-mesh solution of evolution equations like
Eq.~(\ref{EqGeneric}) involves several kinds of errors. These errors
have been described in the context of PDF methods for turbulent
reactive flows~\cite{Pop_95,XuPope_99}. The overall error of the PDF
computation ($P \rightarrow T \rightarrow A$) can be separated into a
deterministic and a random part, the former involving the bias,
spatial and temporal discretisation errors. Numerical errors occur due
to:
\begin{enumerate}
\item[(i)] spatial discretisation, represented by a typical mesh size
  $\Delta x$,
\item[(ii)] finite temporal resolution, determined by the time step
  $\Delta t$, 
\item[(iii)] the use of a finite number of particles, both in the
  whole domain ($N$) and per cell ($N_{pc}$); these are further
  decomposed into the statistical error (of zero average) and the
  bias.
\end{enumerate}

The spatial discretisation error (i) is akin to the classical error in
numerical methods for solving PDEs and depends on the mesh size. In
the present approach (Mac-Kean SDEs), it is inherent
to the use of projection and averaging operators ($A$ and $P$). This
numerical error does not occur when the coefficients of the
SDEs depend only on local values of the state vector. This is the
case, for instance, in numerical computations of stand-alone PDF
methods for homogeneous turbulence~\cite{Pop_95}.
To date, only Xu \& Pope~\cite{XuPope_99}
have addressed this issue for non-linear SDEs and have found some
characteristics of the error for an infinite number of particles per
cell. Note that the spatial discretisation error occurs also in
grid-free methods, such as SPH~\cite{Wel_98}, and is, in this case,
directly related to the smoothing parameter (kernel size).

The temporal discretisation error (ii) is basically the same as in any
numerical method for solving the time evolution of the solutions to
deterministic ODEs or PDEs. Numerical
schemes for the SDEs (operator $T$) must be developed and analysed
with care due to the specificity of stochastic calculus, cf.
Section \ref{sec:cond}.

The statistical error, which is inherent to any Monte Carlo method, is
due to the use of a finite number of particles per cell (samples) to
compute the statistics and is proportional to the inverse of the
square root of $N$, according to the central limit theorem. In
specific applications (e.g., \cite{Min_99b}), the coefficient of
proportionality can be reduced  considerably when appropriate variance
reduction techniques (VRT) are applied~\cite{Pop_95,KW_86}.   

The bias is the difference between the mean value of a quantity for a
finite number $(N)$ of particles and the mean value for infinitely many
particles (all other parameters being unchanged), i.e. for any random
variable $Z$ 
\begin{equation}
  B_Z (N) = \lra{ \lra{Z}_N } - \lra{Z} \quad \text{with}\quad 
\lra{Z}_N = \frac{1}{N}\sum_{i=1}^{N}Z^i,
\end{equation}
where $Z^i$ stand for the different sample values of $Z$. The bias is
thus a deterministic error, important for non-linear stochastic
models~\cite{Pop_95,XuPope_99}. The issue is worth explaining with a
simple example. Consider a random variable $X$ with a certain law
(pdf), say standard Gaussian, i.e. $\lra{X}=0$ and
$\lra{X^2}=1$. We note $X \in {\cal N} (0,1)$. For the mean value
$\lra{X}_N$ computed out of $N$ samples, the central limit theorem
gives (for $N$ sufficiently large)
\begin{equation} \label{XMN}
  \lra{X}_N = \lra{X}_{\infty} + {C} \xi / \sqrt{N},
\end{equation}
where $\xi \in {\cal N} (0,1)$ and $C$ is a proportionality constant 
for the statistical error; obviously $\lra{X}_{\infty}=\lra{X} =
0$. Consider next a function $Y=\lra{X}^2$. Now $Y_{\infty}=0$, but
for a finite number $N$ of samples  $Y_N = {C^2} {\xi}^{2} /N$ and after
averaging $\lra{Y_N} = {C^2}/N$. The bias $B_Y (N) =
\lra{Y_N}-Y_{\infty}$ is thus proportional to $N^{-1}$. In more
general terms~\cite{Pop_95}, for $Y= g(\lra{X})$ we have  $Y_{\infty}=
g(\lra{X}_{\infty})$, and the development into a Taylor series, accounting
for Eq. (\ref{XMN}), yields 
\begin{equation}
\begin{split}
Y_{N} = g(\lra{X}_{N}) = g & \left(\lra{X}_{\infty} +
                \frac{C}{\sqrt{N}}\xi\right) \\
& = Y_{\infty} + \frac{C g'}{\sqrt{N}} \xi + \frac{C^2 g''}{2\,N} {\xi}^2 +
                {\cal O} \left( \frac{g'''}{N^{3/2}} \right) \;.
\end{split}
\end{equation}
After averaging the above, the bias is computed as
\begin{equation}
  B_Y (N) =  C^2\frac{g''}{2\,N} + {\cal O} \left( \frac{g'''}{N^{3/2}}
  \right)\;.
\end{equation}
It depends on the local second derivative of $g$ and is proportional
to $N^{-1}$. In a general case of random fields, the bias
interplays with the spatial error because of the kernel estimation
which is applied to compute averages~\cite{Pop_00,Dre_92}.

%-------------------------------------------------------------------------------
\subsection{Averaging and projection operators} \label{sec:PA}
%-------------------------------------------------------------------------------
We recall that, in the numerical solution process of
Eq.~(\ref{EqGeneric}), moments  of ${\bf Z}(t)$, like $\lra{{\bf Z}}$
and $\lra{Z_i Z_j}$ need to be extracted from the particle data. A
correct computation of these quantities is crucial for the overall
numerical solution, since the moments are put back into the SDE and
serve to advance the particle properties to the next time step of the
simulation, see Section \ref{sec:algo}. These ingredients of the
algorithm, i.e. the computation of mean fields (or averaging)
and their interpolation at (or projection to) particle locations are
well known in the PIC approach~\cite{Hoc_88,Bir_69,Oro_93}. For those
particle models (deterministic equations on a regular mesh), optimum
averaging and projection schemes have been worked out. In the present
case, a new problem is addressed: stochastic models with boundary
conditions typical of fluid mechanics.  Consequently, some new
important numerical features appear.  
\begin{enumerate}
\item[(i)] The first specificity  is related to the computation of
  statistics: in most applications, we need not only the mean values,
  but also (at least) second-order moments present in the evolution
  equations. These moments are usually position-dependent
  (non-homogeneous in space).  
\item[(ii)] A second specificity here (often present in applications
  in fluid mechanics) is that the computational domain is bounded and
  the associated mesh is non-uniform; as argued further, adequate
  boundary conditions may affect the computation of statistics.
\end{enumerate}
Here, attention is focused specifically on the errors due to the
exchange of information  between particles and the mesh, that is how
mean fields (usually the first  and second-order moments) are computed
and projected at particle locations, the main issue being to
investigate whether classical techniques already used in particle
simulations are also suitable for our present particle-mesh
problem.

%-------------------------------------------------------
\subsubsection{Averaging operators, weighting functions} \label{SS_PM_AW}
%-------------------------------------------------------
In order to introduce the numerical issues related to the exchange of
information between the particles and the mesh, let us first discuss
the difference between the ensemble mean (expected value) and the
spatial average. For the case of a deterministic function $\Phi({\bf x})$,
the spatial average $\lra{\,\cdot\,}_\Delta$ (with a characteristic
smoothing length $\Delta$) in the cell centred at ${\bf x}^{[i]}$,
$i=1,\ldots,I$, can be thought of as the integral:
\begin{equation} \label{meanw}
 \lra{{\Phi}^{[i]}}_\Delta = \int
 {\Phi}({\bf x})\,\tilde{w}({\bf x}-{\bf x}^{[i]}) \,d{\bf x}, 
\end{equation}
where ${\Phi}^{[i]}=\Phi({\bf x}^{[i]})$ and $\tilde{w}$ is a given
weighting function (smoothing kernel) satisfying $\int \tilde{w}({\bf
  x}) \,d{\bf x} =1$. For a random field $\Phi(t,{\bf x})$
with a pdf  $f_{\Phi}(t,{\bf x};{\Psi})$, the mean at ${\bf x}^{[i]}$
corresponds to the  probabilistic expectation, i.e.
\begin{equation} \label{meanp}
\lra{{\Phi}^{[i]}}_{} =
\int_{-\infty}^{+\infty} {\Psi}\,f_{\Phi}(t,{\bf x}^{[i]};{\Psi})\,d \Psi,
\end{equation}
where $\Psi$ is a sample-space variable of $\Phi$. In classical Monte
Carlo methods, the pdf $f_{\Phi}(t,{\bf x}^{[i]};{\Psi})$ at each
point ${\bf x}^{[i]}$ is approximated by using a number $N$ of
independent samples of the random variable, say ${\Phi}^{(n)},
n=1,\ldots, N$. In other words, a set of variables ${\Phi}^{(n)}$ is
attached to every particle, located at ${\bf x}^{(n)}$ in the
computational domain (NB: the superscript convention helps to
distinguish between particles ${\bf }^{(n)}$ and cell-related
quantities ${\bf }^{[i]}$).

The mathematical expectation, $\lra{\Phi^{[i]}}_{}$ in
Eq.~(\ref{meanp}), is computed {\it exactly} at ${\bf x}^{[i]}$,
whereas $\lra{\Phi^{[i]}}_\Delta$ in Eq.~(\ref{meanw}) represents the
spatial average {\it centred} on ${\bf x}^{[i]}$; both are equal in
the spatially-homogeneous case only. However, space-dependent moments
cannot be calculated exactly from a Monte Carlo estimation, given a
finite number of particles used in the simulation. In practice, under
a local homogeneity assumption, expectations $\lra{{\Phi}^{[i]}}$ are
approximated by (local) spatial averages
$\lra{\Phi^{[i]}}_{N,\Delta}$, based on a discrete particle set:
\begin{equation} \label{meand}
\lra{{\Phi}^{[i]}}_{} \simeq \lra{{\Phi}^{[i]}}_{N,\Delta} =  
\sum_{n=1}^{N} {\Phi}^{(n)} w({\bf x}^{(n)}-{\bf x}^{[i]}) \;.
\end{equation}
This expression, also known as kernel estimate~\cite{Pop_00}, is
derived  from Eq. (\ref{meanw}) as a quadrature formula; the weights $w$
are the discrete equivalents of $\tilde{w}$, $\sum_{n=1}^{N}
w({\bf x}^{(n)}-{\bf x}^{[i]}) = 1$.

A generalisation of the above discussion to centred moments of any
order is straightforward. Here, the explicit formulation is given for
the variance because of its importance in further considerations,
cf. Sections \ref{SS_CICnu} to \ref{SS_SDE}. By analogy to
Eq. (\ref{meanp}), the local value (at ${\bf x}^{[i]}$) of the
variance $\sigma_\Phi^2$ of the random field $\Phi(t,{\bf x})$ is
\begin{equation} \label{varp}
 (\sigma_\Phi^2)^{[i]} = 
 \int_{-\infty}^{+\infty} ({\Psi}-\lra{{\Phi}^{[i]}}_{})^2 \,
 f_{\Phi}({\Psi}; {\bf x}^{[i]},t)\,d \Psi \;.
\end{equation}
As an extension of formula (\ref{meand}) for the mean value, the
expression  for the variance $\sigma_\Phi^2$ with the use of spatial
averaging becomes
\begin{equation} \label{vard}
({\sigma}_\Phi^2)^{[i]} \simeq ({\sigma}_\Phi^2)^{[i]}_{N,\Delta} =
\sum_{n=1}^{N} ({\Phi}^{(n)}-\lra{\Phi})^2 w({\bf x}^{(n)}-{\bf
  x}^{[i]}),
\end{equation}
where $\lra{\Phi}$ stands either for a cell average
$\lra{{\Phi}^{[i]}}$  or is interpolated at ${\bf x}^{(n)}$. From the
algorithmic standpoint, Eq.~(\ref{vard}) has some
disadvantages. Firstly, the double pass over particles (the mean has
to be computed in advance) results in some computational
overhead. Secondly and more important, for first order weighting
functions (cf. Sections \ref{SS_CICnu} and \ref{SS_CICbc}),
it implies a risky extrapolation of mean fields at the locations
outside the computational domain.  We propose to get around the
difficulty by computing the central moments directly from the standard
(non-centred) moments; for example, the variance of a random variable
$Q$ satisfies $\lra{(Q-\lra{Q})^2}= \lra{Q^2} - \lra{Q}^2$. Yet, some
precaution is needed since such an expression for the variance can not
be guaranteed to remain always non-negative because of round-off
errors.

As far as the choice of the weighting function is concerned, two
methods are widely used and are under investigation in the following:
the NGP (Nearest-Grid-Point) and CIC (Cloud-in-Cell) methods. They
correspond to weighting functions of zero (constant) and first (linear)
order, respectively. These methods have been thoroughly
discussed~\cite{Hoc_88}, as mentioned above, for deterministic systems
solved on uniform meshes and in the particular case of unbounded
domains or bounded domains with periodic boundary conditions. In the
present work, we address the problem for the numerical solutions of
specific stochastic systems (Mac-Kean SDEs) on non-uniform meshes for
bounded domains and non-periodic boundary conditions.

In the NGP method, particle properties are
associated  with the centre of the cell containing the particle 
(for a uniform mesh, it is the grid point nearest to the particle, 
hence the name), and the weighting function is top-hat 
(or piecewise constant, Fig.~\ref{wgh_fun}a), that is
$w({\bf x}^{(n)}-{\bf x}^{[i]})=1/N_{pc}^i$ for particle $n$ in cell
$i$ and $0$ otherwise ($N_{pc}^i$ is the number of particles in cell
$i$). The NGP average is thus found from the sum over all $N_{pc}^i$
particles in a given cell $i$
\begin{equation} \label{NGPm}
\lra{{\Phi}^{[i]}}=\frac{1}{N_{pc}^i}\sum_{n=1}^{N_{pc}^i}{\Phi}^{(n)}.
\end{equation}
In the CIC method, for a uniform mesh in $D$ spatial dimensions, the
weighting function is piecewise-linear (Fig.~\ref{wgh_fun}b):
\begin{equation} \label{CICw}
  w({\bf x})= \prod_{j=1}^D \frac{1}{\Delta_j}
  \max\left\{ 1- |{\bf x}_j| /\Delta_j, 0 \right\},
\end{equation}
where $\Delta_j$ is the width of the cell in direction $j$. The
particle is thus regarded not as a single point but rather as a linear
distribution of properties: a cloud centred at ${\bf x}^{(n)}$, with
a width of $2\Delta_j$. The CIC method is less local than NGP: the
average at a given cell centre is computed not only from the particles
located in this very cell, but also from those in the neighbouring cells. 

Higher order, but less local, weighting functions can be used, for
instance a piecewise-quadratic polynomial (Fig.~\ref{wgh_fun}c) or
a cubic spline as in Smoothed Particle Hydrodynamics
(SPH)~\cite{Mon_92}; quartic and quintic splines as well as Gaussian
kernels  are also quite popular~\cite{Wel_98,Mor_97}.
An alternative to kernel estimators, cf. Eq.\ (\ref{meand})
are somewhat more costly (especially in 2D/3D) formulae with
least-squares or local least-squares approximations, as well as
cross-validated splines~\cite{Pop_85,Dre_92}.

The projection of averaged (cell) values onto the particle  locations
is an interpolation procedure, in a sense akin to averaging,
Eq.~(\ref{meand}), with $^{(n)}$ and $^{[i]}$ replaced by each other
(it is the reverse operation). The consistency of averaging and
projection steps has been a serious concern in PIC
applications. Indeed, it has been demonstrated (see \cite{Hoc_88},
\S 5.2.4) that, in the case of a system of charged particles moving in an
electric field  generated by themselves, unphysical forces may appear
if the averaging scheme is not of order equal to (acceptably also
higher than) the projection scheme. Those authors have also stated that
the CIC method performs better than NGP. 

%----------------------------------------------------
\subsubsection{CIC statistics on a non-uniform mesh}  \label{SS_CICnu}
%----------------------------------------------------
Although the usual presentation of CIC formulae is made for a uniform
mesh, in typical applications of the present mean-field/PDF approach
for polydispersed turbulent two-phase flows, a non-uniform mesh may be
of advantage (wall-bounded flows). A number of difficulties arise in this
generalisation, depending whether the mean density computed from the
particle masses or the mean of a variable attached to the particles is
sought. For the sake of simplicity, we will illustrate the issue in a
1D setting; extension to 2D and 3D is straightforward through the
Cartesian products.

First, consider the computation of fluid density. A particle located
at $x^{(n)}$ where $x^{[i]} \le x^{(n)} < x^{[i+1]}$,
cf.~Fig.~\ref{pm_CIC}(a), will contribute to the mean density
at $x^{[i]}$ and $x^{[i+1]}$. Here, we treat the particle as a
``cloud'' (or a linear distribution of mass) centred at $x^{(n)}$ and
stretched on the interval $\delta x_{i}= \min (\Delta x_{i},\Delta
x_{i+1})$. With this assumption, the particle mass is contained
between $x_0^{[i-1]}$ and $x_0^{[i+1]}$ and does not contribute to
other cell averages (three  possible locations of a ``stretched''
particle are shown in Fig.~\ref{pm_CIC}a). The fraction of the cloud
that belongs to the cell $[i+1]$, $R_x=(x^{(n)}+ \delta
x_{i}/2-x_0^{[i]})/ \delta x_i$, adds to the average at $x^{[i+1]}$,
whereas $(1-R_x)$ adds to the cell average at $x^{[i]}$.
For a boundary cell, the whole mass of particles located close to the
boundary  (between the boundary and the centre of the cell closest to
it) is attributed  to the centre of the cell next to the
boundary. Fig.\ \ref{dens_un_nu} presents 2D computation results of
the r.m.s.\ density (r.m.s.\ deviation from the mean particle density)
a 10$*$10 
mesh, both uniform and non-uniform, using the standard NGP averaging
and the CIC method described above. For the uniform mesh, particle
locations in the domain are generated randomly from the uniform random
distribution.  The non-uniform mesh has been generated using random
numbers from  a uniform distribution on the interval $(0.5\Delta x,
1.5\Delta x)$; in this case, particle locations have been generated
deterministically with a constant number density. To reduce the
statistical error of the r.m.s. density deviation,  it has been computed as an average
of four different runs with different seeding values for pseudorandom
number generator. As expected, the statistical error is higher for the
NGP average and varies as $N_{pc}^{-1/2}$. We note that although the
r.m.s.\ density on the non-uniform mesh is higher, there is no
systematic error for the CIC computation which confirms the
correctness of the method above.

Next, consider the computation of cell-averaged values for any
variable ${\Phi}$ assigned to the particles (a typical example is the
calculation of the mean velocity at a given point from the set of
particle instantaneous velocities). In this case, both the particle
masses and the values of ${\Phi}$ attached to the particles come into
play (it has been mentioned in Appendix \ref{sec:appen} that, in the
frame of the present formalism, one has to resort to mass-weighted
expected values). The first idea could be to use the method described
above for the calculation of the mean density. However, in a basic
test of a linear deterministic function and particles distributed
uniformly on the interval, this method is readily shown not to
retrieve given values of particle variables at cell centres for a
non-uniform grid. Fig.~\ref{pm_CIC}(b) illustrates this point where
cell $i$ is half the size of its neighbours. To compute the average
value of $\Phi$ at $x^{[i+1]}$, the particles labelled 2, 3 and 4
contribute with the respective weights of $1$, $3/4$, and $1/4$,
\begin{equation}
2\,\lra{\Phi^{[i+1]}} = 1 ({\Phi}_0+\Delta\Phi)
+ \frac{3}{4} ({\Phi}_0 - \Delta \Phi)
+ \frac{1}{4} ({\Phi}_0 -3\Delta \Phi).
\end{equation}
For this simple
example, it is readily checked that $\lra{\Phi^{[i+1]}}$ depends on
$\Delta \Phi$ (it should not) and that $\Phi^{[i+1]}\ne {\Phi}_0$. The
reason for this behaviour can be traced into the formula of particle
weights which is not symmetric with respect to $x^{[i+1]}$.
Indeed, the same behaviour can be noticed when non-symmetric CIC weight 
functions are used though they seem to be a natural generalisation of 
Eq.~(\ref{CICw}). This point is also illustrated in Fig.~\ref{pm_CIC}(b)
that includes one such basis function $w(x)$ decreasing linearly
from a maximum at the considered cell centre, here $x^{[i+1]}$, to zero 
at the neighbouring cell centres, here $x^{[i]}$ and $x^{[i+2]}$. 
Using these weighting functions, we readily find again
\begin{equation}
\frac{5}{3}\,\lra{\Phi^{[i+1]}} = \frac{2}{3}({\Phi}_0+\Delta\Phi)
+ \frac{3}{4} ({\Phi}_0 - \Delta \Phi)
+ \frac{1}{4} ({\Phi}_0 -3\Delta \Phi),
\end{equation}
that is $\lra{\Phi^{[i+1]}}\ne {\Phi}_0$, a result that would
explicitly depend  upon the gradient of the function and the size of
the mesh. Instead, we propose the following method that works
correctly for particle variables on a non-uniform mesh. We generalise
Eq.~(\ref{CICw}) by taking $\Delta$ as $\min \{
\Delta x_{i-1}, \Delta x_{i} \}$ to preserve the symmetry of $w(x)$.
This expression gives the correct result for the example of
Fig.~\ref{pm_CIC}(b). An alternative to the above is the use of a
two-stage algorithm~\cite{Dre_92}, satisfactory but arguably more
time-consuming. Again, a 2D computation has been performed for the
r.m.s.\ of a variable assigned to particles from a deterministic
linear profile in space with the same procedure as above for particle
density computations. Both the standard NGP averaging and the CIC
method described above have been used.  Results shown in Fig.\
\ref{ldf_un_nu} confirm the advantages of CIC averaging in this
case. To continue, let us now remove the deterministic assumption for
$\Phi$ and consider averaging of a random variable assigned to
particles
\begin{equation} \label{lrf}
  \Phi^{(n)} = \Phi (x^{(n)}) =  m(x^{(n)}) + s(x^{(n)}) \xi \;,
\end{equation}
where $\xi \in {\cal N} (0,1)$, $m(x)=a_0 +a_1 x$ is a linear function
and  $s(x)=\Delta$ (where $\Delta$ denotes the average mesh size) to study the
effect of the spurious variance resulting from the NGP averaging. A
numerical test has been performed again with the same methodology as
described above. Fig.~\ref{lrf_un_nu_s1} shows the normalised r.m.s.\
of this linear random function (i.e.\ the square root of its
variance computed at the cell centres divided by the prescribed
r.m.s.) computed out of particle locations. A systematic error is
readily noticed for NGP statistics, unless $a_1\Delta \ll {\cal
  O}(1)$.

%------------------------------------------------------
\subsubsection{CIC statistics with boundary conditions} \label{SS_CICbc}
%------------------------------------------------------
In wall-bounded flow applications, the CIC method has to be modified
so that suitable boundary conditions (BC) are properly accounted
for. Dreeben and Pope~\cite{Dre_92} describe their application of CIC
statistics in the PDF method but the treatment of flow boundaries is
not reported there. In Fig.~\ref{pm_1d}, a schematic plot shows
clearly why the CIC averaging requires some care in presence of flow
boundaries (here walls). The NGP statistics are local (in a cell) and
they can thus be computed in border cells without any change. However,
for the CIC method this is not the case. As it transpires from
Fig.~\ref{pm_1d}, the CIC average of the particle density in a border
cell, computed with the weight $w_1(x)$, incorrectly gives a smaller
value than the density in the neighbouring internal cell, computed
with $w_2(x)$. The obvious reason is that less particles contribute to
the mass in the border cell. The problem differs depending upon
whether we consider the particle density field or the mean of any
variable ${\Phi}$ (such as velocity) assigned to particles. To compute
the mean density, for the particles located between the cell centre
and the boundary, it is sufficient to assign all their mass to the
centre of this boundary cell.  For other particle variables, this
treatment is not applicable: even for a linear function $\Phi(x)$, the
higher spatial accuracy (compared to NGP) of CIC is lost.

A working remedy to this situation is the addition of ``ghost'' or
``mirror'' particles outside the actual computational domain in order
to compensate for the incorrect CIC computation at the cell centres
close to the boundaries. The idea of these ``ghost'' particles is
known in the context of the SPH simulation~\cite{Mor_97}. The very
presence of ghost particles with masses equal to their ``hosts''
allows for a correct CIC density computation in boundary cells.  Next,
to compute CIC statistics of any variable $\Phi$ attached  to the
particles, values $\Phi'$ of the variables are to be determined  also
for the mirror ones, as they enter Eq.~(\ref{meand}) with a
corresponding weight. The procedure is relatively straightforward for
the CIC averaging of given functions (either deterministic or random)
where the value of the function at a mirror particle location is
known. However, the main interest for using ghost particles is in
actual particle simulations where precisely these functions are
unknown. The values of the variables attached to the mirror particles
are to be determined directly from those of the host particles through
the application of relevant BCs. For example, if $\Phi$ stands for a
particle velocity component, then the no-slip, impermeable wall
implies $\Phi'=-\Phi$ (Fig.~\ref{pm_1d}) so that at the boundary
$\lra{\Phi}=0$.  Generalisation is possible for more complex, yet
still block-structured, 2D and 3D geometries. Other types of possible
boundary conditions in the PDF method for turbulent flow computations,
e.g. where particle boundary conditions are determined from
a physical reasoning for the near-wall region~\cite{Min_99b}, are also
readily implemented this way.

%---------------------------------------------
\subsubsection{Test case: space-dependent SDE} \label{SS_SDE}
%---------------------------------------------
Let us now analyse the behaviour of a simple but nevertheless
realistic example of the generic SDE, Eq. (\ref{EqGeneric}), where
emphasis is put on the averaging and projection operators. Both the
NGP and CIC techniques are going to be used with the suitable
modifications presented in Sections \ref{SS_CICnu} and
\ref{SS_CICbc}. We consider a one-dimensional Ornstein-Uhlenbeck
process on $0\le x \le 1$ interval
\begin{equation} \label{EqOUx}
d\Phi(t) = - \frac{\Phi(t) - m(x)}{T}\,dt +
\sqrt{\frac{2\,s^2(x)}{T}}\,dW(t),
\end{equation} 
with a fixed timescale $T$; the space-dependence of the mean value
$m(x)$ and the variance $s^2(x)$ is essential to study the spatial
discretisation error, cf.\ Section~\ref{SS_Num_Err}. In
Eq. (\ref{EqOUx}), there is no physical coupling in $x$; it would
occur if convection were added (e.g. $dx=f(\Phi)\,dt$) or if
non-local operators were present (e.g. $\alpha \nabla^2
\lra{\Phi}dt$ terms on the RHS). Here, only a ``numerical'' coupling
exists; it is due to the approximate computation of ensemble averages
through spatial averages, Eq.~(\ref{meand}). 

For a given time step, $\Delta t = t_{l+1} - t_l$, trajectories of the
stochastic process (\ref{EqOUx}) can be integrated  analytically
\cite{Gar_90,Klo_92}, thus avoiding any temporal discretisation error;
this yields (the location $x$ is a parameter)
\begin{equation} \label{EqOU_dt}
\begin{split}
{\Phi}_{l+1} =  {\Phi}_{l}\exp (-\frac{\Delta t}{T} )
+ m(x)&\left[ 1 - \exp (-\frac{\Delta t}{T} )\right] \\ 
& + \sqrt{s^2 (x) \left[1-\exp(-\frac{2\Delta t}{T})\right]}\;\xi,
\end{split}
\end{equation} 
where $\xi$ is a standard Gaussian random variable, $\xi\in{\cal
  N}(0,1)$.

The actual test case consists in using the computed values of 
$\lra{{\Phi}}$ and $\sigma_\Phi^2$ in Eq.~(\ref{EqOU_dt}) in place
of $m(x)$ and $s^2(x)$, respectively. The computed profiles (at
different times) for a quadratic mean $m(x)$ and a constant
variance $s^2(x)$ are presented in
Fig.~\ref{SDE_m_v_CIC_NGP}. Qualitatively, a typical behaviour of CIC
averaging is to modify the shape of the mean $\lra{\Phi}(t,x)$
(ultimately towards a linear profile) and to produce an increase of
the variance in the centre of the computational interval. On the
contrary, the NGP mean value remains basically unchanged, while the
variances in separate cells become increasingly uncorrelated in
time. Indeed, in the CIC method, neighbouring cells are linked through
the averaging procedure, thus the profile of the variance stays
relatively smooth. The NGP computation of Eq.~(\ref{EqOUx}) is local
and as a result, separate cells become independent from one
another, cf. the upper right plot in
Fig.~\ref{SDE_m_v_CIC_NGP}.  These numerical outcomes raise the
question of the existence of a bias (cf.\
Section~\ref{SS_Num_Err}).  A detailed numerical study of this issue
has been performed by analysing the temporal behaviour of the r.m.s.\
of $\Phi$ averaged over all cells, denoted by $\sigma_{\Phi}(t)$ and
given by
\begin{equation}
\sigma_{\Phi}(t)=\frac{1}{I}\sum_{i=1}^I \sigma^{[i]}(t),
\end{equation}
where $\sigma^{[i]}(t)$ are the r.m.s. values obtained in cell
$i$. Indeed, for the NGP method, calculations done in each cell are
uncorrelated and $\sigma^{[i]}(t), i=1,\ldots, I,$ can be regarded as
independent samples. The simplest case to be discussed now is that of
constant mean $m(x)$ and variance $s^2(x)$ in Eq.~(\ref{EqOUx}).
Fig.~\ref{biasM2V0}(a) shows how $\sigma_{\Phi}(t)$ normalised by its
initial value, evolves in time. The procedure, in the NGP
case, is sensitive to the number of particles per cell. In subsequent
time steps, the recomputed variance decreases. The decrease is faster
for a smaller number of particles per cell. This phenomenon is
identified to be related to the bias and is due to the finite value of
$N_{pc}$. Fig.~\ref{biasM2V0}(b) clearly indicates that there is a
bias; however, computations of the variance of ${\sigma}_{\Phi}(t)$
(results presented as error bars in the same figure) show that
the statistical fluctuations of this quantity are significant. With
this precaution in mind, the bias (resulting  from several different
runs for several number of cells, etc.) has been plotted as a function
of the number of particles per cell $N_{pc}$. Results in
Fig.~\ref{biasM2V0_2}(a) indicate that the slope for different runs is
indeed close to the theoretical prediction of the bias, $B_\Phi \sim
N_{pc}^{-1}$ (plotted as a dotted line). The computed probability
distributions of ${\sigma}_{\Phi}$ are shown in
Fig.~\ref{biasM2V0_2}(b).

%-----------------------------------------------------
\subsubsection{NGP or CIC in practical computations?}
%-----------------------------------------------------
In Fig.~\ref{SDE_m_v_CIC_NGP}, the centred second-order moment,
i.e.\ the variance, has been computed using the non-centred
moments in order to avoid problems with the correct statement of
boundary conditions which are necessary for CIC averages. It is
readily seen that the computed profiles of $\lra{\Phi}(x)$ and
$\sigma^2(x)$ tend to become linear (both are fixed to the respective
$m(x)$ and $s(x)$ values at the boundaries $x\!=\!0$ and
$x\!=\!1$). In Fig. \ref{biasM2V0_2}, it is seen that the NGP
computation generates a non-zero spurious variance.

The comparison between the NGP and CIC methods for stochastic
processes reveals that none of the results are entirely
satisfactory. Compared to classical deterministic particle systems,
where the CIC has a known advantage over NGP, new features have been
revealed. NGP is a local and robust method which does not require
specific developments. It respects given mean profiles but leads to
spurious variances. The spurious variances are shown to be related to
a bias which decreases linearly with the number of
particles. On the contrary, when using CIC, developments are needed to
account for non-uniform meshes and for boundary conditions. It is a
less local (higher order) method than NGP but it is more complicated to
implement in the general case. The results obtained in a prototypical
SDE, cf. Section \ref{SS_SDE}, show that the CIC method does not
present advantages in accuracy. Furthermore, in the particular case
considered (no convection), the CIC technique
performs globally worse than the NGP one. Indeed, CIC does not
preserve the mean and shows a systematic error in the variance.

As a consequence, all practical computations performed in the rest of
the present paper (cf. Section \ref{sec:exemples}) will be based upon
the NGP technique.

%-------------------------------------------------------------------------------
\subsection{Accurate schemes for SDE integration}\label{sec:cond}
%-------------------------------------------------------------------------------
It has now been explained how information is exchanged between the
discrete particles and the mesh (operators $P$ and $A$, cf. Section
\ref{sec:algo}).
%modif EP (début)
\ifnum\elscolor>0 \color{blue} \fi
\ifnum\elsitalic>0 \itshape \fi
We recall that this is done here in the case of one-way coupling and
the extension of the numerical schemes, presented in this Subsection, to
the case of two-way coupling will be discussed in Section \ref{sec:twc}.

As shown in Sections \ref{sec:pdf_p} and \ref{sec:model}, and as
explained at the beginning this Section, cf. Eq. (\ref{EqGeneric}),
the SDEs reproducing the dynamics  of the discrete particles are
Mac-Kean SDEs since the coefficients depend not only on the state vector
but also on expected values of functions of ${\bf Z}(t)$. 
\normalfont
\color{black}
%modif EP (fin)
In particle-mesh methods, as
explained above, quantities such as $\lra{f({\bf Z})}$ (where $f$ is a
linear or quadratic function of ${\bf Z}(t)$ in our problem) are
extracted from the particle data and
evaluated at grid points. In Section \ref{sec:PA}, the difficulties
inherent to the projection and averaging procedures have been detailed.

Attention is now focused on the time integration of
the set of SDEs (operator $T$).
The development of a suitable weak numerical scheme for the time
integration of SDEs is a much more difficult task than the
corresponding one for ODEs. Indeed,
SDEs do not obey the rules of classical differential calculus, see
Section \ref{sec:intsto}, and one has to rely on the theory of
stochastic processes \cite{Kle_99}. In that sense, particular
attention has to be payed to the problem of consistency between
discretised equations and the original continuous set of SDEs. It is
recalled that, in the present paper, It\^o's calculus is adopted and
therefore all SDEs are written in the {\em It\^o' sense}, see Section
\ref{sec:intsto}.
%modif EP (début)
\ifnum\elscolor>0 \color{blue} \fi
\ifnum\elsitalic>0 \itshape \fi
This choice has no physical motivation: It\^o's calculus is very
convenient in the development of weak numerical schemes for SDEs
because of the zero mean and isometry properties, cf. Eqs
(\ref{eq:iso_prop}).
\normalfont
\color{black}
%modif EP (fin)

An essential preliminary is to clearly frame the development of the
weak numerical scheme in the general methodology that has been
followed here. We propose, indeed, to describe the problem of
turbulent polydispersed two-phase flows within an engineering context
but with a rigorous treatment of the multiscale character which is a
distinctive feature of these flows, cf. Section \ref{sec:scale}. The
model, Eqs. (\ref{eq:sysEDS}),
contains several characteristic timescales and this system of SDEs
becomes stiff, from a mathematical point of view, whenever one of
these timescales goes to zero. In those cases, various limit systems
are obtained, see Section \ref {sec:scale}, which represent the
asymptotic limits of the physical model. A proper treatment of the
physics of the multiscale aspect imposes to put forward weak
numerical schemes which are consistent with all asymptotic limits when
the different timescales go to zero.

It is worth emphasising that this corresponds to a practical
concern. Indeed, in the numerical simulation of a complex flow, the
timescales may be negligible (much smaller than the time step) in some
areas of the computational domain. For example, in a wall-bounded
flow, the integral timescale of the fluid velocity seen, $T_{L,i}^{*}$,
goes to zero when the distance to the wall decreases. Furthermore,
when dealing with polydispersed particles, or with phenomena involving
evaporation or combustion (when particle diameters decrease in time),
one has often to handle a whole range of particle diameters (say from
$1 \mu m$ to $100-200 \mu m$) and thus a whole range of values for
$\tau_p$ (for example, $10^{-6} {\rm s} \lesssim \tau_p \lesssim
10^{-2} {\rm s}$). It
would be inefficient to carry out computations with a time
step limited by the smallest possible value of $\tau_p$ and/or
$T_{L,i}^{*}$. This is the very reason why in the simulations of
particle dispersion in wall-bounded flows, based on discrete models,
it is necessary  to use different time steps for different classes of
diameters (and thus of $\tau_p$) and to lower the time step in the
wall region.

As a consequence of the above discussion, the constraints which are
required for a suitable weak numerical scheme, can now be summarised,
considering both physical and numerical issues.
Since a particle-mesh method is adopted here, the PDEs for the fluid
are first solved and then the dynamics of the stochastic particles are
computed (see Section \ref{sec:algo}), thus, the scheme has to be 
{\em explicit} 
%modif EP (début)
\ifnum\elscolor>0 \color{blue} \fi
\ifnum\elsitalic>0 \itshape \fi
for the fluid mean fields. By choice, the scheme will also be explicit
for the particle properties. 
\normalfont
\color{black}
%modif EP (fin)
Furthermore, the time step, which has to be the same for
the integration of the PDEs and the SDEs, is imposed by stability
conditions required by the finite volume algorithm solving the
mean-field equations for the fluid. This implies that, since there is
no possibility to control the time step when integrating the SDEs, the
numerical scheme has to be {\em unconditionally stable}. 
At last, since particle localisation on a mesh (needed for projection
and averaging, see Section \ref{sec:num}) is CPU-demanding, the
numerical scheme should minimise these operations.

The constraints required for a suitable weak numerical scheme are:
\begin{enumerate} 
\item[(i)]  {\em the numerical scheme must be explicit, stable and the
    number of calls to particle localisation (sub)routine has to be minimum},
\item[(ii)] {\em the numerical scheme must be consistent with all
    limit systems}.
\end{enumerate}

%-------------------------------------------------------------
\subsubsection{Weak numerical schemes for SDEs} \label{num-sch}
%-------------------------------------------------------------
In Section \ref{sec:formalism}, the correspondence (in a weak sense)
between a set of SDEs and a Fokker-Planck equation (for the associated
law) has been established. In this work, weak numerical schemes shall
be developed for Eqs. (\ref{eq:sysEDS}), i.e. we are
not interested in the exact trajectories of the process but instead in
statistics (the pdf) extracted from the stochastic particles (the real
particles are replaced by stochastic ones which should reproduce the
same statistics). The numerical method proposed in this work is
therefore nothing else than the simulation of an underlying pdf, or in
other words, the equivalent Fokker-Planck equation is solved by
simulating the trajectories of stochastic particles, that is by a
dynamical Monte Carlo method. As briefly
explained in Section \ref{sec:stat}, this non-trivial
numerical procedure, i.e. to resort to SDEs to solve a PDE,
is well suited for PDEs with a large number of degrees of freedom.

Since the It\^o interpretation of stochastic integrals has been
chosen, it is {\em implicitly} assumed, in the discretisation of
the stochastic integral, that $B_{ij}$ should not anticipate the
future, i.e. for each time step $\Delta t=t_{k+1}-t_k$,
$B_{ij}$ should be computed at $t=t_k$. As a result, classical numerical
schemes for ODEs, for example Runge-Kutta schemes, can not be applied
directly. More precisely, careless applications of such schemes for
SDEs can introduce spurious drifts which may not be easy to detect. An
illuminating example of this kind of error  is illustrated in
Ref. \cite{Min_03a}. The key point here is that the numerical
discretisation of the stochastic integral must be in line with its
mathematical definition. 

Let ${\bf Z}^{\Delta t}(t)$ be a numerical approximation of ${\bf Z}(t)$
obtained with a uniform time discretisation, $\Delta t$. A numerical
scheme of order $m$ will converge, in a weak sense, if at time $t=T$
($T$ is called the stopping time), for all sufficiently smooth
functions $f$, there exists a constant $C$ (function of $T$) such that
\begin{equation} \label{eq:order}
\sup_{t \leq T}
\vert \lra{ f[{\bf Z}(t)]-f[{\bf Z}^{\Delta t}(t)] } \vert
\leq C(T) \,(\Delta t)^m.
\end{equation}
Other convergence modes are possible, for example {\em strong
convergence} in the mean-square sense, if one is interested in the
exact trajectories of the process. It is fairly rare that this is the
case for engineering problems. Indeed, in most engineering
applications, one is mainly interested in the expected values
(statistics) of functionals of the variables of interest. For further
discussion on the convergence modes, see the book of Kloeden \& Platen
\cite{Klo_92}. 

%---------------------------------------------------------
\subsubsection{Analytical solution to the system of SDEs} 
%---------------------------------------------------------
In the present work, the weak numerical schemes, with the required
features, are developed based on the analytical solution to Eqs.
(\ref{eq:sysEDS}) {\it with constant coefficients} (independent of time),
the main idea being to derive a numerical scheme by freezing the
coefficients on the integration intervals. This methodology ensures
\textit{stability} and \textit{consistency with all limit systems}:
\begin{enumerate}
\item[-] stability because the form of the equations gives analytical
  solutions with exponentials of the type $\exp(-\Delta t/T)$ where
  $T$ is one of the characteristic timescales ($\tau_p$ and
  $T_{L,i}^*$),
\item[-] consistency with all limit systems by construction, since the
  schemes are based on an analytical solution.
\end{enumerate}
Different techniques shall be used to derive first and second-order
(in time) schemes from the analytical solutions with constant
coefficients. A first-order scheme can be obtained by computing, at
each time step, the variables on the basis of the analytical solutions
(all coefficients are frozen at the beginning of the integration
interval), i.e. a numerical scheme of the {\it Euler} kind is
obtained. A second-order scheme can be derived by resorting to a
predictor-corrector technique where the prediction step is the
first-order scheme.

Before presenting the weak numerical schemes, it is a prerequisite to
give the analytical solutions to system (\ref{eq:sysEDS}), with
constant coefficients (in time). These solutions are
obtained by resorting to It\^o's calculus in combination with the
method of the variation of the constant. For instance, for the fluid
velocity seen, one seeks a solution of the form
$U_{s,i}(t)=H_i(t)\exp(-t/T_i)$, where $H_i(t)$ is a stochastic process
defined by (\textit{from now on the notation is slightly changed:
  $T_{L,i}^*$ is noted $T_i$ for the sake of clarity in the complex
  formulae to come})
%modif EP (début)
\ifnum\elscolor>0 \color{blue} \fi
\ifnum\elsitalic>0 \itshape \fi
\begin{equation} \label{eq:H}
dH_i(t)=\exp(t/T_i)[C_i\,dt + \Check{B}_i\,dW_i(t)],
\end{equation}
that is, by integration on a time interval $[t_0,t]$ ($\Delta t=t-t_0$),
\begin{equation} 
\begin{split}
U_{s,i}(t) = U_{s,i}(t_0)& \exp(-\Delta t/T_i)+ C_i 
             \,T_i\,[1-\exp(-\Delta t/T_i)] \\
+ & \Check{B}_i \exp(-t/T_i)
    \int_{t_0}^{t}\exp(s/T_i)\,dW_i(s), 
\end{split}
\end{equation}
where $\Check{B}_i=B_{ii}$ since $B_{ij}$ is a diagonal matrix,
cf. Eq. (\ref{eq:dUs}).
By proceeding in the same way for the other equations (position and
velocity), the analytical solution is obtained for the entire system,
cf. Table \ref{tab:exa}. The three stochastic integrals, Eqs.
(\ref{eq:gamma_exa}) to (\ref{eq:Omega_exa}) in Table \ref{tab:exa},
are centred Gaussian processes (they are stochastic integrals of
deterministic functions \cite{Kle_99}). These integrals are defined
implicitly, but they can be simplified by integration by parts,
cf. Table \ref{tab:exa}. As explained in Section \ref{sec:fast}, for
the numerical representation of the stochastic integrals, the
knowledge of the covariance matrix (second-order moments) is needed,
see Table \ref{tab:matcov_exa}. Using the isometry property, see
Section \ref{sec:intsto}, the second-order moments, i.e. Eqs.
(\ref{eq:gama2}) to (\ref{eq:GamOme}) in Table \ref{tab:matcov_exa},
can be calculated.
\normalfont
\color{black}
%modif EP (début)
The analytical solutions are now known. Before presenting the
first order scheme, let us verify that the analytical solution given
by Tables \ref{tab:exa} and \ref{tab:matcov_exa} is consistent with
the limit cases obtained in Section \ref{sec:limcases}, i.e. Eqs.
(\ref{eq:limc1}) to (\ref{eq:limc4}).

\vspace*{2mm} \underline{\textbf{Limits systems of analytical
    solution:}} in 
limit case 1, where the discrete particles behave as fluid particles,
the limit system is given by Eq. (\ref{eq:limc1}). When $\tau_p
\rightarrow 0$, Eq. (\ref{eq:Upa_exa}) becomes
\begin{equation} \label{eq:Upcasi}
U_{p,i}(t)= U_{s,i}(t_0)\exp(-\Delta t/T_i) + C_i\,T_i\,\exp(-\Delta
t/T_i) + \Gamma_i(t),
\end{equation}
and for the stochastic integral $\Gamma_i(t)$, one has
\begin{equation} 
\left\{\begin{split}
& \lra{\Gamma_i^2(t)} \xrightarrow[\tau_p \to 0]{}
\frac{\Check{B}_i^2\,T_i}{2}[1-\exp(-2\Delta t/T_i)] =
\lra{\gamma_i^2(t)},\\
& \lra{\Gamma_i(t)\,\gamma_i(t)} \xrightarrow[\tau_p \to 0]{}
\lra{\gamma_i^2(t)}.
\end{split}\right.
\end{equation}
The last two equations indicate that $\Gamma_i(t) \to \gamma_i(t)$
when $\tau_p \to 0$. By comparing Eq. (\ref{eq:Upcasi}) to
Eq. (\ref{eq:Ufa_exa}) with $\Gamma_i(t)=\gamma_i(t)$, the results of
Eq. (\ref{eq:limc1}) are retrieved, i.e. ${\bf U}_p(t)={\bf U}_s(t)$.

In limit case 2, the fluid velocity seen, ${\bf U}_s(t)$, is a fast
variable which is eliminated. The results obtained in Table
\ref{tab:exa} and \ref{tab:matcov_exa} with $T_i \to 0$ and
$\Check{B}_i\,T_i={\rm cst}$, give 
\begin{equation} \label{eq:Up_casii}
\begin{split}
U_{p,i}(t) = U_{p,i}(t_0)\exp(-\Delta t/\tau_p) +
[& \lra{U_i}+\mc{A}_i\,\tau_p] [1-\exp(-\Delta t/\tau_p)] \\ +
& \sqrt{\frac{\Check{B}_i^2\,T_i^2}{2\tau_p}
[1-\exp(-2\Delta t/\tau_p)]}\;\;{\mc G}_{p,i},
\end{split}
\end{equation}
where ${\mc G}_{p,i}$ is a ${\mc N}(0,1)$ vector (composed of
independent standard Gaussian random variables) and we recall that 
$\lra{U_i}=\lra{U_i(t,{\bf x}_p(t)}$. It can be rapidly
verified, by applying It\^o's calculus, that Eq. (\ref{eq:Up_casii})
is the solution to system (\ref{eq:limc2}) when the coefficients are
constant.

In limit case 3, both the fluid velocity seen and the velocity of the
discrete particles become rapid variables. When $\tau_p \to 0$ and
$T_i \to 0$ with $\Check{B}_i\,T_i={\rm cst}$, Eq. (\ref{eq:xpa_exa}) becomes
\begin{equation} \label{eq:xp_casiii}
x_{p,i}(t) = x_{p,i}(t_0) + [\lra{U_i}+\mc{A}_i\,\tau_p]\,\Delta t
+ \sqrt{\Check{B}_i^2\,T_i^2\,\Delta t} \;\;{\mc G}_{x,i},
\end{equation}
which is the solution to Eq. (\ref{eq:limc3}) when the coefficients
are constant (${\mc G}_{x,i}$ is a ${\mc N}(0,1)$ vector).

In limit case 4, when $T_i\rightarrow 0$ (with no condition on $\Check{B}_i
T_i$) the system becomes deterministic, the results are in agreement
with Eq. (\ref{eq:limc4}). When $T_i\rightarrow 0$, Eqs.
(\ref{eq:xpa_exa}) to (\ref{eq:Ufa_exa}) become
\begin{equation} \label{eq:xU_casiv}
\left\{\begin{split}
& U_{s,i}(t)=\lra{U_i}, \\
& U_{p,i}(t) = U_{p,i}(t_0)\exp(-\Delta t/\tau_p)
  + [\lra{U_i}+\mc{A}_i\,\tau_p][1-\exp(-\Delta t/\tau_p)],\\
& x_{p,i}(t) = x_{p,i}(t_0) + \tau_p[1-\exp(-\Delta t/\tau_p)]
U_{p,i}(t_0) \\
& \qquad \qquad + [\lra{U_i}+\mc{A}_i\,\tau_p]\{\Delta t -
\tau_p[1-\exp(-\Delta t/\tau_p)]\},
\end{split}\right.
\end{equation}
which is the analytical solution to system (\ref{eq:limc4}) when the
coefficients are constant. 

%-------------------------------------------------------
\subsubsection{Weak first order scheme} \label{sec:sch1}
%-------------------------------------------------------
The derivation of the weak first order scheme is now rather
straightforward since the analytical solutions to system
(\ref{eq:sysEDS}) 
with constant coefficients have been already calculated. Indeed, the
Euler scheme (which is a weak scheme of order $1$ \cite{Klo_92}) is
simply obtained by freezing the coefficients at the beginning of the
time intervals $\Delta t = [t_n,t_{n+1}]$. Let $Z_i^n$ and $Z_i^{n+1}$
be the approximated values of $Z_i(t)$ at time $t_n$ and $t_{n+1}$,
respectively. The Euler scheme is then simply written by using the
results of Tables \ref{tab:exa} and \ref{tab:matcov_exa} as shown in
Table \ref{tab:sch1}. Before showing that the scheme is consistent
with all limit cases, some clarifications must be given. Here, the
limit systems are considered in the {\em discrete sense}. The
observation timescale $dt$ has now become the time step $\Delta
t$. The timescales $\tau_p$ and $T_i$ do not go to zero, as in the
continuous sense (Section \ref{sec:scale}), but their values,
depending on the history of the particles, can be smaller or greater
than $\Delta t$. The continuous limits, i.e. Eqs.
(\ref{eq:limc1}) to (\ref{eq:limc4}), represent a mathematical limit,
whereas in the discrete formulation, as we shall see just below, the limit
systems correspond to a numerical solution where the ratios $\Delta
t/T_i$ and $\Delta t/\tau_p$ become large (the limit systems are
obtained by Taylor expansions).

In limit case 1, when $\tau_p \to 0$ in the continuous sense and
$\tau_p \ll \Delta t \ll T_i$ in the discrete sense, the numerical
scheme gives $U_{p,i}^{n+1}=U_{s,i}^{n+1}$, see Table \ref{tab:sch1},
which is consistent with the results of Section \ref{sec:limcases}.

In limit case 2, in the continuous sense $T_i \to 0$ and $\Check{B}_i T_i
={\rm cst}$, that is the fluid velocity seen ${\bf U}_s(t)$ becomes a fast
variable which is eliminated. In the discrete case, ${\bf U}_s(t)$ is
simply observed at a timescale which is great compared to its memory, 
that is $T_i \ll \Delta t \ll \tau_p$, and the numerical scheme yields
(see Table \ref{tab:sch1})
\begin{equation}
U_{s,i}^{n+1}=\lra{U_i^n}
+ \sqrt{\frac{[\Check{B}_i^n]^2 T_i^n}{2}}\,{\mc G}_{1,i},
\end{equation}
where $\lra{U_i^n}=\lra{U_i(t_n,{\bf x}_p^n)}$.
The fluid velocity seen becomes a Gaussian random variable, a result
which is physically sound since ${\bf U}_s(t)$ is observed at time steps
which are greater than its memory. This result is in line with that
of the model problem presented in Section \ref{sec:fast}.
Furthermore, by Taylor expansion,
it can be shown that the numerical scheme is consistent with
Eq. (\ref{eq:Up_casii}).

In limit case 3, that is when $1 \ll \Delta t/T_i$ and $1 \ll \Delta
t/\tau_p$ (discrete case), one obtains for the velocity of the
particles and for the fluid velocity seen (see Table \ref{tab:sch1})
\begin{equation} \label{eq:Up_casiii}
\left\{\begin{split}
& U_{p,i}^{n+1} = \lra{U_i^n} + \mc{A}_i^n\,\tau_p^n +
  \sqrt{\frac{[\Check{B}_i^n]^2}{2}}
 \frac{T_i^n}{T_i^n+\tau_p^n} (\sqrt{T_i^n}\;{\mc G}_{1,i} +
 \sqrt{\tau_p^n}\;{\mc G}_{2,i}), \\ 
& U_{s,i}^{n+1} = \lra{U_i^n} +
\sqrt{\frac{[\Check{B}_i^n]^2\,T_i^n}{2}}\;\;{\mc G}_{1,i}.
\end{split}\right.
\end{equation}
Once again, $U_{p,i}(t)$, and $U_{s,i}(t)$, which were eliminated in
the continuous case, do not disappear. They become Gaussian random
variables, a result which is physically sound since these two random
variables are observed at time steps which are greater than their
respective memories. Moreover, by Taylor expansion, one can show that
the numerical scheme is consistent with Eq. (\ref{eq:xp_casiii}).

In limit case 4, $T_i =0$, and the flow becomes laminar. It can be
easily shown that the numerical scheme is consistent with Eqs.
(\ref{eq:xU_casiv}). For instance, one has for the fluid velocity
$U_{s,i}^{n+1}=\lra{U_i^n}$, cf. Table \ref{tab:sch1}.

The previous results show that the Euler scheme presented in Table
\ref{tab:sch1} is consistent with all limit cases. Therefore, the
scheme gives numerical solutions which are physically sound,
i.e. a consistent representation of the multiscale character
of the model is obtained.

%--------------------------------------------
\subsubsection{Weak second order scheme} \label{sec:sch2}
%--------------------------------------------
Most of the time, dynamical Monte Carlo methods are used with
first-order schemes only, for example in nuclear or particle
physics. In those cases, the time-step value is not a very important
factor, and attention is rather focused on obtaining accurate
statistics. On the contrary, in industrial fluid mechanics
applications with complex geometries and strong inhomogeneities in the
flow, a high-order accuracy in time can be critical in order to avoid
prohibitively small time steps resulting in huge computational time. Such
an example will be presented in Section \ref{sec:cyclone}.

From a formal point of view, weak high-order schemes for a set of SDEs
can be derived for any desired accuracy, though this is much more
complicated than for ODEs. Such high-order schemes are generally based
on truncated stochastic Taylor expansions, see for example Refs.
\cite{Klo_92,Tal_96}. These techniques can not be applied directly in
our particular case since neither the unconditional stability nor the
consistency in limit cases can be obtained.

%modif EP (début)
\ifnum\elscolor>0 \color{blue} \fi
\ifnum\elsitalic>0 \itshape \fi
\vspace*{2mm} \underline{\textbf{Property of the system of SDEs:}} the
diffusion  matrix of system (\ref{eq:SDEp}) has a
singular property of crucial importance here \cite{Min_03}. In the
present case, this nine-dimensional matrix can be written, using block
notation, as (we recall that ${\bf Z}(t)=({\bf x}_p(t),{\bf
  U}_p(t),{\bf U}_s(t))$)
\begin{equation}
{\bf \sigma }(t,{\bf Z}(t)) =
\begin{bmatrix}
\quad 0 & \quad 0 & 0 \\
\quad 0 & \quad 0 & 0 \\
\quad 0 & \quad 0 & {\bf B}_s(t,{\bf x}_p(t))
\end{bmatrix},
\end{equation}
where each block represents a three-dimensional matrix. Indeed, from
Eq. (\ref{eq:dUs}), it can be noticed that $B_{s,ij}$ depends only on
time, position, ${\bf x}_p$, and the mean value of the relative
velocity,  $\lra{{\bf U}_r}$. Therefore, the only variable of the
state vector on which $B_{s,ij}$ depends is position, because 
$\lra{{\bf U}_r}(t,{\bf x}_p(t))$ is a mean field. The fact that
$\sigma_{ij}$ depends neither on ${\bf U}_p$ nor on ${\bf U}_s$
implies that quantities such that  $\partial \sigma_{ij}/\partial z_k$
are non-zero only when $1 \leq k \leq 3$ and $6 \leq i,j \leq 9$. For
these values of $k$ and $j$, one has $\sigma_{kj}=0$. 

Thus, the diffusion matrix $\sigma_{ij}$ has the following singular
property
\begin{equation} \label{eq:commu}
\sum_k \sum_j \sigma_{kj}\,\frac{\partial \sigma_{ij}}{\partial
z_k}=0, \quad \forall \; i.
\end{equation}

\vspace*{2mm} \underline{\textbf{General idea:}} let us consider the
following model problem
\begin{equation} \label{eq:frak}
dX_i(t) = \mathfrak{A}_i({\bf X}(t))\,dt + 
                 \sum_j \mathfrak{B}_{ij}({\bf X}(t))\,dW_j(t),
\end{equation}
where $\mathfrak{B}_{ij}$ verifies property (\ref{eq:commu}). It can
be shown, for example by stochastic Taylor expansions \cite{Klo_92},
that a predictor-corrector scheme of the type 
\begin{equation} \label{eq:idee}
\left\{ \begin{split}
& \td{X}_i^{n+1} = X_i^n + \mathfrak{A}_i^n\,\Delta t + 
                       \sum_j \mathfrak{B}_{ij}^n\,\Delta W_j, \\
& X_i^{n+1} = X_i^n + \frac{1}{2}\left(\mathfrak{A}_i^n 
                    + \td{\mathfrak{A}}_i^{n+1}\right)\Delta t
                + \sum_j \frac{1}{2}\left(\mathfrak{B}_{ij}^n 
          + \td{\mathfrak{B}}_{ij}^{n+1}\right)\,\Delta W_j,
\end{split}\right.
\end{equation}
is a weak second-order scheme
($\td{\mathfrak{A}}_i^{n+1}=\mathfrak{A}_i(\td{\bf{X}}^{n+1})$,
$\td{\mathfrak{B}}_{ij}^{n+1}=\mathfrak{B}_{ij}(\td{\bf{X}}^{n+1})$, 
$\Delta t=t_{n+1}-t_n$ and $\Delta W_j=W_j^{n+1}-W_j^n$). This result is
true, once again, only when the diffusion matrix verifies
property (\ref{eq:commu}), cf. Ref. \cite{Min_03a}. If this property is not
verified, the problem is more complex and other terms are needed to
enforce second-order accuracy, see for example Talay \cite{Tal_95}.
Since the predictor step of the scheme above is the Euler scheme
(already developed in Section \ref{sec:sch1}), the remaining task
consists in finding a suitable correction step which ensures the
fulfilment of the constraints listed above. 
\normalfont
\color{black}
%modif EP (fin)

\vspace*{2mm} \underline{\textbf{Derivation of the numerical scheme:}}
how should 
the coefficients of the predictor step, $\td{\mathfrak{A}}_i^{n+1}$ and
$\td{\mathfrak{B}}_{ij}^{n+1}$, be computed? The main idea here is to generate
a correction step based on the analytical solutions by considering
that the acceleration terms vary linearly with time. This idea
originates from considerations related to Taylor series
expansions. The numerical solution obtained from the analytical
solution with constant coefficients is an approximation of first-order
accuracy. Mathematically, the solution is given in terms of the
integral of acceleration terms. Thus, one can state that the solution
based on the zero-th order (constant terms) development of the
acceleration terms gives a first-order approximation in time. By
analogy, it can be guessed that approximating the acceleration terms
by piecewise linear functions in time yields a second-order
approximation in time.

Let us introduce the following notation: $\td{U}_{p,i}^{n+1}$ and
$\td{U}_{s,i}^{n+1}$ stand for the predicted velocities and
$\td{T}_i^{n+1}$ and $\td{\tau_p}^{n+1}$ are the predicted time
scales. The values of the fields related to the fluid taken at
$(t_{n+1},\,{\bf x}_p^{n+1})$ are denoted, for example,
$\lra{U_{i}^{n+1}}$ or $\lra{P^{n+1}}$. As far as the computation of
the mean fields extracted from the discrete particles are concerned,
it is worth emphasising that none of them are computed at
$(t_{n+1},\,{\bf x}_p^{n+1})$, because the scheme would become
implicit, i.e. fields such as the expected value of the particle
velocity are computed from the predicted velocities. For example,
one has
\begin{equation}
C_i(t_{n+1},\,{\bf x}_p^{n+1}) = C_i^{n+1}=
\frac{ \lra{U_i^{n+1}}}{\td{T}_i^{n+1}}
+ f(\lra{\td{{\bf U}}_p^{n+1}},\lra{{\bf U}^{n+1}},\lra{P^{n+1}}).
\end{equation}

Let us first consider the fluid velocity seen. The analytical
solution to system (\ref{eq:sysEDS}) when the coefficients are
constant is, by applying  the rules of It\^o's calculus
\begin{equation} \label{eq:Us_sch2}
U_{s,i}(t) = U_{s,i}(t_0)\,\exp(-\Delta t/T_i)
+ \int _{t_0}^{t}C_i(s,{\bf x}_p)\exp[(s-t)/T_i]\,ds + \gamma_i(t),
\end{equation}
where the temporal coefficients (the timescales) are considered
constant, while the term $C_i(s,{\bf x}_p)$ is retained in the
integral. Following the previous ideas, let us suppose that
$C_i(s,{\bf x}_p)$ varies linearly on the integration interval
$[t_0,t]$, that is ($\Delta t=t-t_0$)
\begin{equation} \label{eq:Ci_int}
C_i(s,{\bf x}_p(s)) = C_i(t_0,{\bf x}_p(t_0)) + \frac{1}{\Delta t}
[C_i(t,{\bf x}_p(t))-C_i(t_0,{\bf x}_p(t_0))](s-t_0).
\end{equation}
By inserting Eq. (\ref{eq:Ci_int}) into Eq. (\ref{eq:Us_sch2}),
integration gives
\begin{equation}
\begin{split}
& U_{s,i}(t) = U_{s,i}(t_0)\,\exp(-\Delta t/T_i)
+ [T_i\,C_i(t_0,{\bf x}_p(t_0))]\,A_2(\Delta t,T_i) \\
& \qquad + [T_i\,C_i(t,{\bf x}_p(t))]\,B_2(\Delta t,T_i)+ \gamma_i(t),
\end{split}
\end{equation}
where the functions $A_2(\Delta t,x)$ and $B_2(\Delta t,x)$ are given
by ($x$ is a positive real variable)
\begin{equation}
\left\{\begin{split}
& A_2(\Delta t,x)=-\exp(-\Delta t/x)+[1-\exp(-\Delta t/x)][\Delta t/x],\\
& B_2(\Delta t,x)=1-[1-\exp(-\Delta t/x)][\Delta t/x]. 
\end{split}\right.
\end{equation}
Accounting for the time dependence of the coefficients, i.e. $T_i$, it
is proposed to write the following correction step, which is in line
with the treatment of the acceleration terms,
\begin{equation} \label{eq:Us_pred}
\begin{split}
U_{s,i}^{n+1} = & 
\frac{1}{2}\,U_{s,i}^n\,\exp(-\Delta t/T_i^n)
+ \frac{1}{2}\,U_{s,i}^{n}\,\exp(-\Delta t/\td{T}_i^{n+1}) \\
+ & A_2(\Delta t,\,T_i^n) \,[T_i^n C_i^n]
+ B_2(\Delta t,\,\td{T}_i^{n+1}) \,[\td{T}_i^{n+1} C_i^{n+1}]
+ \td{\gamma}_i^{n+1},
\end{split}
\end{equation}
where a consistent formulation for the stochastic integral
$\td{\gamma}_i^{n+1}$ is needed. The same procedure is used,
i.e. the diffusion matrix $B_{ij}$ is linearised and
integration is carried out. The final expression is
\begin{equation}
\td{\gamma}_i^{n+1}= \sqrt{[B_i^{*}]^2\frac{\td{T}_i^{n+1}}{2}
               [1-\exp(-2\Delta t/\td{T}_i^{n+1})]}\;\; {\mc G}_{1,i},
\end{equation}
where ${\mc G}_{1,i}$ is the same ${\mc N}(0,1)$ random variable used
in the simulation of $\gamma_i^n$ in the Euler scheme and where
$B_i^{*}$ is defined by
\begin{equation} \label{sigma1}
\begin{split}
\left[1-\exp(-2\,\Delta t/\td{T}_i^{n+1})\right] & B_i^{*} = \\
A_2(2\,\Delta t,\,\td{T}_i^{n+1})\,& \sqrt{(\Check{B}_i^n)^2} +
B_2(2\,\Delta t,\,\td{T}_i^{n+1})\,\sqrt{ (\td{\Check{B}}_i^{n+1})^2}.
\end{split}
\end{equation}
Here, some explanations must be given. During integration,
another step is necessary in order to achieve the closed form
presented in Eq. (\ref{sigma1}). Indeed, two parts derive from the
integration by parts carried out when  $B_{ij}$ varies linearly.
The first term is an analytical function, while the second term is still
a stochastic integral, therefore the global integral can be written
formally  $\td{\gamma}_i^{n+1} = \delta_1 + \delta_2$.
It has been considered that a projection of this second integral
on the first remains of second-order accuracy for the global scheme.
Therefore, the following hypothesis has been used,
$\delta_2 \approx (\lra{\delta_1 \delta_2}/\lra{\delta_1^2})\delta_1$.

In the case of the velocity of the particles, the same approach
followed for the fluid velocity seen is adopted. Let us start from the 
exact solution with constant coefficients for ${\bf U}_p(t)$. By
resorting to the rules of It\^o's calculus, one can write
\begin{equation}
\begin{split}
U_{p,i}(t) = U_{p,i}(t_0)& \exp(-\Delta t/\tau_p) + \\
& \frac{1}{\tau_p}\exp(-\Delta t/\tau_p)\int_{t_0}^{t}\exp(s/\tau_p)
[U_{s,i}(s)+\tau_p\,\mc{A}_i(s,{\bf x}_p)]\,ds,
\end{split}
\end{equation}
and by inserting Eq. (\ref{eq:Us_sch2}) in the previous equation,
one has
\begin{equation} \label{eq:depart}
\begin{split}
& U_{p,i}(t) = \; U_{p,i}(t_0)\exp(-\Delta t/\tau_p)
  + U_{s,i}(t_0)\,\theta_i
    [\exp(-\Delta t/T_i)-\exp(-\Delta t/\tau_p)]  + \Gamma _i(t) \\
& + \frac{1}{\tau_p}\exp(-t/\tau_p)\int _{t_0}^{t}\exp(s/\tau_p)
    \left[\exp(-s/T_i) \int_{t_0}^{s} C_i(u,{\bf x}_p)\,\exp(u/T_i)\,du
    + \tau_p\,\mc{A}_i(s,{\bf x}_p)\right]\,ds.
\end{split}
\end{equation}
Two deterministic integrals must be treated in
Eq. (\ref{eq:depart}). A multiple one, involving $C_i(t,{\bf x}_p)$ and
a simple one with the acceleration term $\mc{A}_i(t,{\bf x}_p)$. Both
integrals are handled as done previously for the fluid velocity seen,
that is, it is assumed that both accelerations vary linearly on the
integration interval, see for example Eq. (\ref{eq:Ci_int}) for
$C_i(t,{\bf x}_p)$. By integration by parts of both integrals, one
finds after some derivations
\begin{equation}
\begin{split}
U_{p,i}(t) = & \; U_{p,i}(t_0)\exp(-\Delta t/\tau_p) \\
& + U_{s,i}(t_0)\,\theta_i
    [\exp(-\Delta t/T_i)-\exp(-\Delta t/\tau_p)] \\
& + [T_i\,C_i(t_0,{\bf x}_p(t_0))]\,A_{2c}(\tau_p,T_i)
  + [T_i\,C_i(t,{\bf x}_p(t))]\,B_{2c}(\tau_p,T_i) \\
& + [\tau_p\,\mc{A}_i(t_0,{\bf x}_p(t_0))]\,A_2(\Delta t,\tau_p)
  + [\tau_p\,\mc{A}_i(t,{\bf x}_p(t))]\,B_2(\Delta t,\tau_p) \\
& + \Gamma_i(t),
\end{split}
\end{equation}
where the functions $C_{2c}(x,y)$, $A_{2c}(x,y)$ and $B_{2c}(x,y)$ are
given by ($x$ and $y$ are two positive real variables)
\begin{equation}
\left\{\begin{split}
& C_{2c}(x,y) = [y/(y-x)][\exp(-\Delta t/y)-\exp(-\Delta t/x)], \\
& A_{2c}(x,y) =
- \exp(-\Delta t/x) + [(x+y)/\Delta t][1-\exp(-\Delta t/x)] \\
& \hskip 4.0cm - (1+y/\Delta t)\,C_{2c}(x,y), \\
& B_{2c}(x,y) =  1 - [(x+y)/\Delta t][1-\exp(-\Delta t/x)]
+ (y/\Delta t)C_{2c}(x,y).
\end{split}\right.
\end{equation}
In analogy with the expression proposed for the fluid velocity seen,
cf. Eq. (\ref{eq:Us_pred}), the following correction step is
proposed,
\begin{equation} \label{eq:Up_sch2}
\begin{split}
U_{p,i}^{n+1} =
    \frac{1}{2}\,U_{p,i}^n\, & \exp(-\Delta t/\tau_p^n)
  + \frac{1}{2}\,U_{p,i}^n\,\exp(-\Delta t/\td{\tau}_p^{n+1}) \\
& + \frac{1}{2}\,U_{s,i}^n\,C_{2c}(\tau_p^n,\,T_i^n)
  + \frac{1}{2}\,U_{s,i}^n\,C_{2c}(\td{\tau}_p^{n+1},\,\td{T}_i^{n+1}) \\
& + A_{2c}(\tau_p^n,\,T_i^n) \,[T_i^n C_i^n]
  + B_{2c}(\td{\tau}_p^{n+1},\,\td{T}_i^{n+1})\,[\td{T}_i^{n+1}C_i^{n+1}] \\
& + A_2(\Delta t,\tau_p^n)[\tau_p^n\,\mc{A}_i^n]
  + B_2(\Delta t,\td{\tau}_p^{n+1})[\td{\tau}_p^{n+1}\,\mc{A}_i^{n+1}]
  + \td{\Gamma}_i^{n+1}.
\end{split}
\end{equation}
For the simulation of the stochastic integral, one has 
(${\mc G}_{2,i}$ is the ${\mc N}(0,1)$ random variable used in the
simulation of $\Gamma_i^n$ in the Euler scheme, see Table
\ref{tab:sch1})
\begin{equation} 
\td{\Gamma}_i^{n+1} =
\frac{\lra{ \td{\Gamma}_i^{n+1}\td{\gamma}_i^{n+1} } }
     {\lra{(\td{\gamma}_i^{n+1})^2}}\,\td{\gamma}_i^{n+1}+
\sqrt{\lra{(\td{\Gamma}_i^{n+1})^2}-
      \frac{[\lra{\td{\Gamma}_i^{n+1}\td{\gamma}_i^{n+1}}]^2}
           {\lra{(\td{\gamma}_i^{n+1})^2}} }\;{\mc G}_{2,i},
\end{equation}
where the second order moments
$\lra{(\td{\Gamma}_i^{n+1})^2}$ and
$\lra{\td{\Gamma}_i^{n+1}\td{\gamma}_i^{n+1}}$ are computed from
Eqs. (\ref{eq:Gama2}) and (\ref{eq:gaGam}), respectively, by inserting 
the suitable timescales and diffusion matrix, that is
$\tau_p^{n}$, $\td{T}_i^{n+1}$ and $B^{*}_i$. This completes
the weak second order scheme. 

It can be shown, by means of stochastic Taylor expansion
\cite{Klo_92}, that the present scheme is a weak scheme of order
$2$ in time for system (\ref{eq:sysEDS}). It is worth emphasising that
no correction is done on position, ${\bf x}_p(t)$, since the prediction
is already of order $2$.  The complete scheme is summarised in Table
\ref{tab:sch2c}.

\vspace*{2mm} \underline{\textbf{Limit cases:}} in limit case 1, when
$1 \ll  \Delta t/\tau_p$, one has $A_{2c}(\tau_p,T_i) \to A_2(\Delta
t,T_i)$, 
$B_{2c}(\tau_p,T_i) \to B_2(\Delta t,T_i)$ and
$C_{2c}(\tau_p,T_i) \to \exp(-\Delta t/T_i)$. For the stochastic
integral, one can show that
$\td{\Gamma}_i^{n+1} \to \td{\gamma}_i^{n+1}$. Inserting these
results in Eq. (\ref{eq:Up_sch2}) yields $U_{p,i}^{n+1} =
U_{s,i}^{n+1}$, which is consistent with Eq. (\ref{eq:limc1}). This
result is a second order scheme for ${\bf U}_p(t)$, and
therefore the scheme remains of order $2$ in limit case 1.

When $1 \ll \Delta t/T_i$ and $\Check{B}_i\,T_i ={\rm cst}$
(limit case 2), one has $A_{2c}(\tau_p,T_i) \to A_2(\Delta t,\tau_p)$
and $B_{2c}(\tau_p,T_i) \to B_2(\Delta t,\tau_p)$, which gives for the
numerical correction of the velocity of the particles
\begin{equation} \label{eq:lim_uti}
\begin{split}
U_{p,i}^{n+1} & =
    \frac{1}{2}\,U_{p,i}^n\, \exp(-\Delta t/\tau_p^n)
  + \frac{1}{2}\,U_{p,i}^n\,\exp(-\Delta t/\td{\tau}_p^{n+1}) \\
& + A_2(\Delta t,\tau_p^n)[\lra{U_i^n}+\tau_p^n\,\mc{A}_i^n]
  + B_2(\Delta t,\td{\tau}_p^{n+1})[\lra{U_i^{n+1}}+
  \td{\tau}_p^{n+1}\,\mc{A}_i^{n+1}] \\
& + \td{\Gamma}_i^{n+1}.
\end{split}
\end{equation}
For the simulation of the stochastic integral, one can prove by
looking at the limit values (when $1 \ll \Delta t/T_i$ and
$\Check{B}_i\,T_i ={\rm cst}$) in Eqs. (\ref{eq:gama2}),
(\ref{eq:Gama2}) and (\ref{eq:gaGam}) that (here ${\mc G}^{\prime}_{p,i}$
is a ${\mc N}(0,1)$ random variable)
\begin{equation} \label{eq:gamp}
\td{\Gamma}_i^{n+1} \to \sqrt{\frac{[B^*_i\,\td{T}_i^{n+1}]^2}
{2\,\tau_p^{n}}[1-\exp(-2\Delta t/\tau_p^{n})]}\;\;{\mc
G}^{\prime}_{p,i},
\end{equation}
which is in accordance with Eq. (\ref{eq:Up_casii}). Unfortunately,
it can be established, again by Taylor stochastic expansion, that the
scheme is not of second order in time for system (\ref{eq:limc2}), but 
of first order. This is due to the treatment of the correction step
for the stochastic integral $\Gamma_i(t)$ where $\tau_p^n$ has been
retained in order to avoid anticipation and inconsistent numerical
expressions of the It\^o integral.
As far as the fluid velocity seen is concerned, one has
\begin{equation} \label{eq:lim_uti1}
U_{s,i}^{n+1} = \lra{U_i^{n+1}} +
\sqrt{\frac{[B_i^{*}]^2\,\td{T_i}^{n+1}}{2}}\;{\mc G}_{1,i},
\end{equation}
which is in line with the previous result.
 This scheme is of second order, but the whole
scheme is not. Indeed, as mentioned above, the scheme is only of first
order for the velocity of the particles.

When both the fluid velocity seen and the velocity of the particles
become fast variables (limit case 3), that is when
$1 \ll \Delta t/T_i$, $1 \ll \Delta t/\tau_p$ and
$\Check{B}_i\,T_i ={\rm cst}$, one can write for the velocity of 
the particle, for example from Eq. (\ref{eq:lim_uti}) with $1 \ll
\Delta t/\tau_p$,
\begin{equation} 
U_{p,i}^{n+1} = \lra{U_i^{n+1}}+
\td{\tau}_p^{n+1}\,\mc{A}_i^{n+1} +
\sqrt{\frac{[B^*_i\,\td{T}_i^{n+1}]^2}{2\,\tau_p^{n}}}\;
{\mc G}^{\prime}_{p,i}.
\end{equation}
For the fluid velocity seen, Eq. (\ref{eq:lim_uti1}) is
unchanged. These results are consistent with the expressions of
Section \ref{sec:sch1}. In limit case 3, the numerical scheme
for the position of the  particles is equivalent to the Euler scheme
written previously and is of first order in time. 

When the flow becomes laminar, that is when $T_i \to 0$ with no
condition on the product $\Check{B}_i\, T_i$, one has the following
limits: $A_2(\Delta t,T_i) \to 0$, $B_2(\Delta t,T_i) \to 1$ and
$\gamma_i(t) \to 0$, which gives for the fluid velocity seen,
$U_{s,i}^{n+1}=\lra{U_i^{n+1}}$. For the velocity of the
particles, the coefficients have the following limits:
$A_{2c}(\tau_p,T_i) \to A_2(\Delta t,\tau_p)$, $B_{2c}(\tau_p,T_i) \to
B_2(\Delta t,\tau_p)$ and $C_{2c}(\tau_p,T_i) \to 0$ which gives
together with the limit $\Gamma_i(t) \to 0$,
\begin{equation}
\begin{split}
& U_{p,i}^{n+1} =
    \frac{1}{2}\,U_{p,i}^n\, \exp(-\Delta t/\tau_p^n)
  + \frac{1}{2}\,U_{p,i}^n\,\exp(-\Delta t/\td{\tau}_p^{n+1}) \\
& + A_2(\Delta t,\tau_p^n)[\lra{U_i^n}+\tau_p^n\,\mc{A}_i^n]
  + B_2(\Delta t,\td{\tau}_p^{n+1})
    [\lra{U_i^{n+1}}+\td{\tau}_p^{n+1}\,\mc{A}_i^{n+1}].
\end{split}
\end{equation}
It can be shown, by regular Taylor expansion, that this
scheme, together with the prediction step (Euler scheme) is a
second order scheme for system (\ref{eq:limc4}).

In summary, a weak second-order scheme for system (\ref{eq:sysEDS})
has been derived. This scheme satisfies all conditions listed in
Section \ref{sec:cond}. However, second order convergence is not
obtained in limit cases 2 and 3. In this latter case, the first order
convergence is inherent to the spirit of the scheme, that is a single
step to compute position ${\bf x}_p(t)$ (in order to minimise the number
of particle localisations in the algorithm).

%modif EP (début)
\ifnum\elscolor>0 \color{blue} \fi
\ifnum\elsitalic>0 \itshape \fi
%===============================================================================
\section{Specific and open issues} \label{sec:twc}
%===============================================================================
The main objective of this paper is to present a consistent and
rigorous numerical method for the computations of polydispersed
turbulent two-phase flows using a mean-field/PDF approach. The
mathematical framework and the models used in this approach have
been put forward (Sections \ref{sec:formalism} and
\ref{sec:pdf_models}) and, a general methodology, for the derivation
of weak numerical schemes for the set of SDEs describing the dynamics
of the stochastic particles, has been given in the context of
particle-mesh methods (Sections \ref{sec:scale} and \ref{sec:num0}).

In this general methodology, the derivation of the weak numerical
schemes has been performed only in the case of one-way coupling. As
mentioned before, this is not a limitation of the methodology and it
is simply the status of the developments so far. The extension of the
present results to two-way coupling is now discussed. Two issues are
addressed: 
\begin{enumerate}
\item[(i)] the computations of the source terms in the PDEs describing
  the dynamics of the fluid mean fields, i.e. when two-way coupling is
  accounted for, that is when the particle mass fraction is high
  enough and the influence of the particles on the fluid mean fields
  must be taken into account, particle source terms are added to the
  fluid equations, cf. Eqs. (\ref{eq:feq_Uf}) and (\ref{eq:feq_ufuf}),
\item[(ii)] the extension of the methodology introduced in Sections
  \ref{sec:scale} and \ref{sec:num0}, i.e. when
  two-way coupling is considered, an acceleration is added to the SDE
  describing the dynamics of the fluid velocity seen,
  cf. Eqs. (\ref{eq:Ap_s}) and (\ref{eq:SDEp}), and the structure of
  the system of SDEs is changed.
\end{enumerate}
The first point is a specific issue, that is a practical solution is
given for the computational procedure of the source terms. The second
point is considered as an open issue since only explanations on the
procedure to follow, for the extension of the weak numerical schemes
to two-way coupling, are provided.

After the treatment of the two-way coupling issues and before
showing computational examples, some possible improvements and some
open questions related to the numerical method will be discussed. The
list of open questions related to the present numerical
(particle-mesh) method is long. Here, attention is focused on two
open issues:
\begin{enumerate}
\item[(iii)] the formulation of boundary conditions in wall-bounded
  flows,
\item[(iv)] the development of new numerical methods based on the
  present one. 
\end{enumerate}
We start now with aspects related to two way coupling, i.e. issues (i)
and (ii).
\normalfont
\color{black}
%modif EP (fin)

%-------------------------------------------------------------------------------
\subsection{Computation of the source terms} \label{sec:source}
%-------------------------------------------------------------------------------
As can be seen from the equations given in Table \ref{tab:RANS},
i.e. Eqs. (\ref{eq:feq_Uf}) and (\ref{eq:feq_ufuf}), two
source terms are present when two-way coupling is considered. 
The first one, say ${\bf S}_U$,
represents the exchange of momentum between the discrete particles and
the fluid. In the present paper, the only force exerted by the fluid
on the discrete particles is the drag force, see Eqs. (\ref{exa_part_eqns}),
and by reaction the force exerted by the particles on the fluid is the
reverse drag force. The mean momentum source term is expressed by
\begin{equation}
(S_U)_i = \chi \lra{ \frac{U_{p,i}-U_{s,i}}{\tau_p} },
\end{equation}
where $\chi=(\alpha_p\rho_p)/(\alpha_f\rho_f)$. From the particle
equation of motion, cf. Eqs. (\ref{eq:SDEp}), the drag term is
equal to the discrete particle acceleration (when gravity is first
subtracted) and ${\bf S}_U$ can be re-expressed as
\begin{equation}
(S_U)_i = -\chi \lra{ \frac{ dU_{p,i} }{dt} }.
\end{equation}
%modifs JPM(debut)
\ifnum\elscolor>0 \color{blue} \fi
\ifnum\elsitalic>0 \itshape \fi
From the discrete point of view, if we use the NGP technique, as 
explained in Section \ref{SS_PM_AW}, for the sake of simplicity 
(since most of what is presented below concern particle instantaneous 
quantities that can be put within the CIC formalism), this source term 
is the sum of the reverse drag force due to the discrete particles 
that are found in a given fluid cell. Then, the total fluid momentum 
in a cell $[k]$, whose volume is ${\cal V}_f^{[k]}$ can be written as
\begin{equation}
\alpha_f\rho_f {\cal V}_f^{[k]}\,(S_U)_i^{[k]}= \sum_{l=1}^{N_k} m_p^l 
\frac{ (U_{p,i}^l)^{n+1} - (U_{s,i}^l)^{n}}{\Delta t},
\end{equation}
\normalfont
\color{black}
%modifs JPM(fin)
where $m_p^l$ and $U_{p,i}^l$ stand for the mass and the velocity of
the discrete particle labelled $l$, respectively. The sum is performed
over the $N_k$ particles that are located in cell $[k]$ at iteration
$n$ ($t=n\Delta t$).

The source term for the fluid Reynolds-stress equations, see
Eq. (\ref{eq:feq_ufuf}), raises new questions and its numerical
evaluation is an interesting example of the specificities of
stochastic calculus, cf. Section \ref{sec:intsto}. For the
discussion of its expression, we limit ourselves to the simplified
case of a stationary one-dimensional system and to the source term,
$S_k$, for the fluid kinetic energy $k$. The system of SDEs that we
consider is
\begin{equation} 
\left\{\begin{split}
dU_p(t) =& \frac{ U_s(t) - U_p(t)}{\tau_p}\, dt, \\
dU_s(t) =& -\frac{U_s(t)}{T}\, dt -
\chi\left(\frac{U_s(t)-U_p(t)}{\tau_p}\right)\, dt + \sqrt{K}\,
dW(t),
\end{split}\right.
\end{equation}
and the fluid kinetic source term which represents the work performed
by the drag force is 
\begin{equation}
S_k = \chi\lra{U_s\left(\frac{U_p-U_s}{\tau_p}\right)}.
\end{equation}
%modifs JPM(debut)
\ifnum\elscolor>0 \color{blue} \fi
\ifnum\elsitalic>0 \itshape \fi
For this simplified case, and when the coefficients of the equations
are constant, we can derive the analytical expression of the
second-order moments.
Indeed, for a long-enough time after the initial conditions, the
stochastic process ${\bf Z}(t)=\{ U_p(t),U_s(t)\}$ (since
here $x_p(t)$ is irrelevant) reaches a stationary state and 
$\lra{ U_p^2 }, \lra{ U_p\,U_s}$ and $\lra{U_s^2}$ become constant. 
Therefore, using It\^o's calculus, we have that
\begin{equation} 
\left\{\begin{split}
&d\, \lra{ U_p^2 }=2\, \lra{ U_p\, dU_p }=0, \\
&d\, \lra{ U_p\,U_s}= \lra{ U_p\, dU_s + U_s\, dU_p }=0, \\
&d\, \lra{ U_s^2 }= 2\, \lra{ U_s\, dU_s } + K\, dt =0.
\end{split}\right.
\end{equation}
The first two equations yield the equilibrium formulae for the
second-order moments
\begin{equation} 
\left\{\begin{split}
&\lra{ U_p^2 }=\lra{ U_p\, U_s }, \\
&\lra{ U_p\,U_s}=\lra{U_s^2}\, \frac{1}{ 1 + \tau_p/T},
\end{split}\right.
\end{equation}
while the third one gives the expression of the diffusion coefficient
$K$ to maintain a constant value of the fluid kinetic energy
\begin{equation} 
K= 2\, \lra{U_s^2}\, \left( \frac{1}{T} + \frac{\chi}{T+\tau_p} \right).
\end{equation}
Using these formulae, the (equilibrium) analytical expression of the 
kinetic source term can be written as
\begin{equation} \label{expression_Sk}
S_k = - \chi \, \lra{U_s^2}\, \frac{1}{T + \tau_p} .
\end{equation}
This source term is always negative which indicates that the drag
force, which is indeed a friction force, induces a loss of energy in the
fluid energy budget. This is valid for the total energy budget, whereas
if we consider the fluid energy spectrum and its modulation by particles,
particles may enhance turbulence at some lenghtscales (or wave numbers)
due, for example, to wakes generated behind the particles. In the
present model, we consider only $S_k$ which is the integrated value
of the exchange term over the whole spectrum, and, if we leave out the
(possible) energy injected from particles by their initial conditions,
the total energy gained by the particles comes from the fluid and the
fluid kinetic energy source term is negative. 
\normalfont
\color{black}
%modifs JPM(fin)
However, when $\tau_p \to 0$, that is when the
discrete particles behave as fluid elements (but with a constant mass
fraction, $\chi$), we expect the kinetic source term to vanish ($S_k
\to 0$) since we consider a stationary case. Yet, from
Eq.~(\ref{expression_Sk}), it is seen that the limit is
\begin{equation} 
S_k \xrightarrow[\tau_p \to 0]{} -\chi \,\lra{ U_s^2}\,\frac{1}{T}.
\end{equation}
This spurious non-zero limit for vanishing particle characteristic
timescale can be traced back to the Langevin model and is related to
the fact that acceleration is indeed replaced by a white-noise
term~, cf. Chapter 6.8 in Ref. \cite{Min_01}. 

Nevertheless, it is possible to retrieve the correct limit in the
numerical evaluation of $S_k$ by resorting to a discretisation based
on the Stratonovich definition, see Section \ref{sec:intsto}. The
first step is to write the source term with the particle acceleration
as
\begin{equation} 
S_k = - \chi \lra{ U_s\, \frac{dU_p}{dt} }.
\end{equation}
Therefore, if we consider the integration of the source term in a time
interval, we get  
\begin{equation} 
S_k \, dt = - \chi \lra{ U_s\, dU_p }
\end{equation}
and, in a formal sense, when $\tau_p \to 0$, we expect the source term
to become
\begin{equation} 
S_k\, dt \xrightarrow{} - \chi \, \lra{U_s\, dU_s },
\end{equation}
since in that case $U_p \to U_s$, cf. Eq. (\ref{eq:limc1}) in
Section \ref{sec:limcases}. Now, from Section \ref{sec:intsto}, we
know that the above expression can have different meanings. If we
decide to regard the term $\lra{U_s\, dU_s }$ as being defined in the
It\^o sense, as it should be in order to be consistent with the algebra
retained throughout the paper, we would find the non-zero limit given
above. Yet, if for this expression of the source term, we decide to
consider it as being defined in the Stratonovich sense, then
\begin{equation} 
S_k \,dt \xrightarrow[\tau_p \to 0]{} - \chi \, \lra{U_s \circ dU_s }=
-\frac{\chi}{2} \, \lra{ d (U_s)^2 },
\end{equation}
which is indeed zero since we are in a stationary case. 

%modifs JPM(debut)
\ifnum\elscolor>0 \color{blue} \fi
\ifnum\elsitalic>0 \itshape \fi
The difference between the two stochastic calculus is only presented here
since it provides a useful guideline. In the present case, it is seen that
the interest of the Stratonovich expression is that the formal quantity
$dU_s/dt$ can still be handled as if it were a normal derivative (and, in
our case, the limit of $dU_p/dt$ when $\tau_p \to 0$). 
\normalfont
\color{black}
%modifs JPM(fin)
This suggests therefore to express the kinetic source term numerically,
in a fluid cell $[k]$ at time $t=n\, \Delta t$ with $N_k$ particles,
as
\begin{equation}
\alpha_f\rho_f {\cal V}_f^{[k]} \, S_k^{[k]}= - \sum_{l=1}^{N_k} m_p^l 
\frac{1}{2}\left( (U_s^l)^{n+1} + (U_s^l)^n \right) 
\frac{ (U_p^l)^{n+1} - (U_p^l)^n }{\Delta t} .
\end{equation}
From the properties of the numerical schemes developed in the previous
Sections, we have that $(U_p^l)^n \to (U_s^l)^n$ when $\tau_p \to
0$. Thus, in that limit
\begin{equation}
\left.\begin{split}
\alpha_f\rho_f {\cal V}_f^{[k]}\, S_k^{[k]} 
& \xrightarrow{} - \sum_{l=1}^{N_k} m_p^l\,
\frac{1}{2}\left( (U_s^l)^{n+1} + (U_s^l)^{n} \right) 
\frac{ (U_s^l)^{n+1} - (U_s^l)^{n} }{\Delta t} \\
& = - \sum_{l=1}^{N_k} m_p^l\,
\frac{1}{2}\left( [(U_s^l)^{n+1}]^2 - [(U_s^l)^{n}]^2 \right)
\simeq \alpha_p\rho_p {\cal V}_f^{[k]}\, \lra{ [(U_s^l)^{n+1}]^2 - [(U_s^l)^{n}]^2 }
\end{split}\right.
\end{equation}
and, when the stationarity of $U_s$ is indeed enforced numerically,
this term is zero. Finally, going back to the exact fluid Reynolds
stress equations, we propose to express the numerical source terms as
\begin{equation}
\begin{split}
\alpha_f\rho_f {\cal V}_f^{[k]}\, S_{R_{ij}}^{[k]}= 
- \sum_{l=1}^{N_k} m_p^l & \left\{
\frac{1}{2}\left[ (U_{s,j}^l)^{n+1} + (U_{s,j}^l)^{n} \right] 
\frac{ (U_{p,i}^l)^{n+1} - (U_{p,i}^l)^{n} }{\Delta t} +  \right.\\
& \hskip 0.5cm \left. \frac{1}{2}\left[ (U_{s,i}^l)^{n+1} + (U_{s,i}^l)^{n} \right] 
\frac{ (U_{p,j}^l)^{n+1} - (U_{p,j}^l)^{n} }{\Delta t} \right\}.
\end{split}
\end{equation}

%modifs JPM(debut)
\ifnum\elscolor>0 \color{blue} \fi
\ifnum\elsitalic>0 \itshape \fi
As mentioned at the beginning of this section, most of the arguments that
have been presented concern the discrete evaluation of each particle term
$U_s(dU_p/dt)$. The expression proposed above in the NGP formulation can
be used directly within the CIC technique. 
Another interesting question is to ask to what cell (or cells) the different
source terms should be assigned. Indeed, within one time step, particles
may cross several fluid cells and the source terms, say ${\bf S}_U$ and
${\bf S}_R$ which represent the total momentum and energy exchange terms,
should be distributed between the different cells crossed by the particles.
One possibility is to apply to each fluid cell crossed by a particle, the
different reverse expressions, say $dU_p/dt$ and $U_s(dU_p/dt)$ in proportion
of the time spent in that cell (the residence time) which is then a fraction 
of the time step. This is probably the most precise expression and the most
accurate discrete formulation, but it implies to keep track of the different
fluid cells along the particle trajectory within one time step. In a complex
geometry and unstructured meshes, given present localisation algorithms, this
is not an easy task and it induces computational overloads. For these reasons,
at the moment, it is proposed to evaluate the total source terms from the
particles that were located in that cell at the beginning of the time step.
This evaluation has been applied in the various computational examples
presented in Section \ref{sec:exemples}. It can be seen as a
first-order spatial approximation or based on an implicit assumption
that the particle Courant number remains of order one in most cases.  
\normalfont
\color{black}
%modifs JPM(fin)

%modif EP (début)
\ifnum\elscolor>0 \color{blue} \fi
\ifnum\elsitalic>0 \itshape \fi
%-------------------------------------------------------------------------------
\subsection{Extension of the weak numerical schemes} \label{sec:extention}
%-------------------------------------------------------------------------------
When two-way coupling is accounted for, the SDE describing the
dynamics of the fluid velocity seen is supplemented with an
acceleration term, cf. Eqs. (\ref{eq:Ap_s}) and (\ref{eq:SDEp}), in
order to account for the influence of the discrete particles on the
statistics of the fluid velocity sampled along the trajectory of a
discrete particle. This supplementary acceleration changes drastically
the nature of the equation system and one has (the equation for
position is omitted for the sake of clarity)
\begin{equation}
\left\{
\begin{split}
dU_{p,i}(t) & = -\frac{1}{\tau_p}U_{p,i}\, dt +
              \frac{1}{\tau_p}U_{s,i}\, dt + g_i\,dt,\\
dU_{s,i}(t) & = \frac{\chi}{\tau_p}U_{p,i}\,dt 
           -\left(\frac{1}{T_i}+\frac{\chi}{\tau_p}\right)U_{s,i}\,dt
           + C_i\, dt + \sum_j B_{ij}\, dW_j(t),
\end{split} \right .
\end{equation}
that is, the SDE for the fluid velocity seen, ${\bf U}_s(t)$, depends
explicitly on the velocity of the discrete particle, ${\bf U}_p(t)$. 
This dependence complicates the analysis of the system, in particular
the limit systems when the time scales go to zero, cf. Section
\ref{sec:scale}. If one is able to find the limit systems in the
continuous sense, the extension of the numerical schemes can be
obtained in the same way as presented in Section
\ref{sec:cond}. However, in order to calculate the analytical solution
with constant coefficients, one has to express the following matrix in
diagonal or triangular form (depending on the roots of the
characteristic polynomial)
\begin{equation}
\begin{bmatrix}
-1/\tau_p   & 1/\tau_p \\
\chi/\tau_p & -(1/T_i + \chi/\tau_p)
\end{bmatrix}.
\end{equation}
Once this is done, the previous analysis can be used, but in the frame
of much more complex algebra. Once the analytical solution is obtained
for the eigensystem, one has to go back to the original system (state
vector) with some transformation matrix (which is formed by the
eigenvectors).

This difficulty is actually not a typical feature of two-way
coupling. As a matter of fact, in the one-way coupling case, if the
alternative model is chosen, cf. Section \ref{sec:pdf_p}, the drift
term is written in terms of the local instantaneous
velocities. Therefore, an acceleration which has the same form as
${\bf A}_{p \rightarrow s}$ is introduced
\begin{equation}
\sum_j \frac{1}{T_{ij}}(U_{p,j}-U_{s,j}),
\end{equation}
where the time scales of the mean flow, $T_{ij}$, are given by 
$T_{ij}^{-1}= \partial \lra{U_i}/\partial x_j$. As a consequence, if
such a model is used for the drift term, the problems inherent to the
derivation of the present schemes with two-way coupling are already
encountered for the one-way coupling case. There is, to our knowledge,
no specific work in the literature dealing with this subject.
\normalfont
\color{black}
%modif EP (fin)

%-------------------------------------------------------------------------------
\subsection{Boundary conditions in wall-bounded flows} \label{sec:boundary}
%-------------------------------------------------------------------------------
%modif EP (début)
\ifnum\elscolor>0 \color{blue} \fi
\ifnum\elsitalic>0 \itshape \fi
In the present work, for the computational examples, cf. Section
\ref{sec:exemples}, the wall boundary conditions for the system of
SDEs are treated as follows: for the discrete particle velocity,
${\bf U}_p(t)$, an elastic wall-particle collision is applied whereas
for the fluid velocity seen,
${\bf U}_s(t)$, we build on ideas from turbulent single-phase flows
\cite{Min_99b,War_04} in order to ensure consistency when $\tau_p \to 0$. 
As a matter of fact, 
in one-point PDF models for single-phase turbulent flows, cf. Section
\ref{sec:pdf_f}, or in one-point PDF models for the discrete
particles, cf. Section \ref{sec:pdf_p}, the derivation of boundary
conditions for fluid or discrete particles, when solid boundaries are
present, has not received the needed attention. Here, for the sake of
simplicity, we make our point by considering, as an example, only the
motion of fluid particles.
\normalfont
\color{black}
%modif EP (fin)

In the framework of PDF methods for single-phase turbulent flows,
boundary conditions of the \textit{wall function} kind have been
studied and proposed \cite{Min_99b,Pop_97b}. This solution has been
investigated rigorously from the mathematical and physical (based on
the knowledge of the phenomenology of the near-wall region) points of
view. In practice, the numerical treatment is developed in analogy
with the wall-function approach used in RANS computations, a method
which is perfectly in line with one-point high-Reynolds
number PDF models. However, in some engineering applications where a
precise description of the near-wall region is needed, it may be of
interest to replace the wall-function boundary conditions with a direct 
particle-wall interaction, i.e. ${\bf U}_f(t)={\bf 0}$ at the
wall. According to Section \ref{sec:pdf_f}, a one-point PDF model for
single-phase turbulent flows reads
\begin{equation} \label{eq:model_f}
\left\{\begin{split}
& dx_{f,i}(t)=U_{f,i}(t)\, dt, \\
& dU_{f,i}(t)= \mathfrak{A}_i(t,{\bf x}_f(t),{\bf U}_f(t))\,dt + 
\sum_j \mathfrak{B}_{ij}(t,{\bf x}_f(t))\,dW_{j}(t),
\end{split}\right.
\end{equation}
and it has been shown in Section \ref{sec:sch2} that a possible weak
second-order scheme is (when $\mathfrak{B}_{ij}$ verifies property 
(\ref{eq:commu}))
\begin{equation}
\left\{\begin{split} 
& \td{x}_{f,i}^{n+1} = x_{f,i}^n + U_{f,i}^n\, \Delta t \\
& \td{U}_{f,i}^{n+1} = U_{f,i}^n + \mathfrak{A}_i^n\,\Delta t 
                 + \mathfrak{B}_{ij}^n\,\Delta W_i(t), \\
& U_{f,i}^{n+1} = U_{f,i}^n + \frac{1}{2}\left( \mathfrak{A}_i^n + 
        \td{\mathfrak{A}}_i^{n+1} \right)\,\Delta t + \sum_j
\frac{1}{2} \left( \mathfrak{B}_{ij}^n + 
        \td{\mathfrak{B}}_{ij}^{n+1} \right) \, \Delta W_j(t),
\end{split}\right.
\end{equation}
where
$\td{\mathfrak{A}}_i^{n+1}=\mathfrak{A}_i(t+\Delta
t,\td{x}_{f,i}^{n+1},\td{U}_{f,i}^{n+1})$ and
$\td{\mathfrak{B}}_{ij}^{n+1}=\mathfrak{B}_{ij}(t+\Delta
t,\td{x}_{f,i}^{n+1})$. This scheme is used for our present discussion
and it is different from the one developped above, cf. Table \ref{tab:sch2c}.
With this \textit{stochastic} framework in mind, some open questions remain.
\begin{enumerate}
\item[(i)]Is it possible to propose a general form of wall boundary
  conditions for the fluid particles, independently of our particular
  Langevin model?  
\item[(ii)] What boundary condition ensures that the impermeability
  condition is valid just at the wall (as in the real world)? 
\item[(iii)] What is the order of accuracy of the boundary conditions
  in the frame of our numerical schemes?
\item[(iv)] Is it possible to propose high-order (second-order) boundary
  conditions?
\end{enumerate} 
The above questions might seem easy to answer at first glance, but
the subtleties of stochastic calculus make these open issues
difficult to solve. At present, only one proposition has been made
\cite{Pop_98}. In that work, the authors have proposed to impose a
zero velocity to the particles reaching the wall during a time step
and to move them in space by symmetry at the wall. The proposed
treatment is sensible, but it presents some shortcomings, i.e. it
remains dependent on the particular model used (a Wiener process was
used in the equation for position to account for viscous
effects in the vicinity of the wall), and the order of accuracy is not
given.  A mathematical approach to this problem can be found in the
book of \"Ottinger \cite{Ott_96}.

The scarcity of the literature on this subject calls for future
rigorous development in the formulation of boundary conditions at the
wall in one-point PDF methods. A good illustration of the lack of
knowledge will be presented in Section \ref{sec:exemples} for a
computational example of particle deposition phenomena. 

%-------------------------------------------------------------------------------
\subsection{New hybrid methods}
%-------------------------------------------------------------------------------
In the present work, a hybrid method has been used: the fluid
is described with a mean-field (RANS) approach whereas the statistics
(the pdf) of the discrete particles are reproduced by
introducing stochastic particles (SDEs).

In stand-alone methods for one-point PDF models for single-phase
turbulent flows, it is known that the bias (cf. Section
\ref{SS_Num_Err}) is the main concern in the control of the numerical
error \cite{XuPope_99}. If this is also the case for the numerics put
forward for one-point PDF models for discrete particles, every idea
improving this shortcoming is welcome. A solution could be to resort
to VRT that have been developed in
disparate fields. A possibility could be to resort to a hybrid algorithm
for the numerical treatment of the discrete particles where some
variables could be solved by a mean-field method and others by a PDF
method. In such configurations, duplicate fields usually arise, and
consistency conditions must be imposed. These consistency conditions
give the opportunity of introducing VRT. Indeed, the mean
variables computed from a mean-field method are by construction not
biased. If the PDF method contains the evolution in time of the
corresponding instantaneous variables, the operation of centering the
moments extracted from the PDF approach with the ones computed from
the mean-field algorithm leads to an excellent reduction of variance
\cite{Pop_95}. Moreover, it would be helpful to find a criterion,
in the frame of domain decomposition, in order to use the mean-field
or PDF algorithms where it is most appropriate. For example, in some
parts of the flow where the knowledge of some mean-fields is
sufficient, one would resort to mean-field algorithms whereas in other
regions, where the physics are complex and the pdf is needed, one
would use a PDF algorithm. In such an approach, the central issue
becomes the consistency at the boundaries between the contiguous
domains. Some work in that sense has been carried out in the field of
Direct Monte Carlo Simulations (DSMC) \cite{Gar_99}.

At last, in the present work, the proposed particle algorithm (the
numerical algorithm for the set of SDEs, cf. Section \ref{sec:algo})
is compatible with other approaches for the fluid. It is one of
the strong points of this numerical
method. Therefore, it is conceptually possible to think about some
other configuration and in particular to a LES/PDF one, i.e. the set
of SDEs is provided with filtered fluid fields instead of mean fluid
fields. Even though, in such a configuration, an increase of the
computational effort is expected, the quality of the results, in
particular, for cases where RANS models are known to be inadequate,
could be improved, cf. the computational example for particle deposition 
in Section \ref{sec:exemples}. In such an algorithm, the challenge is
to reconstruct the subgrid scale fluid velocity along the discrete
particle trajectories \cite{Boi_00,Bel_04,Poz_04a}. A possibility is to
use PDF methods; it has been attempted in single-phase flows
\cite{Pop_02} and it remains to be developed for dispersed two-phase
flows.

%===============================================================================
\section{Computational examples} \label{sec:exemples}
%===============================================================================
Three numerical computations of polydispersed turbulent two-phase
flows are now presented. The first one (\textit{swirling flow}) is
chosen in order to show that significant improvements in the
computing efficiency can be achieved by using, for the integration of
the set of SDEs, a second-order scheme instead of a first-order
scheme. The second computation (\textit{bluff-body flow}) demonstrates
the ability of the models to capture the main physics of the flow and
the specificity of PDF models from which valuable information
can be extracted. The third example (\textit{particle deposition})
illustrates the ability of PDF models to treat flows were complex
physics are involved.

%modif EP (début)
\ifnum\elscolor>0 \color{blue} \fi
\ifnum\elsitalic>0 \itshape \fi
In the first and third examples, i.e. swirling flow and particle
deposition, both flows are dilute enough so that only one-way coupling
is under consideration. The numerical schemes presented in Section
\ref{sec:cond} can therefore be used directly. In the second example,
bluff-body flow, the suspension is rather dense and one has to take
into account two way-coupling. Numerically, for the integration in
time of the set of SDEs, this is done by resorting
to the first-order scheme and by treating the coupling term,
cf. Eq. (\ref{eq:Ap_s}), as an explicit source term. Collisions, that
might occur in some restricted areas of the computational domain,
are not taken care of. It is, however, fully possible to treat
collisions between discrete particles in the frame of the present
PDF approach. This has been discussed elsewhere \cite{Min_01} on a
theoretical basis and the inherent numerical developments remain to be
done.
\normalfont
\color{black}
%modif EP (fin)
%-------------------------------------------------------------------------------
\subsection{Swirling flow} \label{sec:cyclone}
%-------------------------------------------------------------------------------
In this particular example, no comparison with experimental data is
attempted since the purpose of the computations is to show the
benefits of using second-order schemes instead of first-order schemes
for the integration of the set of SDEs.

%----------------------------------
\subsubsection{Experimental setup}
%----------------------------------
The turbulent polydispersed two-phase flow under
investigation corresponds to a gas-solid flow (air and solid
particles) in a cyclone of the Stairmand type \cite{Sta_51}, see
Fig. \ref{fig:cyclone}. Cyclone separators are devices used to
separate particles from gas flows. The gas flow inside the cyclone has
quite complicated patterns, that is a reverse swirling flow with quite
high rotational velocities. The swirl is created by the tangential
inlet, but it is well-known from experiments, that the gas flow
exhibits a double helix structure. The flow spirals downwards (with a
constant intensity) to the vortex finder (exit tube of the cyclone at
the bottom) where it reverses and spirals upwards in a cylindrical
volume having roughly 
the diameter of the exit. In such a device, the separation between air
and particles is not due to gravity but to the effect of the double
helix. Indeed particles entering the device are entrained towards the
outer wall (by centrifugal forces) where they flow downwards to the
exit (the axial velocity of the gas is oriented downwards at the
walls).

The efficiency of a cyclone is characterised by its selectivity
curve. This curve expresses the ratio (in mass) of captured particles
as a function of their diameter. For very small particles, this curve
goes to zero efficiency as their inertia decreases ($\tau_p \rightarrow 0$),
i.e. particles tend to behave as fluid elements. On the
contrary, large particles are all collected and the efficiency
converges to $1$. Between these two asymptotic cases, the cyclone
efficiency is an increasing function of the particle diameter.

%------------------------------------
\subsubsection{Numerical simulations}
%------------------------------------
The simulated cyclone \cite{Boy_83,Boy_86} has a diameter 
$D=0.2\,{\rm m}$,
the glass particles have  a density of $\rho_p=2500\;{\rm kg/m^3}$ and
diameters ranging between $0.5$ and $5\;{\rm \mu m}$. The gas (air at
ambient temperature $\simeq 293 \,{\rm K}$) is injected with a constant
velocity of $30 \,{\rm m\cdot s^{-1}}$.
In the present simulations, the flow is dilute enough (the mass of
particles  per unit of gas is quite low) not to consider two-way
coupling effects  (the particles have no influence on the flow
field). In addition, it is well established from experiments that such
flows are stationary. Consequently, the flow field is computed
in advance and all particles are tracked in a \textit{frozen}
field. The prediction of the flow field is rather challenging given
the complex structure of the flow. In this work, a second-order
turbulence model (Rotta model \cite{Pop_00}, which is consistent with
the form of the SDEs) was used with a fine grid (approximately
$4\cdot10^{5}$ nodes) in order to obtain mesh-independent calculations. Figures
\ref{fig:gas_axe} and \ref{fig:gas_rad} show the axial and radial mean
velocity profiles of the air flow at two different heights: it can be
seen that the numerical results are in good agreement with the
measurements.

Particles are then tracked in this frozen field. The diameter range of
the glass particles has been discretised as follows,
$d_p=[0.5;1;1.5;2;3;4;5]\;{\rm \mu m}$. For each class (diameter of
particles), a number $N_{pc}$ of particles is released (this number is 
identical for each class). The computation stops when all particles
have left the computational domain.

%-------------------------------------
\subsubsection{Results and discussion}
%-------------------------------------
A numerical study has been carried out to show that the results (the
obtained selectivity curves) are independent of the time step, $\Delta
t$, and the number of particles per class, $N_{pc}$. For instance, for
the second-order scheme, the computations show, for two different time
steps, that roughly $400$ particles per class are necessary to obtain
selectivity curves which do not depend on $N_{pc}$, see
Fig. \ref{fig:res_2}. The time step of $\Delta t=10^{-4}\;{\rm s}$
guarantees that the results do not depend on the time discretisation,
Fig. \ref{fig:res_2}. For the first-order scheme,
similar results are obtained : $\Delta t=5\cdot10^{-6}\;{\rm s}$ and
$N_{pc}=400$ ensure that the results depend neither on the time step
nor on the number of particles per class, Fig. \ref{fig:res_2}.

It is then observed that, for this particular flow, there is a great
difference between the respective time steps for the first and second
order schemes, see Fig. \ref{fig:res_2}, and this for the same
numerical results (selectivity curve). In fact, with a second-order
scheme, the time step can be multiplied by a factor $20$ compared to a 
first-order scheme. If one accounts for the computer time (the
computation of a time step takes approximately $30\,\%$ extra time
compared to the first-order scheme), there is a gain in CPU time by a
factor $15$ by using a second-order scheme. Therefore, in this case,
it is seen that the complexity of the second-order scheme is balanced
by the reduction of computing time.

%modif EP (début)
\ifnum\elscolor>0 \color{blue} \fi
\ifnum\elsitalic>0 \itshape \fi
It can be stated that, for flows where velocities and the curvature of
the trajectories of the discrete particles are important, it is
recommended to use a second-order scheme rather than a first-order
one, unless one is ready to pay the computational price. This is
clearly seen in this computational example where the time step of the
first-order scheme is extremely small: as a matter of fact, in such a
flow, a precise prediction of the particle velocity is needed because
the numerical error on this quantity amounts at simulating an
additional centrifugal force.
\normalfont
\color{black}
%modif EP (fin)

%-------------------------------------------------------------------------------
\subsection{Bluff body flow}
%-------------------------------------------------------------------------------
In this second example, the numerical results are compared to
experimental data in order to show the ability of the present approach 
to reproduce the main trends of complex turbulent polydispersed two
phase-flows. Other results are displayed to enhance the specificity of
the mean-field/PDF approach, that is the type of information which can
be extracted.

%--------------------------------------------------------
\subsubsection{Experimental setup} \label{sec:experiment}
%--------------------------------------------------------
The 'Hercule' experimental setup \cite{Ish_99,Bor_01} is
characteristic of pulverised coal combustion where primary air and
coal are injected in the centre and secondary air is introduced on the
periphery, Fig. \ref{fig:hercule}. This is a typical bluff-body flow
where the gas (air at ambient temperature 
and atmospheric pressure) is injected both in the inner region (jet) and the
outer cylinders (exterior). The ratio between the gas velocity in the inner region,
$U_j$, and the gas velocity in the outer region, $U_e$, is low enough so
that a recirculation zone downstream of the injection is created. Two
honeycombs are used in the experimental setup in order to stabilise
the flow so that no swirl is present. Solid particles (glass spheres)
are then injected from the inner cylinder with a given mass flow
rate. The injected glass spheres have a density $\rho_p=2470\,{\rm kg/m^3}$
and a known diameter distribution, typically between $d_p=20\,{\rm \mu m}$
and $d_p=110\,{\rm \mu m}$ around a mass-weighted average of $d_p\sim
65\,{\rm \mu m}$. A polydispersed turbulent two-phase flow which is
\textit{stationary} and \textit{axi-symmetric} is then
obtained. Moreover, two-way coupling takes place since the particle
mass loading, $\phi=\chi\,\lra{U_p}/\lra{U_f}$, at the inlet is high
enough.
Experimental data is available for radial profiles of different
statistical quantities at five axial distances downstream of the
injection, Fig. \ref{fig:hercule} (axial profiles along the axis of
symmetry have also been measured).
The 'Hercule' experimental setup is a very interesting test case for
polydispersed turbulent two-phase flow modelling and numerical
simulations since most of the different aspects encountered in such
flows are present. The particles are dispersed by the turbulent flow
but in return modify it. Furthermore, the existence of a
recirculation zone (with two stagnation points, $S_1$ and $S_2$ in
Fig. \ref{fig:hercule}) where particles interact with negative axial fluid
velocities constitutes a much more stringent test case compared to
cases where the fluid and the particle mean velocities are of the same
sign (the problem is then mostly confined to radial dispersion
issues). These features are displayed in Fig. \ref{fig:hercule} where mean
streamlines are shown (solid lines for the fluid and dashed lines for
the particles). For the fluid, there is a rather  large recirculation
zone with stagnation points. For the particles, depending on their
inertia, several behaviours can be observed: some particles do not
'feel' the recirculation zone and leave the test section
immediately. Others are partially influenced and change direction
before leaving the apparatus, whereas some particles follow closely
the recirculation pattern. This will be seen in the
results showing the pdf of the particle residence time at different
locations in the flow.

%----------------------------------------------------------
\subsubsection{Numerical simulation} \label{sec:simulation}
%----------------------------------------------------------
A two-dimensional, single block, non-Cartesian, non-uniform mesh (142 x 3
x 75 nodes for the $(x,r,\theta)$ coordinate system) has been
generated in accordance with the axi-symmetric property of the flow. 
It was carefully checked that the results are not too sensitive
either to the time and spatial discretisations or to the number of
stochastic particles ($\Delta t=10^{-3}\,{\rm s}$ and $N=14000$
particles). A second-order turbulence model (Rotta model, which is
consistent with the form of the SDEs) was used. The projection and
averaging operators were approximated with a NGP technique
\cite{Hoc_88}. For further details on the numerical computations,
see Ref. \cite{Min_01r}.

In the simulations, the following procedure is adopted. The
single-phase flow case is first computed until the stationary state
is reached. By
doing so, it is possible to check that the prediction of the flow
field, without the particles, is accurate. Then the discrete particles
are introduced until the stationary state is obtained again. At that
point, the number of particles in the flow is roughly constant (it
fluctuates around a mean value). From this state, computations are
continued to extract the statistics which are compared to the
experimental values. The last computation is performed to allow time
averaging on the ensemble averages so that the statistical noise can
be reduced to a minimum (VRT).

%---------------------------------------------------------
\subsubsection{Results and discussion} \label{sec:results}
%---------------------------------------------------------
Figures \ref{fig:vitz}, \ref{fig:vitx} and \ref{fig:vitzf} show that,
in this particular flow, there is almost no difference between the
predictions with the first and second-order schemes. The main
differences take place in regions where only few particles are present
(large diameters) and consequently the statistical results contain some 
noise (see the ragged behaviour of the curves).

Three sets of results are given: (i) radial profiles of the particle
axial velocity, Fig. \ref{fig:vitz}, (ii) radial profiles of the
particle radial velocity, Fig. \ref{fig:vitx} and (iii) radial
profiles of the fluid axial velocity, Fig. \ref{fig:vitzf}.
All sets of numerical results
compare relatively well with the experimental data, in terms of the
shape of the curve and of the magnitude which are observed.
The recirculation zone ($x=0.16 \,{\rm m}$) is well predicted as indicated by
the velocity profiles. All (mean and fluctuating) velocities go to
zero when no particles are present. The widths of the numerical curves
indicate that the predicted radial dispersion of the particles is
also in line with experimental findings.

In Figs. \ref{fig:vitz} to \ref{fig:vitzf}, only first and second-order
moments of the variables of interest have been displayed (mean and
fluctuating velocities for the fluid and the particles). This
information could have been obtained by resorting to classical
mean-field equations. However, in many engineering
applications (for example combustion), it is necessary to know the
distribution of the residence time of the particles at a certain
time. In other words, one would like to know for all the particles
found in a certain zone how much time they have spent inside the
domain, or even if they have previously entered a marked region. This
kind of information is not available in a mean-field model whereas in
the present hybrid approach, this information is directly provided
without additional costs: the pdf of the variables attached to each
particle, which contains far more information than a few moments, is
explicitly computed.

A typical example of this type of information is given in
Fig. \ref{fig:res_time}. In the first plot, on the left-hand side, a
snap shot of the local instantaneous positions of the particles is
given, where particles are coloured by their residence time. Two
distributions are extracted, one in a cell close to the inlet and the
other one in a cell close to the outlet. The pdf in the cell near the
inlet clearly shows the recirculation pattern: the distribution is
highly peaked, which represents particles which have just entered the
domain, but a small number of particles have a quite high residence
time, i.e. they recirculate. At the outlet, since the
particles have different trajectories in the domain, a continuous
spread in residence time is observed. In the region near the inlet,
more information can be gathered, for example the local instantaneous
axial  velocity, see the RHS graph in
Fig. \ref{fig:res_time}. Most particles have the same axial velocity,
around $4\,{\rm m/s}$ which is actually the inlet velocity. These particles
correspond to the peak observed in the residence time: they have just
entered the domain and travel directly to this region. The rest of the
particles have a smaller axial velocity but with much wider
fluctuations: these particles correspond to particles which are
recirculating.

At last, it is often argued that the mean-field/PDF approach is
time-consuming. As a matter of fact, in this computation, the time
spent by both solvers (the mean-field solver for the fluid and the
PDF solver for the particles) has been compared for the
\textit{same number of computational elements} (mesh points and
particles, respectively). It is found that the PDF solver is
slightly faster: this is not really surprising since the mean-field
fluid solver implies the use of a full second-order turbulence model
(6 coupled PDEs).

%-------------------------------------------------------------------------------
\subsection{Pipe flow: deposition}
%-------------------------------------------------------------------------------
In this last example of numerical applications with the mean-field/PDF
approach, a flow where complex physics are involved,
\textit{i.e. particle deposition}, is under investigation. Particle
deposition from a turbulent flow on walls is a phenomenon which is
observed in many engineering applications (for example thermal and
nuclear systems, cyclone separators, spray cooling) and also in
various environmental situations. Given the large number of possible
applications, a lot of interest has been devoted to this subject
and many studies have been carried out in the last decades.

%--------------------------------------------------------
\subsubsection{Experimental setup}
%--------------------------------------------------------
Different experiments have been conducted to observe deposition
in turbulent flows. In most of them, attention is focused on the
deposition velocity~\cite{Liu_74,McC_77} which is defined as 
$k_p = m_p/\bar{C}$, 
where  $m_p$ is the mass flux and $\bar{C}$ is the bulk mean particle
concentration. This deposition rate, often presented as 
the dimensionless deposition velocity $k_p/u^*$, is a function
of  the dimensionless particle relaxation time,
$\tau_p^+$, defined as
\begin{equation} \label{taup}
\tau_p^+ = S^+\frac{u^*}{U_{p0}} = \frac{d_p^2 \,\rho_p\, U_{p0}
  \,u^*}{18\,\mu_f\,\nu_f}\,\frac{u^*}{U_{p0}} = \frac{d_p^2 \,\rho_f^2
  \,u^{*2}}{18\,\mu_f^2}\,\frac{\rho_p}{\rho_f},
\end{equation}
where $S^+$ is the dimensionless stopping distance, $U_{p0}$ is the
particle initial velocity and $u^*$ is the friction velocity. $u^*$ is
evaluated with the Blasius formula, $u^* = [0.03955\,Re^{0.25}]^{0.5}\,U_m$, 
where $U_m$ is the bulk mean velocity. The deposition velocity is the
key point in many engineering applications where one seeks the law
that gives $k_p/u^*$ as a function of $\tau_p^+$, that is as a
function of the particle diameter. Recently, several experimental
studies and DNS studies of particle deposition have been presented,
for example~\cite{Eat_94,Van_98}, and have improved the understanding
of the physical mechanisms at play. In particular, a lot of information
has been obtained on the dynamical structures of wall-bounded
flows, like the coherent structures which manifest themselves in the
near-wall region.

In the present computational example, the principal interest is to
show the advantage of solving the set of SDEs with a numerical scheme
consistent with all limit cases, see Sections \ref{sec:scale},
\ref{sec:sch1} and \ref{sec:sch2}. Indeed, in pipe flows with particle
diameters ranging from $~1\,{\rm \mu m}$ to $100\,{\rm \mu m}$, all
limit cases can be encountered:
\begin{enumerate}
\item[1.] for the smallest particles ($\tau_p \to 0$ in the continuous
  sense and $\tau_p \ll \Delta t$ in the discrete sense), limit case 1
  is obtained,
\item[2.] in the near wall region, i.e. $T_{L,i}^* \ll \Delta
  t$, for example in the peak-production region where turbulent kinetic
  energy is maximal, one has $B_{ij}\,T_{L,i}^* ={\rm cst}$, a situation
  which is characteristic of limit case 2,
\item[3.] in the same region as above with small particles, one has
  limit case 3,
\item[4.] in the close vicinity of the wall, i.e. $T_{L,i}^*
  \ll \Delta t$, with no condition on the other moments, the flow is
  laminar, that is limit case 4.
\end{enumerate}
The need to cope with limit cases is the result of practical
considerations. Indeed, if the numerical scheme were not consistent
with the limit cases, it would be very inefficient to carry out
computations with a time step limited by the smallest timescale.

In the present work, numerical simulations corresponding to the
experimental setup of Liu and Agarwal~\cite{Liu_74} are
presented, i.e. the deposition of particles ($920 \,{\rm kg/m^3}$ in
density and diameters in the range $1.4$ to $68.5 \,{\rm \mu m}$) in a
vertical pipe flow at a Reynolds number of $10^4$.

%------------------------------------
\subsubsection{Numerical simulation}
%-------------------------------------
In order to describe the particle phase, $10^4$ stochastic particles
(distributed in 10 diameter classes, cf. Table \ref{tab:sej1}) are
released in a frozen field, i.e. the flow is stationary
and dilute enough so that only one-way coupling is under
consideration. Two frozen fields are computed with a standard
$k-\epsilon$ turbulence model and with a $R_{ij}-\epsilon$ (Rotta model)
both with wall-function boundary conditions. The computations are
performed with a 2D mesh, 100 x 20 x 3 nodes (the flow is
axisymmetric).

To compute the deposition velocity, the fraction of particles
remaining in the flow, $F$, is evaluated as a function of the axial
position $x$ \cite{Mati_00}. $F$ is calculated as the number of
particles that reach the sampling cross-section, divided by the total
number of released particles. The particle deposition velocity is then
computed as follows \cite{Mati_00}
\begin{equation}
k_p = \frac{U_m D_h}{4(x_2 - x_1)}\,ln\left(\frac{F_1}{F_2}\right),
\end{equation}
where $D_h$ is the diameter of the pipe and $F_i$ is the value of $F$
at a given sampling cross section labelled $i$ (axial position $x_i$). 

Numerical tests have been performed to check that the numerical
results are independent of the values of the numerical parameters, in
particular the number of particles, $N$, and the time-step, $\Delta
t$. It was checked beforehand that the numerical prediction of the
fluid field is grid-independent.

Both numerical schemes (first and second order) were tested with
different time steps, cf. Fig.~\ref{dt}. All computations
were then performed with the weak second-order scheme and a time step
of $10^{-4}\,{\rm s}$. Indeed, Fig. \ref{dt} shows that both schemes give
similar results and that a time step $\Delta t = 10^{-4}\,{\rm s}$ ensures
that the computations are independent of $\Delta t$. The independence
of the deposition velocity for the whole range of particle diameters
with respect to the time step illustrates the benefits of a numerical
scheme which is consistent in all limit cases: for instance, the
values of the particle relaxation timescales given in Table~
\ref{tab:sej1} cover three orders of magnitude (limit case
1). Computations can anyway be carried out with by using the same
constant time-step for all classes and in the whole domain.

An analysis of the statistical error has also been carried out. Since
particle deposition velocities are calculated by a Monte Carlo method,
it is important to check that the number of particles (which represents
samples of the pdf) is sufficiently large so that statistical error
is reasonably small. In Fig.~\ref{part}, results obtained with three different
values of $N_{pc}$ (the number of particles used for each class of
diameter) are presented ($N_{pc} = 500, 1000$ and $5000$).
Fig.~\ref{part} shows that there is no clear difference between the
results obtained with different values of $N_{pc}$. As a matter of fact,
it seems that 500 particles for each class of diameter is
enough. Nevertheless, for all following simulations, the value of $N_{pc} =
1000$ particles for each class of diameter has been chosen, in order
to reduce statistical noise.

%----------------------------------------
\subsubsection{Results and discussion}
%----------------------------------------
The numerical results are now compared to experiments and some
sensitivity tests are conducted. Some proposals are put forward for
the features which seem to be the most significant ones for a good
representation of deposition phenomena. Possible improvements
of the computational method shall be exposed.

In Fig. \ref{turb_dep}, results obtained with the general PDF model,
Eqs.~(\ref{exa_part_eqns}) and (\ref{eq:dUs}), and two turbulence
models ($k - \epsilon$ and $R_{ij}-\epsilon$), are displayed. The
difference between the simulations performed with the two different
turbulence models is negligible: this is not too surprising, since
for turbulent pipe flow, both models give similar mean fluid velocity
profiles. The standard PDF model is integrated with wall-function
conditions for the fluid and pure-deposition boundary conditions for
the particles. These results are coherent with those obtained in an
analogous configuration by Schuen \cite{Sch_87}.

Figure \ref{turb_dep} shows that, for heavy particles ($\tau_p^+ > 10$),
the model predictions are in good agreement with experiments whereas
for light particles ($\tau_p^+ < 10$), the deposition velocities are
strongly overestimated (they remain at the same level as that of the
heavy particles). Therefore, the model is not suitable for simulations
of deposition phenomena in the range $\tau_p^+ < 10$. This statement is
consistent with experimental and DNS findings~\cite{Nar_03}: heavy particles are
slightly affected by near-wall boundary layer and more especially by
the specific features of the local instantaneous
turbulent structures in the near-wall region. On the contrary, for
light particles, the physical mechanism of deposition changes, with a
growing importance of turbulent structures and near-wall physics. 
In the current PDF model, near wall physics are mainly described by
wall-function boundary conditions which may be sufficient for heavy
particles deposition but not for light particle deposition.

Wall-functions give a reasonable approximation of the mean fluid
velocity profile in the logarithmic region, but in any case they do
not describe the viscous sub-layer. Therefore, a question arises: is
the prediction of small particle deposition velocity sensitive to
changes in the fluid mean-field profiles? This matter was recently
investigated for other Lagrangian models~\cite{Mati_00}. Following the
same reasoning, simulations have been carried out with a given frozen
field (consequently, wall-function boundary conditions are
suppressed). The frozen field can be obtained, in this particular case,
either from analytical solutions for the mean fluid fields ($\lra{{\bf
    U}},k,\lra{\epsilon})$~\cite{Mon_75} and/or from DNS data.
In Fig. \ref{viscous}, two frozen fields are tested. In the first field,
the axial mean fluid velocity $\lra{U_{i}}$ is given by the
law-of-the-wall equations (the values $k$ and $\epsilon$ are that of
the computations). In the second field, $\lra{U_{i}}$ is still given
by the law-of-the-wall, and the turbulent kinetic energy, $k$, and the
turbulent dissipation rate, $\epsilon$, are curve-fitted to the DNS
data that can be found in the work of Matida et
al.~\cite{Mati_00}. Thus, in the second field, mean fluid profiles are
exact. Figure \ref{viscous} shows that an exact frozen field hardly
improves the results.  An explanation might be that the eventual
effect of the exact mean fluid profiles are concentrated in a very
thin region. The most important quantity is expected to be the
turbulent kinetic energy, which goes to zero at the wall and should
affect mainly light particles.  Nevertheless, $k$ diminishes only 
from $y^+ \approx 10$, where it has its maximum (peak production).
The resulting effect is not easy to be foreseen and it may be
negligible with respect to the overall effect of migration of
particles towards the wall due to the net mean flow.

In order to further support the argument above, mean near-wall
residence times of deposited particles have been computed in the layer
$y^+ < 30$, for each class of diameters. Indeed, this quantity has 
been found to properly distinguish different deposition
mechanisms~\cite{Nar_03}. A rough description of the physics of deposition
is that heavy particles, which are slightly influenced by near-wall
structures, deposit with small near-wall residence times by the so
called {\it free-flight} mechanism. On the contrary, light particles
are trapped and driven by turbulent structures and depose with large
near-wall residence times, this mechanism is called {\it diffusional}.
The lighter the particles are, the more important the diffusional
mechanism is. In Table \ref{tab:sej1}, the results obtained for each
class of diameters are given for a simulation corresponding to the
exact frozen field.  For the sake of clarity, the residence time is
always expressed in non-dimensional form (it is normalised with the
viscous timescale, $\nu_f/(u^*)^2$, $t^+=t\,(u^*)^2/\nu_f$). Table
\ref{tab:sej1} shows that all particles deposit after small near-wall
residence time, that is by free-flight mechanism. Moreover, since the
residence time slightly increases with $d_p$, the motion of particles
is influenced by the migratory flux. The force exerted on the
particles by the fluid being inversely proportional to $d_p$, light
particles reach the walls faster than the coarse ones. Therefore, in
absence of a representation of turbulent coherent structures (which
should be able to trap particles in the near-wall region and which
should describe correctly the mechanisms of deposition), the sole mean
fluid profiles are not the main mechanism.

Two possibilities exist to improve the prediction of deposition
phenomena and in particular of light particles:
\begin{enumerate}
\item[(i)] some phenomenological model can be introduced, based on the
  present knowledge of deposition physics. In this type of approach
  some hypotheses are made in accordance with experimental
  findings. Some parameters may be present and may be fitted in order
  to find good comparison with experiments. This approach ought to
  verify if the hypotheses made are correct or not, and, thus, ought
  to show which are the dominant aspects not covered by the standard
  model.
\item[(ii)] it is also possible to propose extensions of the present
  PDF model that reproduce correctly the variations of the fluid
  statistical moments (such as $\lra{{\bf U}}$, $ R_{ij}$,
  $\epsilon$, etc.) throughout the near-wall region, including the
  viscous sublayer~\cite{War_04}. Such a model might lead to
  improvement for the deposition velocity. However, it would require a
  very refined mesh in the near-wall region, considering the high
  value of the Reynolds number. Furthermore, only the statistics of
  the fluid would be well reproduced and, as mentioned
  above, it is believed that the contribution of the specific
  features of coherent structures should be considered for small
  particle deposition. 
\end{enumerate}
Therefore, for practical purposes, the first proposition has been
retained~\cite{Chi_03}. This subject is not further developed here
and it is left as a challenging open issue.

%===============================================================================
\section{Conclusions and perspectives}
%===============================================================================
In this paper, we have presented a comprehensive review of the
numerical methods involved for the computation of polydispersed
turbulent two-phase flows using a particle stochastic method based on
Langevin equations. The present mean-field/PDF model is one among a
host of other so-called Euler/Lagrange models. However, it is worth
putting forward two main specific aspects of the present framework. 
\begin{enumerate}
\item[(i)]The usual term Euler/Lagrange refers to the point of view
  adopted for the description of the two phases: an Eulerian point of
  view for the fluid phase and a Lagrangian one for the particle
  phase. This terminology can be misleading and does not clearly
  identify the physics involved, the level of information contained in
  the statistical description and the numerical tools which are
  adopted. For example, the so-called Eulerian equations for the fluid
  phase can be directly obtained through the two-particle stochastic
  formulation sketched in Section \ref{sec:two_point} and Appendix
  \ref{sec:appen}, and its numerical solution may involve Lagrangian
  ideas (for instance, in the method of characteristics for the
  discretisation of the convection terms). And, at least in
  theory, the PDF equation could also be solved using a mesh and an
  Eulerian description in phase space. 
  In the present work, the complete model is called a
  \emph{mean-field/PDF approach}. This refers directly to the level of
  information contained in the description: the fluid phase is
  described by a limited number of statistical moments (in practice,
  at most two for the fluid velocity) while the particle phase is
  characterised by the PDF of the variables retained for its
  description. The complete model is therefore a \emph{hybrid
    model}. The hybrid nature of the model is then reflected in the
  numerical approach developed in Section \ref{sec:num0}. The fluid
  mean fields are computed as the solution of PDEs, involving a mesh
  and, say, classical Finite Volume schemes. As explained in Section
  \ref{sec:formalism}, the PDF equation is solved, in a weak sense, by
  a particle Monte Carlo method where the particles should be seen as
  instantaneous realizations of this PDF rather than real
  particles. The numerical approach is thus an hybrid
  \emph{PDE/Particle Stochastic} method, or a \emph{Mesh/Particle
    Stochastic} method; its specificities have been discussed at length in
  Section \ref{sec:num0}. The various details treated in that Section
  can be improved, but the important point is that they are developed
  once a clear framework about the complete hybrid method is first set
  forth. This framework is a necessary guideline.   

\item[(ii)]Drawing on these first remarks, the second aspect of the
  present formulation is the fact that there is a separation between
  the theoretical construction of the model and its numerical
  solution. Indeed, given the correspondence (in a weak sense) between
  the PDF equation and particle stochastic equations, as explained in
  Section \ref{sec:formalism}, it may be tempting to develop directly
  the model in discrete time. This amounts to treating without
  distinction the model and its numerical scheme. This may be
  confusing  and may not help to identify the actual issues. In the
  present work, the theoretical stochastic model is developed and
  first written in continuous time. This requires knowledge of the
  mathematical background of stochastic diffusion, but actually this
  effort is a valuable investment and the formulation in terms of the
  particle trajectories of the stochastic process in continuous time
  simplifies the situation and the numerical developments.
\end{enumerate}

The central point of the complete work presented in Section
\ref{sec:pdf_models} concerns the multiscale character of the
stochastic theoretical model which is then reflected in the numerical
developments of Section \ref{sec:num0}. Placed between the
mathematical background and the numerical implementations, this
Section illustrates the interplay between mathematical formulation,
physical modelling and numerical developments. The mathematical
manipulation of the system of equations reveals the property of the
model: different limits are continuously reached depending upon
the values of the observation timescale with respect to the different
physical timescales. These limits correspond to natural physical
diffusive limits and they point out that some physical variables are
not real white-noise terms, but that their effects may be regarded as
such \emph{at a certain scale}. Using the numerical time step as the
observation timescale, this physical property appears in turn as a basis
for the development of the numerical schemes which, while being
explicit and stable, can satisfy these limits without any constraint
or threshold on the time step.

Finally, the main purpose of this work has been to propose a
consistent and specific framework for the simulation of polydispersed
two-phase flows based on Langevin stochastic equations. Together with
the presentation of the theoretical aspects~\cite{Min_01}, it provides
a comprehensive description of the model and of the numerical
ideas. It does not pretend to be the ultimate word in this field and much work
remains to be done. The Langevin equations still require new
developments~\cite{Min_01} and, on the numerical side, boundary
conditions must be properly addressed, see Section
\ref{sec:boundary}. Yet, it is hoped that the present framework paves the
way for the improvement of current methods as well as for the
formulation of new ideas. In particular, new hybrid methods may
benefit from these first steps, by trying to go further into a mixed
mean-field/PDF approach within the description of the particle phase
itself. This will require a good understanding of the consistency
between the mean field equations satisfied by particle statistical
properties and the instantaneous stochastic equations for the
trajectories, of their mathematical manipulation and of the issues
involved with particle/mesh exchange of information.

%===============================================================================
\addcontentsline{toc}{section}{Acknowledgements}
\section*{Acknowledgements}
%===============================================================================
Dr. Chibbaro's research was supported by a Marie Curie Transfer of
Knowledge fellowship of the European Community programme, Contract
No. MTKD-CT-2004-509849. The authors would like to thank Dr. Mehdi
Ouraou for his fruitful collaboration in numerical simulations and
Prof. Denis Talay for his precious help in developing numerical
schemes.

%===============================================================================
\pagebreak
\include{appendix}
%===============================================================================

%+++++++++++++++++++++++++++++++++++++++++++++++++++++++++++++++++++++++++++++++
\newpage
\addcontentsline{toc}{section}{References}
\bibliographystyle{unsrt}
\bibliography{%
\ch{article},%
\ch{book},%
\ch{reports},%
\ch{thesis},%
\ch{proceed},%
\ch{math},%
\ch{fluidisation},%
\ch{deposition}}
%+++++++++++++++++++++++++++++++++++++++++++++++++++++++++++++++++++++++++++++++

%+++++++++++++++++++++++++++++++++++++++++++++++++++++++++++++++++++++++++++++++
\addcontentsline{toc}{section}{List of symbols}
\include{symbols}
\addcontentsline{toc}{section}{Figures}
\include{figures}
\addcontentsline{toc}{section}{Tables}
\include{tables}

\ifnum\elsevier>0 
\listoffigures
\listoftables
\fi
%+++++++++++++++++++++++++++++++++++++++++++++++++++++++++++++++++++++++++++++++

\end{document}

%% file: appendix.tex
%===============================================================================
\appendix
\section{Appendix: two-point description} \label{sec:appen}
\ifnum\elsevier>0 
\numberwithin{equation}{section}
\fi
%===============================================================================
Here, some additional information is given on the construction of the
two-point description, i.e., the form of the acceleration term to be
added in the Langevin equation describing the velocity increments
along the trajectory of a fluid particle. Once this is done, it is
briefly explained how the mean-field (RANS) equations for the fluid
can be extracted from this two-point description.

%-------------------------------------------------------------------------------
\subsection{Model for a two-way coupling term}
%-------------------------------------------------------------------------------
In the exact local instantaneous equations for the fluid (the
Navier-Stokes equations), a formal treatment of the force exerted on
the fluid by the discrete particles implies the use of a distribution
(or density of force) acting on the fluid located in the neighbourhood
of the discrete particles in order to express the resulting
acceleration on nearby fluid particles. This accurate treatment, which
would result in a multi-point treatment of the discrete phase, is
outside the scope of the present work. Here, in the frame of the
one-point approach, the influence of the discrete particles on the
fluid is expressed directly in the SDEs, Eqs. (\ref{eq:SDEfp}), with
stochastic tools.

As explained in Section \ref{sec:pdf_p}, for 
${\bf A}_{p \rightarrow s}$, the underlying force corresponds to the
exchange of momentum between the fluid and the particles (drag
force). The acceleration acting on the fluid element surrounding a
discrete particle can be obtained as the sum of all elementary
accelerations (due to the neighbouring particles), i.e., at the
discrete particle location ${\bf x}_p$, the elementary acceleration
$({\bf U}_p-{\bf U}_s)/\tau_p$ is multiplied by $\chi=(\alpha_p
\rho_p)/(\alpha_f \rho_f)$, that is the probable mass of particles
divided by the probable mass of fluid (since the total force is
distributed only on the fluid phase).

%modif EP (début)
\ifnum\elscolor>0 \color{blue} \fi
\ifnum\elsitalic>0 \itshape \fi
For ${\bf A}_{p \rightarrow f}$, the problem of finding a suitable
stochastic model is slightly more difficult since the drag force can
only be defined in terms of variables attached to the discrete
particles (which are not defined at the location of a fluid
particle). As a consequence, the influence of the neighbouring
discrete particles on the fluid particle located, at time $t$,
at ${\bf x}={\bf x}_f(t)$, is ensured by considering
that ${\bf A}_{p \rightarrow f}$ is a random variable given by
\begin{equation}
{\bf A}_{p \rightarrow f} =
\left\{\begin{split}
& 0 \quad \text{with probability} \quad 1-\alpha_p(t,{\bf x}), \\
& {\bf \Pi}_p \quad \text{with probability} \quad \alpha_p(t,{\bf x}), 
\end{split}\right.
\end{equation}
where ${\bf \Pi}_p$ is a random variable which plays the role of an
ersatz of the Eulerian random variable which is formed from the
discrete particles at ${\bf x}={\bf x}_p(t)$
\begin{equation} \label{eq:Apf}
{\bf \Pi}_p \equiv -\frac{\rho_p}{\rho_f}\frac{{\bf U}_s-{\bf U}_p}{\tau_p}.
\end{equation}
This random term mimics the reverse force due to the discrete
particles and is only non zero when the fluid particle is in the close
neighbourhood of a discrete particle. In addition, it is required
that, at ${\bf x}={\bf x}_f(t)$, ${\bf \Pi}_p$ and ${\bf U}(t,{\bf x})$
are correlated so that 
\begin{equation}
\left\{\begin{split}
& \lra{{\bf \Pi}_p} = -(\rho_p/\rho_f)
                       \lra{({\bf U}_s-{\bf U}_p)/\tau_p},\\
& \lra{{\bf \Pi}_p {\bf U}}=-(\rho_p/\rho_f) 
           \lra{({\bf U}_s-{\bf U}_p){\bf U}_s/\tau_p}.
\end{split}\right.
\end{equation}
For further explanations on the modelling of two-way coupling, see
Refs.~\cite{Min_01,Pei_02}.
\normalfont
\color{black}
%modif EP (fin)

%-------------------------------------------------------------------------------
\subsection{Mean-field (RANS) equations for the fluid}
%-------------------------------------------------------------------------------
In sample space, Eqs. (\ref{eq:SDEfp}) are equivalent (in a weak
sense) to a general Fokker-Planck equation for the two-point
Lagrangian pdf, $p^r(t;{\bf z}_f,{\bf z}_p)$, cf. the correspondence
between Eq. (\ref{sde}) and Eq. (\ref{fokker-planck}). It can be shown
\cite{Min_01} that the Fokker-Planck equation verified by 
$p^r(t;{\bf z}_f,{\bf z}_p)$ is also verified by the two-point Eulerian
mass density function (mdf) and therefore by one of its marginals,
$F^E_f(t,{\bf x};{\bf V}_f)$. This mdf is given by 
$F^E_f(t,{\bf x};{\bf V}_f)=\rho_f \, p^E_f(t,{\bf x};{\bf V}_f)$
where $p^E_f$ is the Eulerian distribution function of the fluid. The
knowledge of the PDE verified by an Eulerian quantity allows us, using
classical tools of kinetic theory \cite{Lib_98,Cha_70}, to write field
equations for the velocity moments of the fluid: the PDE verified by
$F^E_f$ is multiplied by a given function of 
${\bf V}_f$, ${\cal H}({\bf V}_f)$. Applying the following operator
\begin{equation}
\alpha_f(t,{\bf x})\,\rho_f\,
\lra{{\cal H}({\bf U}(t,\mb{x})}= \int {\cal H}({\bf V}_f) 
F_k^E(t,\mb{x};{\bf V}_f)\,d{\bf V}_f
\end{equation}
to this PDE gives field equations for any 
$\lra{{\cal H}({\bf U}(t,{\bf x})}$. By replacing ${\cal H}$ by
${\cal H}=1$, ${\cal H}=V_{f,i}$ and ${\cal H}=V_{f,i} V_{f,j}$, the
continuity equation, the momentum equations and the Reynolds-stress
equations are obtained, respectively \cite{Pei_02}. These equations
are given in Table \ref{tab:RANS},
i.e. Eqs. (\ref{eq:feq_alphaf})-(\ref{eq:feq_ufuf}).

%%% Local Variables: 
%%% mode: latex
%%% TeX-master: t
%%% End: 

%% file: symbols.tex
%-------------------------------------------------------------
\section*{List of symbols}
%-------------------------------------------------------------
\begin{longtable}[htbp]{p{1.5cm}p{8cm}p{2cm}}
$A_i$ & drift vector & \\
${\bf A}_{p \rightarrow f}$ & acceleration in Langevin eq. for the
fluid & ${\rm m\cdot s^{-2}}$ \\
${\bf A}_{p \rightarrow s}$ & acceleration in Langevin eq. for the
particles & ${\rm m\cdot s^{-2}}$ \\
$A_{f,i}$ & drift vector defined by Eq. (\ref{eq:SDEf}) & ${\rm m\cdot s^{-2}}$ \\
$A_{s,i}$ & drift vector defined by Eq. (\ref{eq:dUs}) & ${\rm m\cdot s^{-2}}$ \\
${\cal A}_i$ & acceleration defined by Eq. (\ref{eq:sysEDS}) & ${\rm m\cdot s^{-2}}$ \\
${\cal A}_i^n$ & approximated value of ${\cal A}_i$ at $t_n$ & ${\rm m\cdot s^{-2}}$ \\
${\cal A}_i^{n+1}$ & predicted value of ${\cal A}_i$ at $t_{n+1}$ & ${\rm m\cdot s^{-2}}$ \\ 
$\mathfrak{A}_i$ & drift vector defined by Eqs. (\ref{eq:frak}) and
(\ref{eq:model_f}) &  \\
$\mathfrak{A}_i^n$ & approximated value of $\mathfrak{A}_i$ at $t_n$ & \\
$\td{\mathfrak{A}}_i^{n+1}$ & predicted value of $\mathfrak{A}_i$ at $t_{n+1}$ & \\ 
$A_1$ & coefficient defined in Table \ref{tab:sch1}& ${\rm s}$ \\
$A_2$ & function defined in Table \ref{tab:sch2c}& \\
$A_{2c}$ & function defined in Table \ref{tab:sch2c}& \\
$b_i$ & coefficient for $\td{k}$, $b_i=T_L/T_{L,i}^*$ & \\
$B$ & diffusion coefficient in Eq. (\ref{eq:langevin}) & ${\rm m\cdot s^{-3/2}}$ \\
$B_{\cdot}(N)$ & bias for variable $\cdot$ & \\
$B_{ij}$ & diffusion matrix & ${\rm m\cdot s^{-3/2}}$ \\
$\Check{B}_i$ & diagonal elements of $B_{ij}$, i.e. $\Check{B}_i=B_{ii}$ & 
 ${\rm m\cdot s^{-3/2}}$ \\
$\Check{B}_i^n$ & approximated value of $\Check{B}_i$ at $t_n$ & ${\rm m\cdot
s^{-3/2}}$ \\
$\td{\Check{B}}_i^{n+1}$ & predicted value of $\Check{B}_i$ at $t_{n+1}$ &
${\rm m\cdot s^{-3/2}}$ \\
$B_i^*$ & approximated value defined by Eq. (\ref{sigma1}) & ${\rm m\cdot s^{-3/2}}$ \\
$\mathfrak{B}_{ij}$ & diffusion matrix defined by Eqs. (\ref{eq:frak})
and (\ref{eq:model_f}) &  \\
$\mathfrak{B}_{ij}^n$ & approximated value of $\mathfrak{B}_{ij}$ at $t_n$ & \\
$\td{\mathfrak{B}}_{ij}^{n+1}$ & predicted value of $\mathfrak{B}_{ij}$ at $t_{n+1}$ & \\ 
$B_{f,ij}$ & diffusion matrix defined by Eq. (\ref{eq:SDEf}) & ${\rm m\cdot s^{-3/2}}$ \\
$B_{s,ij}$ & diffusion matrix defined by Eq. (\ref{eq:dUs}) & ${\rm m\cdot s^{-3/2}}$ \\
$B_1$ & coefficient defined in Table \ref{tab:sch1}& ${\rm s}$ \\
$B_2$ & function defined in Table \ref{tab:sch2c}& \\
$B_{2c}$ & function defined in Table \ref{tab:sch2c}& \\
$C$ & proportionality constant, cf. Eq. (\ref{XMN}) & \\
$C(T)$ & proportionality constant, cf. Eq. (\ref{eq:order}) & \\
$C_{D}$ & drag coefficient & \\
$C_i$ & acceleration defined by Eq. (\ref{eq:sysEDS}) & ${\rm m\cdot s^{-2}}$ \\
$C_i^n$ & approximated value of $C_i$ at $t_n$ & ${\rm m\cdot s^{-2}}$ \\
$C_i^{n+1}$ & predicted value of $C_i$ at $t_{n+1}$ & ${\rm m\cdot s^{-2}}$ \\
$C_{\beta}$ & coefficient for $T_{L,i}^*$, cf. Eq. (\ref{eq:Csanady}) & \\
$C_{0}$ & Kolmogorov constant, cf. Eq. (\ref{eq:cov}) & \\
$C_1$ & coefficient defined in Table \ref{tab:sch1}& ${\rm s}$ \\
$C_{2c}$ & function defined in Table \ref{tab:sch2c}& \\
$d_{p}$ & particle diameter & ${\rm m}$ \\
$dt$ & time increment or observation time scale & ${\rm s}$ \\
$D$ & diffusion coefficient or cyclone diameter & ${\rm m^{2}\cdot s}$ or ${\rm m}$ \\
$D_{ij}$ & positive-definite matrix, ${\bf D}={\bf B}{\bf B}^T$ & \\
$D_1$ & coefficient defined in Table \ref{tab:sch1}& \\
$E_1$ & coefficient defined in Table \ref{tab:sch1}& \\
$f$ & function & \\
$f_{\Phi}$ & pdf of $\Phi$ & \\
$\mc{F}^E_f$ & fluid Eulerian mass density function & \\
$g$ & function & \\
${\bf g}$ & gravitational acceleration & ${\rm m\cdot s^{-2}}$ \\
$G_{ij}$ & return-to-equilibrium matrix, cf. Eq. (\ref{eq:SDEf}) & ${\rm s^{-1}}$ \\
$G_{ij}^a$ & anisotropy matrix, cf. Eq. (\ref{eq:SDEf}) & ${\rm s^{-1}}$ \\
${\cal G}_{p,i}$ & standard Gaussian random variable, Eq. (\ref{eq:Up_casii}) &  \\
${\cal G}_{p,i}^{\;'}$ & standard Gaussian random variable, Eq. (\ref{eq:gamp}) &  \\
${\cal G}_{x,i}$ & standard Gaussian random variable, Eq. (\ref{eq:xp_casiii})&  \\
${\cal G}_{1,i}$ & standard Gaussian random variable, cf. Table \ref{tab:sch1} &  \\
${\cal G}_{2,i}$ & standard Gaussian random variable, cf. Table \ref{tab:sch1} &  \\
${\cal G}_{3,i}$ & standard Gaussian random variable, cf. Table \ref{tab:sch1} &  \\
${\bf H}(t)$ & stochastic process, cf. Eq. (\ref{eq:H}) &  \\
$\mc{H}({\bf V}_f)$ & function of ${\bf V}_f$ &  \\
$I_x(t)$ & stochastic integral in model problem,
cf. Eq. (\ref{eq:IxIu}) & ${\rm m}$ \\
$I_U(t)$ & stochastic integral in model problem,
cf. Eq. (\ref{eq:IxIu}) & ${\rm m\cdot s^{-1}}$ \\
$k$ & turbulent kinetic energy & ${\rm m^{2}\cdot s^{-2}}$ \\
$\td{k}$ & modified turbulent kinetic energy & ${\rm m^{2}\cdot s^{-2}}$ \\
$k_p$ & deposition velocity & ${\rm m\cdot s^{-1}}$ \\
$m(x)$ & deterministic function, cf. Section \ref{SS_PM_AW} & \\
$N$ & total number of discrete particles or samples & \\
$N_k$ & number of discrete particles in cell $k$ & \\
$N_{pc}$ & number of discrete particles per cell or per class & \\
$N_{pc}^i$ & number of discrete particles in cell $i$ & \\
$p$ & probability density function (pdf) for ${\bf Z}(t)$ & \\
$p^E_f$ & Eulerian fluid distribution function & \\
$p^r$ & probability density function (pdf) for ${\bf Z}^r(t)$ & \\
$P(t,{\bf x})$ & local instantaneous pressure field & ${\rm Pa}$ \\
$\lra{P^n}$ & approximated value of $P(t,{\bf x})$ at $(t_n,{\bf x}_p^n)$, & ${\rm Pa}$ \\
$\lra{P^{n+1}}$ & predicted value of $P(t,{\bf x})$ at $(t_{n+1},{\bf x}_p^{n+1})$ & ${\rm Pa}$ \\
$P_{ij}$ & coefficients defined in Table \ref {tab:sch1}, $(i,j) \in
   (1,2,3)^2$ & ${\rm m}$ or ${\rm m\cdot s^{-1}}$ \\ 
$Re_p$ & Reynolds number (discrete particles) &  \\
$R_{ij}$ & Reynolds stress tensor & ${\rm m^{2}\cdot s^{-2}}$ \\
$s(x)$ & deterministic function, cf. Section \ref{SS_PM_AW} & \\
$S_U$ & source term in Eq. (\ref{eq:feq_Uf}) & ${\rm m\cdot s^{-2}}$ \\
$S_k$ & trace of source term tensor in Eq. (\ref{eq:feq_ufuf}) & ${\rm m^{2}\cdot s^{-3}}$ \\
$S_{R_{ij}}$ & source term in Eq. (\ref{eq:feq_ufuf}) & ${\rm m^{2}\cdot s^{-3}}$ \\
$t$ & time & ${\rm s}$ \\
$t^+$ & dimensionless time, $t^+=t\,(u^*)^2/\nu_f$ & \\
$T$ & characteristic time scale & ${\rm s}$ \\
$T_E$ & fluid Eulerian integral time scale & ${\rm s}$ \\
$T_{i}$ & fluid seen integral time scale (in Section \ref{sec:num0}) & ${\rm s}$ \\
$T_{i}^n$ & approximated value of $T_i$ at $t_n$ & ${\rm s}$ \\
$\td{T}_{i}^{n+1}$ & predicted value of $T_i$ at $t_{n+1}$ & ${\rm s}$ \\
$T_L$ & fluid Lagrangian integral time scale & ${\rm s}$ \\
$T_{L,i}^*$ & fluid seen integral time scale & ${\rm s}$ \\
${\bf u}(t,{\bf x})$ & fluctuating fluid velocity field & ${\rm m\cdot s^{-1}}$ \\
$u^*$ & friction velocity & ${\rm m\cdot s^{-1}}$ \\
$U(t)$ & velocity in model problem, cf. Eq. (\ref{eq:langevin}) & ${\rm m\cdot s^{-1}}$ \\
${\bf U}(t,{\bf x})$ & local instantaneous fluid velocity field & ${\rm m\cdot s^{-1}}$ \\
$\lra{U_i}$ & mean fluid velocity field at $(t,{\bf x})$ or $(t,{\bf x}_p(t))$ 
& ${\rm m\cdot s^{-1}}$ \\
$\lra{U_i^n}$ & approximated value of $\lra{U_i}$ at $(t_n,{\bf x}_p^n)$ & ${\rm m\cdot s^{-1}}$ \\
$\lra{U_i^{n+1}}$ & predicted value of $\lra{U_i}$ at $(t_{n+1},{\bf x}_p^{n+1})$ 
& ${\rm m\cdot s^{-1}}$ \\
${\bf U}_{f}(t)$ & velocity of fluid particles & ${\rm m\cdot s^{-1}}$ \\
$U_{f,i}^n$ & approximated value of ${\bf U}_f(t)$ at $t_n$ & ${\rm m\cdot s^{-1}}$ \\
$\td{U}_{f,i}^{n+1}$ & predicted value of ${\bf U}_f(t)$ at $t_{n+1}$ &
   ${\rm m\cdot s^{-1}}$ \\
$U_{f,i}^{n+1}$ & approximated value of ${\bf U}_f(t)$ at $t_{n+1}$ & ${\rm m\cdot s^{-1}}$ \\
${\bf U}_{p}(t)$ & velocity of the discrete particles & ${\rm m\cdot s^{-1}}$ \\
$U_{p,i}^n$ & approximated value of ${\bf U}_p(t)$ at $t_n$ & ${\rm m\cdot s^{-1}}$ \\
$\td{U}_{p,i}^{n+1}$ & predicted value of ${\bf U}_p(t)$ at $t_{n+1}$ &
   ${\rm m\cdot s^{-1}}$ \\
$U_{p,i}^{n+1}$ & approximated value of ${\bf U}_p(t)$ at $t_{n+1}$ & ${\rm m\cdot s^{-1}}$ \\
${\bf U}_{r}(t)$ & particle relative velocity & ${\rm m\cdot s^{-1}}$ \\
${\bf U}_{s}(t)$ & fluid velocity seen & ${\rm m\cdot s^{-1}}$ \\
$U_{s,i}^n$ & approximated value of ${\bf U}_s(t)$ at $t_n$ & ${\rm m\cdot s^{-1}}$ \\
$\td{U}_{s,i}^{n+1}$ & predicted value of ${\bf U}_s(t)$ at $t_{n+1}$ &
   ${\rm m\cdot s^{-1}}$ \\
$U_{s,i}^{n+1}$ & approximated value of ${\bf U}_s(t)$ at $t_{n+1}$ & ${\rm m\cdot s^{-1}}$ \\
${\bf V}_f$ & sample space value for ${\bf U}_f(t)$ & ${\rm m\cdot s^{-1}}$ \\
${\cal V}_f^{[k]}$ & volume of fluid in cell $[k]$ & ${\rm m^{3}}$ \\
$w({\bf x})$ & weighting function (continuous form) & \\
$\td{w}({\bf x})$ & weighting function (discrete form) & \\
$W_{i}(t)$ & Wiener process & ${\rm s^{1/2}}$ \\
$x(t)$ & position in model problem, cf. Eq. (\ref{eq:langevin}) & ${\rm m}$ \\
${\bf x}_{f}(t)$ & position of the fluid particles & ${\rm m}$ \\
${\bf x}_{p}(t)$ & position of the discrete particles & ${\rm m}$ \\  
$x_{p,i}^n$ & approximated value of ${\bf x}_p(t)$ at $t_n$ & ${\rm m}$ \\
$x_{p,i}^{n+1}$ & approximated value of ${\bf x}_p(t)$ at $t_{n+1}$ & ${\rm m}$ \\
${\bf X}(t)$ & stochastic process or deterministic variable, $X$ & \\
$y^+$ & dimensionless distance from wall & \\  
${\bf y}$ & sample space value of ${\bf Y}(t)$ & \\  
$Y(t)$ & stochastic process or deterministic variable, $Y$ & \\  
${\bf Y}(t)$ & set of external variables & \\  
${\bf z}$ & sample space value of ${\bf Z}(t)$ & \\  
${\bf z}_f^i$ & sample space value of ${\bf Z}_f^i(t)$ & \\  
${\bf z}_p^i$ & sample space value of ${\bf Z}_p^i(t)$ & \\  
${\bf Z}(t)$ & state vector & \\  
$Z_{f,j}^i$ & variable $j$ for fluid particle $i$ & \\  
$Z_{p,j}^i$ & variable $j$ for discrete particle $i$ & \\  
${\bf Z}^r(t)$ & reduced state vector & \\  
\end{longtable}
%---------------------------
\subsection*{Greek letters}
%---------------------------
\begin{longtable}[htbp]{p{1.5cm}p{8cm}p{2cm}}
$\alpha_{f}(t,{\bf x})$ & volume fraction of fluid & \\
$\alpha_{p}(t,{\bf x})$ & volume fraction of particles & \\
$\beta_{i},\beta$ & constants defined in Eq. (\ref{eq:Csanady}) & \\
$\gamma_i(t)$ & stochastic process defined by
Eq. (\ref{eq:gammaN_exa}) & ${\rm m\cdot s^{-1}}$ \\
$\gamma_i^n$ & approximated value of $\gamma_i(t)$ at $t_n$ & ${\rm m\cdot s^{-1}}$ \\
$\td{\gamma}_i^{n+1}$ & predicted value of $\gamma_i(t)$ at $t_{n+1}$ & ${\rm m\cdot s^{-1}}$ \\
$\Gamma_i(t)$ & stochastic process defined by Eq. (\ref{eq:GammaN_exa}) & ${\rm m\cdot s^{-1}}$ \\
$\Gamma_i^n$ & approximated value of $\Gamma_i(t)$ at $t_n$ & ${\rm m\cdot s^{-1}}$ \\
$\td{\Gamma}_i^{n+1}$ & predicted value of $\Gamma_i(t)$ at $t_{n+1}$ & ${\rm m\cdot s^{-1}}$ \\
$\delta(\cdot)$ & Dirac delta function & \\ 
$\delta_{ij}$ & Kronecker's symbol & \\ 
$\Delta t$ & time step & ${\rm s}$ \\
$\Delta x$ & characteristic cell size & ${\rm m}$ \\
$\epsilon(t,{\bf x})$ & dissipation rate of $k$ & ${\rm m^{2}\cdot s^{-3}}$ \\
$\varepsilon(t)$ & energy dissipation for fluid particles & ${\rm m^{2}\cdot s^{-3}}$ \\
$\eta(t)$ & Gaussian white noise & \\
$\theta_i$ & ratio, $\theta_i=T_i/(T_i-\tau_p)$ & \\
$\theta_i^n$ & approximated value of $\theta_i$ at $t_n$ & \\
$\mu_f$ & dynamic viscosity of fluid & ${\rm Pa\cdot s}$ \\
$\nu_f$ & kinematic viscosity of fluid & ${\rm m^{2}\cdot s^{-1}}$ \\
$\xi$ & standard Gaussian random variable, cf. Eq. (\ref{XMN}) & \\
$\xi_x$ & standard Gaussian random variable, cf. Eq. (\ref{eq:xi}) & \\
$\xi_U$ & standard Gaussian random variable, cf. Eq. (\ref{eq:xi}) & \\
${\bf\Pi}_p(t)$ & random acceleration defined by Eq. (\ref{eq:Apf}) & ${\rm m\cdot s^{-2}}$ \\
$\rho_f$ & density of fluid & ${\rm kg.m^{-3}}$ \\
$\rho_p$ & density of discrete particles & ${\rm kg.m^{-3}}$ \\
${\bds \sigma}$ & diffusion matrix in Eq. (\ref{EqGeneric}) & ${\rm m\cdot s^{-3/2}}$ \\
$\sigma[\;\cdot\;]$ & standard deviation of $\cdot$ & \\
$\sigma_{\Phi}^2$ & variance of $\Phi$ & \\
$\tau$ & characteristic time scale & ${\rm s}$ \\
$\tau_p$ & particle relaxation time  & ${\rm s}$ \\
$\tau_p^+$ & dimensionless particle relaxation time  & \\
$\tau_p^n$ & approximated value of $\tau_p$ at $t_n$ & ${\rm s}$ \\
$\td{\tau_p}^{n+1}$ & predicted value of $\tau_p$ at $t_{n+1}$ & ${\rm s}$ \\
$\tau_{\eta}$ & Kolmogorov time scale & ${\rm s}$ \\
$\Phi$ & random variable or deterministic function &  \\
$\chi$ & ratio, $\chi=\alpha_f\,\rho_f/\alpha_p\,\rho_p$ & \\
$\Psi$ & sample space value of $\Phi$ &  \\
$\Omega_i(t)$ & stochastic process defined by
Eq. (\ref{eq:OmegaN_exa}) & ${\rm m}$ \\
$\Omega_i^n$ & approximated value of $\Omega_i(t)$ at $t_n$ & ${\rm m}$ \\
\end{longtable}
%------------------------
\subsection*{Subscripts}
%------------------------
\begin{longtable}[htbp]{p{1.5cm}p{8cm}p{2cm}}
$f$ & continuous phase (fluid) & \\
$p$ & discrete phase (particles) & \\
$s$ & fluid properties sampled along particle trajectories & \\
\end{longtable}
%------------------------
\subsection*{Superscripts}
%------------------------
\begin{longtable}[htbp]{p{1.5cm}p{8cm}p{2cm}}
$[i]$ & variable calculated at cell centre $i$ & \\
$[k]$ & variable calculated in cell $k$ & \\
$n$ & approximated values at $t=t_n$ & \\
$n+1$ & approximated values at $t=t_n+\Delta t$ & \\
$(n)$, $(N)$ & variables calculated at particle locations & \\
$r$ & reduced information & \\
$T$ & transpose of a matrix & \\
$[x]$ & variables calculated on the mesh/at cell centres & \\
$+$ & dimensionless quantities & \\
$\;\widetilde{.}$ & predicted quantities (numerical schemes) & \\
\end{longtable}
%------------------------------
\subsection*{Special notation}
%------------------------------
\begin{longtable}[htbp]{p{1.5cm}p{8cm}p{2cm}}
$\{\,\cdot\,\}$ & set of variables & \\ 
$\lra{\,\cdot\,}$ & mathematical expectation & \\ 
$\lra{\,\cdot\,}_N$ & mean value, i.e. $(1/N)\sum_{i=1}^N \cdot$ & \\ 
$\lra{\,\cdot\,}_{\Delta}$ & spatial average & \\ 
$\lra{\,\cdot\,}_{N,\Delta}$ & approximation of $\lra{\,\cdot\,}$,
spatial average on $N$ samples & \\ 
$\lra{\,\cdot\,}_{\infty}$ & $=\lra{\,\cdot\,}_N$ with $N \to \infty$,
i.e. $\lra{\,\cdot\,}$ & \\ 
$\lra{\,\cdot\,|\,\cdot\,}$ & conditional expectation & \\ 
$\mid \,\cdot\, \mid$ & norm of a vector & \\ 
$\partial$ & partial derivative & \\ 
$\textbf{U}$ & bold style for vector notation & \\ 
$D\cdot/Dt$ & $\partial\cdot/\partial t+\lra{U_{i}}\partial \cdot/\partial x_{i}$ & \\ 
$d\cdot(t)$ & time increment, e.g. $d{\bf U}_f(t)={\bf U}_f(t+dt)-{\bf U}_f(t)$ & \\ 
\end{longtable}
%------------------------------
\subsection*{Abbreviations}
%------------------------------
\begin{longtable}[htbp]{p{1.5cm}p{8cm}p{2cm}}
CIC & Cloud In Cell & \\
CPU & Central Processing Unit & \\
cst & a given constant & \\ 
CTE & Crossing Trajectory Effect & \\ 
DNS & Direct Numerical Simulation & \\ 
DSMC & Direct Simulation Monte Carlo & \\ 
LES & Large Eddy Simulation & \\
NGP & Nearest Grid Point & \\ 
ODE & Ordinary Differential Equation & \\ 
PDE & Partial Differential Equation & \\ 
pdf/PDF & Probability Density Function & \\ 
PIC & Particle In Cell & \\ 
RANS & Reynolds-Averaged Navier-Stokes & \\ 
RHS & Right-Hand Side & \\ 
r.m.s. & root-mean square & \\ 
RSM & Reynolds Stress Models & \\ 
SDE & Stochastic Differential Equation & \\ 
SPH & Smoothed Particle Hydrodynamics & \\ 
VRT & Variance Reduction Technique & \\ 
\end{longtable}

%%% Local Variables: 
%%% mode: latex
%%% TeX-master: "titi"
%%% End: 

%% file: figures.tex
%===============================================================================
% Figures
%===============================================================================
\ifnum\elsevier<0 
\setlength\abovecaptionskip{5pt}
\setlength\belowcaptionskip{0pt}
\fi

%------------------------------
% Figure 1
%------------------------------
\begin{figure}[H]
\centerline{ (a)}
\vspace{0.5mm}
\centerline{ \epsfig{file=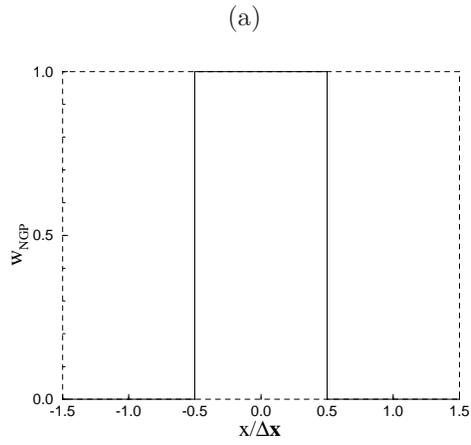,width=7cm} } \vspace{0.1cm}
\centerline{ (b)}
\vspace{0.5mm}
\centerline{ \epsfig{file=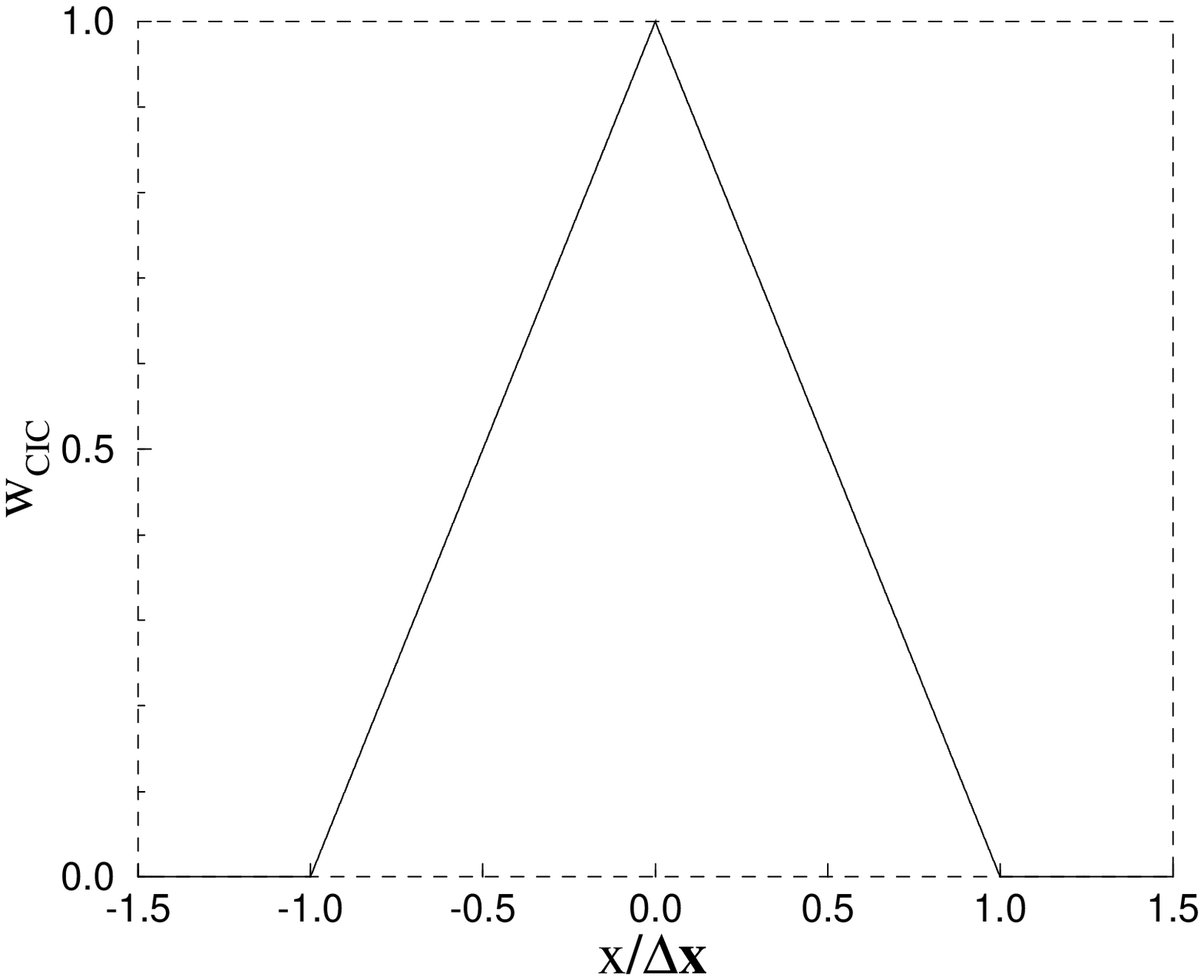,width=7cm} } \vspace{0.1cm}
\centerline{ (c)}
\vspace{0.5mm}
\centerline{ \epsfig{file=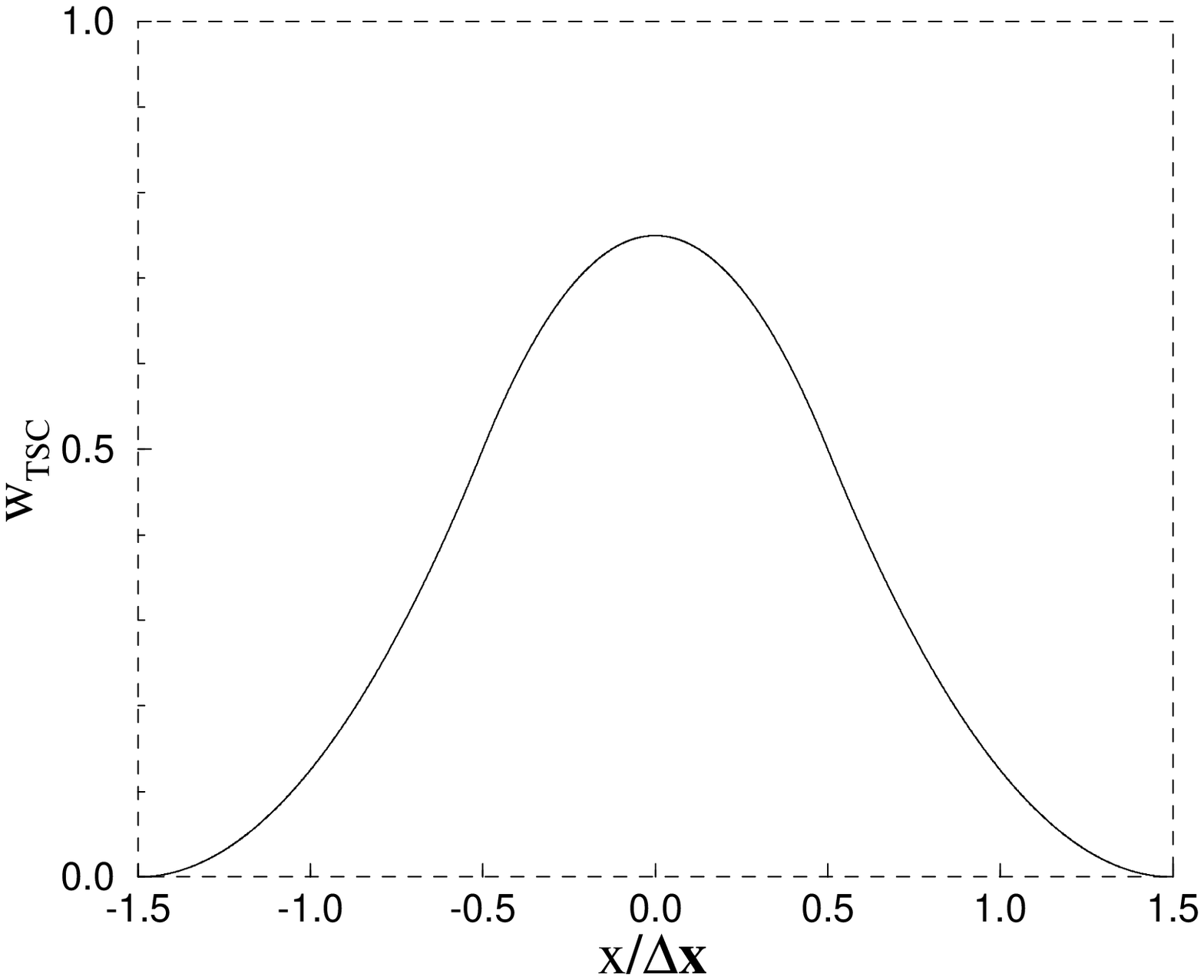,width=7cm} }
\caption{Weighting functions of different orders used for averaging:
  (a) top hat (constant) or Nearest-Grid-Point (NGP); (b) linear or
  Cloud-In-Cell (CIC); (c) piecewise quadratic.}
\label{wgh_fun}
\end{figure}

%------------------------------
% Figure 2
%------------------------------
\afterpage{\clearpage} \newpage
\begin{figure}[H]
\centerline{ (a)}
\vspace{0.5mm}
\centerline{ \epsfig{file=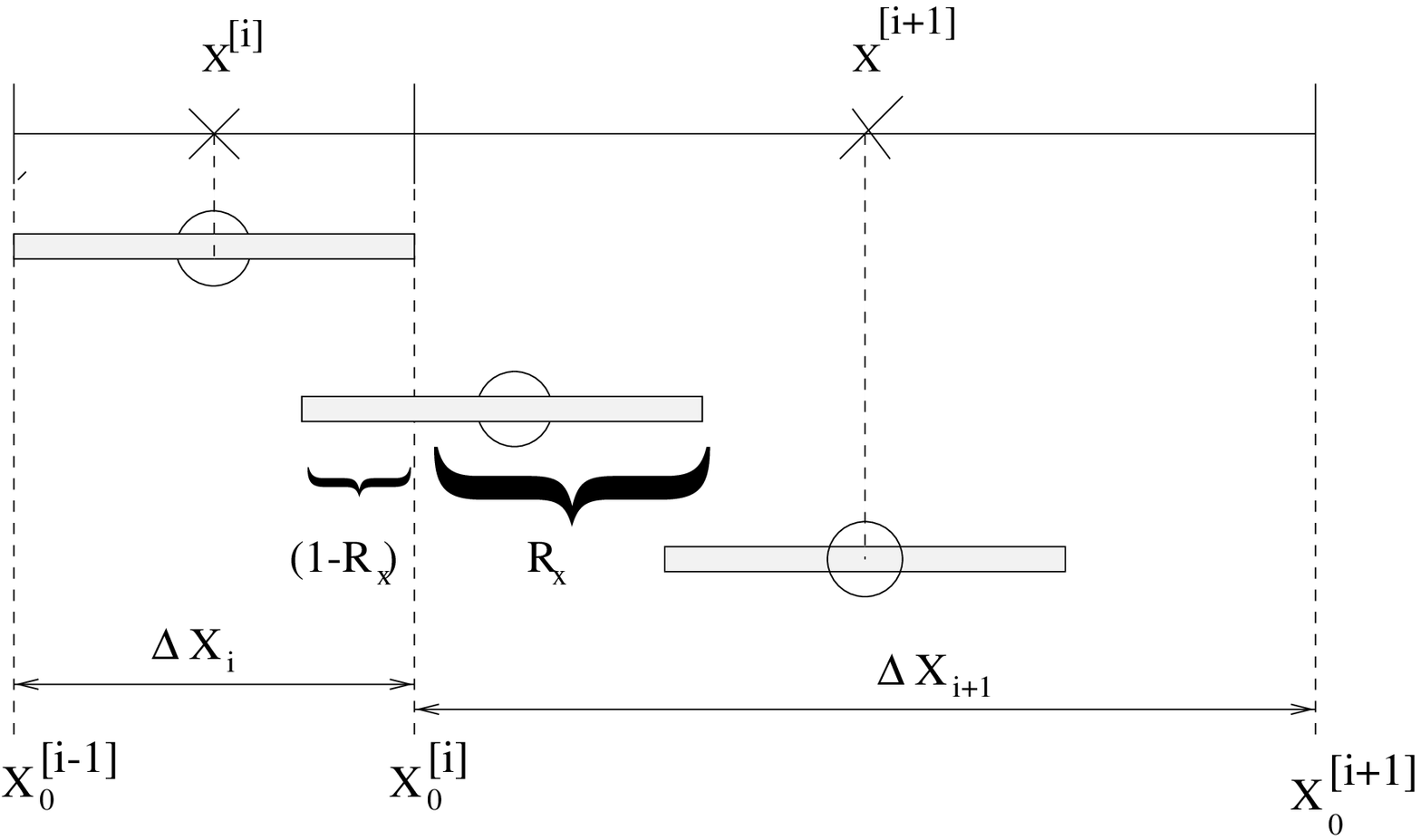,width=9cm}}
\vspace{1cm}
\centerline{ (b)}
\vspace{0.5mm}
\centerline{ \epsfig{file=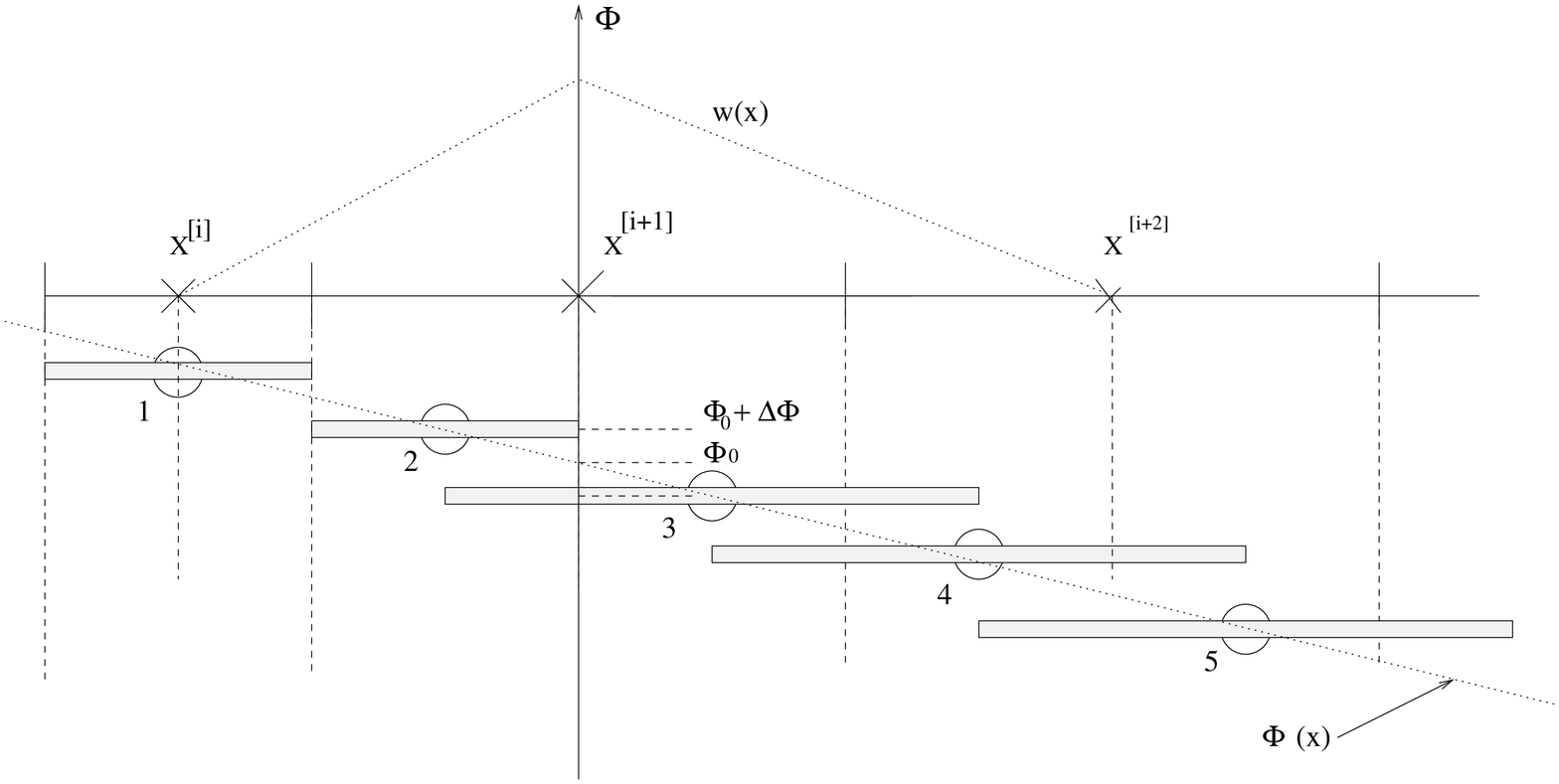,width=14cm}}
\vspace{1cm}
\caption{CIC scheme used to compute cell averages for (a) particle
  density and (b) a linear function attached to particles (the linear
  function is defined by its slope $\Delta \Phi$ and $\Phi_0$ which is
  the value of $\Phi$ at $x=x^{[i+1]}$). The cells are characterised
  by the cell boundary coordinates
  $x_0^{[i-1]},x_0^{[i]},x_0^{[i+1]},\dots$ and by the coordinates of
  the cell centres $x^{[i-1]},x^{[i]},x^{[i+1]},\dots$. $\Delta x_i$
  and $\Delta x_{i+1}$ represent the cell sizes.}
\label{pm_CIC}
\end{figure}

%------------------------------
% Figure 3
%------------------------------
\afterpage{\clearpage} \newpage
\begin{figure}[h]
\centerline{ \hspace*{1cm} (a) \hspace*{6cm} (b)}
\vspace{0.5mm}
\centerline{ \epsfig{file=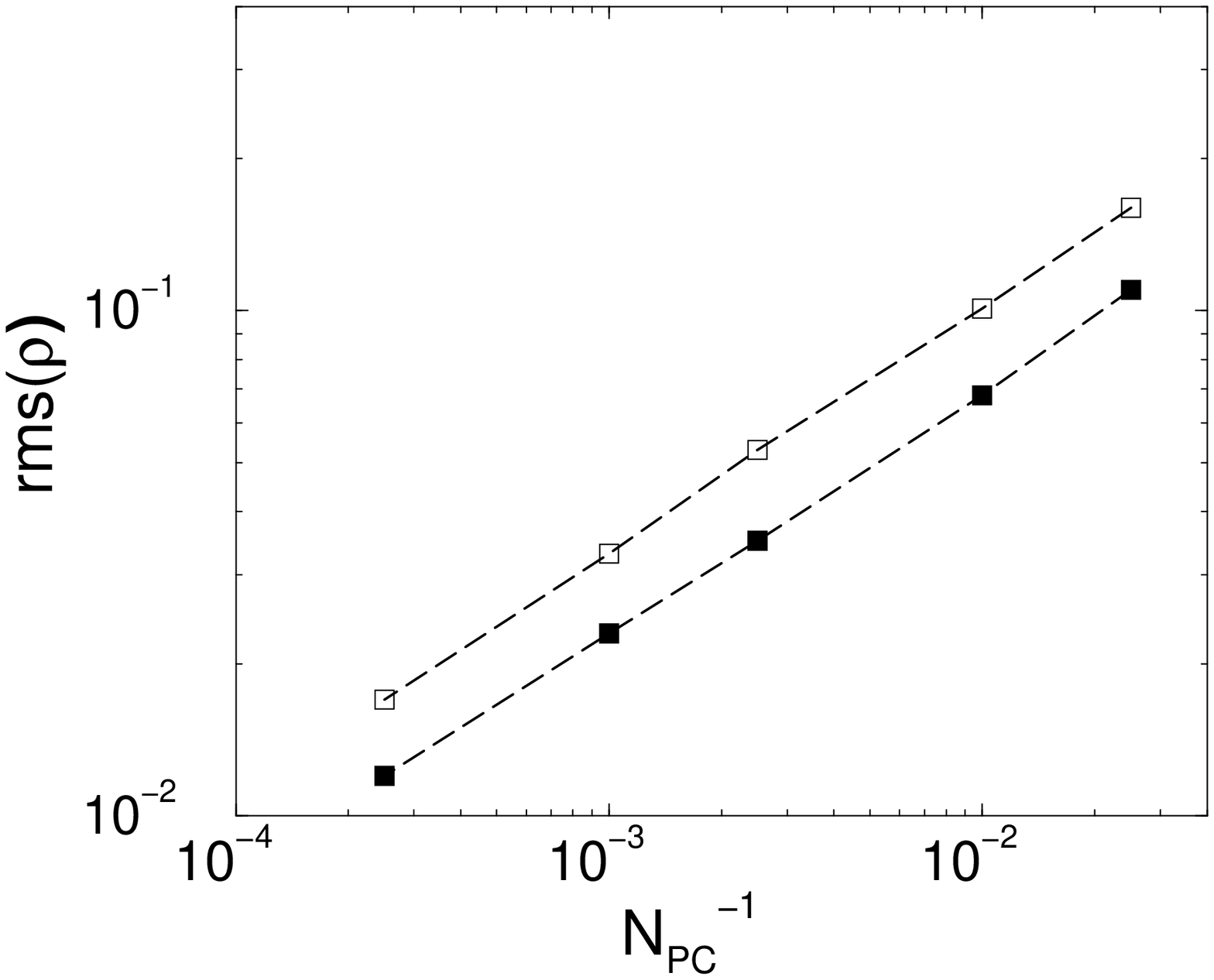,width=6.5cm}
\hskip 0.5cm \epsfig{file=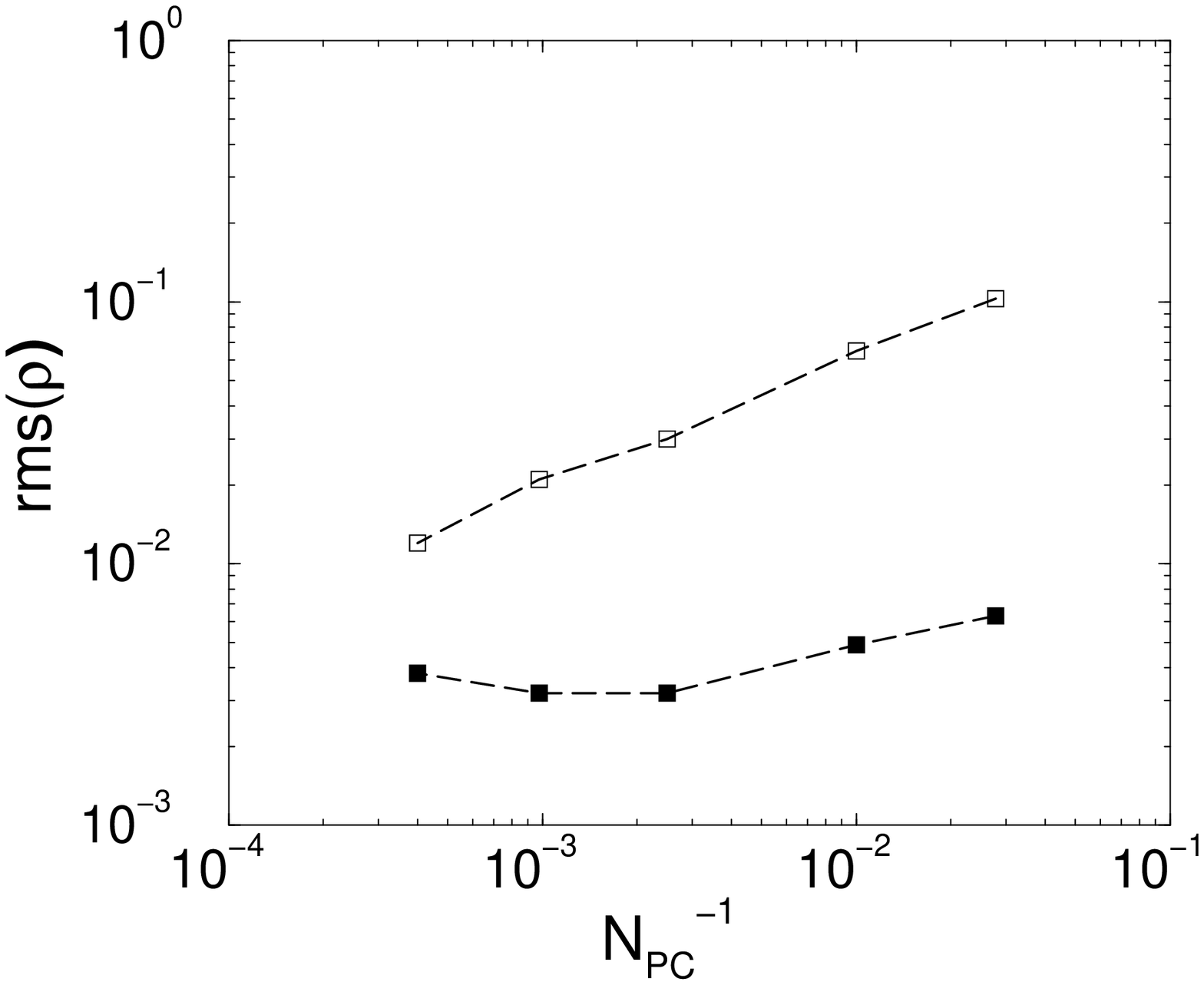,width=6.5cm} }
\vspace{3mm}
\caption{Computations of the r.m.s. of particle density on a mesh
  using NGP and CIC averaging : (a) uniform mesh, (b) non-uniform
  mesh. NGP method ($\square$), CIC method ($\blacksquare$).}
\label{dens_un_nu}
\end{figure}

%------------------------------
% Figure 4
%------------------------------
\afterpage{\clearpage} \newpage
\begin{figure}[h]
\centerline{ \hspace*{1cm} (a) \hspace*{6cm} (b)}
\vspace{0.5mm}
\centerline{ \epsfig{file=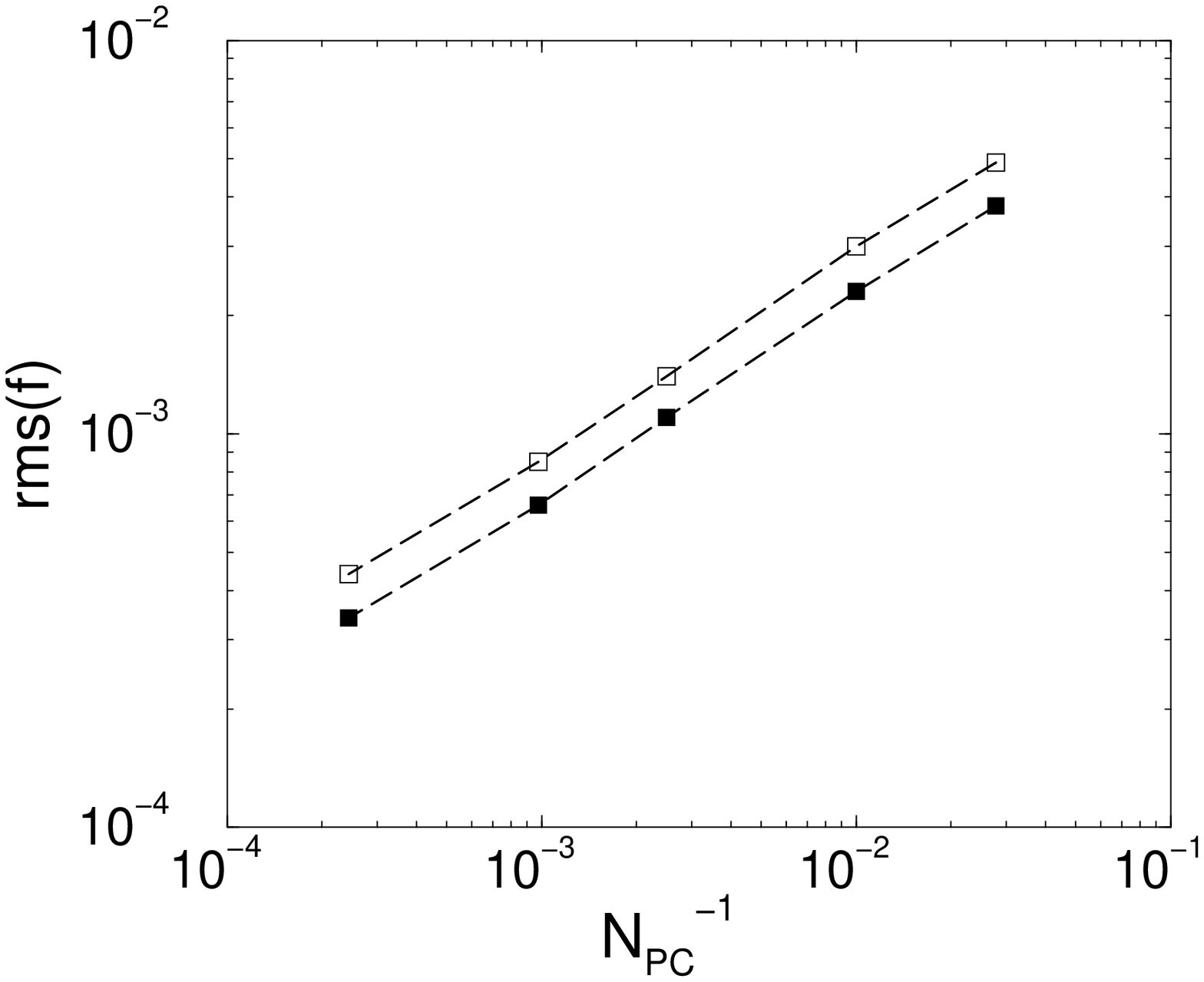,width=6.5cm}
\hskip 0.5cm \epsfig{file=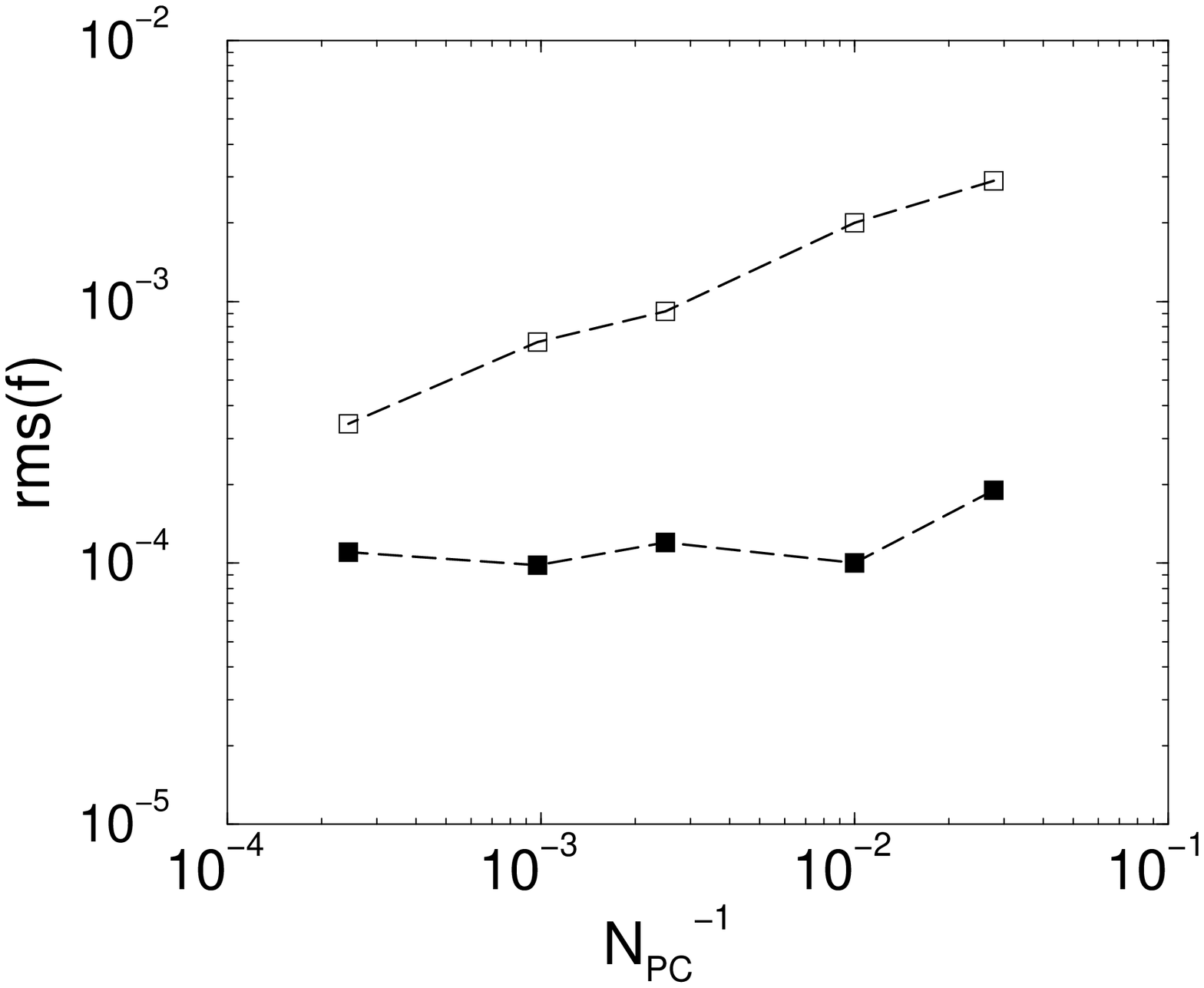,width=6.5cm} }
\vspace{3mm}
\caption{The r.m.s. value of the mean of a linear deterministic function
  computed on a mesh using NGP and CIC averaging : (a) uniform mesh, (b)
  non-uniform mesh.  NGP method ($\square$), CIC method ($\blacksquare$).}
\label{ldf_un_nu}
\end{figure}

%------------------------------
% Figure 5
%------------------------------
\afterpage{\clearpage} \newpage
\begin{figure}[H]
\centerline{ \hspace*{1cm} (a) \hspace*{6cm} (b)}
\vspace{0.5mm}
 \centerline{ \epsfig{file=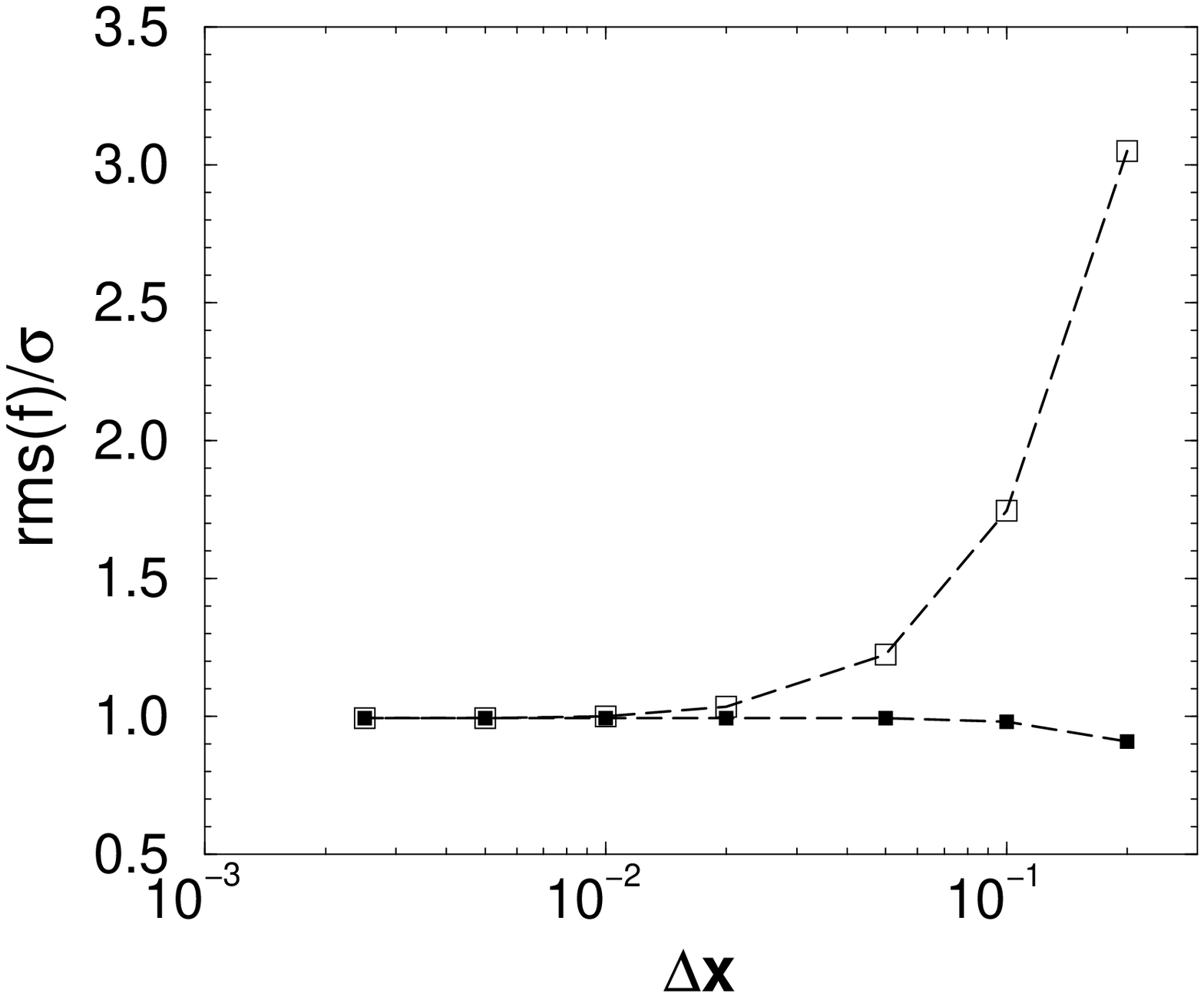,width=6.5cm}
 \hskip 0.5cm \epsfig{file=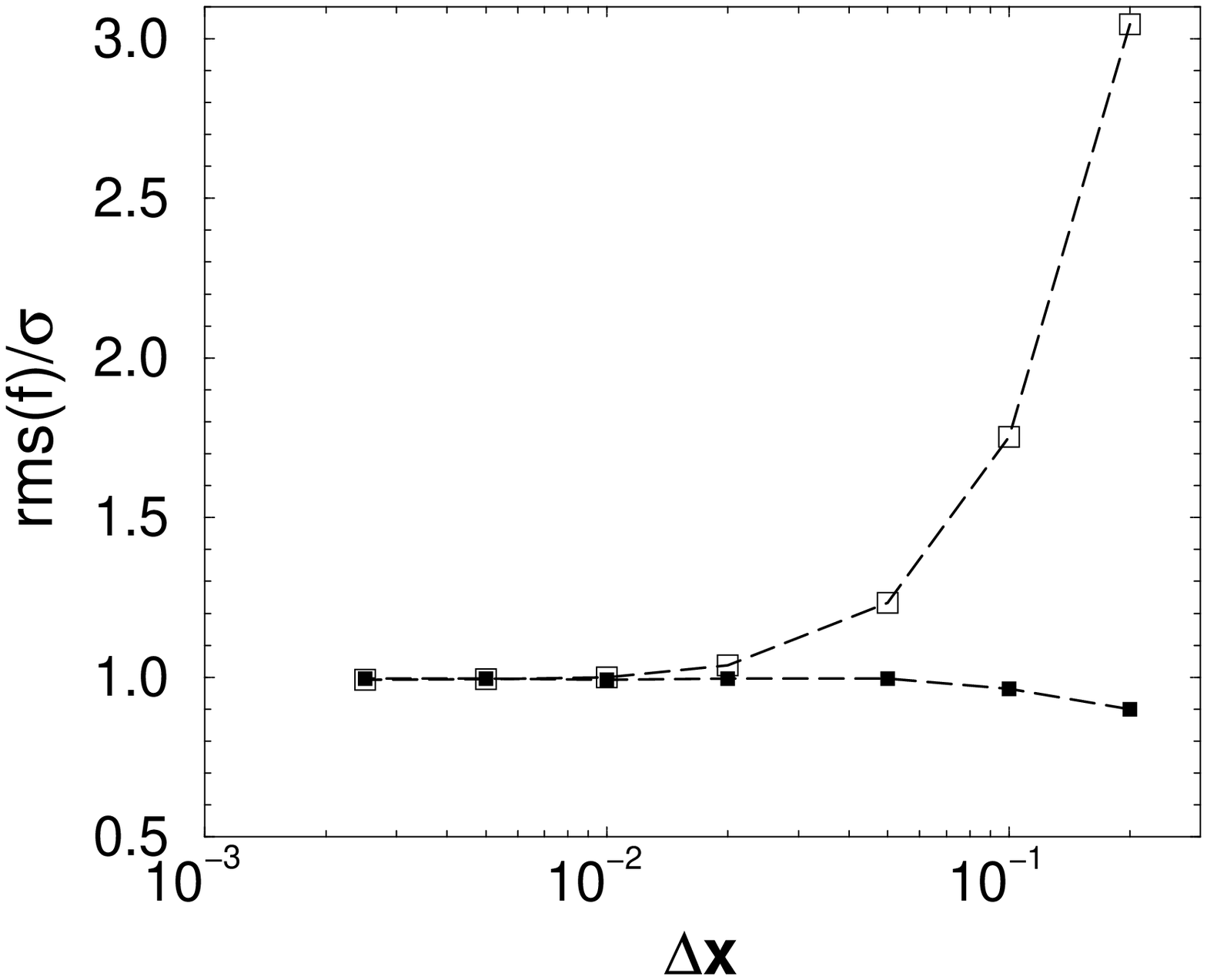,width=6.5cm} }
\vspace{3mm}
\caption{Normalised r.m.s.\ of a linear random function 
  computed on a mesh using NGP and CIC averaging: (a) uniform mesh, (b)
  nonuniform mesh. NGP method ($\square$), CIC method ($\blacksquare$).}
\label{lrf_un_nu_s1}
\end{figure}

%------------------------------
% Figure 6
%------------------------------
\afterpage{\clearpage} \newpage
\begin{figure}[H]
\centering \epsfig{file=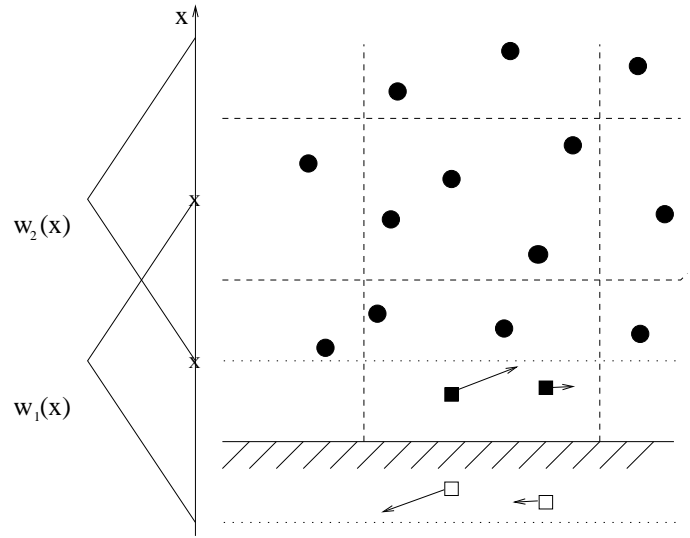,width=9cm}
\vskip 3mm
\caption{Cells and particles -- schematic plot.
         Mirror particles ($\square$), corresponding to ($\blacksquare$),
         are added outside of the computational
         domain. The values of variables attached to them (like velocity)
         correspond to those of their ``host'' particles in border cells.
         Dashed lines delimit cells and dotted lines indicate where
         mirror particles are needed.}
\label{pm_1d}
\end{figure}

%------------------------------
% Figure 7
%------------------------------
\afterpage{\clearpage} \newpage
\begin{figure}[H]
\centering \epsfig{file=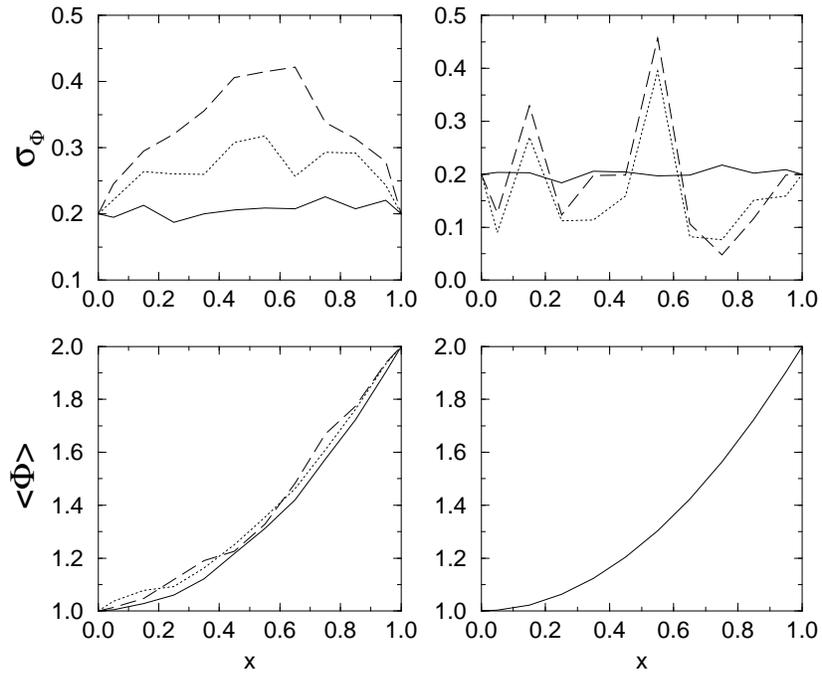,width=11cm}
\caption{Temporal evolution of the mean (lower plots),
  $\lra{{\Phi}}$, and variance (upper plots), $(\sigma_\Phi^2)$,
    profiles for quadratic initial mean profile and constant initial
    non-zero variance. Left plots: CIC; right plots: NGP. Three
    successive time instants (solid, dotted and dashed lines).}
\label{SDE_m_v_CIC_NGP}
\end{figure}

%------------------------------
% Figure 8
%------------------------------
\afterpage{\clearpage} \newpage
\begin{figure}[H]
\centerline{ \hspace*{1cm} (a) \hspace*{6cm} (b)}
\vspace{0.5mm}
\centerline{\epsfig{file=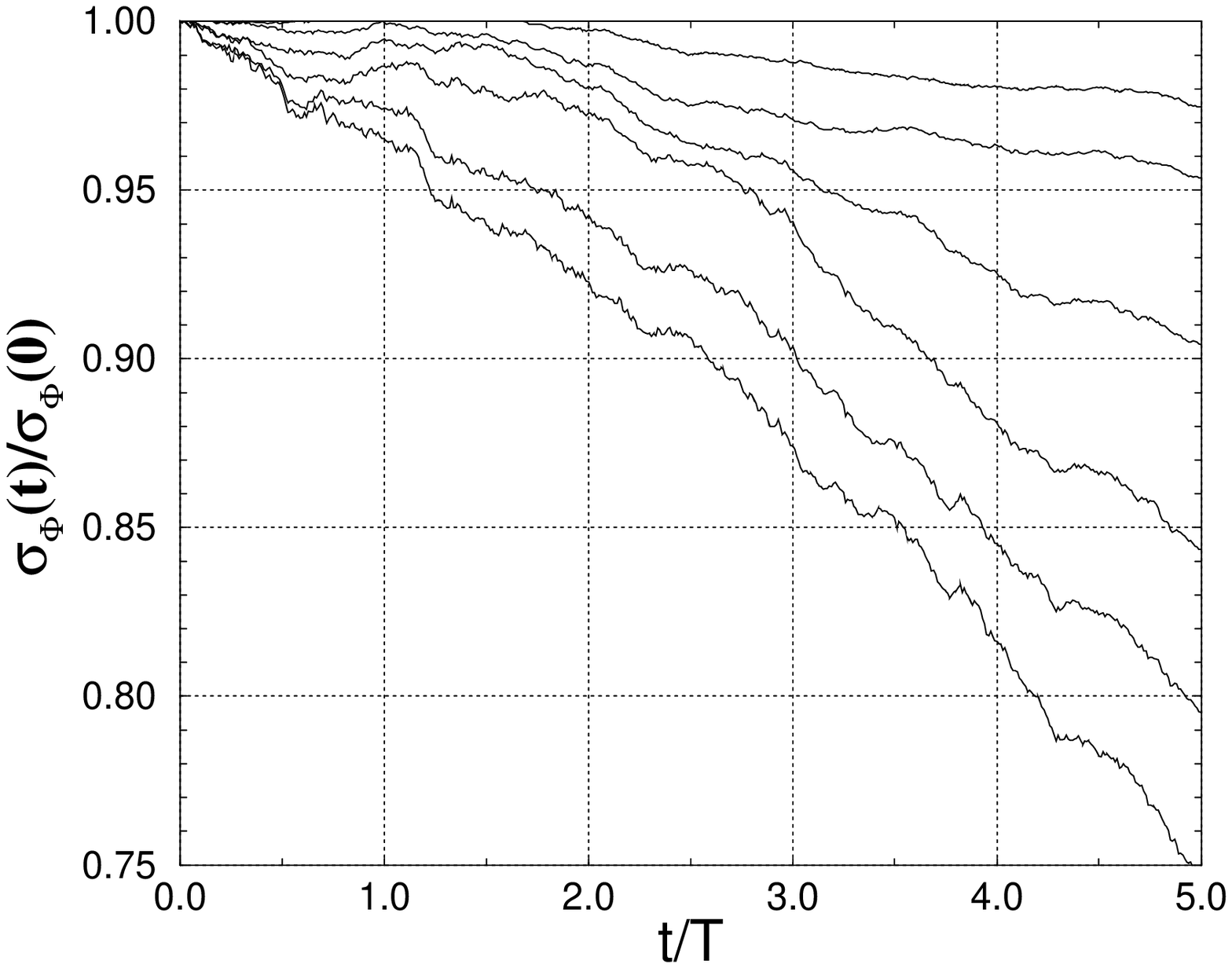,width=7cm}
\hskip 0.2cm  \epsfig{file=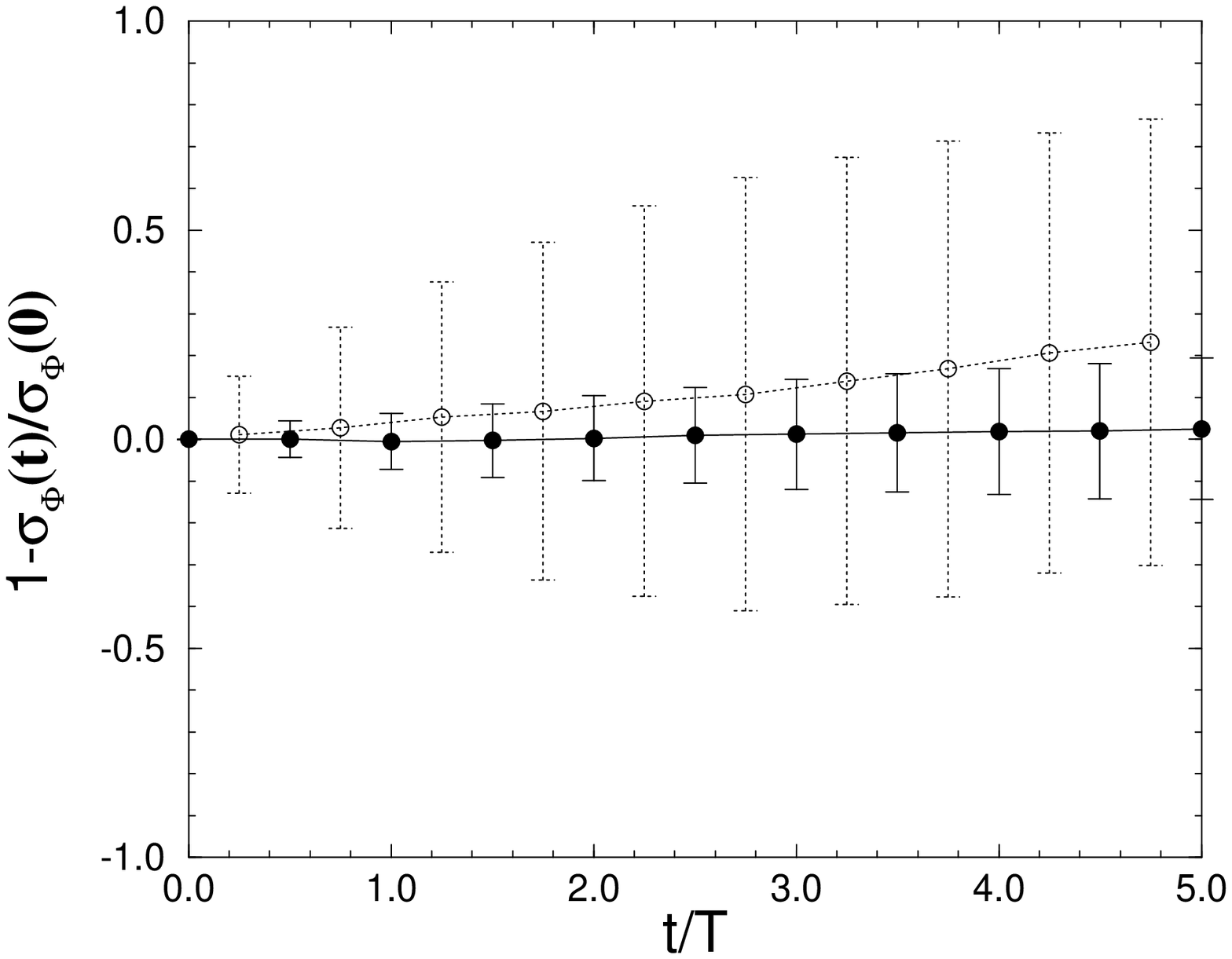,width=7cm} }
\caption{ Temporal evolution of the r.m.s. of $\Phi$ in scaled
   time; NGP computation over 1000 cells.
   (a) Results for $N_{pc}=20,25,40,80,160,$ and $320$, respectively
   (from the lowest to the highest curve);
   (b) the r.m.s.\ of $\Phi$ with its standard deviation for
   $N_{pc}=25$ (dotted line with $\circ$) and
   $N_{pc}=320$ (solid line with $\bullet$).}
\label{biasM2V0}
\end{figure}

%------------------------------
% Figure 9
%------------------------------
\afterpage{\clearpage} \newpage
\begin{figure}[H]
\centerline{ \hspace*{1cm} (a) \hspace*{6cm} (b)}
\vspace{0.5mm}
\centerline{\epsfig{file=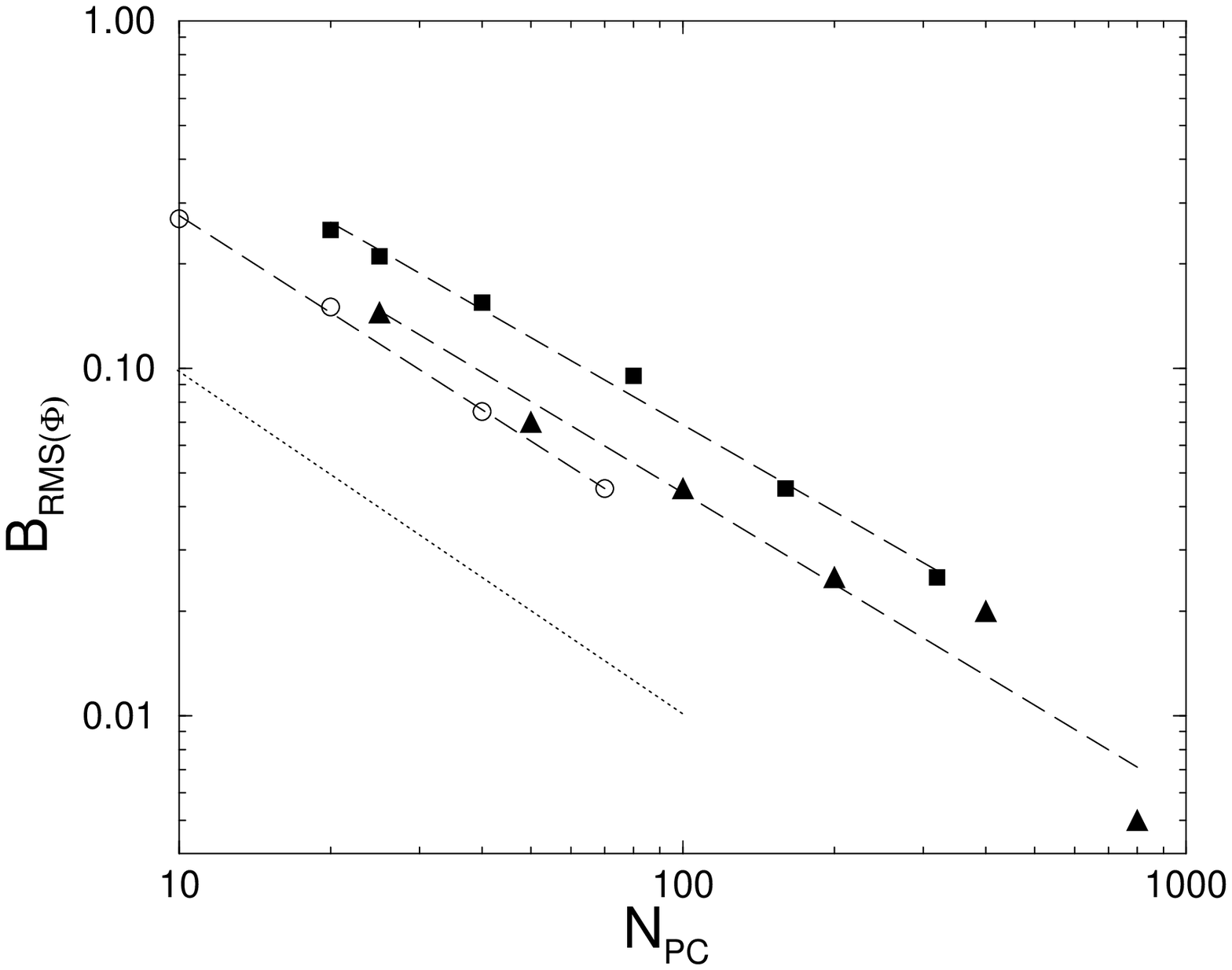,width=7cm}
\hskip 0.5cm  \epsfig{file=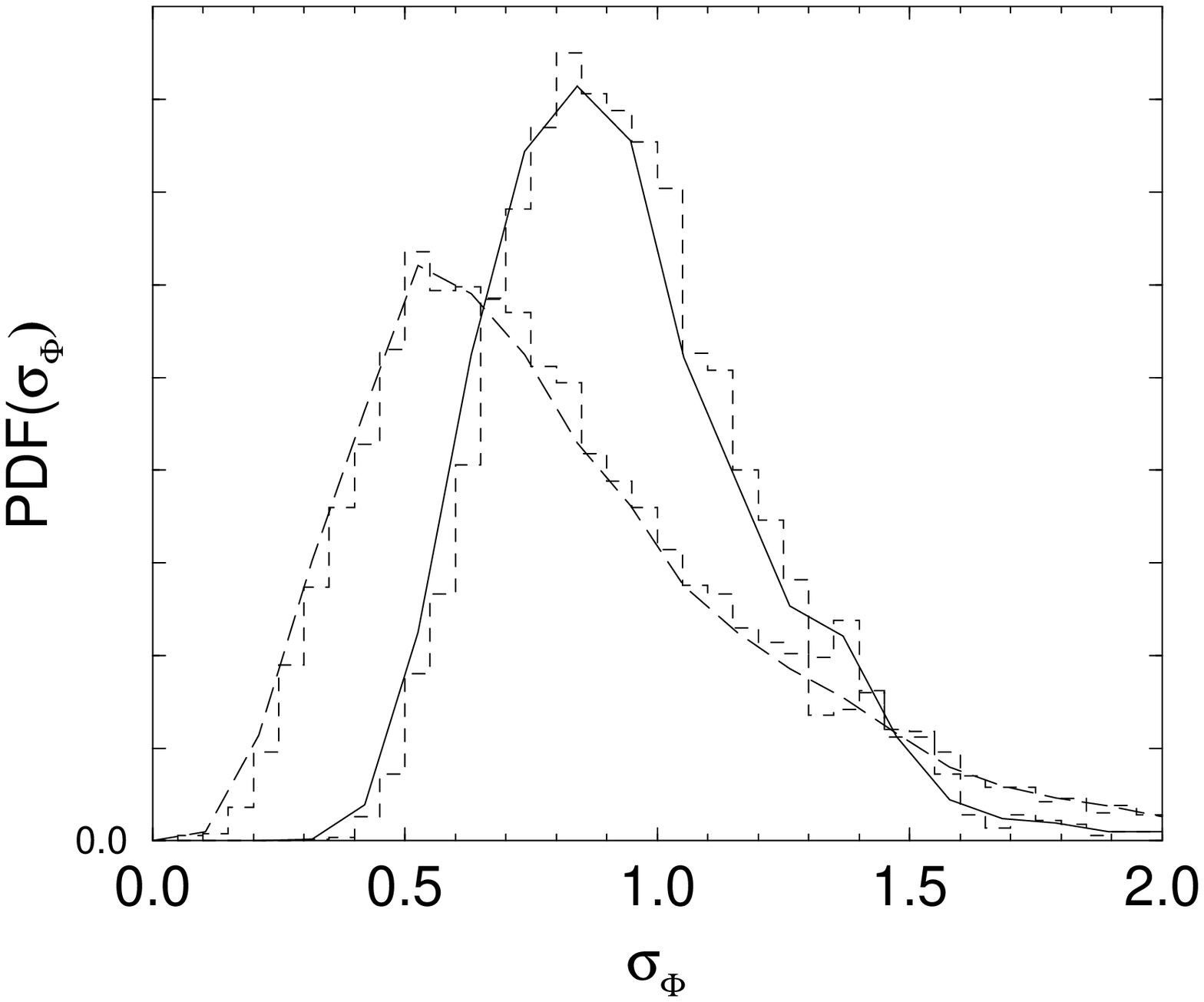,width=7cm} }
\caption{(a) Analysis of the bias of the r.m.s.\ of $\Phi$
   as a function of the number of particles per cell $N_{pc}$
   for three different NGP computations; dotted line: -1 slope.
(b) PDF of the r.m.s. of $\Phi$ at $t/T=3$ in the NGP computation 
over 5000 cells. Solid line: $N_{pc}\!=\!20$, dashed line: $N_{pc}\!=\!70$.
Results are smoothed histograms (dashed lines).}
\label{biasM2V0_2}
\end{figure}

%-------------------------------
% Figure 10 : géométrie cyclone
%-------------------------------
\afterpage{\clearpage} \newpage
\begin{table}[H]
\centering
\begin{tabular}{|c|c|c|c|c|c|c|}
\hline
$De/D$ & $H/D$ & $h/D$ & $a/D$ & $b/D$ & $S/D$ & $B/D$ \\
\hline\hline
$1/2$ & $4$ & $3/2$ & $1/2$ & $1/5$ & $1/2$ & $9/25$ \\
\hline
\end{tabular}
\end{table}
\vspace*{-5mm}
\begin{figure}[H]
\centering
\includegraphics[height=10cm]{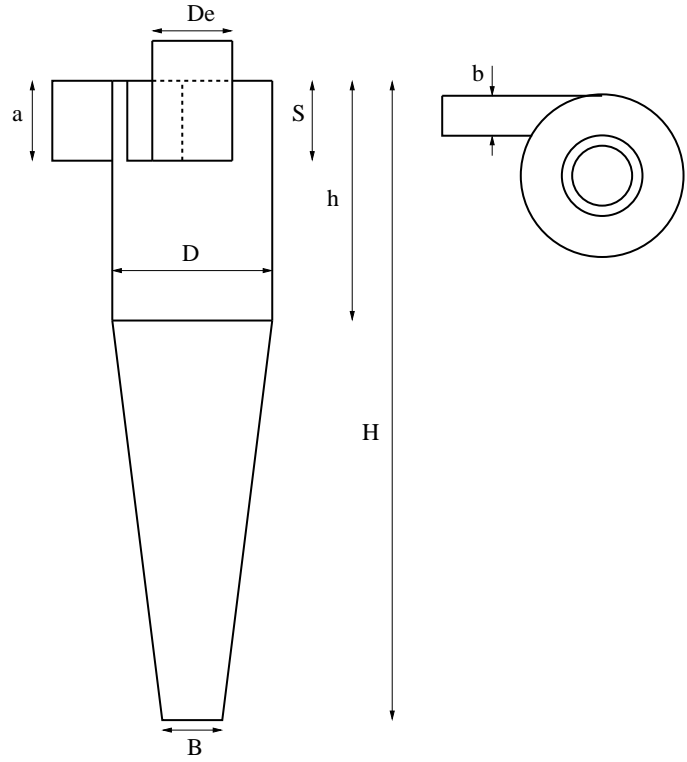}
\caption{Definition of the geometry of a cyclone of the Stairmand
  design. All dimensions are given as a function of the diameter $D$
  of the cyclone.} 
\label{fig:cyclone}
\end{figure}

%--------------------------------
% Figure 11 : résultats cyclone
%--------------------------------
\afterpage{\clearpage} \newpage
\setlength{\unitlength}{1cm}
\begin{figure}[H]
\centering
\includegraphics[width=0.8\textwidth]{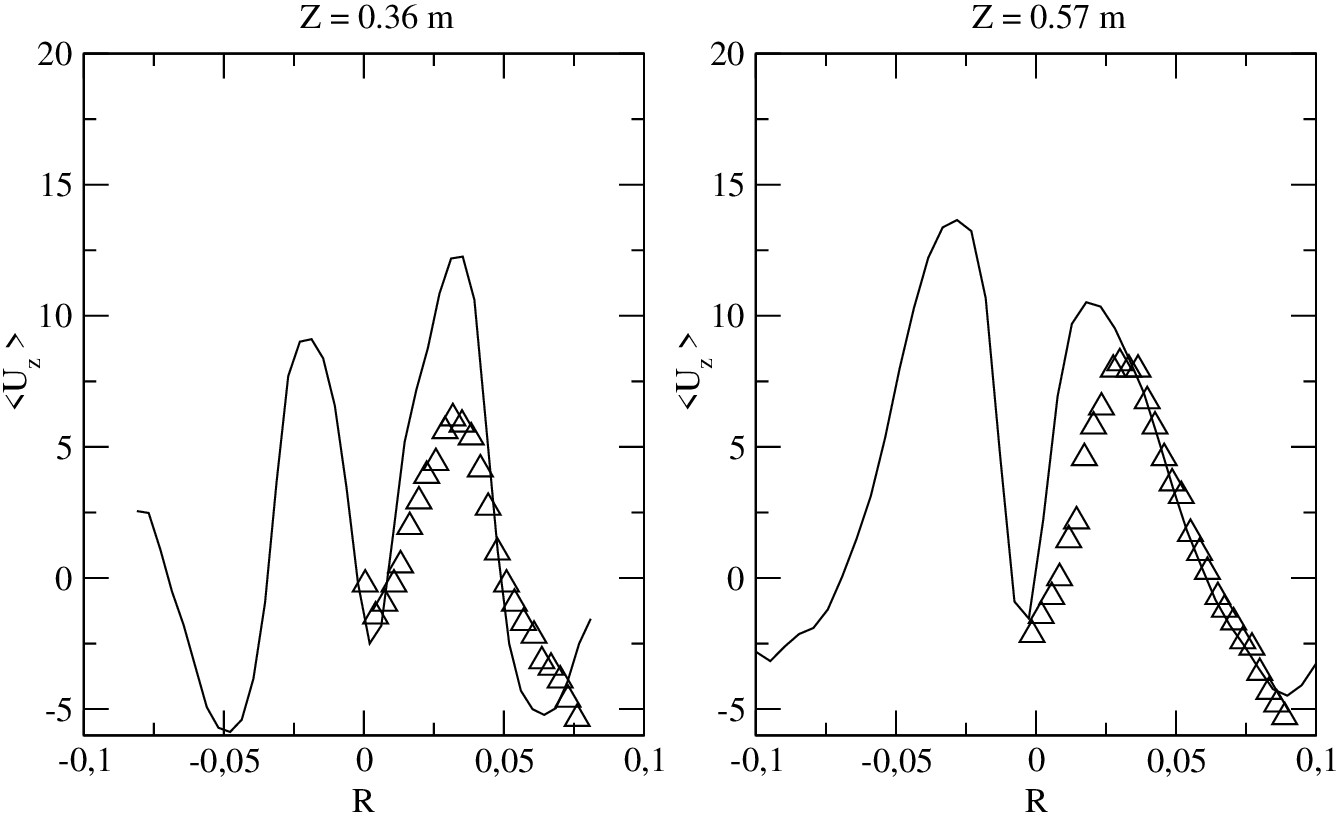}
\caption{Gas mean axial velocity profiles at two different heights:
  $z=0.36\,{\rm m}$ (left) and $z=0.57\,{\rm m}$ (right). The velocities are plotted as
  functions of the radius $R$. Continuous lines:
  computations. Experimental data: $\triangle$.}
\label{fig:gas_axe}
\end{figure}

%--------------------------------
% Figure 12 : résultats cyclone
%--------------------------------
\afterpage{\clearpage} \newpage
\begin{figure}[H]
\centering
\includegraphics[width=0.8\textwidth]{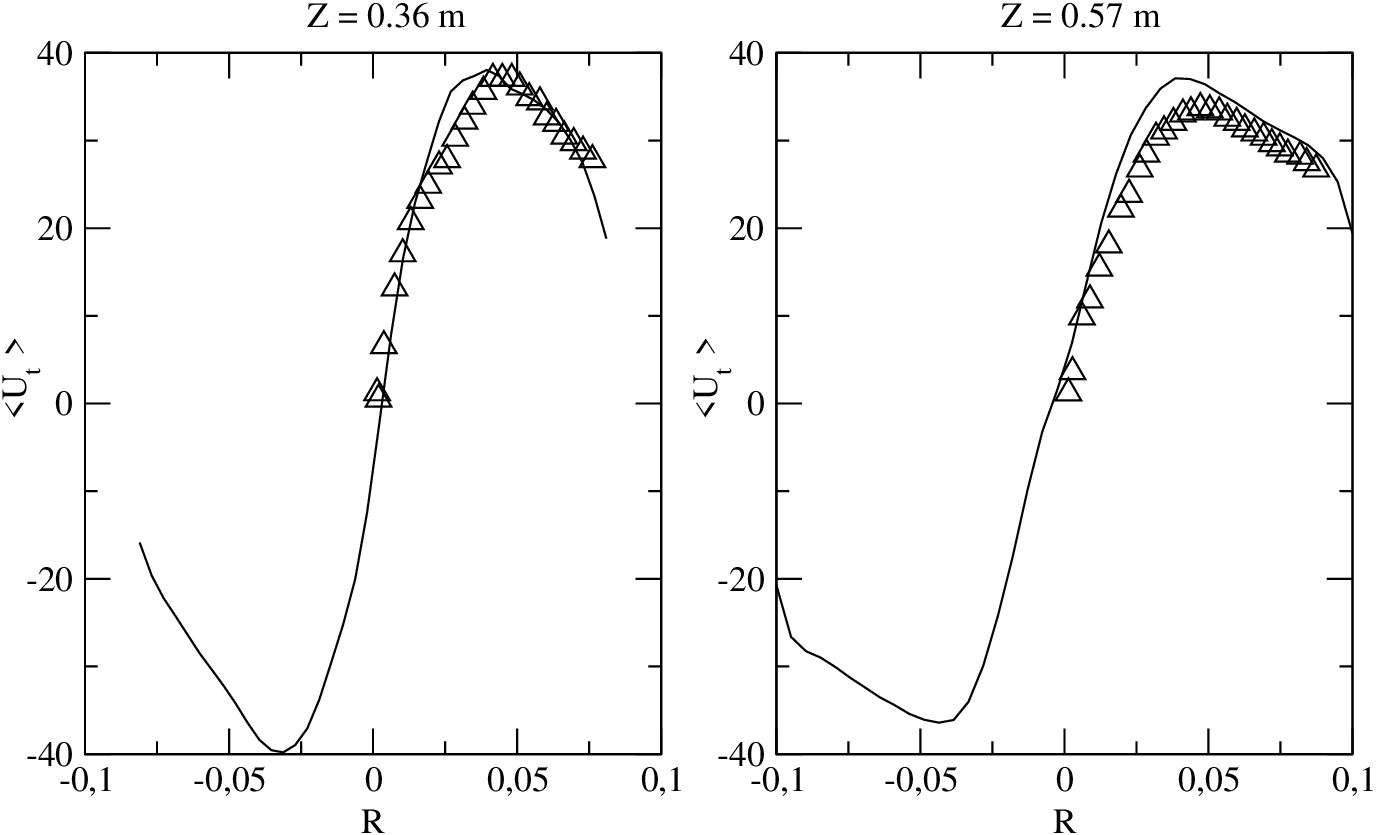}
\caption{Gas mean tangential velocity profiles at two different heights:
  $z=0.36\,{\rm m}$ (left) and $z=0.57\,{\rm m}$ (right). The velocities are plotted as
  functions of the radius $R$. Continuous lines:
  computations. Experimental data: $\triangle$.}
\label{fig:gas_rad}
\end{figure}

%--------------------------------
% Figure 13 : résultats cyclone
%--------------------------------
\afterpage{\clearpage} \newpage
\begin{figure}[H]
\begin{minipage}[c]{1.0\textwidth}
\begin{minipage}[c]{0.5\textwidth}
\centering
\includegraphics[width=0.92\textwidth]{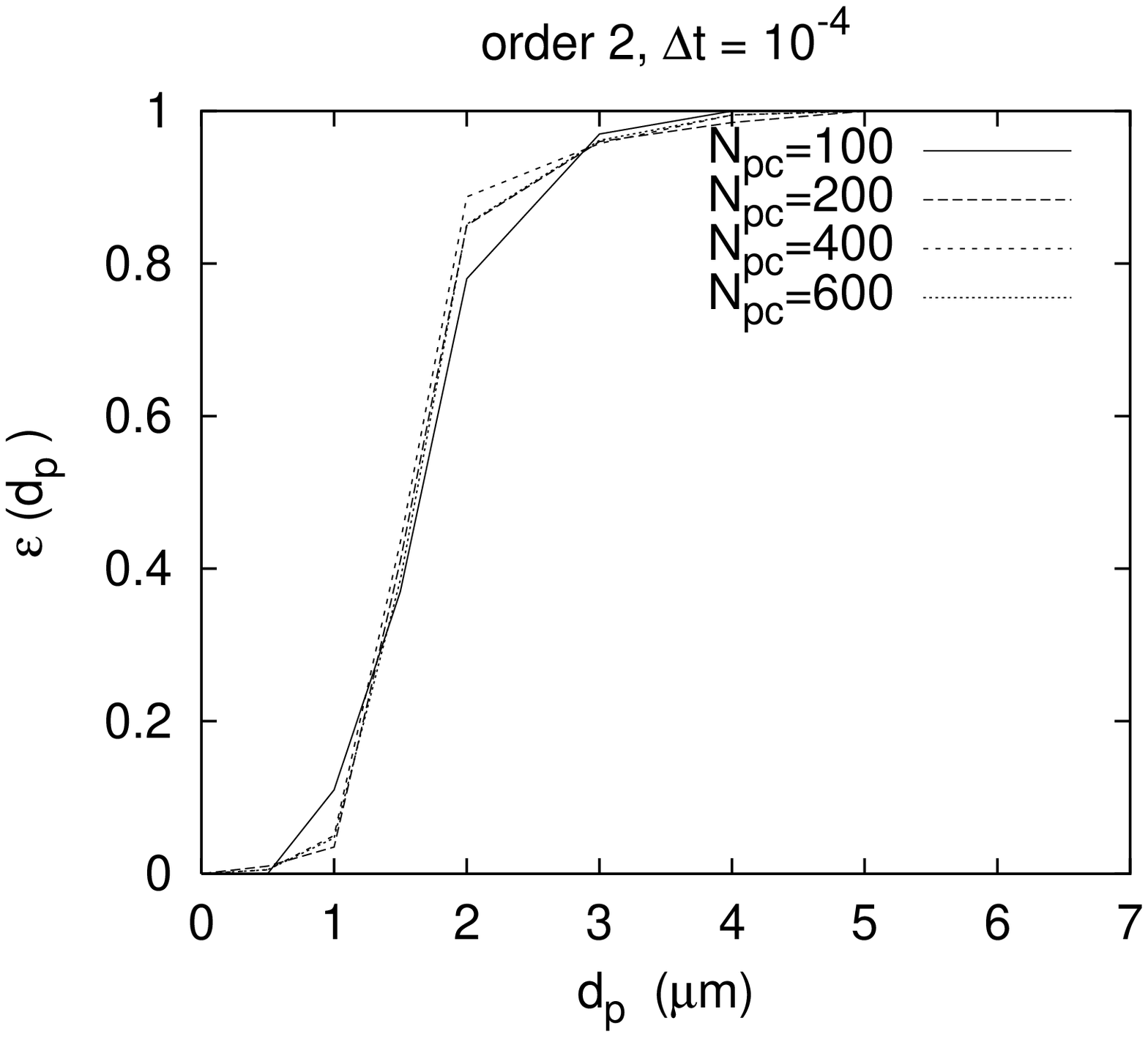}
\end{minipage}
\begin{minipage}[c]{0.5\textwidth}
\centering
\includegraphics[width=0.92\textwidth]{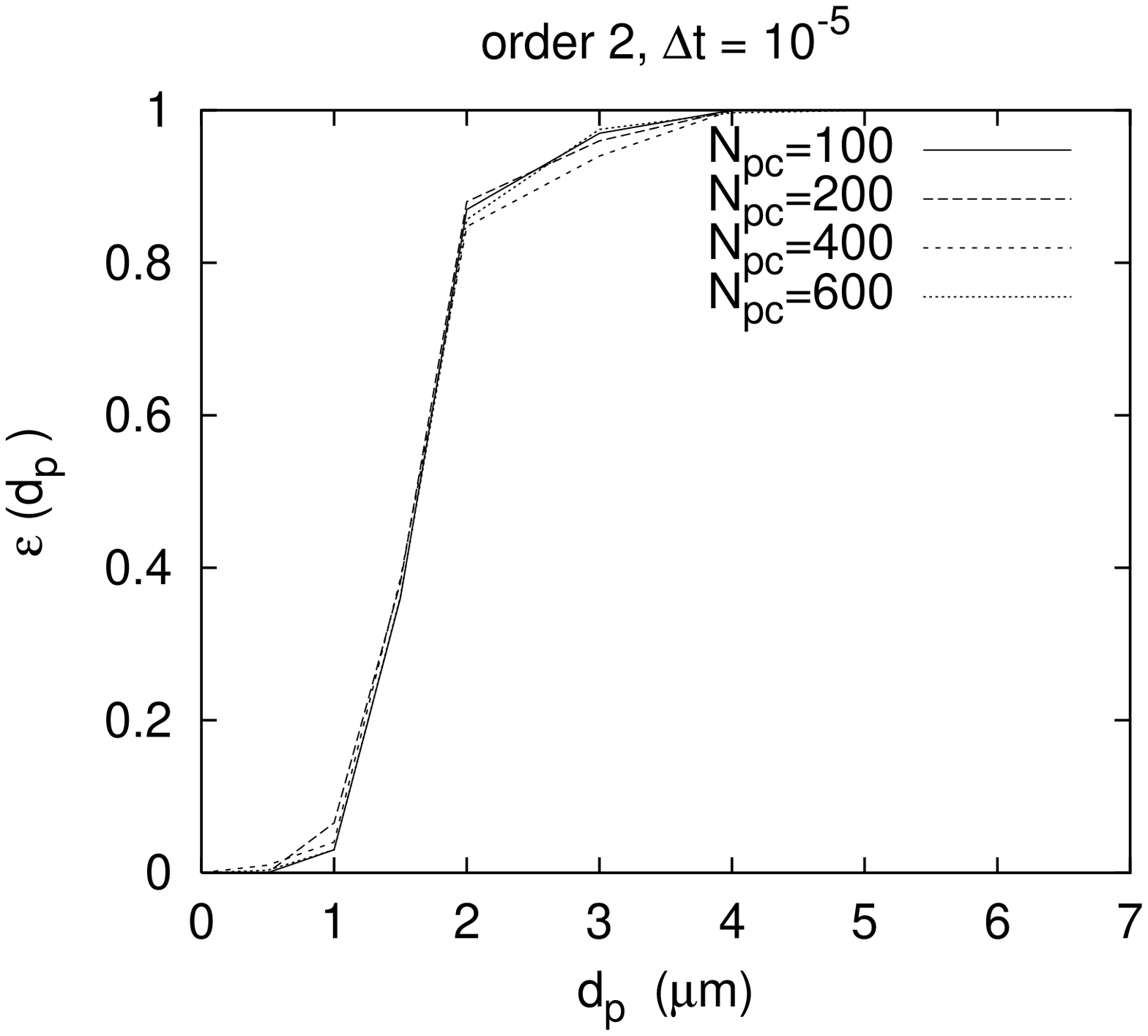}
\end{minipage}
\begin{minipage}[c]{0.5\textwidth}
\centering
\includegraphics[width=0.92\textwidth]{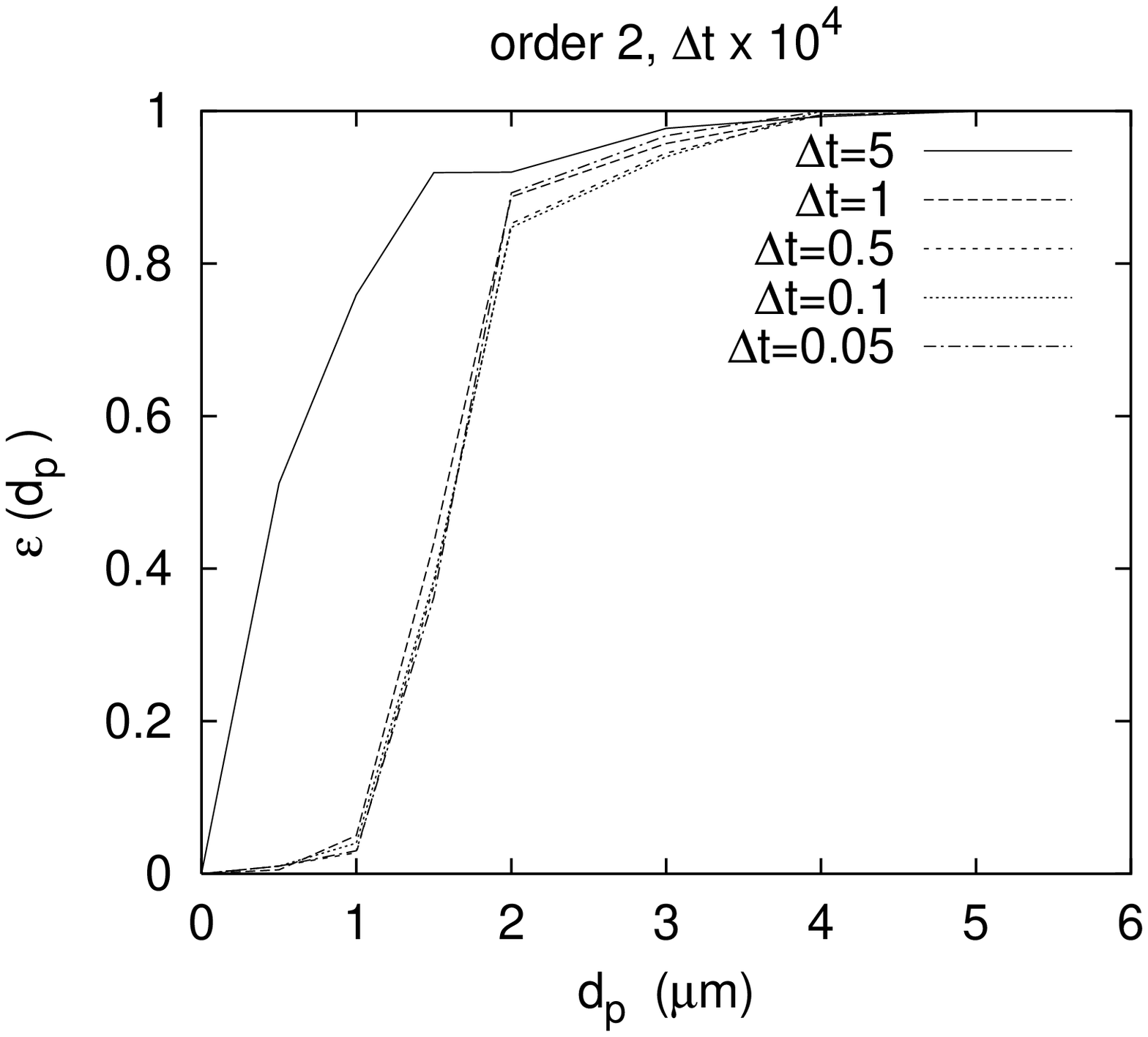}
\end{minipage}
\begin{minipage}[c]{0.5\textwidth}
\centering
\includegraphics[width=0.92\textwidth]{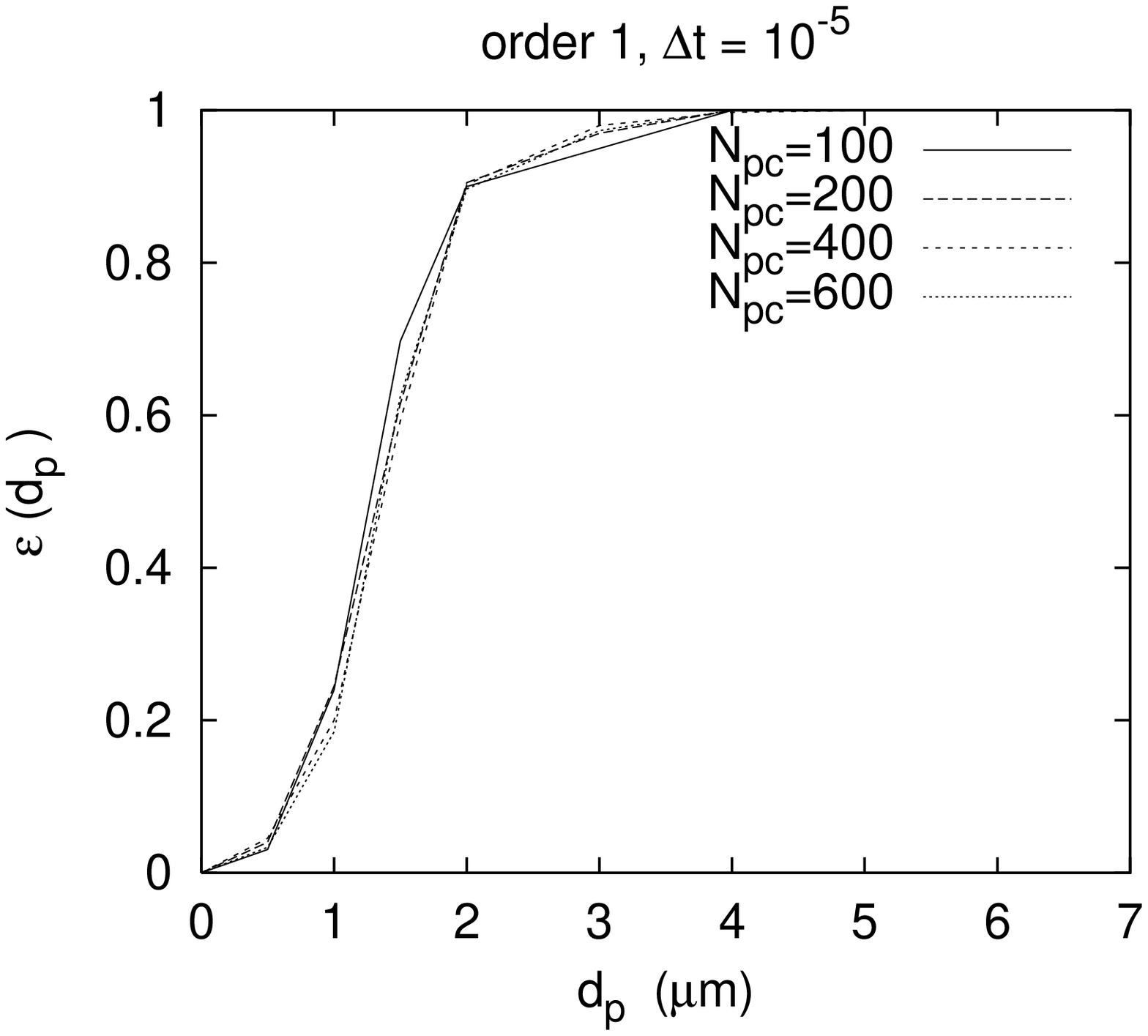}
\end{minipage}
\begin{minipage}[c]{0.5\textwidth}
\centering
\includegraphics[width=0.92\textwidth]{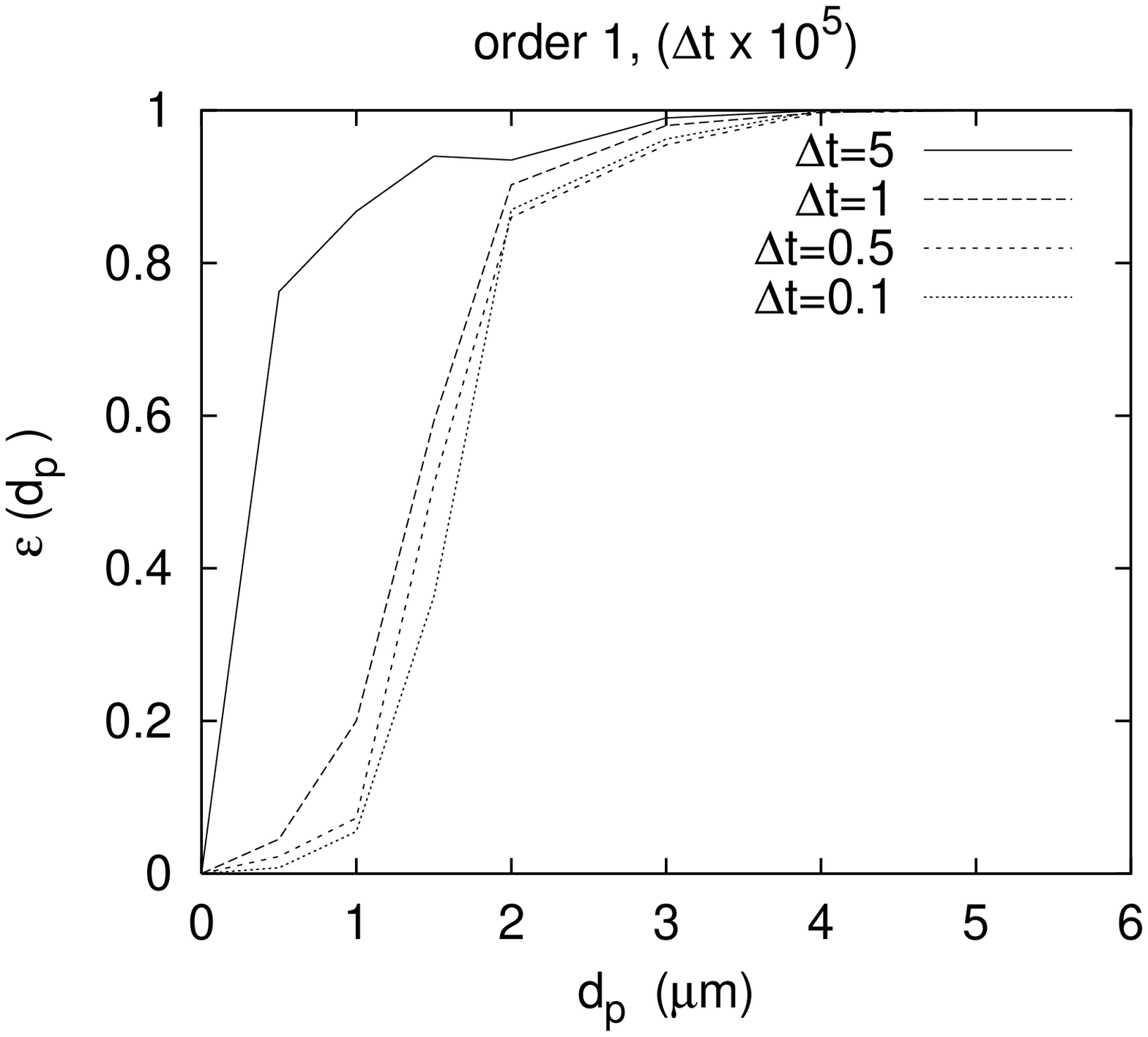}
\end{minipage}
\begin{minipage}[c]{0.5\textwidth}
\centering
\includegraphics[width=0.92\textwidth]{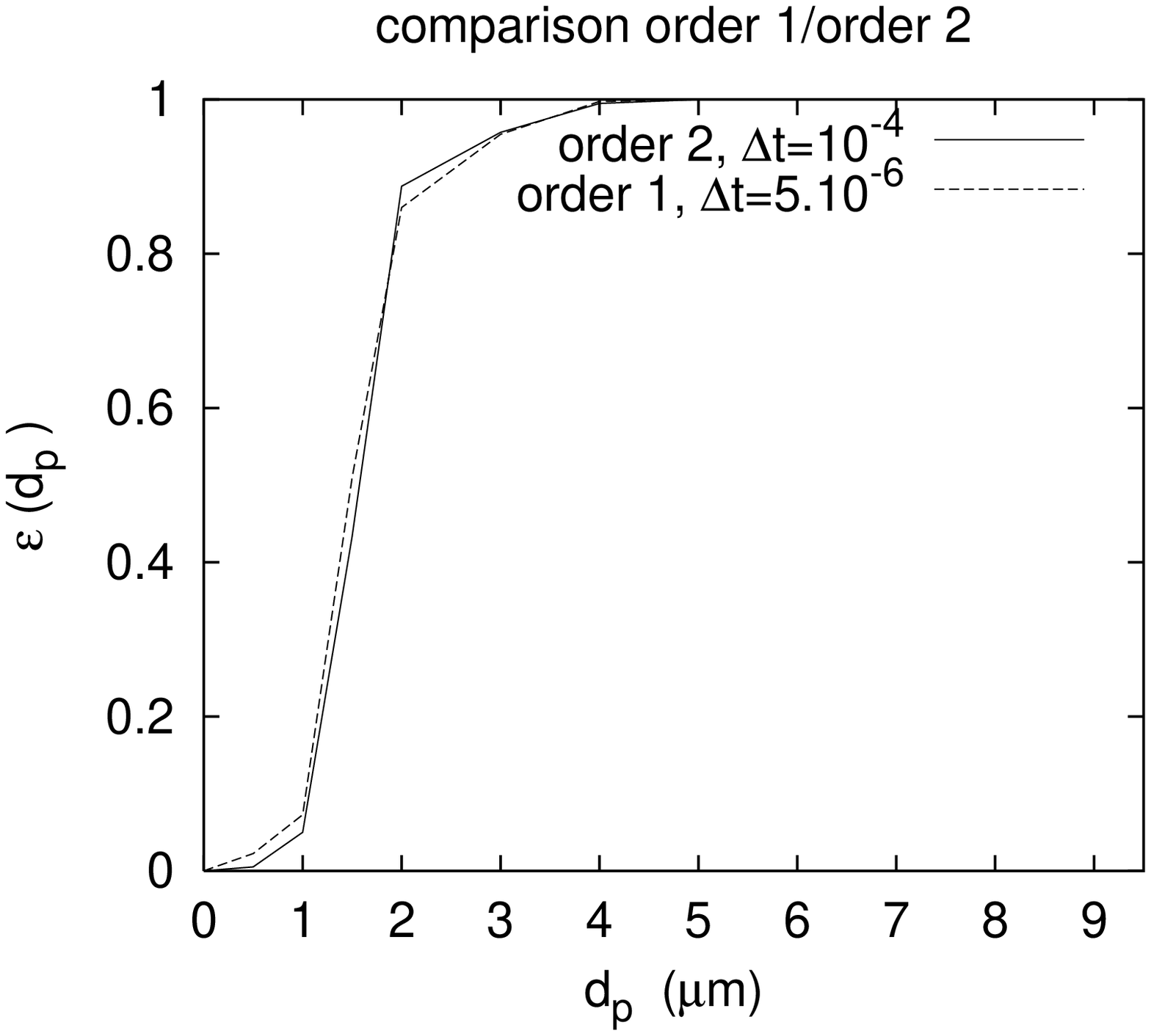}
\end{minipage}
\end{minipage}
\caption{Sensitivity analysis for the second order and first
order schemes. The sensitivity analysis is carried out for the number
of particles per class, $N_{pc}$, and the time step, $\Delta t$ (with
$N_{pc}=400)$. At last, comparison of the time steps for the first and
second order schemes for identical numerical results (bottom right
corner). All numerical results represent the selectivity curve which
gives the efficiency of the cyclone, $\epsilon$ as a function of the
particle diameter, $d_p$, \textit{i.e.} $\epsilon(d_p)$.}
\label{fig:res_2}
\end{figure}

%------------------------------
% Figure 14 : géométrie Hercule
%------------------------------
\afterpage{\clearpage} \newpage
\begin{figure}[H]
\centering
\includegraphics[height=12cm]{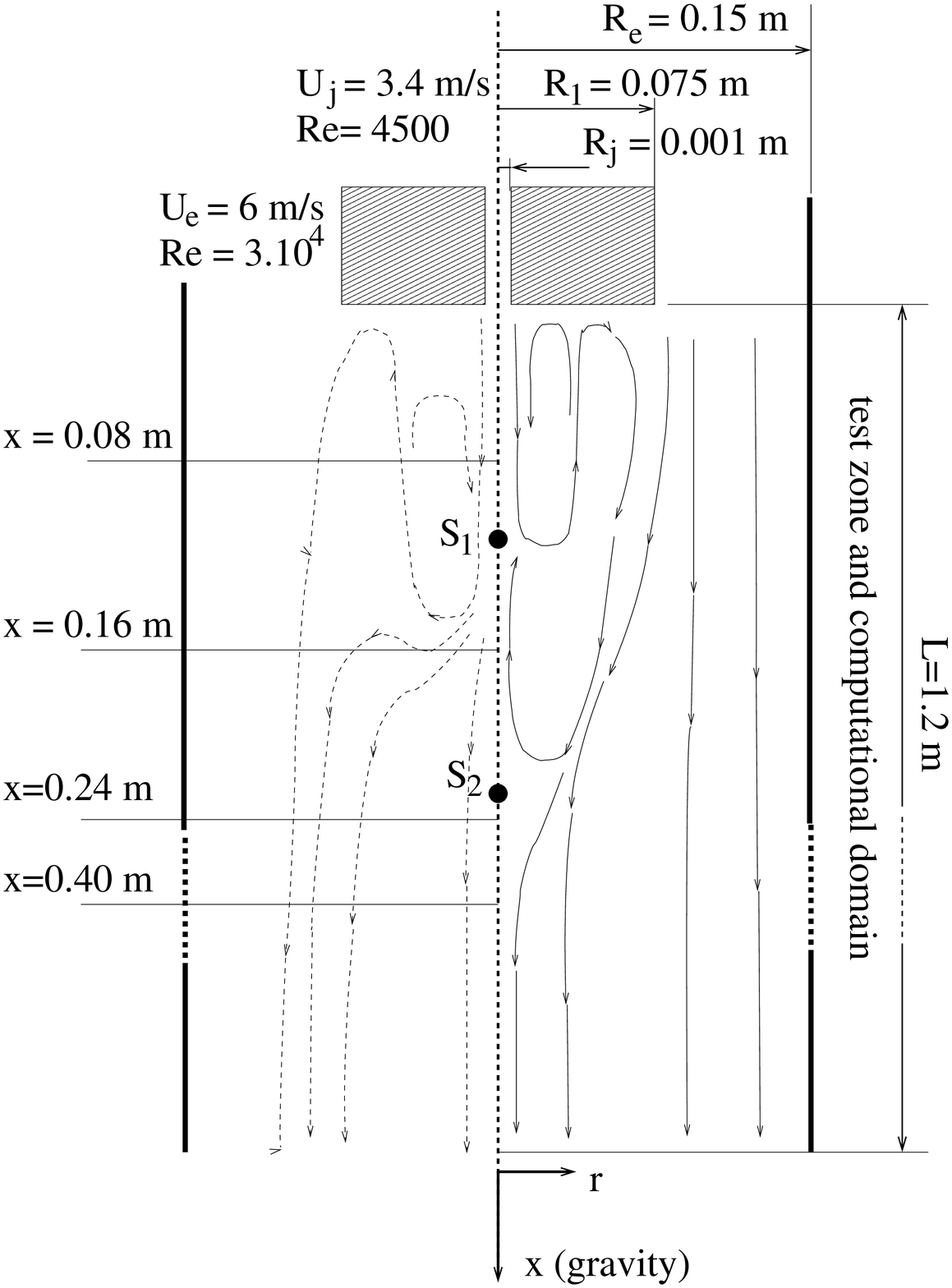}
\caption{The 'Hercule' experimental setup. The mean streamlines are
shown for the fluid (solid lines) and the particles (dashed
lines). Two stagnation points in the fluid flow can be observed
($S_1$ and $S_2$). Experimental data is available for radial profiles
of different statistical quantities at five axial distances downstream
of the injection ($x=0.08, 0.16,0.24,0.32, 0.40 \,{\rm m}$) (experimental
data is also available on the symmetry axis).}
\label{fig:hercule}
\end{figure}

%------------------------------
% Figure 15 : résultats Hercule
%------------------------------
% VITESSE AXIALE PARTICULES
\afterpage{\clearpage} \newpage
\begin{figure}[H]
 \begin{minipage}[c]{1.0\textwidth}
 \begin{minipage}[c]{0.5\textwidth}
   \centering
   \includegraphics[width=0.92\textwidth]{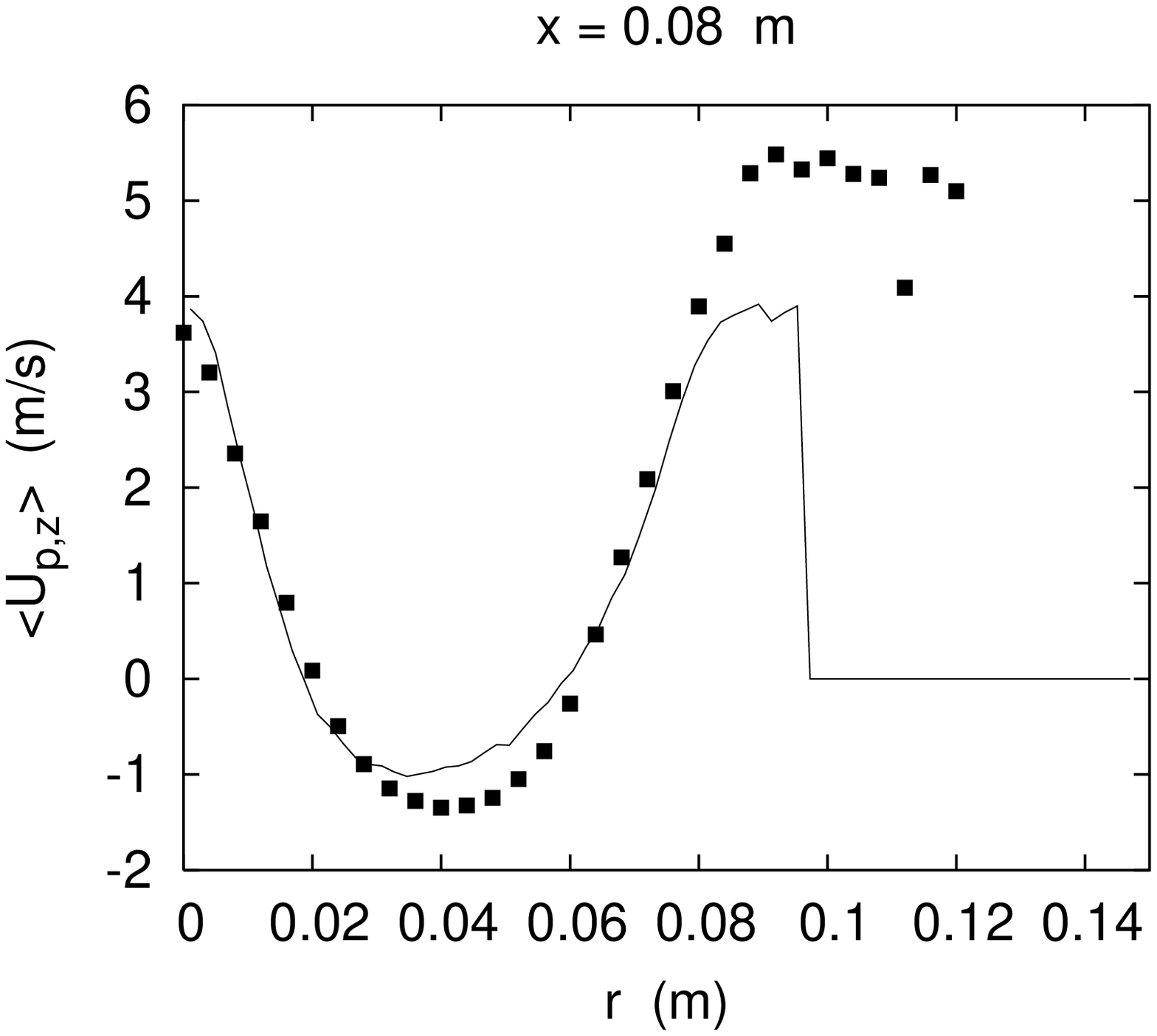}
 \end{minipage}
 \begin{minipage}[c]{0.5\textwidth}
   \centering
   \includegraphics[width=0.92\textwidth]{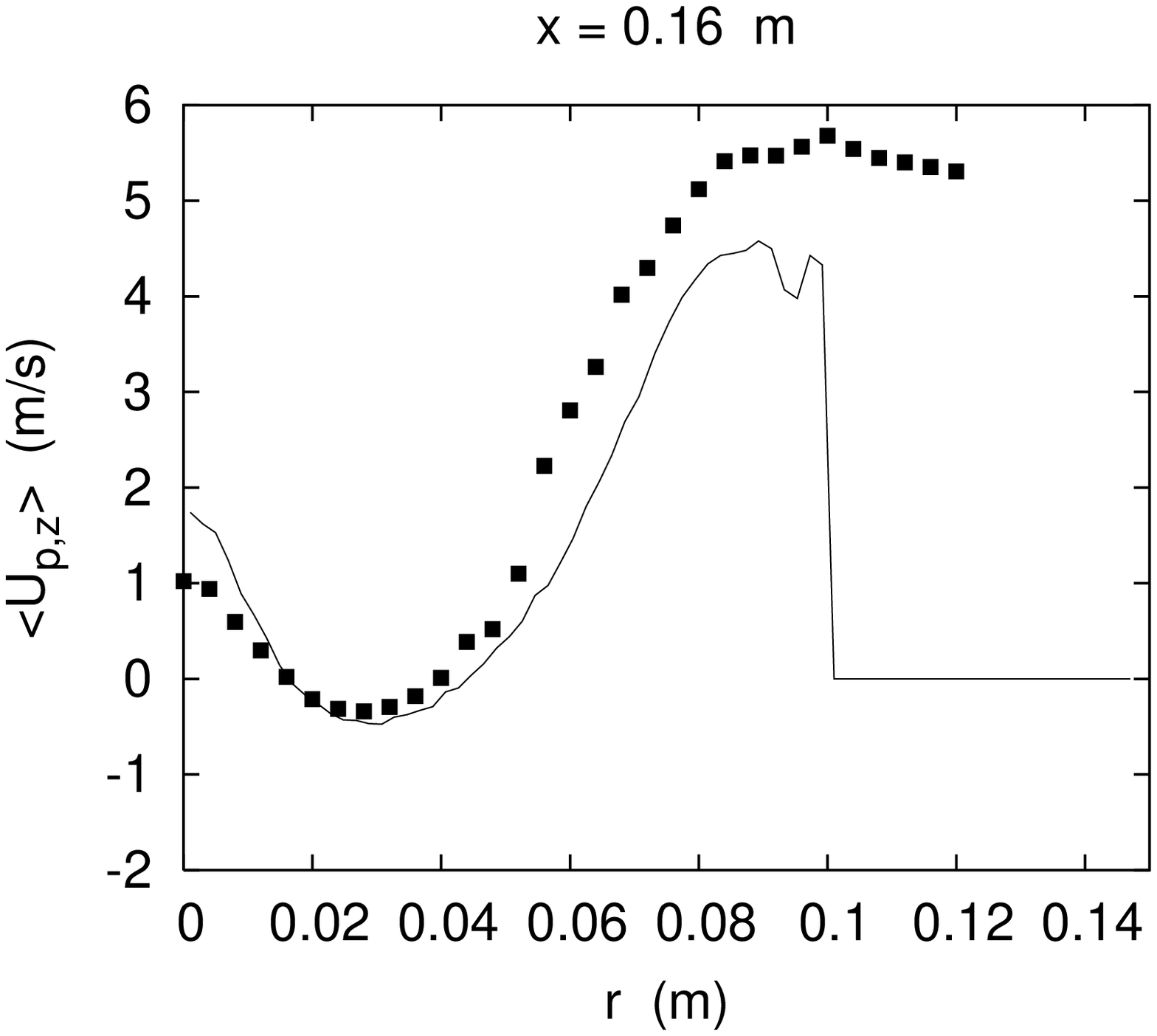}
 \end{minipage}
 \begin{minipage}[c]{0.5\textwidth}
   \centering
   \includegraphics[width=0.92\textwidth]{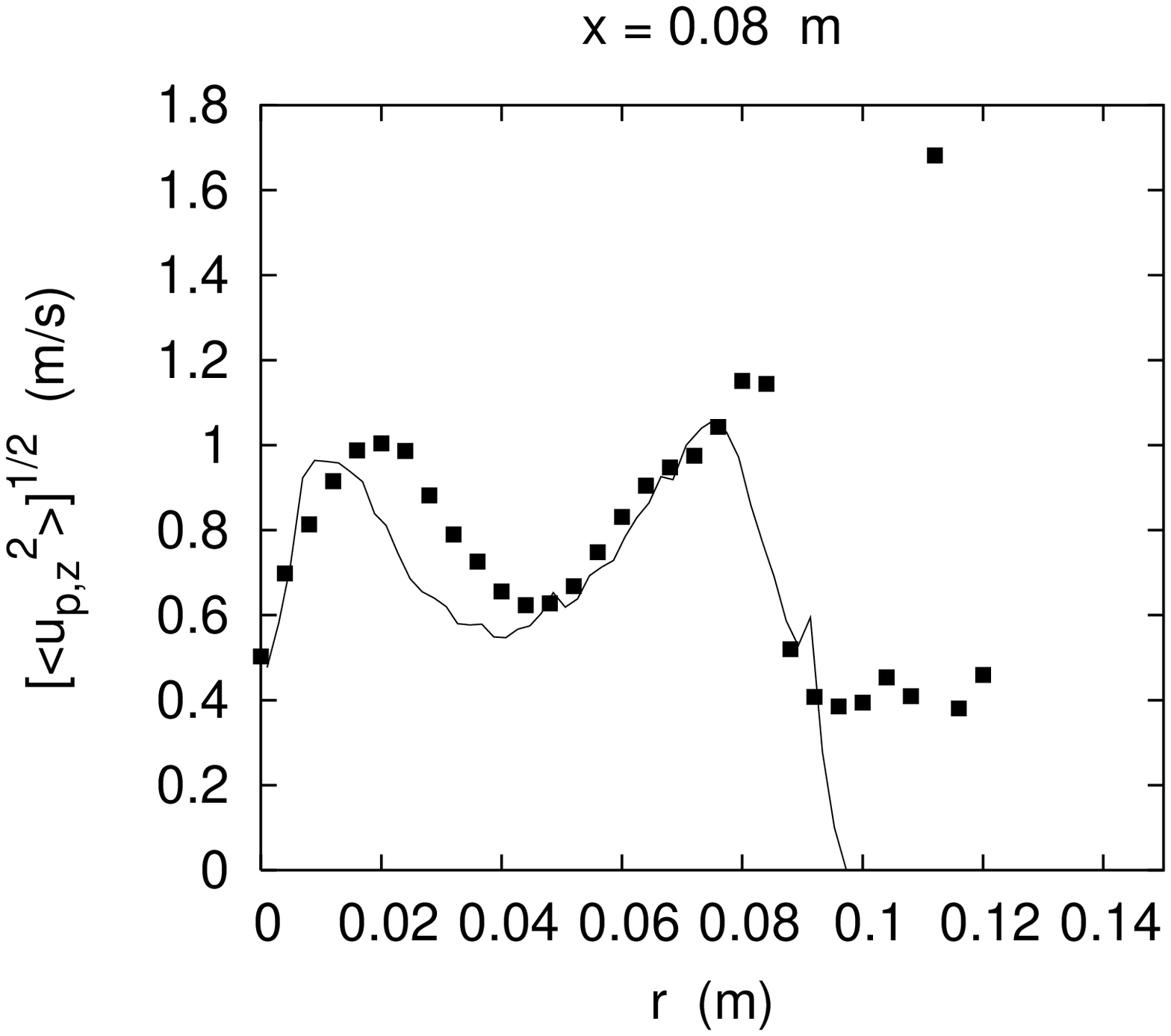}
 \end{minipage}
 \begin{minipage}[c]{0.5\textwidth}
   \centering
   \includegraphics[width=0.92\textwidth]{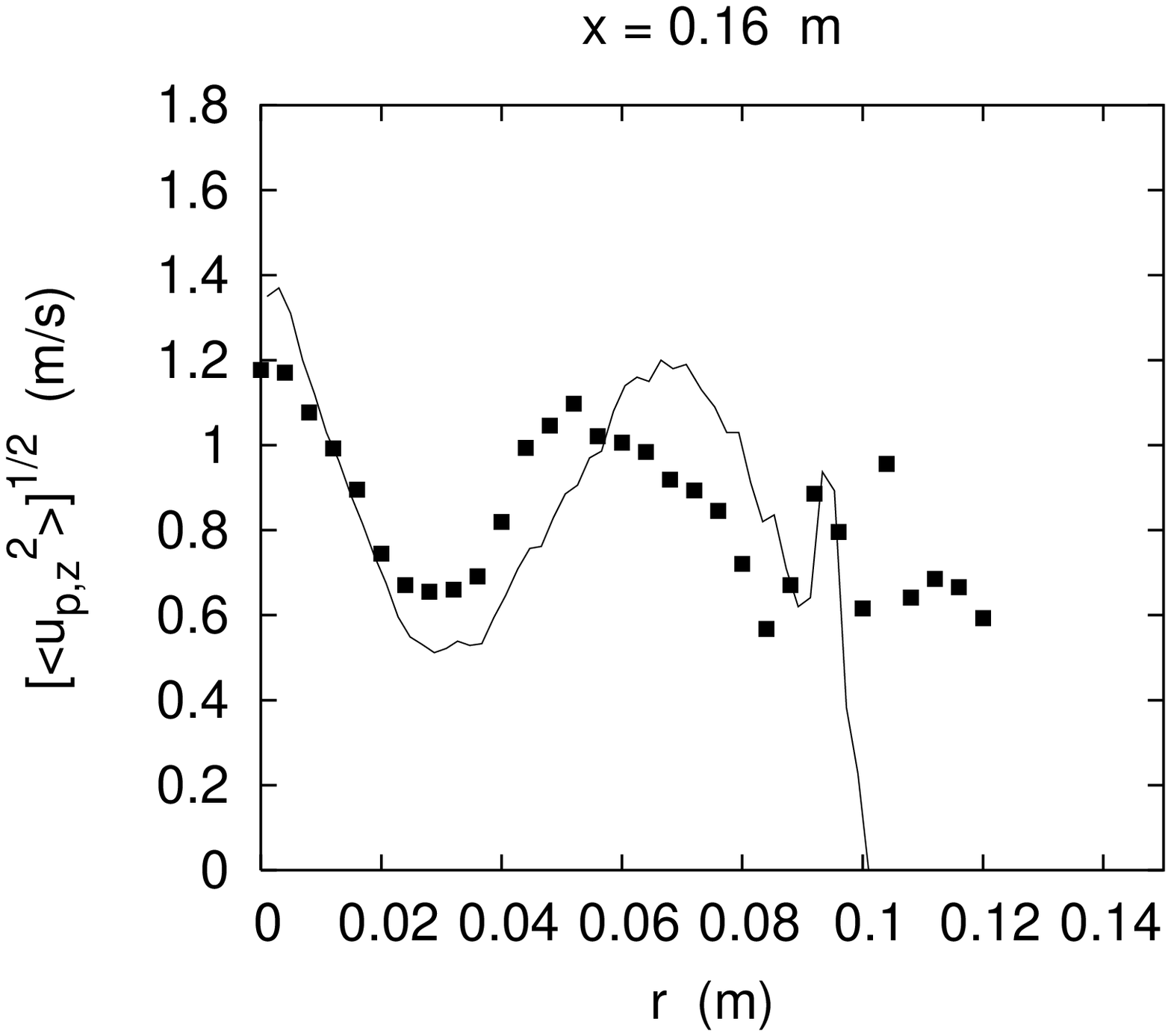}
 \end{minipage}
 \end{minipage}
 \caption{Radial profiles of the particle axial velocity at
 $x=0.08\,{\rm m}$ and $x=0.16\,{\rm m}$. Mean velocities (top) and fluctuating
 velocities (bottom). Continuous lines: computations (first order
 scheme). Experimental data: $\blacksquare$.}
\label{fig:vitz} 
\end{figure}

%------------------------------
% Figure 16 : résultats Hercule
%------------------------------
% VITESSE RADIALE PARTICULES
\afterpage{\clearpage} \newpage
\begin{figure}[H]
 \begin{minipage}[c]{1.0\textwidth}
 \begin{minipage}[c]{0.5\textwidth}
   \centering
   \includegraphics[width=0.92\textwidth]{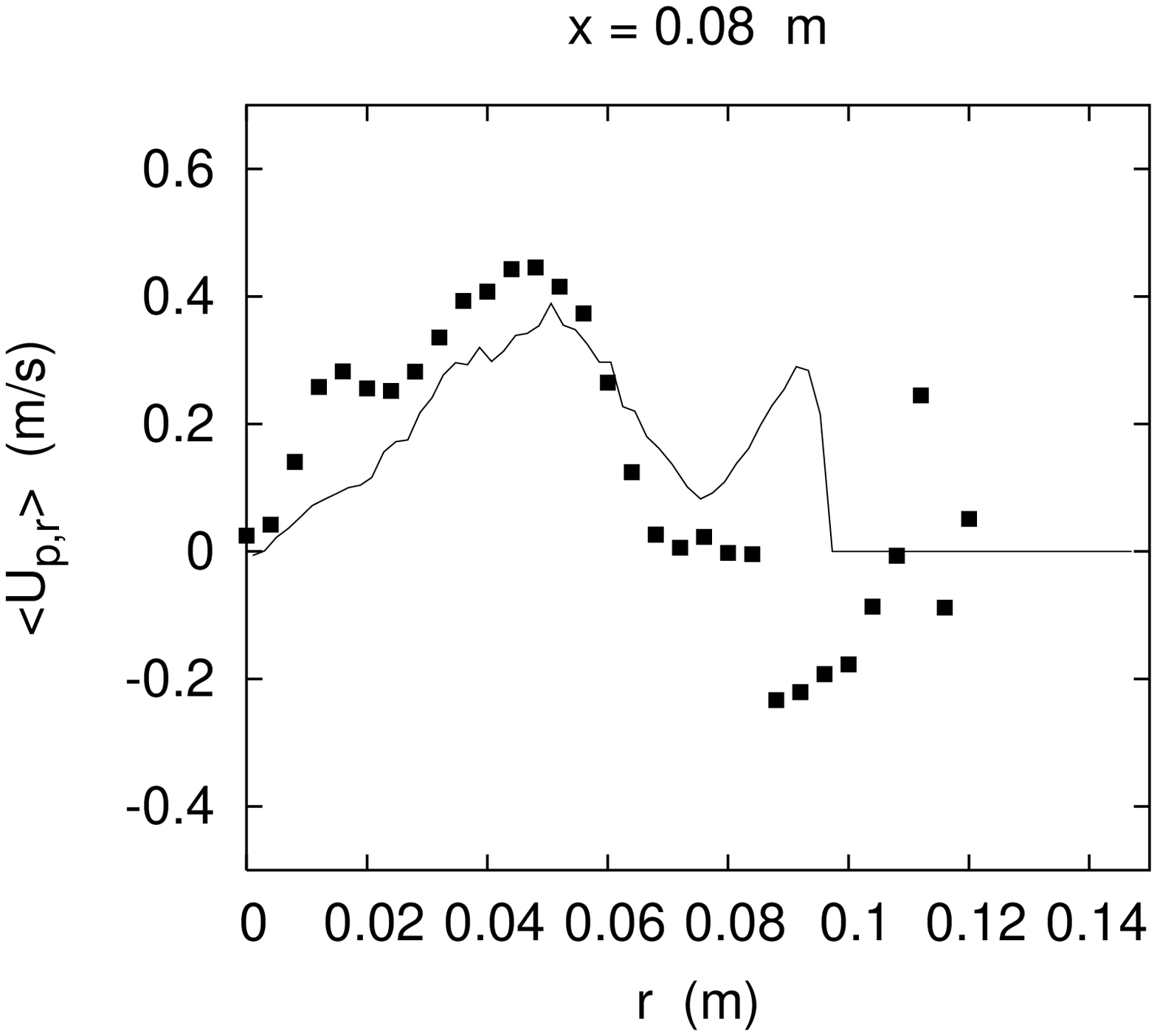}
 \end{minipage}
 \begin{minipage}[c]{0.5\textwidth}
   \centering
   \includegraphics[width=0.92\textwidth]{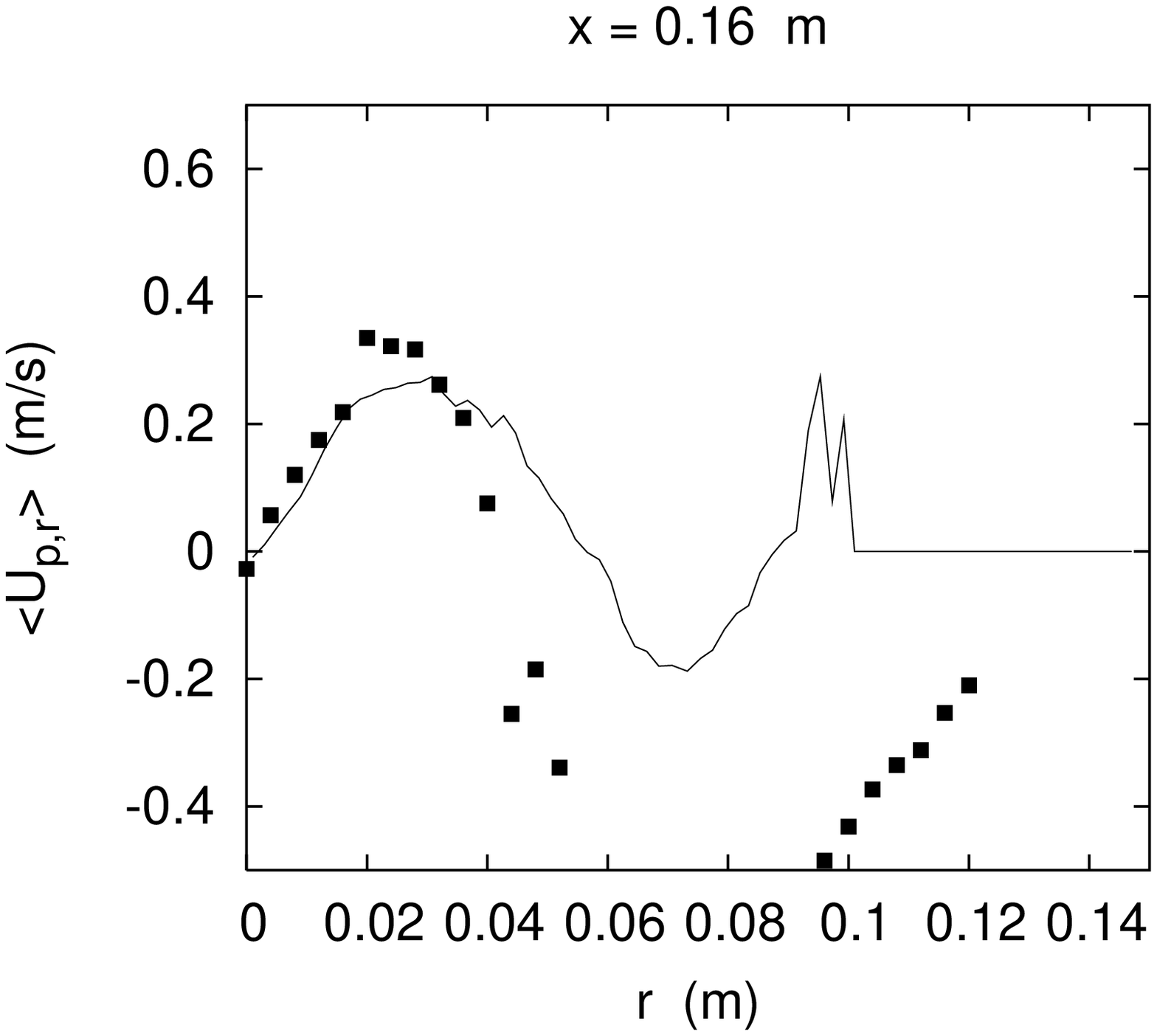}
 \end{minipage}
 \begin{minipage}[c]{0.5\textwidth}
   \centering
   \includegraphics[width=0.92\textwidth]{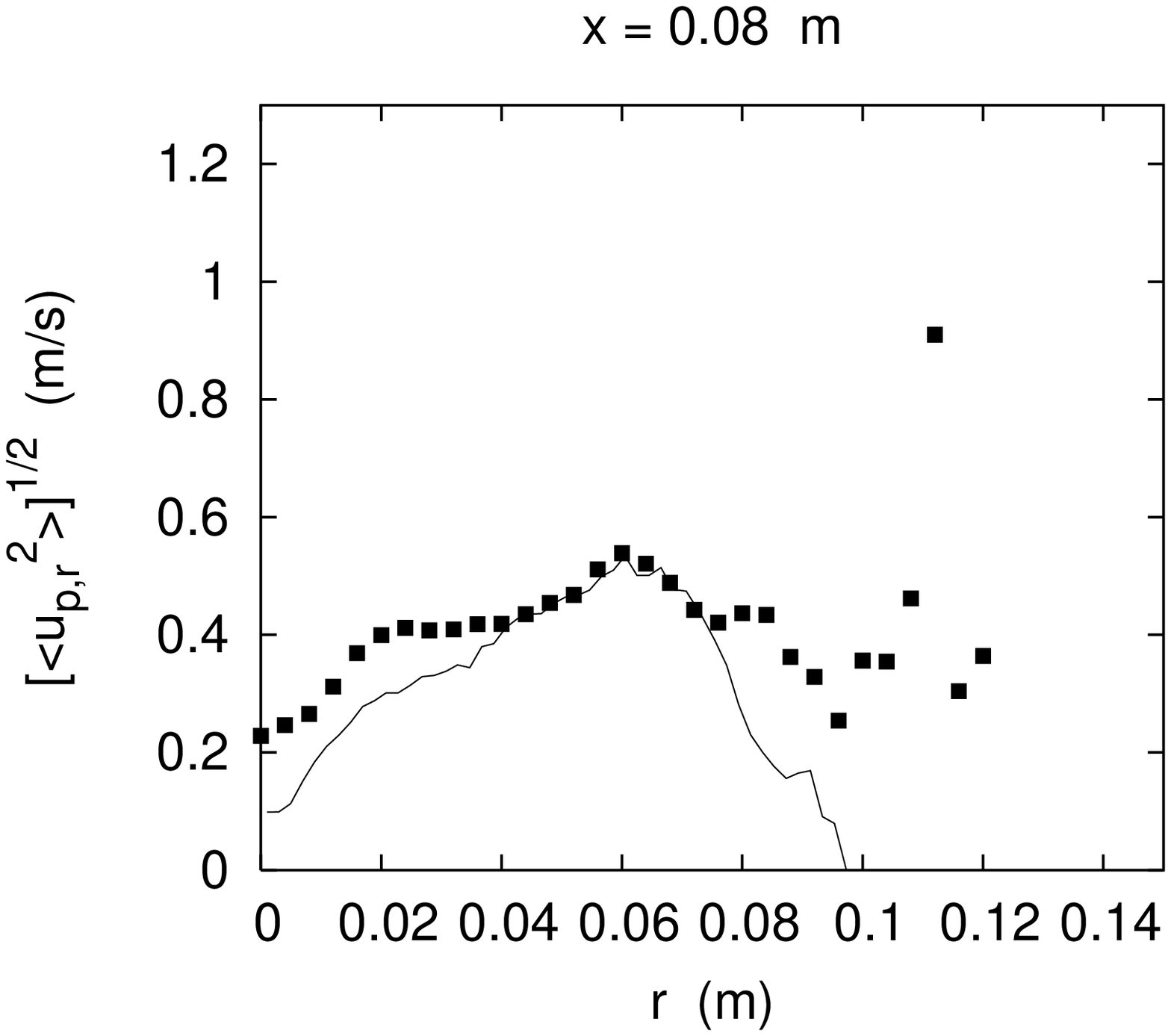}
 \end{minipage}
 \begin{minipage}[c]{0.5\textwidth}
   \centering
   \includegraphics[width=0.92\textwidth]{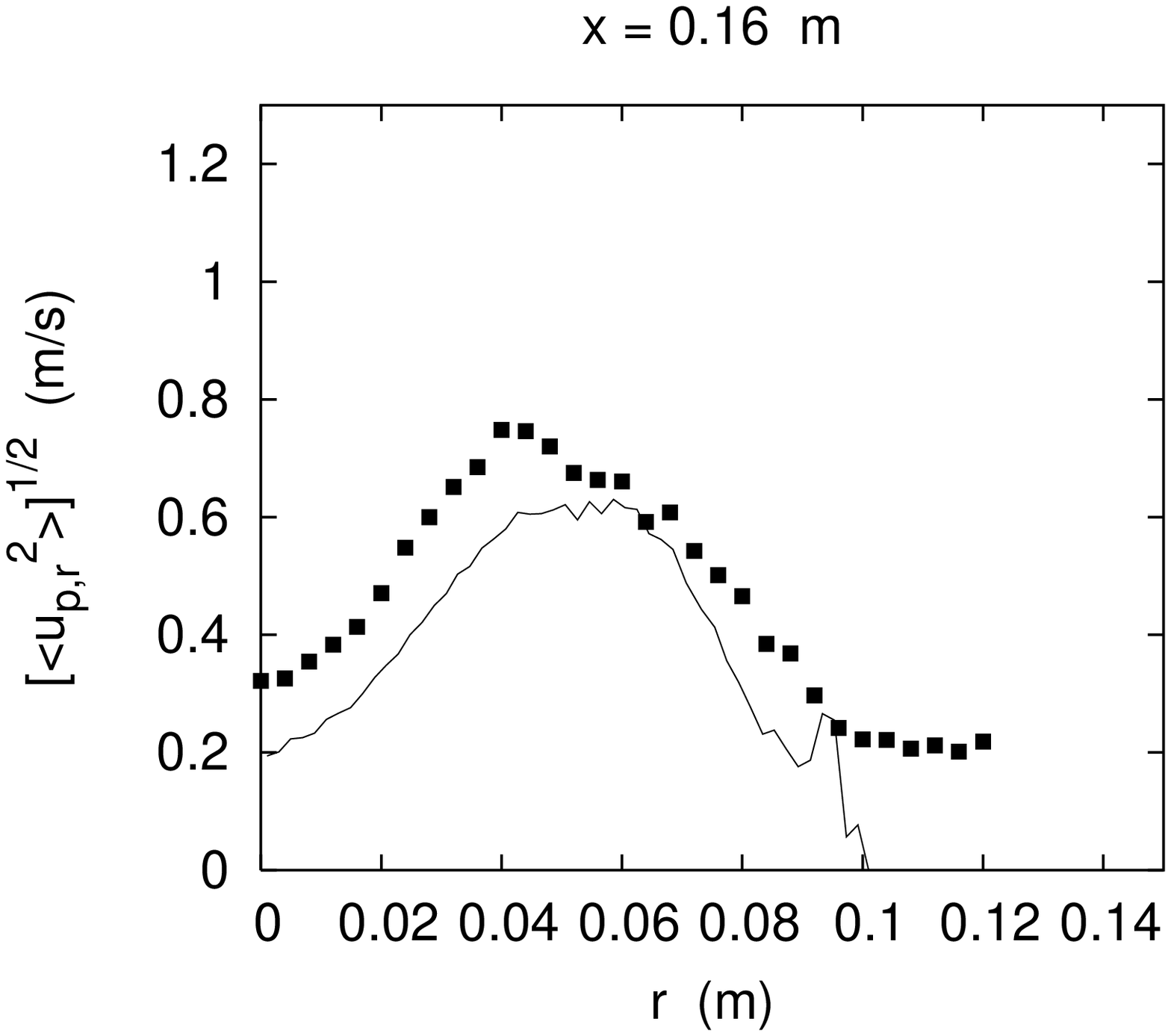}
 \end{minipage}
 \end{minipage}
 \caption{Radial profiles of the particle radial velocity at
 $x=0.08\,{\rm m}$ and $x=0.16\,{\rm m}$. Mean velocities (top) and fluctuating
 velocities (bottom). Continuous lines: computations (first order
 scheme). Experimental data: $\blacksquare$.}
\label{fig:vitx}
\end{figure}

%------------------------------
% Figure 17 : résultats Hercule
%------------------------------
% VITESSE AXIALE FLUIDE
\afterpage{\clearpage} \newpage
\begin{figure}[H]
 \begin{minipage}[c]{1.0\textwidth}
 \begin{minipage}[c]{0.5\textwidth}
   \centering
   \includegraphics[width=0.92\textwidth]{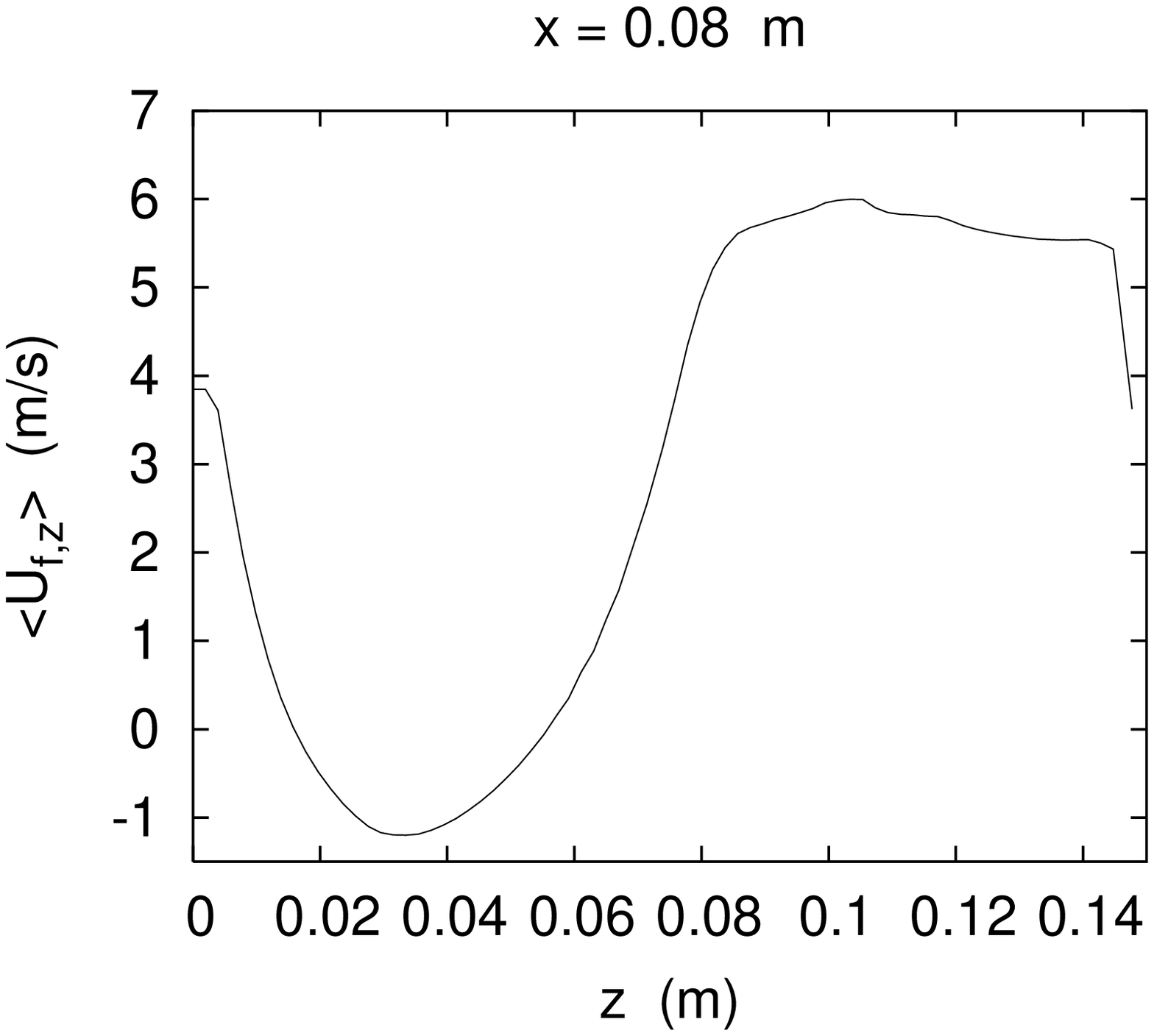}
 \end{minipage}
 \begin{minipage}[c]{0.5\textwidth}
   \centering
   \includegraphics[width=0.92\textwidth]{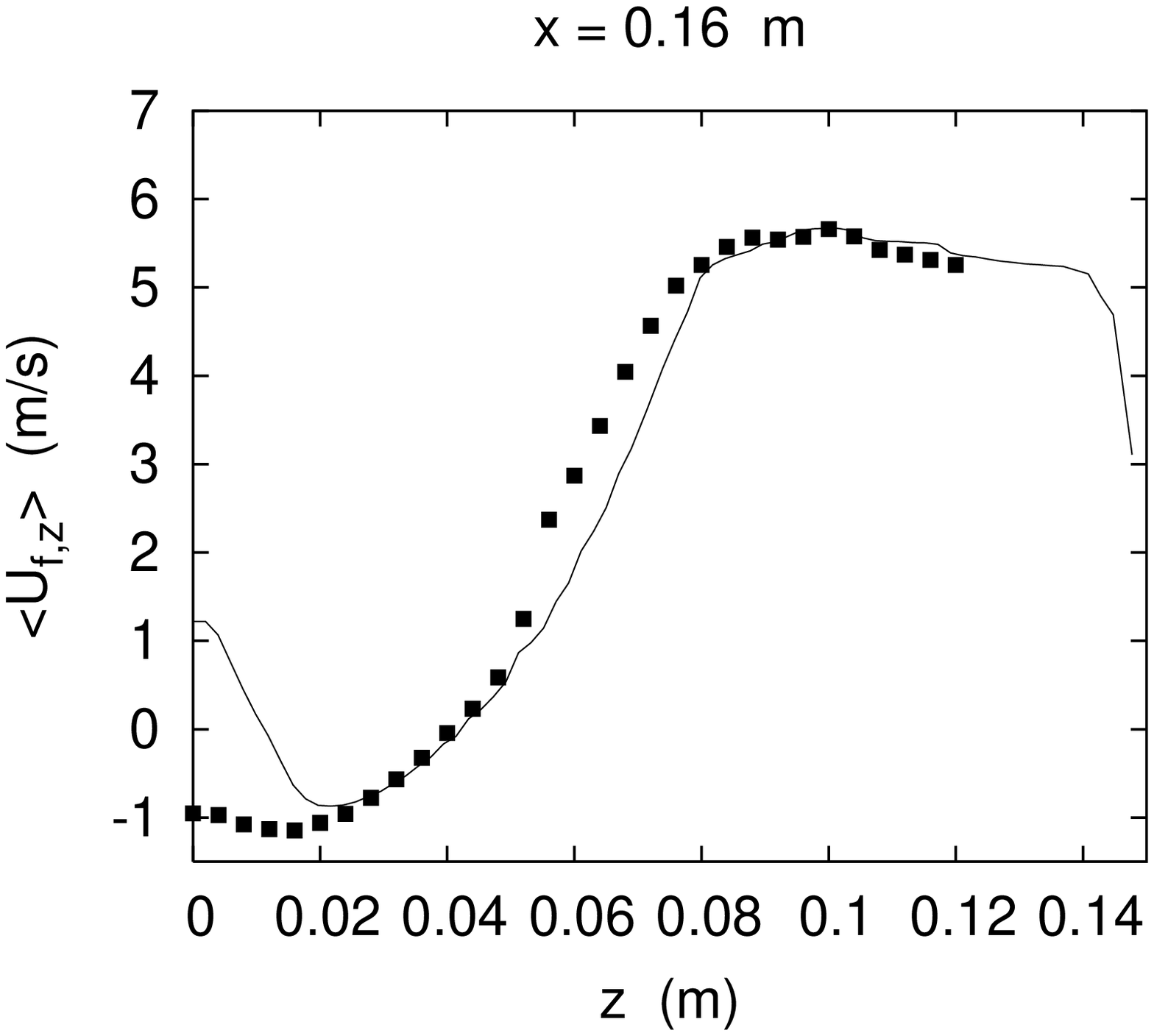}
 \end{minipage}
 \end{minipage}
 \caption{Radial profiles of the fluid axial velocity at
 $x=0.08\,{\rm m}$ and $x=0.16\,{\rm m}$ (mean velocities only). Continuous lines:
 computations (first order scheme). Experimental
 data: $\blacksquare$.} 
\label{fig:vitzf}
\end{figure}

%------------------------------
% Figure 18 : résultats Hercule
%------------------------------
%\afterpage{\clearpage} \newpage
\begin{figure}[H]
 \begin{picture}(0,22)
 \put(-1,0.5){\includegraphics[width=22cm]{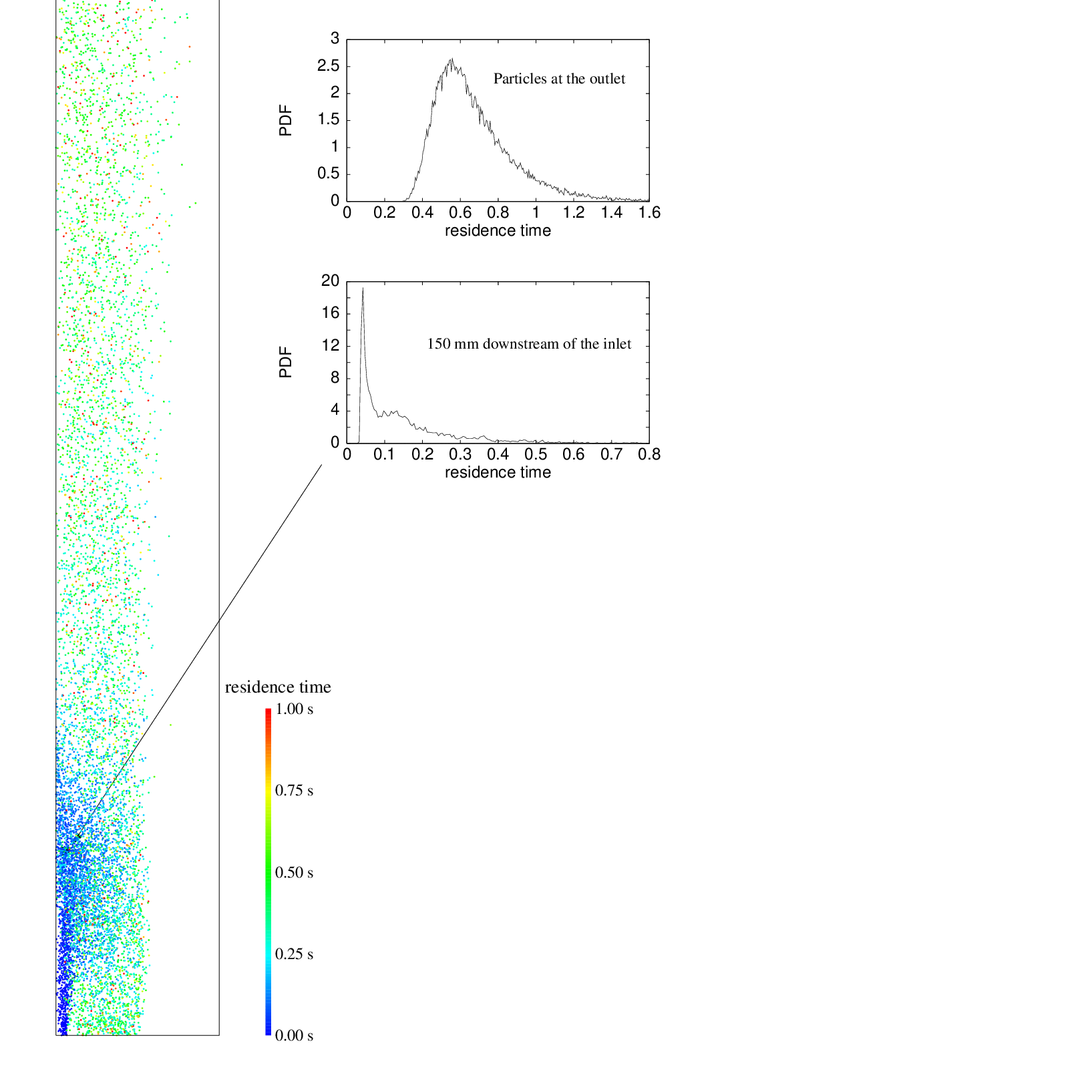}}
 \put(7.8,2.0){\includegraphics[angle=90,width=0.33\textwidth]{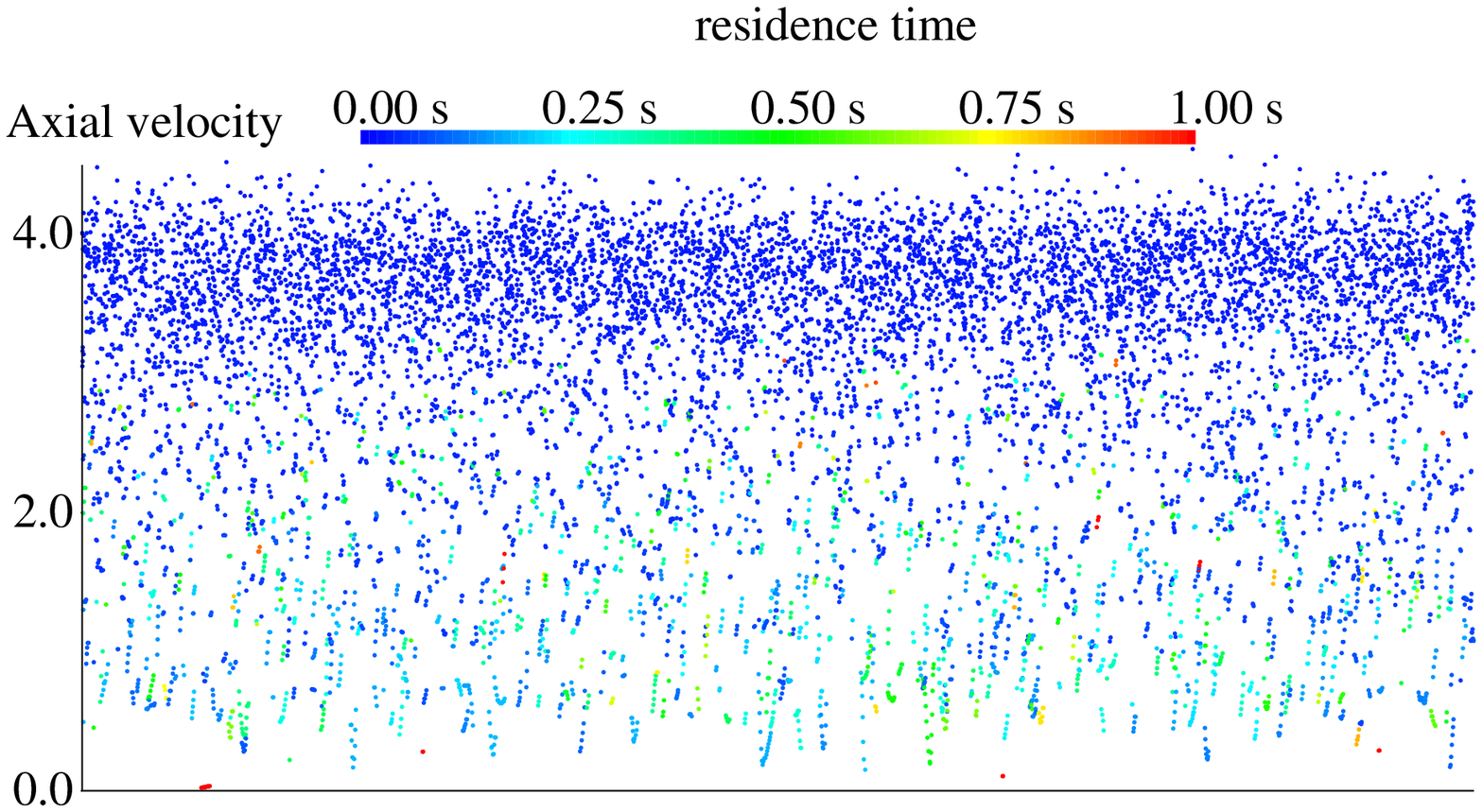}} 
 \end{picture}
 \vskip -1cm 
%modif EP (début)
\ifnum\elsitalic>0 \color{blue} \fi
 \caption{Left-hand side: snapshot of the flow, at a given time, where
   the particles are coloured by their residence time in the
  computational domain. Right-hand side (upper corner): probability density
  functions of the particle residence time inside the domain (close to
  the injection and close to the outlet of the domain). Right-hand
  side (bottom corner): axial velocity (horizontal axis) of the particles
  (coloured by their residence time in the computational domain)
  close to the inlet at different time steps (vertical axis).}
%modif EP (fin)
\label{fig:res_time}
\color{black}
\end{figure}

%----------------------------------
% Figure 19 : résultats déposition
%----------------------------------
\afterpage{\clearpage} \newpage
\begin{figure}[H]
\begin{center}
\epsfig{file=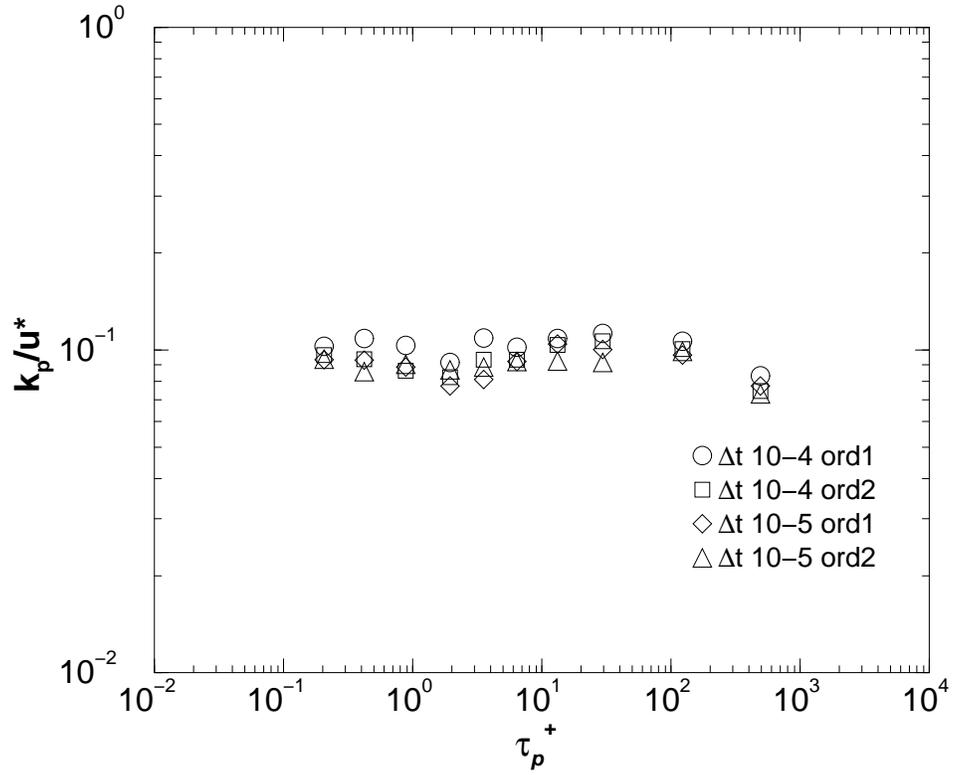,height=11.5truecm}
\caption{Sensitivity analysis : deposition velocity computed with
 different time steps $\Delta t$ and with the first and second order
 schemes. $\Delta t=10^{-4}\,{\rm s}$ ($\circ$) and $\Delta t=10^{-5}\,{\rm s}$
 ($\diamond$) with the first order scheme. $\Delta t=10^{-4}\,{\rm s}$
 ($\square$) and $\Delta t=10^{-5}\,{\rm s}$ ($\triangle$) with the second
 order scheme.}
\label{dt}
\end{center}
\end{figure}

%----------------------------------
% Figure 20 : résultats déposition
%----------------------------------
\afterpage{\clearpage} \newpage
\begin{figure}[H]
\begin{center}
\epsfig{file=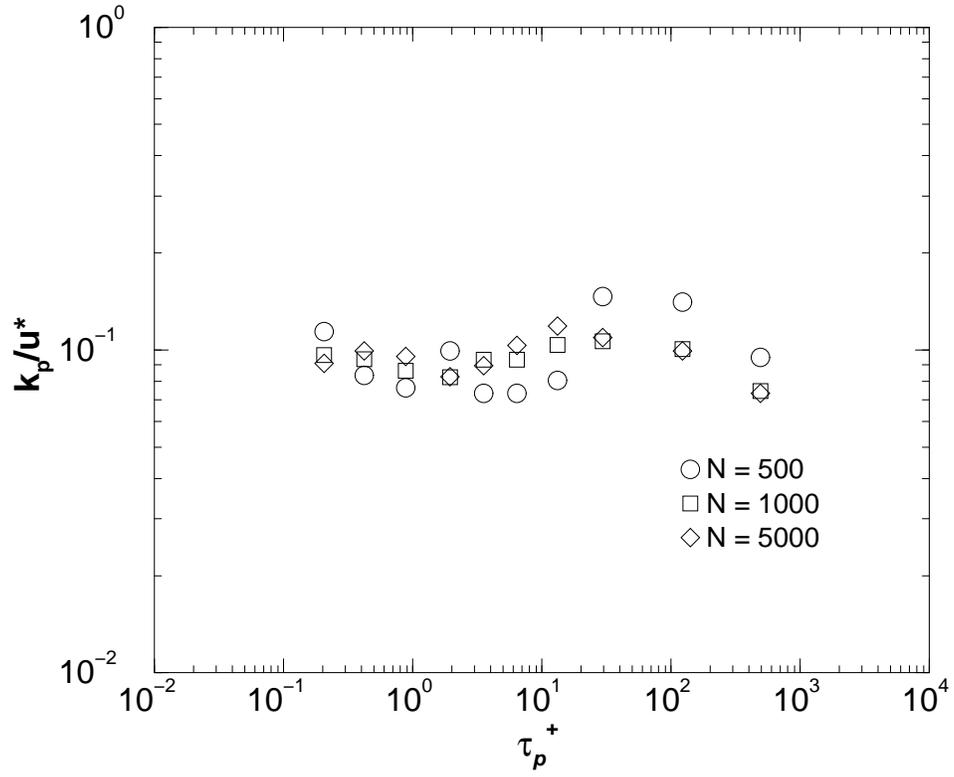,height=11.5truecm}
\caption{Sensitivity analysis : deposition velocity computed with
  different number of particles per class $N_{pc}$ for the same fluid
  mean-fields ($N_{pc}=500$
  ($\circ$), $N_{pc}=1000$ ($\square$) and $N_{pc}=5000$
  ($\diamond$)).}
\label{part}
\end{center}
\end{figure}

%----------------------------------
% Figure 21 : résultats déposition
%----------------------------------
\afterpage{\clearpage} \newpage
\begin{figure}[H]
\begin{center}
\epsfig{file=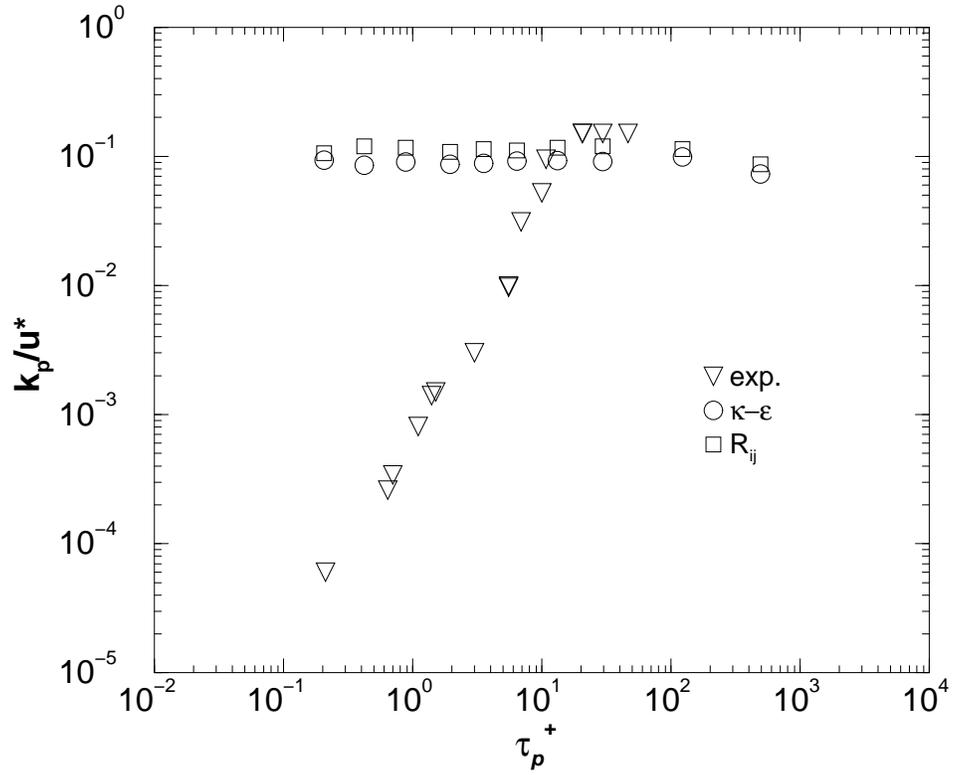,height=11.5truecm}
\caption{Deposition velocity with different mean fields for the
  fluid. Fluid mean-fields calculated with a
  $k-\epsilon$ model ($\circ$). Fluid mean-fields computed with a
  $R_{ij}-\epsilon$ model ($\square$). Experimental data
  ($\triangledown$).}
\label{turb_dep}
\end{center}
\end{figure}

%----------------------------------
% Figure 22 : résultats déposition
%----------------------------------
\afterpage{\clearpage} \newpage
\begin{figure}[H]
\begin{center}
%\vspace{5cm}
\rotatebox{0}{\epsfig{file=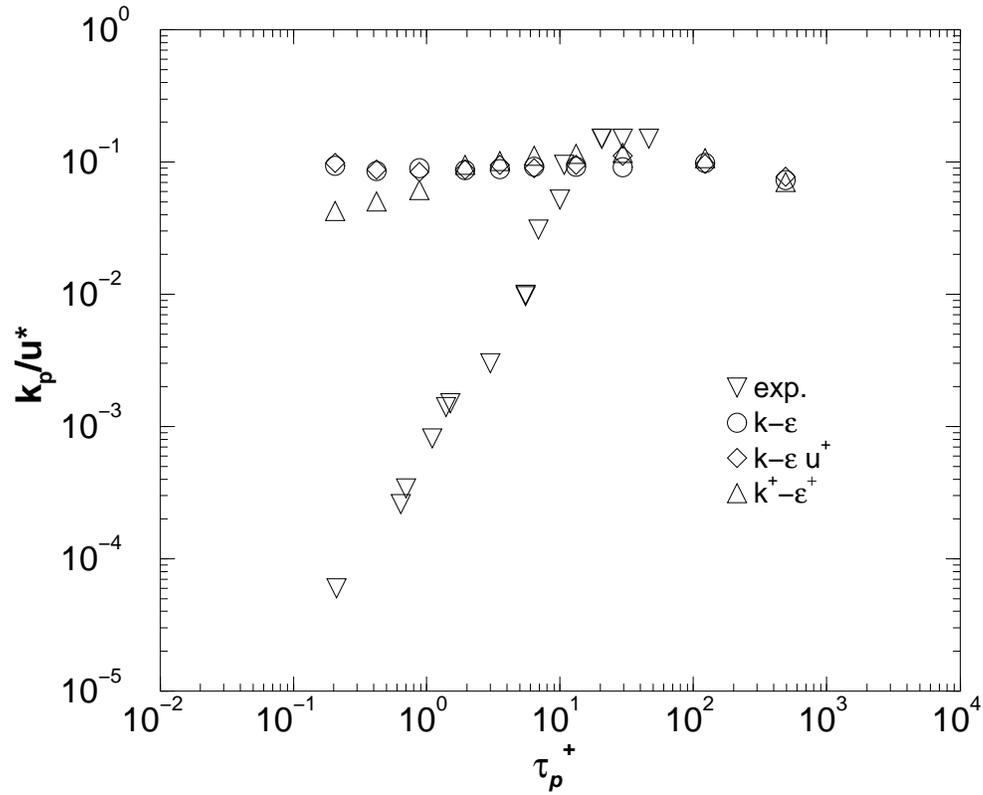,height=10.5truecm}}
\caption{Deposition velocity with different mean fields for the
  fluid. First, mean field obtained by computation with a standard
  $k-\epsilon$ model ($\circ$). Second, mean field where
  $\lra{U_i}$ is computed with the law-of-the-wall equations and
  where the values $k$ and $\epsilon$ are that of the computations
  ($\diamond$). Third, $\lra{U_i}$ is still given by the
  law-of-the-wall, and $k$ and $\epsilon$ are curve-fitted to DNS
  data~\cite{Mati_00} ($\triangle$). Experimental data
  ($\triangledown$).}
\label{viscous}
\end{center}
\end{figure}

%%% Local Variables: 
%%% mode: latex
%%% TeX-master: "figure"
%%% End: 

%% file: tables.tex
%+++++++++++++++++++++++++++++++++++++++++++++++++++++++++++++++++++++++++++++++
%
%                       TABLES
%
%+++++++++++++++++++++++++++++++++++++++++++++++++++++++++++++++++++++++++++++++
\setcounter{table}{0}
%++++++++++
% TABLE 0
%++++++++++
\ifnum\elsevier<0 
\setlength\abovecaptionskip{0pt}
\setlength\belowcaptionskip{10pt}
\fi
\begin{table}[htbp]
\caption{Complete mean-field (RANS)/PDF model.}
\hrule
\begin{align}
& \text{\underline{Mean-field (RANS) equations for the
    fluid.}} \notag \\
& \hspace*{5mm} \text{Continuity equation:} \notag \\
& \hspace*{10mm} \frac{D}{Dt}(\alpha_f\rho_f) = - \alpha_f\rho_f
\frac{\partial \lra{U_{i}}}{\partial x_i} \label{eq:feq_alphaf} \\
& \hspace*{5mm} \text{Momentum equation:} \notag \\
& \hspace*{10mm} \frac{D}{Dt}\lra{U_{i}}= -\frac{1}{\rho_f}
\frac{\partial\lra{P}}{\partial x_i}
-\frac{1}{\alpha_f\,\rho_f}\frac{\partial}{\partial
  x_j}(\alpha_f\,\rho_f\lra{u_{i} u_{j}}) +
\chi\lra{(U_{p,i}-U_{s,i})/\tau_p} \label{eq:feq_Uf} \\
& \hspace*{5mm} \text{Reynolds stress equation:} \notag \\
& \hspace*{10mm} \frac{D}{Dt}\lra{u_{i} u_{j}}  =
-\frac{1}{\alpha_f\,\rho_f}\frac{\partial}{\partial
  x_k}(\alpha_f\,\rho_f \lra{u_{i}u_{j}u_{k}})
- \lra{u_{i} u_{k}}\frac{\partial\lra{U_{j}}}{\partial x_k}
 - \lra{u_{j} u_{k}}\frac{\partial\lra{U_{i}}}{\partial x_k} \notag \\
& \hspace*{10mm} + G_{jk}\lra{u_{i} u_{k}}+ G_{ik}\lra{u_{j}
  u_{k}} + C_0 \lra{\epsilon}\delta_{ij} \notag \\ 
& \hspace*{10mm} + \chi \lra{\frac{1}{\tau_p}\left[(U_{p,i}-U_{s,i})U_{s,j} +
  (U_{p,j}-U_{s,j})U_{s,i}\right] } \label{eq:feq_ufuf} \\
& \hspace*{5mm} \text{with \quad} \frac{D}{Dt}=\frac{\partial}{\partial t} +
  \lra{U_{k}}\frac{\partial}{\partial x_k } \quad \text{and}
  \quad \chi=\frac{\alpha_p\,\rho_p}{\alpha_f\,\rho_f} \notag \\ \notag \\ 
& \text{\underline{SDEs for the discrete particles.}}\notag 
\end{align}
\begin{equation}\notag
\left\{\begin{split}
& dx_{p,i}(t)= U_{p,i}\,dt, \\
& dU_{p,i}(t)= \frac{U_{s,i}-U_{p,i}}{\tau_p}\,dt + g_i\,dt, \\
& dU_{s,i}(t) = A_{s,i}\,dt + A_{p \rightarrow s,i}\,dt +
  B_{s,ij}\,dW_j(t),
\end{split}\right.
\end{equation}
\begin{align}
& \hspace*{5mm} A_{s,i}=-\frac{1}{\rho_f}\frac{\partial \lra{P}
  }{\partial x_i} + \left( \lra{U_{p,j}} - \lra{U_{j}} \right)
\frac{\partial \lra{U_{i}}}{\partial x_j}
-\frac{1}{T_{L,i}^*}\left( U_{s,i}-\lra{U_{i}} \right) \notag \\
& \hspace*{5mm} B_{s,i}^2 = \lra{\epsilon}\left( C_0b_i
    \tilde{k}/k + \frac{2}{3}( b_i \tilde{k}/k -1) \right) \notag \\
& \hspace*{5mm} A_{p \rightarrow s,i} = -\chi (U_{s,i}-U_{p,i})/\tau_p \notag \\
& \hspace*{5mm} T_{L,i}^{*}= T_L/\sqrt{ 1 + \beta_i^2 \displaystyle
  \frac{\vert\lra{{\bf U}_r}\vert^2}{2k/3}}, \quad b_i= T_L/T_{L,i}^*,
\quad  \tilde{k}= \frac{3}{2}
\frac{\sum^3_{i=1}b_i\lra{u_{i}^2}}{\sum^3_{i=1}b_i} \notag \\ \notag
\end{align}
\hrule
\label{tab:RANS}
\end{table}

%++++++++++
% TABLE I
%++++++++++
\afterpage{\clearpage}
\begin{table}[htbp]
\caption{Analytical solutions to system (\ref{eq:sysEDS}) for
time-independent coefficients.}
\hrule
\begin{align}
& x_{p,i}(t) = x_{p,i}(t_0)
  + U_{p,i}(t_0)\tau_p  [1-\exp(-\Delta t/\tau_p)]
  + U_{s,i}(t_0)\,\theta_i \{T_i[1-\exp(-\Delta t/T_i)] \notag \\
&  \hspace*{5mm} + \tau_p[\exp(-\Delta t/\tau_p)-1]\} 
  + [C_i\,T_i]
    \{\Delta t-\tau_p[1-\exp(-\Delta t/\tau_p)] \notag \\
&  \hspace*{5mm} - \theta_i (T_i[1-\exp(-\Delta t/T_i)]+
      \tau_p[\exp(-\Delta t/\tau_p)-1])\} 
     + \Omega_i(t) \label{eq:xpa_exa} \\
&  \hspace*{5mm} \text{\quad with \quad} \theta_i = T_i/(T_i-\tau_p)
  \notag \\
& U_{p,i}(t) = U_{p,i}(t_0)\exp(-\Delta t/\tau_p)
  + U_{s,i}(t_0)\,\theta_i
    [\exp(-\Delta t/T_i)-\exp(-\Delta t/\tau_p)] \notag \\
& \hspace*{5mm} + [C_i\,T_i]
    \{[1-\exp(-\Delta t/\tau_p)]-\theta_i
      [\exp(-\Delta t/T_i)-\exp(-\Delta t/\tau_p)]\} \notag \\
& \hspace*{5mm} + \Gamma_i(t) \label{eq:Upa_exa} \\
& U_{s,i}(t) = U_{s,i}(t_0)\exp(-\Delta t/T_i)
  + C_i\,T_i[1-\exp(-\Delta t/T_i)]
  + \gamma _i(t) \label{eq:Ufa_exa} \\ \notag \\
%------------
& \text{\underline{The stochastic integrals $\gamma _i(t),\;\Gamma
_i(t),\;\Omega _i(t)$ are given by:}}\notag \\
& \quad \gamma _i(t) = \Check{B}_i\exp(-t/T_i)
  \int _{t_0}^{t} \exp(s/T_i)\,dW_i(s), \label{eq:gamma_exa}\\
& \quad \Gamma _i(t) = \frac{1}{\tau_p}\exp(-t/\tau_p)
  \int _{t_0}^{t}\exp(s/\tau_p)\,\gamma _i(s)\,ds, \label{eq:Gamma_exa}\\
& \quad \Omega _i(t) = \int _{t_0}^{t}\Gamma
_i(s)\,ds. \label{eq:Omega_exa} \\ \notag \\
%------------
& \text{\underline{By resorting to stochastic integration by parts, $\gamma
_i(t),\;\Gamma _i(t),\;\Omega _i(t)$ can be written:}}\notag \\
& \quad \gamma _i(t) = \Check{B}_i\,\exp(-t/T_i)
\,I_{1,i}, \label{eq:gammaN_exa}\\
& \quad \Gamma _i(t) = \theta_i \,\Check{B}_i\,
 [\exp(-t/T_i)\,I_{1,i} -\exp(-t/\tau_p)\,I_{2,i}], \label{eq:GammaN_exa}\\
& \quad \Omega _i(t) = \theta_i \,\Check{B}_i\,
   \{ (T_i-\tau_p)\,I_{3,i} \notag \\ 
& \hskip 4.0cm -[T_i \exp(-t/T_i)\,I_{1,i} -
   \tau_p \exp(-t/\tau_p) \,I_{2,i}]\}, \label{eq:OmegaN_exa} \\ 
& \text{with} \quad I_{1,i} = \int_{t_0}^{t}\exp(s/T_i)\,dW_i(s),
\quad I_{2,i}= \int_{t_0}^{t}\exp(s/\tau_p)\,dW_i(s) \notag \\
& \hskip 7.5cm \text{and} \quad I_{3,i}=\int_{t_0}^{t}dW_i(s).\notag
\end{align}
\hrule
\label{tab:exa}
\end{table}

%+++++++++++
% TABLE II
%+++++++++++
\afterpage{\clearpage}
\begin{table}[htbp]
\caption{Derivation of the covariance matrix for constant coefficients.}
\hrule
\begin{align}
& \lra{\gamma _i^2(t)} = \Check{B}_i^2 \,\frac{T_i}{2}
  \left[1-\exp(-2\Delta t/T_i)\right] \quad \text{where} \quad
  \Check{B}_i^2 = B_{ii}^2 \label{eq:gama2} \\ \notag \\
& \lra{\Gamma _i^2(t)} =
    \Check{B}_i^2 \,\theta_i ^2
   \left\{\frac{T_i}{2}[1-\exp(-2\Delta t/T_i)]
  - \frac{2\tau_p T_i}{T_i+\tau_p}
    [1-\exp(-\Delta t/T_i)\exp(-\Delta t/\tau_p)] \right . \notag \\
&+ \left . \frac{\tau_p}{2}[1-\exp(-2\Delta t/\tau_p)] \right\}
  \label{eq:Gama2} \\ \notag \\
& \frac{1}{\Check{B}_i^2 \,\theta_i ^2}\lra{\Omega _i^2(t)} =
  (T_i-\tau_p)^2\Delta t + \frac{T_i^3}{2}[1-\exp(-2\Delta t/T_i)]
  + \frac{\tau_p^3}{2}[1-\exp(-2\Delta t/\tau_p)] \notag \\
&- 2T_i^2(T_i-\tau_p)[1-\exp(-\Delta t/T_i)] 
+ 2\tau_p^2(T_i-\tau_p)[1-\exp(-\Delta t/\tau_p)] \notag \\
&- 2\frac{T_i^2\tau_p^2}{T_i+\tau_p}
    [1-\exp(-\Delta t/T_i)\exp(-\Delta t/\tau_p)] \\ \notag \\
& \lra{\gamma _i(t)\,\Gamma _i(t)} =
    \Check{B}_i^2 \, \theta_i \,T_i
\left\{ \frac{1}{2}[1-\exp(-2\Delta t/T_i)]
  - \frac{\tau_p}{T_i+\tau_p}[1-\exp(-\Delta t/T_i)
    \exp(-\Delta t/\tau_p)]\right\} \label{eq:gaGam} \\ \notag \\
& \lra{\gamma _i(t)\,\Omega _i(t)} =
  \Check{B}_i^2 \, \theta_i \,T_i
  \left\{(T_i-\tau_p)[1-\exp(-\Delta t/T_i)]
  - \frac{T_i}{2}[1-\exp(-2\Delta t/T_i)] \right . \notag \\
&  + \left . \frac{\tau_p^2}{T_i+\tau_p}[1-\exp(-\Delta t/T_i)
    \exp(-\Delta t/\tau_p)]\right\} \\ \notag \\
& \frac{1}{\Check{B}_i^2 \, \theta_i ^2}
  \lra{\Gamma _i(t)\,\Omega _i(t)} =
  (T_i-\tau_p)\{T_i[1-\exp(-\Delta t/T_i)]-
               \tau_p[1-\exp(-\Delta t/\tau_p)]\}  \notag \\
&  - \frac{T_i^2}{2}[1-\exp(-2\Delta t/T_i)] 
  - \frac{\tau_p^2}{2}[1-\exp(-2\Delta t/\tau_p)] \notag \\
& + T_i \tau_p \,[1-\exp(-\Delta t/T_i)\exp(-\Delta t/\tau_p)] \label{eq:GamOme}
\end{align}
\hrule
\label{tab:matcov_exa}
\end{table}

%+++++++++++
% TABLE III
%+++++++++++
\afterpage{\clearpage}
\begin{table}[htbp]
\caption{Weak first-order scheme (Euler scheme)}
\hrule
\begin{align}
& \text{\underline{Numerical integration of the system:}}\notag \\
& \quad x_{p,i}^{n+1} = x_{p,i}^n + A_1\,U_{p,i}^n + B_1\,U_{s,i}^n
  + C_1\,[T_i^n C_i^n] + \Omega _i^n,
  \notag \\
& \quad U_{s,i}^{n+1} = U_{s,i}^n\, \exp(-\Delta t/T_i^n)
              + [T_i^n C_i^n] [1-\exp(-\Delta t/T_i^n)]
              + \gamma _i^n,\notag \\
& \quad U_{p,i}^{n+1} = U_{p,i}^n\, \exp(-\Delta t/\tau_p^n)
              + D_1\,U_{s,i}^n + [T_i^n C_i^n](E_1-D_1)
              + \Gamma _i^n. \notag \\ \notag \\
%----------------
& \text{\underline{The coefficients $A_1,\;B_1,\;C_1,\;D_1$ and $E_1$ are
  given by:}}\notag \\ 
& \quad A_1 = \tau_p^n\,[1-\exp(-\Delta t/\tau_p^n)],\notag\\
& \quad B_1 = \theta_i ^n\,[T_i^n(1-\exp(-\Delta t/T_i^n)-A_1]
      \quad \text{with}\quad \theta_i ^n = T_i^n/(T_i^n-\tau_p^n),\notag\\
& \quad C_1 = \Delta t - A_1 - B_1, \notag \\
& \quad D_1 = \theta_i ^n [\exp(-\Delta t/T_i^n)-\exp(-\Delta
  t/\tau_p^n)],\notag\\ 
& \quad E_1 = 1 - \exp(-\Delta t/\tau_p^n).\notag \\ \notag \\
%----------------
& \text{\underline{The stochastic integrals $\gamma _i^n,\;\Omega
_i^n,\;\Gamma_i^n$ are simulated by:}}\notag \\
& \quad \gamma_i^n = P_{11}\,{\mc G}_{1,i},\notag \\
& \quad \Omega_i^n = P_{21}\,{\mc G}_{1,i}+ P_{22}\,{\mc G}_{2,i} \notag \\
& \quad \Gamma_i^n = P_{31}\,{\mc G}_{1,i}+ P_{32}\,{\mc G}_{2,i}+
                     P_{33}\,{\mc G}_{3,i}, \notag\\ 
& \quad \text{where ${\mc G}_{1,i},\;{\mc G}_{2,i},\;{\mc G}_{3,i}$ are
  independent ${\cal N}(0,1)$ random variables.} \notag\\ \notag \\
%----------------
& \text{\underline{The coefficients
 $P_{11},\;P_{21},\;P_{22},\;P_{31},\;P_{32},\;P_{33}$ are defined
 as:}}\notag\\
& \quad P_{11} = \sqrt{\lra{(\gamma_i^n)^2}}, \notag \\
& \quad P_{21} = \frac{\lra{\Omega_i^n\gamma_i^n}}{\sqrt{\lra{(\gamma_i^n)^2}}}, 
  \quad P_{22} = \sqrt{\lra{(\Omega_i^n)^2}-
  \frac{\lra{\Omega_i^n\gamma_i^n}^2}{\lra{(\gamma_i^n)^2}}},\notag \\
& \quad P_{31} = \frac{\lra{\Gamma_i^n\gamma_i^n}}{\sqrt{\lra{(\gamma_i^n)^2}}},
  \quad P_{32} = \frac{1}{P_{22}}(\lra{\Omega_i^n\Gamma_i^n}-P_{21}P_{31}),
  \quad P_{33} = \sqrt{\lra{(\Gamma_i^n)^2}-P_{31}^2-P_{32}^2)}.\notag
\end{align}
\hrule
\label{tab:sch1}
\end{table}

%+++++++++++
% TABLE IV
%+++++++++++
\afterpage{\clearpage}
\begin{table}[htbp]
\caption{Weak second-order scheme}
\hrule
\begin{align}
& \text{\underline{Prediction step:}}
\quad \text{Euler scheme, see Table \ref{tab:sch1}}.\notag \\
\ifnum\elsevier<0 \notag \\ \fi
%----------------
& \text{\underline{Correction step:}} \notag \\
& \quad U_{p,i}^{n+1} =
    \frac{1}{2}\,U_{p,i}^n\, \exp(-\Delta t/\tau_p^n)
  + \frac{1}{2}\,U_{p,i}^n\,\exp(-\Delta t/\td{\tau }_p^{n+1})\notag \\ 
& \qquad \qquad \qquad \qquad + \frac{1}{2}\,U_{s,i}^n\,C_{2c}(\tau_p^n,\,T_i^n)
  + \frac{1}{2}\,U_{s,i}^n\,C_{2c}(\td{\tau}_p^{n+1},\,\td{T_i}^{n+1})
\notag \\ 
& \qquad \qquad \qquad \qquad + A_{2c}(\tau_p^n,\,T_i^n) \,[T_i^n C_i^n]
  + B_{2c}(\td{\tau}_p^{n+1},\,\td{T}_i^{n+1})\,[\td{T}_i^{n+1} C_i^{n+1}]\notag \\ 
& \qquad \qquad \qquad \qquad + A_2(\Delta t,\tau_p^n)[\tau_p^n\,{\cal A}_i^n]
  + B_2(\Delta t,\td{\tau_p}^{n+1})[\td{\tau_p}^{n+1}\,{\cal A}_i^{n+1}]
  + \td{\Gamma}_i^{n+1}, \notag \\
& \quad U_{s,i}^{n+1} = \frac{1}{2}\,U_{s,i}^n\,\exp(-\Delta t/T_i^n)
+ \frac{1}{2}\,U_{s,i}^{n}\,\exp(-\Delta t/\td{T}_i^{n+1})
+ A_2(\Delta t,\,T_i^n) \,[T_i^n C_i^n]\notag \\ 
& \qquad \qquad \qquad \qquad 
+ B_2(\Delta t,\,\td{T}_i^{n+1}) \,[\td{T}_i^{n+1} C_i^{n+1}]
+ \td{\gamma}_i^{n+1}.\notag \\ \ifnum\elsevier<0 \notag \\ \fi
%----------------
& \text{\underline{The coefficients $A_2,\;B_2,\;A_{2c},\;B_{2c}$ et $C_{2c}$ are
  defined as:}}\notag \\ 
& \quad A_2(\Delta t,x) = -\exp(-\Delta t/x)
+ [1-\exp(-\Delta t/x)][\Delta t/x], \notag\\
& \quad B_2(\Delta t,x) = 1-[1-\exp(-\Delta t/x)][\Delta t/x],\notag \\ 
& \quad A_{2c}(x,y) =
- \exp(-\Delta t/x) + [(x+y)/\Delta t][1-\exp(-\Delta t/x)]
- (1+y/\Delta t)\,C_{2c}(x,y), \notag \\
& \quad B_{2c}(x,y) =  1 - [(x+y)/\Delta t][1-\exp(-\Delta t/x)]
+ (y/\Delta t)\,C_{2c}(x,y), \notag\\
& \quad C_{2c}(x,y) = [y/(y-x)][\exp(-\Delta t/y)-\exp(-\Delta
  t/x)].\notag \\ \ifnum\elsevier<0 \notag \\ \fi
%----------------
& \text{\underline{The stochastic integrals
        $\td{\gamma}_i^{n+1}\;$ \text{and} $\;\td{\Gamma}_i^{n+1}$
        are simulated as follows:}}\notag \\
& \quad \td{\gamma}_i^{n+1}= \sqrt {\frac{[B_i^{*}]^2\td{T}_i^{n+1}}{2}
               [1-\exp(-2\Delta t/\td{T}_i^{n+1})]}\; {\mc G}_{1,i},\notag \\
& \quad \text{with} \quad \left[1-\exp(-2\,\Delta t/\td{T}_i^{n+1})\right]\,B_i^{*} =
    A_2(2\,\Delta t,\,\td{T}_i^{n+1})\,\sqrt{(\Check{B}_i^n)^2} + \notag \\
& \hskip 8.0cm B_2(2\,\Delta t,\,\td{T}_i^{n+1})\,\sqrt{(\td{\Check{B}}_i^{n+1})^2}. \notag \\
& \quad \td{\Gamma}_i^{n+1} =
  \frac{\lra{ \td{\Gamma}_i^{n+1}\td{\gamma}_i^{n+1} } }
     {\lra{(\td{\gamma}_i^{n+1})^2}}\,\td{\gamma}_i^{n+1}+
  \sqrt{\lra{(\td{\Gamma}_i^{n+1})^2}-
      \frac{\lra{\td{\Gamma}_i^{n+1}\td{\gamma}_i^{n+1}}^2}
           {\lra{(\td{\gamma}_i^{n+1})^2}} }\;{\mc G}_{2,i} \notag \\ \notag \\
& \quad \text{with}\quad \lra{ \td{\Gamma}_i^{n+1}\td{\gamma}_i^{n+1} } =
  \lra{\Gamma_i\gamma_i}(\tau_p^n,\,\td{T}_i^{n+1},\,B^{*}_i)
  \quad \text{and}\quad \lra{(\td{\Gamma}_i^{n+1})^2} =
  \lra{\Gamma_i^2}(\tau_p^n,\,\td{T}_i^{n+1},\,B^{*}_i). \notag
\end{align}
\hrule
\label{tab:sch2c}
\end{table}

\afterpage{\clearpage}
\begin{table}[htb]
\caption{Mean near-wall ($y^+ < 30$) residence time for different
  diameters (the simulation is carried out with the exact frozen
  field). The residence time is given in non-dimensional form (it is
  normalised with the viscous time scale, $\nu_f/(u^*)^2$, that is
  $t^+=t\,(u^*)^2/\nu_f$).}
\begin{center}
\begin{tabular}{|l|c|c|}
\hline 
$\tau^+_p$  & $d_p$ ($\mu m$) & $t^+$ (wall units) \\
\hline \hline
$0.2$  & $1.4$ & 29.5 \\
\hline
$0.4$  & $2.0$ & 29.9 \\
\hline
$0.9$  & $2.9$ & 28.7  \\
\hline
$1.9$  & $4.3$ & 30.9     \\
\hline
$3.5$  & $5.8$ & 31.5     \\
\hline
$6.4$  & $7.8$ & 31.6   \\
\hline
$13.2$  & $11.2$ & 35.4  \\
\hline
$29.6$ & $16.8$  & 40.6   \\
\hline
$122.7$  & $34.2$  & 55.3 \\
\hline
$492.2$  & $68.5$  & 96.8 \\
\hline
\end{tabular}
\label{tab:sej1}
\end{center}
\end{table}

%%% Local Variables: 
%%% mode: latex
%%% TeX-master: "tables"
%%% End: 